# A geometric theory of waves and its applications to plasma physics

Daniel Edward Ruiz

A Dissertation
Presented to the Faculty
of Princeton University
in Candidacy for the Degree
of Doctor of Philosophy

Recommended for Acceptance
by the Department of
Astrophysical Sciences
Program in Plasma Physics
Adviser: Dr. Ilya Y. Dodin

September 2017

# Abstract

Waves play an essential role in many aspects of plasma dynamics. For example, they are indispensable in plasma manipulation and diagnostics. Although the physics of waves is well understood in the context of relatively simple problems, difficulties arise when studying waves that propagate in inhomogeneous or nonlinear media. This thesis presents a new systematic wave theory based on phase-space variational principles. In this dissertation, waves are treated as geometric objects of a variational theory rather than formal solutions of specific PDEs. This approach simplifies calculations, highlights the underlying wave symmetries, and leads to improved modeling of wave dynamics. Specifically, this dissertation presents two important breakthroughs that were obtained in the general theory of waves.

The first main contribution of the present dissertation is an extension of the theory of geometrical optics (GO) in order to include polarization effects. Even when diffraction is ignored, the GO ray equations are not entirely accurate. This occurs because GO treats wave rays as classical particles described by their position and momentum coordinates. However, vector waves have another degree of freedom, their polarization. As a result, wave rays can behave as particles with spin and show polarization dynamics, such as polarization precession and polarization-driven bending of ray trajectories. In this thesis, the theory of GO is reformulated as a first-principle Lagrangian wave theory that governs both mentioned polarization phenomena simultaneously. The theory was applied successfully to several systems of interest, such as relativistic spin-1/2 particles and radio-frequency waves propagating in magnetized plasmas.

The second main contribution of this thesis is the development of a phase-space method to study basic properties of nonlinear wave–wave interactions. Specifically, a general theory is proposed that describes the ponderomotive refraction that a wave can experience when interacting with another wave. It is also shown that phase-space methods can be useful to study problems in the field of wave turbulence, such as the nonlinear interaction of high-frequency waves with large-scale structures. Overall, the results obtained can serve as a basis for future studies on more complex nonlinear wave–wave interactions, such as modulational instabilities in general wave ensembles or wave turbulence.

# Acknowledgements

I would first like to express my sincerest gratitude to my thesis advisor, Ilya Dodin. It was a real pleasure to work with him. I am very grateful that he introduced me to wave physics and that together we built a very productive research program. I shall always be impressed by Ilya's mastery of many subjects in physics, and I hope to reach that level of wisdom in the future. After I leave Princeton, I am going to miss our coffee breaks and our afternoon discussions on various topics, which ranged from general physics, highlights in the arXiv, and house-maintenance projects. I shall be forever indebted to Ilya's efforts to teach me how to perform high-quality research and how to communicate the findings clearly.

I am grateful to my readers, John Krommes and Hong Qin. Their comments and suggestions were very valuable in improving the overall quality of the thesis. Thank you, John, for suggesting future opportunities for research that could be explored with the techniques developed in this dissertation.

Some of the work presented in this thesis would have not been completed without the help of my collaborators. In this regard, I would like to thank Jeff Parker, Leland Ellison, and Eric Shi. It was a pleasure to discuss the technical aspects of the interactions between drift waves and zonal flows with Jeff. I am also grateful for the help that I received from Leland and Eric in developing the numerical algorithms to solve our equations. I feel fortunate to have worked with them. Not only were the final results that we obtained more complete, but the projects were also more amusing when working alongside them.

I am grateful for the scientific education that I have received before arriving to Princeton. I would like to thank the faculty of the Physics Department at Tecnológico de Monterrey. Specially, I thank Professors Carlos Hinojosa, Francisco Rodríguez, Alfonso Serrano, Julio Gutiérrez, and Raúl Hernández who inspired my interest in physics. I would also like to thank the faculty of the Physics Department at École Polytechnique. In particular, I would like to thank Professors Jean-Marcel Rax, Victor Malka, and Julien Fuchs for the introductory courses in plasma physics.

I would like to express my gratitude to the faculty at the Princeton Plasma Physics Program for the wonderful training that I received in plasma physics. Special thanks to John Krommes, Matt Kunz, Jong-Kyu Park, Hong Qin, Amitava Bhattacharjee, and Bill Tang for their time and support. I would also like to thank Nat Fisch for his deep insights and for the various discussions that I had with him. I cannot thank Yevgeny Raitses enough. I worked with Yevgeny as a summer intern in the Hall Thruster Laboratory. He was the first one to introduce me to research in plasma physics. His love and dedication to the field lead me to pursue a PhD degree in this field.

The Princeton Plasma Physics Program would not be the same without the service and dedication of Barbara Sarfaty and Beth Leman. One one hand, Barbara welcomed me two times to Princeton: first as a student summer intern and then as a first-year graduate student. On the other hand, I leave Princeton under Beth's watch of the program. Both of them were always very kind to me and were always willing to help me out. I thank them for their time and support.

I would also like to extend my gratitude to my friends and coworkers who made my life at Princeton so enjoyable. I thank Peter Bolgert for our running and weight-lifting sessions and for attending my wedding. I thank Eric Shi for our interesting and sometimes never-ending conversations in the office and for his help on various computer-related problems that I had. I shall always be indebted to Yao Zhou and Seth Davidovits for the numerous rides to Princeton Junction and Jersey City during the winter season or when it was raining outside. It was awesome to share rooms with the legendary Vasily Geyko during the APS meetings. It was a pleasure to play with Noah Mandel and Leland Ellison at the green table. Lunch-time bike rides with David Pfefferlé were awesome. It was always enjoyable to have morning coffee with Vinicius Duarte and Alex Glasser. It was a pleasure to "sing" Italian with Luca Commisso in the hallways of the theory department and to discuss difficult and always challenging physics questions with Eero Hirvijoki, Kenan Qu, and Josh Burby. I am grateful to have met Amanda Sieman during my first few days in Princeton; she is an awesome friend. Special thanks to the theory nubbers, with whom I shared office space in the last few years. I also thank the rest of the graduate students of the program, especially those in my cohort: Peter Bolgert, Charles Swanson, Yuan Shi, Jacob Schwartz, Lee Gunderson, and Jonathan Ng. It was an honor and pleasure to endure prelims and generals with you.

I would like to thank my friends outside Princeton, who have supported me during the last few years. In particular, thanks to Sammie Laycock, Adrián Ramos, Mauricio Silva, Raúl López, Kendra Rajchel and family, Guillaume Aoust, Loïc Binan, Samy Jazaerli, Sabine Corcos, Samuel Rosat, Chiara Altomare, Paolo Carrozzo, and Manuele Aufiero. I feel very lucky to have met you guys at different points in my life and to call you friends.

Quisiera escribir estos pensamientos para mis papás, Alicia and Miguel. Este trabajo no hubiera sido posible sin el apoyo constante que ustedes me han brindado desde siempre. Gracias al amor y al cariño de ustedes y Anabel, yo soy la persona que soy. Te agradezco, papá, porque cultivaste en mí el interés por la ingeniería y las ciencias. En cierta manera, este trabajo es fruto de los pequeños problemas de geometría y matemáticas que me dabas cuando era pequeño. Te agradezco, mamá, por tu apoyo incondicional en la escuela y en la natación. Además, te doy gracias por la formación cívica y ética que me has brindado. Los quiero mucho a los dos y espero que se sientan orgullosos y satisfechos que, al fin, he terminado la escuela.

Finally, I would like to thank Alison, the love of my life. Five years ago, we moved to New Jersey to finally date normally after too many years of long-distance dating. Since then, we got married and lived in our beloved community of the Ironbound for the past years. You are an amazing wife and partner. You are a great example of service and dedication for the welfare and progress of others. Thank you for understanding me and for listening to my weird analogies when trying to explain the physics problems that I was working on. I thank you for the wonderful moments that we have lived together, for your support, and for your love.

Although this might seem odd, I would like to thank this country for the wonderful opportunities that I have received. In this regard, the research presented in this dissertation was financially supported by the U.S. Government. Specifically, this work was supported by U.S. DOE through Contract Nos. DE-AC02-09CH11466 and DE-AC52-07NA27344, by the NNSA SSAA Program through DOE Research Grant No. DE-NA0002948, by the U.S. DTRA through Research Grant No. HDTRA1-11-1-0037, and by the U.S. DOD NDSEG Fellowship through Contract No. 32-CFR-168a.

A mis queridos papás, Alicia y Miguel,

y a mi linda esposa, Alison.

# Contents

# List of Tables

# List of Figures

# Summary of Notation

The following notation is used throughout this thesis. The symbol "$\doteq$" denotes definitions, "c. c." denotes "complex conjugate," and "h. c." denotes "Hermitian conjugate." Unless otherwise specified, natural units are used in this work so that the speed of light equals one ($c = 1$), and so does the Planck constant ($\hbar = 1$). Four-dimensional spacetime indices are denoted by lowercase Greek letters ($\mu, \nu, ...$), taking the values 0, 1, 2, 3. The spacetime coordinate is denoted as $x^\mu = (x^0, \mathbf{x})$, with $x^0 = t$. In the three-dimensional space, Latin indices span from 1 to 3 and denote the spatial variables, i.e., $\mathbf{x} = (x^1, x^2, x^3)$, or equivalently, $\mathbf{x} = (x, y, z)$. The spatial derivatives are denoted by $\partial_i \doteq \partial/\partial x^i$. The Einstein summation convention is always implicitly assumed for all types of repeated indices that appear only on one side of an equation. The Minkowski spacetime metric $g^{\mu\nu} = g_{\mu\nu} = \text{diag}[+,-,-,-]$ is adopted and is used to raise and lower Lorentz indices so that $a^\mu = g^{\mu\nu} a_\nu$. Hence, the inner product between two four-vectors $a$ and $b$ is given by $a \cdot b \doteq a^\mu b_\mu = a^0 b^0 - \mathbf{a} \cdot \mathbf{b}$. The Dirac bra–ket notation is used to denote $|\,\Psi\,\rangle$ as a state of a Hilbert space. In Euler–Lagrange equations, the notation "$\delta a$ :" denotes that the corresponding equation was obtained by extremizing the action functional with respect to $a$.

Tables of frequently used symbols and acronyms are given below.

Table 1: Frequently used symbols

| **Common symbols** | |
|---|---|
| $a^\mu = (a^0, \mathbf{a})$ | Spacetime four-vector |
| $\mathbf{a}^i = (a^1, a^2, a^3)$ | Spatial vector |
| $g^{\mu\nu} = g_{\mu\nu} = \mathrm{diag}[+,-,-,-]$ | Minkowski spacetime metric |
| $\delta^\mu_\nu$ | Kronecker symbol |
| $a \cdot b = a^0 b^0 - \mathbf{a} \cdot \mathbf{b}$ | Scalar product of two four-vectors |
| $\mathbf{a} \cdot \mathbf{b}$ | Scalar product of two spatial vectors |
| $\partial_\mu = \partial/\partial x^\mu = (\partial_t, \boldsymbol{\nabla})$ | Spacetime derivative |
| $\boldsymbol{\nabla}_i = \partial_i = (\partial/\partial x^1, \partial/\partial x^2, \partial/\partial x^3)$ | Spatial derivative |
| $\mathrm{d}^4x = \mathrm{d}x^0\,\mathrm{d}^3x$ | Four-dimensional volume element |
| $\mathrm{d}^3x = \mathrm{d}x^1\,\mathrm{d}x^2\,\mathrm{d}x^3$ | Three-dimensional volume element |
| $\mid \Psi \rangle, \langle \Psi \mid$ | In the Dirac notation, "ket" and "bra" states of a Hilbert space |
| $\widehat{\mathcal{A}}$ | In the Dirac notation, operator in a Hilbert space |
| $\widehat{\mathcal{A}}^\dagger, \widehat{\mathcal{A}}^\mathrm{T}, \widehat{\mathcal{A}}^{-1}$ | Hermitian conjugate, transpose, inverse |
| $[\widehat{\mathcal{A}}, \widehat{\mathcal{B}}]_- = \widehat{\mathcal{A}}\widehat{\mathcal{B}} - \widehat{\mathcal{B}}\widehat{\mathcal{A}}$ | Commutator |
| $[\widehat{\mathcal{A}}, \widehat{\mathcal{B}}]_+ = \widehat{\mathcal{A}}\widehat{\mathcal{B}} + \widehat{\mathcal{B}}\widehat{\mathcal{A}}$ | Anti-commutator |
| $\widehat{\mathcal{A}}_H = (\widehat{\mathcal{A}} + \widehat{\mathcal{A}}^\dagger)/2$ | Hermitian part of the operator $\widehat{\mathcal{A}}$ |
| $\widehat{\mathcal{A}}_A = (\widehat{\mathcal{A}} - \widehat{\mathcal{A}}^\dagger)/(2i)$ | Anti-Hermitian part of the operator $\widehat{\mathcal{A}}$ |
| $\mathsf{W}[\widehat{\mathcal{A}}]$ | Weyl transformation of the operator $\widehat{\mathcal{A}}$ |
| $A(x,p) = \mathsf{W}[\widehat{\mathcal{A}}]$ | Weyl symbol of the operator $\widehat{\mathcal{A}}$ |
| $A(x,p) \star B(x,p)$ | Moyal product |
| $\{\{A, B\}\} = -i(A \star B - B \star A)$ | Moyal sine bracket |
| $[[A, B]] = (A \star B + B \star A)$ | Moyal cosine bracket |
| $\{A, B\}$ | Eight-dimensional Poisson bracket |
| $\gamma^\mu$ | Dirac gamma matrices |
| $\not{a} = a_\mu \gamma^\mu$ | Feynmann's slashed notation |
| $\delta(x^i)$ | Dirac delta function |
| $\delta^4(x) = \delta(x^0)\delta^3(\mathbf{x})$ | Four-dimensional Dirac delta function |
| $\delta^3(\mathbf{x}) = \delta(x^1)\delta(x^2)\delta(x^3)$ | Three-dimensional Dirac delta function |
| $\mathrm{Tr}[M]$ | Trace of the matrix $M$ |

Table 2: Frequently used acronyms

| **Acronyms** | |
|---|---|
| ACT | action conservation theorem |
| BMT | Bargmann–Michel–Telegdi |
| CE2 | second-order cumulant expansion |
| DW | drift wave |
| ELE | Euler–Lagrange equation |
| EM | electromagnetic |
| FW | Foldy–Wouthuysen |
| GO | geometrical optics |
| HF | high frequency |
| MW | modulating wave |
| OC | oscillation center |
| ODE | ordinary differential equation |
| PDE | partial differential equation |
| PW | probe wave |
| RF | radio frequency |
| SG | Stern–Gerlach |
| WKE | wave kinetic equation |
| XGO | extended geometrical optics |
| ZF | zonal flow |

# Part I

# A primer on variational theories of waves

# Chapter 1

# Introduction

## 1.1 Motivation

Waves play an essential role in many aspects of plasma dynamics, and they are also indispensable for the manipulation and diagnostics of plasmas (Stix, 1992; Swanson, 2003). In the particular context of magnetic confinement fusion, several important applications of waves include plasma heating (Stix, 1958; Stix and Palladino, 1958), current drive (Fisch, 1978, 1987), mode stabilization (Reiman, 1983; Westerhof, 1987), and alpha channeling (Fisch and Herrmann, 1994). These applications of waves represent important breakthroughs in the mission of obtaining reliable nuclear fusion energy. Although much work has been done in the theory of these and other plasma-wave effects in the past, the field is still full of challenges.

Historically, fundamental theories of waves in plasmas have focused on relatively simple problems such as the propagation of waves in stationary homogeneous plasmas of infinite extent. In these ideal cases, wave physics can be studied by applying the standard Laplace and Fourier transforms to Maxwell's equations and other linearized partial differential equations (PDEs) that describe the plasma response (Stix, 1992). However, realistic problems are indeed more complex to deal with; for example, waves usually propagate in inhomogeneous plasmas, can undergo mode conversion, and are subject to nonlinear phenomena. Describing the dynamics of waves in these non-ideal situations quickly becomes cumbersome or even impossible to treat using the standard analytical approaches.

To describe waves in nonideal plasmas, one possible solution could be to use modern-day supercomputers in order to numerically solve the equations describing the wave dynamics (Heuraux *et al.*, 2015). Although this alternative might seem straightforward, the task still remains daunting and far from trivial. As an example, full-wave simulations of radio-frequency waves in tokamaks can take several hours (or even days) to complete using modern-day supercomputers (Jaeger *et al.*, 2001). In consequence, full-wave simulations remain prohibitively costly for practical purposes, especially those involving parameter scans.

To further advance our understanding of wave dynamics, we must resort to reduced modeling of the wave equations. Reduced modeling involves the development of approximate models that describe a dynamical system of interest by using a smaller number of dynamical variables or by eliminating certain phenomena that evolve at different time scales. Although reduced models are less accurate than the original equations of motion from which they are based, reduced models are valuable since they are often simpler to solve numerically and, in many cases, can explicitly show the physical effects of interest that would otherwise remain buried in the original equations of motion. Developing new theoretical approaches to reduced modeling of waves remains one of the key frontiers of plasma science. Contributing to this area is the main theme of the present dissertation.

Importantly, not all reduced models are created equal. Reduced modeling of waves in plasmas has been traditionally based on simplifying and manipulating systems of PDEs. However, studying waves in nonideal media is not always feasible with this approach because PDE-based methods do not recognize basic principles of wave dynamics such as the conservation of the wave action (Whitham, 2011). With the advent of variational methods in plasma theory (Low, 1958; Newcomb, 1962; Littlejohn, 1983) and in physics in general (Feldmeier and Schnack, 2000; Grabowski, 2014; Seliger and Whitham, 1968; Whitham, 2011; Joannopoulos *et al.*, 2008), it is becoming increasingly appreciated that dynamical reductions can be made more accurate if they exploit basic principles that underlie the problem of interest; for example, the existence of symmetries in the system can often lead to the discovery of conserved invariants (Marsden and Ratiu, 1999). In this regard, dynamical reductions made within variational principles are inherently advantageous.[1] In the context of wave physics, the advantages of variational methods can be summarized as follows.

- All the physics of nondissipative waves is contained in a single object: the system action functional. Variational methods are modular; in other words, physical effects can be considered or neglected by simply inserting or removing terms in the system action functional.

- In contrast to PDE-based approaches, where dropping terms from the equations can often break the conservation properties of the system (unless one is extremely careful), variational methods are more robust. With these methods, one approximates only the action functional in order to obtain equations of motion that automatically conserve key invariants. This makes the method attractive for reduced modeling (Whitham, 2011).

[1] Dissipation, which is not readily captured by variational principles, can be handled *ad hoc* or by using extended variational techniques such as those discussed by Dodin *et al.* (2017).

- Possible symmetries of a system and their corresponding conservation laws can all be extracted by using Noether's theorem, regardless of specific details of the underlying system. This helps simplify the analysis of problems.

In light of the arguments above, this dissertation seeks: (i) to develop a fundamental theory of waves in plasmas (and, as a spin-off, in other media too) that realizes the aforementioned advantages of the variational approach, and (ii) to study, using this new machinery, some fundamental wave phenomena in selected physical systems. Although the main motivation of this research program is to better understand wave dynamics in plasmas, this thesis will not be solely focused on this particular class of waves. Instead of treating waves as formal solutions of specific PDEs, this dissertation considers waves as abstract objects of a Lagrangian theory. This research methodology will lead to the discovery of fundamental properties that are universal to all types of waves. The discovered principles will then be applied to study specific problems involving waves in plasmas and will advance our understanding of plasma physics. Additionally, this work will make use of phase-space methods to describe wave dynamics (Tracy *et al.*, 2014). Studying waves from the geometrical phase-space perspective will allow one to borrow many of the tools commonly used in classical Hamiltonian mechanics and apply these tools to the study of waves. This variational phase-space description of waves will help simplify calculations, highlight the underlying wave symmetries, and lead to improved reduced modeling of wave dynamics.

## 1.2 Thesis objectives

Two major topics will be covered in this dissertation. The first topic is focused on the effects of wave polarization on the propagation of linear vector waves. The second topic is concentrated on the basic principles of nonlinear wave–wave interactions. A brief description of these main research themes, as well as the corresponding research objectives, is given below.

**Polarization effects on the propagation of linear vector waves**

Among the various reduced models for waves (Whitham, 2011; Tracy *et al.*, 2014), geometrical optics (GO) is an approximation used to describe the propagation of waves in the limit of small wavelengths. Even when one ignores diffraction, the well-known ray equations in GO are not entirely accurate. This occurs because GO treats wave rays as particles, which are solely described by their position and momentum coordinates. However, vector waves have another degree of freedom, namely, their polarization. As a result, wave rays can behave as particles with spin and can show polarization dynamics, such as polarization precession and polarization-driven bending of ray trajectories.

This thesis presents an extension of GO in order to include polarization effects. Specifically, the main objectives regarding this topic include the following.

- Extend and reformulate GO as a first-principle Lagrangian theory that self-consistently includes the effects of wave polarization. (This model will be referred to as *extended geometrical optics*, or XGO.)
- Study the basic phenomena related to the wave polarization, or wave spin, of vector waves.
- Explain the connection between the transfer of wave quanta [also known as mode conversion (Friedland *et al.*, 1987)] and the precession of the wave polarization.
- Apply XGO to study polarization effects on the propagation of waves in plasmas; in particular, determine the regime in which the polarization-driven bending of ray trajectories is important for ray tracingof radio-frequency (RF) waves in tokamaks.
- Apply XGO to study the dynamics of quantum spinning particles; in particular, obtain a classical point-particle model for the relativistic spin-1/2 particle; also, compare the resulting model with other previously developed classical models for spinning particles.

### Nonlinear wave–wave interactions

Waves of small amplitude can often be described using linear models. Within the linear approximation, waves propagate independently from one another so their dynamics are decoupled. However, in the nonlinear regime waves can interact with one another. Such nonlinear wave–wave interactions can lead to interesting effects.

To study such effects is the second major theme of this thesis. Specifically, I shall focus on the following two phenomena. (i) Waves can experience a ponderomotive refraction when propagating in a medium that is modulated by a second wave (Dodin and Fisch, 2014). This effect is similar to how charged particles experience time-averaged ponderomotive forces in high-frequency fields. Understanding this ponderomotive effect on waves could lead to novel methods for manipulating waves. (ii) In some cases, an ensemble of high-frequency incoherent waves can collectively interact and spontaneously form large-scale structures. This process, which is often referred to as the zonostrophic instability, is repeatedly observed in planetary atmospheres and tokamak plasmas (Farrell and Ioannou, 2003; Fujisawa, 2009). In magnetic confinement fusion, this topic has recently gained attention as it can provide a means of suppressing turbulence and improving plasma confinement.

This dissertation presents a study of the previously mentioned topics through the lens of the phase-space perspective. The main objectives include the following:

- Develop a general variational theory that describes the ponderomotive refraction of waves.

- Explain the connection between the newly developed ponderomotive theory of waves and the theory of wave dispersion.

- Using the previously obtained methods, derive a ponderomotive model for the relativistic spin-1/2 electron, which includes the leading-order classical dynamics and the effects due to the particle spin.

- Study the collective dynamics of incoherent high-frequency waves in the particular context of zonal-flow formation in turbulence.

## 1.3 Thesis overview and main contributions

This dissertation is organized in five parts. Part I, composed of Chapters 1 to 3, gives the main motivations for the current research and introduces the basic concepts and methodologies that will be used throughout the dissertation. Part II, which comprises Chapters 4 to 6, is focused on polarization effects on the propagation of linear vector waves. Part III is composed of Chapters 7 to 9 and presents some recent discoveries regarding nonlinear wave–wave interactions. Part IV, which only includes Chapter 10, presents the final remarks and conclusions. Finally, Part V, which is composed of Appendices A to C, gives a general overview of the Weyl symbol calculus and presents some auxiliary calculations. A brief summary of each Chapter is given below.

**Part I: A primer on variational theories of waves**

In Chapter 2, I present a general overview of variational formulations commonly used in wave dynamics. The variational formulations based on the physical-space, abstract, and phase-space representations are introduced. In order to present the phase-space representation of wave dynamics, a brief survey of the main properties of the Weyl symbol calculus is given in this Chapter. Examples are shown on how to apply the different variational formulations.

In order to introduce some of the basic methodologies used in the thesis, I present a brief discussion of the GO approximation in Chapter 3. First, the main assumptions underlying the GO approximation are given. Then, starting from the phase-space representation of the action functional, I derive the GO equations for linear scalar waves propagating in a nondissipative medium. For completeness, both eikonal waves and incoherent waves are treated. For eikonal waves, the resulting GO equations are the well-known Whitham's equations for wave dynamics. For incoherent waves, the governing equation is the wave kinetic equation. Finally, a brief comment is also presented on how to extend GO in order to include diffraction effects. Illustrative examples are also given throughout the Chapter.

**Part II: Extending geometrical optics to vector waves**

As mentioned previously, the GO ray equations are completely characterized by the wave coordinates and momenta. However, unlike scalar waves, multi-component (vector) waves have another degree of freedom, namely, their polarization. This additional degree of freedom manifests itself as an effective "wave spin" that can be assigned to the rays and affects the wave dynamics accordingly. Two well-known manifestations of polarization dynamics are mode conversion and the precession of the wave polarization. However, another less-known polarization effect is the polarization-driven bending of ray trajectories. In Chapter 4, I present an extension and reformulation of GO as a first-principle Lagrangian theory whose action includes the aforementioned effects of wave polarization. I call such a theory *extended geometrical optics* (XGO).

In Chapter 5, I apply XGO theory to obtain the first-ever point-particle Lagrangian model for the relativistic spin-1/2 particle. The starting point of the theory is the first-principle Dirac action that describes the relativistic spin-1/2 quantum particle. By applying the procedure given in Chapter 4, I obtain a point-particle phase-space Lagrangian, whose Hamiltonian captures spin dynamics such as the Stern–Gerlach force and the particle spin precession. This new model is then compared with previously obtained classical theories for spinning particles.

In Chapter 6, I apply the developed XGO theory to study polarization effects on RF waves propagating in magnetized plasma. Analytical results are presented for the case of waves propagating in weakly magnetized plasma. It is shown that polarization effects are manifested as a precession of the wave polarization and as a polarization-driven bending of the GO ray trajectories. Numerical simulations are presented for the case of waves in strongly magnetized plasma. In particular, I investigate the regime in which polarization effects are important in wave ray tracing in tokamaks.

**Part III: Nonlinear wave–wave interactions**

In Chapter 7, I apply the machinery developed in the previous Chapters to study another class of problems, which is as follows. As is well known, charged particles can experience time-averaged ponderomotive forces in high-frequency fields. In a similar way, linear waves can also experience time-averaged refraction in modulated media. In this Chapter, I present a general variational theory that can describe this refraction, or *ponderomotive effect on waves*. The same theory also shows that any wave is, in fact, a polarizable object that contributes to the linear dielectric tensor of the ambient medium. Examples are given on how to calculate the ponderomotive Hamiltonians of quantum particles and photons within a number of models.

In Chapter 8, I present a ponderomotive model of a relativistic spin-1/2 electron interacting with a high-frequency EM field. The starting point of the theory is the action for the quantum Dirac electron.

After applying the techniques developed in Chapters 4 and 7, I derive a reduced model that describes the time-averaged dynamics of the relativistic spin-1/2 electron in a laser pulse propagating in vacuum. The pulse is allowed to have an arbitrarily large amplitude provided that radiation damping and pair production are negligible. The model captures the Bargmann–Michel–Telegdi spin dynamics, the Stern–Gerlach spin–orbital coupling, the conventional ponderomotive forces, and the interaction with large-scale background fields. Predictions of this ponderomotive model are compared with the non-averaged point-particle model given in Chapter 5.

In Chapter 9, I present a study of the collective behavior of incoherent high-frequency waves in the particular context of zonal-flow (ZF) formation in wave turbulence. In recent years, the appearance of ZFs in drift-wave (DW) turbulence inside tokamaks has captured much attention. Among the various reduced models, the wave kinetic equation (WKE) is widely used to study this phenomenon. However, this formulation neglects the exchange of enstrophy between DWs and ZFs and also ignores effects beyond the GO limit. In this Chapter, I present a new theory that captures both of these effects while still treating DW quanta ("driftons") as particles in phase space. This formulation can be considered as a phase-space representation of the second-order cumulant expansion (CE2). In the GO limit, this formulation features additional terms missing in the traditional WKE that ensure exact conservation of the total enstrophy of the system, in addition to the total energy.

**Part IV: Discussion and conclusions**

In Chapter 10, I summarize the main results obtained in this dissertation. Potential opportunities for future research are outlined.

**Part V: Appendices**

To understand the results presented in this thesis, some basic knowledge of the Weyl symbol calculus will be required. In this regard, in Appendix A I include a summary of the basic properties of the Weyl symbol calculus that are frequently used throughout this dissertation.

In Appendix B, I present a construction of the mathematical formalism behind the Weyl symbol calculus. Proofs of the various properties of the Weyl symbol calculus are given with considerable detail in order to facilitate a beginner's introduction to the subject.

In Appendix C, I present some auxiliary calculations needed to derive some of the results presented in this dissertation.

# Chapter 2

# Variational principles of wave dynamics

This Chapter presents a general overview on variational formulations commonly used to study wave dynamics. The variational formulations based on the physical-space, abstract, and phase-space representations are introduced. Also, a brief survey on the Weyl symbol calculus is given in order to introduce the phase-space representation of wave dynamics. Examples are given on how to apply the different variational formulations.

## 2.1 Introduction

### 2.1.1 Motivation

The development of reduced models for wave dynamics has enjoyed great progress ever since the seminal works by Whitham on the propagation of linear and nonlinear dissipationless waves (Whitham, 1961, 1965a,b, 2011). The cornerstone of Whitham's method is that general nondissipative wave equations can be cast into a transparent and relatively universal variational formulation. Variational methods are particularly advantageous as they lead to wave equations in a manifestly conservative form by extremizing a single scalar object, the action functional. The fact that wave equations can be approximated robustly and self-consistently by approximating a single functional makes the method particularly attractive for the asymptotic analysis of wave equations. Furthermore, variational methods have been convenient for the development of new reduced wave theories.[1]

[1] See, for example, Dewar (1970, 1972), Hayes (1973), and Dodin (2014b).

In Whitham's variational approach (Whitham, 1961), the wave fields are functions of the four-dimensional spacetime coordinates. This description of the fields is called the *physical-space representation* and is often used for studying Maxwell's equations for electromagnetism or the Navier–Stokes equations for fluid dynamics. However, other representations of the field dynamics are also possible. For example, from classical mechanics it is well known that the dynamics of a physical system can be described by using either the physical-space perspective (Lagrangian approach) or the phase-space perspective (Hamiltonian approach) (Goldstein *et al.*, 2002). Although these representations are largely equivalent, the Hamiltonian approach is considered to be a more "geometric" theory and can often lead to new insights to old problems (Marsden and Ratiu, 1999; Abraham and Marsden, 1987). Drawing from this analogy, one could wonder if further advances on wave theory can be achieved by using a *variational phase-space approach* for describing waves.

It turns out that there exists a natural *phase-space representation* of waves. In this description, the wave fields depend on both position and momentum coordinates; i.e., they are functions of the *ray phase space*. The pioneering ideas of this approach were developed in the field of quantum mechanics by Hermann Weyl, Eugene Wigner, José Moyal, and others (Weyl, 1931; Wigner, 1932; Moyal, 1949). Years later, the Berkeley plasma theory group, led by Allan Kaufman, contributed to major breakthroughs in classical wave dynamics by using the variational phase-space approach.[2] Some contributions include the development of a covariant phase-space formulation of wave dynamics (Kaufman *et al.*, 1987), a wave-kinetic equation for describing incoherent fields (McDonald, 1988), and a general approach to linear mode conversion (Friedland *et al.*, 1987). In summary, during the past years a more profound understanding of wave dynamics has been achieved by using the phase-space representation of wave dynamics. This approach will be reviewed in the present Chapter.

### 2.1.2 Overview

This Chapter presents a general overview on the different variational formulations of wave dynamics. This Chapter is organized as follows. In Sec. 2.2, the variational formulations in the physical-space and abstract representations are introduced. In Sec. 2.3, a brief survey of the Weyl symbol calculus is given. Then, Sec. 2.4 presents the variational principle in the the phase-space representation. In Sec. 2.5, conclusions and final remarks are given.

[2] For the interested reader, Tracy and Brizard (2009) present an excellent historical account of the many contributions of Allan Kaufman to the field of wave dynamics.

## 2.2 Basic formulation

### 2.2.1 Physical-space representation

Let us begin by introducing the variational formulation of wave dynamics in the physical-space representation. Consider a wave field, either classical or quantum, as a complex-valued vector $\Psi(x)$ that depends solely on the spacetime coordinate $x = (t, \mathbf{x})$.[3] This field has an arbitrary (yet finite) number of components $N$ so that

$$\Psi(x) = \begin{pmatrix} \Psi^1(x) \\ \vdots \\ \Psi^N(x) \end{pmatrix}, \tag{2.1}$$

where $\Psi^n(x)$ denotes the $n$th wave component.[4] In the absence of external sources and parametric resonances,[5] the dynamics of any linear wave $\Psi(x)$ can be cast in a general form as (Dodin, 2014a):

$$\int \mathrm{d}^4x' \, \mathcal{D}(x, x')\Psi(x') = 0, \tag{2.2}$$

where $\mathcal{D}(x, x')$ is a $N \times N$ matrix kernel that depends on two spacetime coordinates $x$ and $x'$. Here $\mathcal{D}(x, x')$ describes the effects of the underlying medium on the propagating wave $\Psi(x)$. Unless otherwise noted, this thesis uses the Minkowski metric $g^{\mu\nu} = g_{\mu\nu} = \mathrm{diag}\,[+, -, -, -]$. Also, $\mathrm{d}^4x \doteq \mathrm{d}x^0\,\mathrm{d}x^1\,\mathrm{d}x^2\,\mathrm{d}x^3$ is the four-dimensional volume element, and the integrals are taken over spacetime.

***Example.*** — Equation (2.2) encompasses both partial differential equations and integro-differential equations. For example, in the case of a Schrödinger particle in a potential $V(x)$, the kernel $\mathcal{D}(x, x')$ is a scalar whose expression is given by

$$\mathcal{D}(x, x') = \left( i \frac{\partial}{\partial t'} - \frac{1}{2m} \frac{\partial}{\partial \mathbf{x}'} \cdot \frac{\partial}{\partial \mathbf{x}'} + V(x) \right) \delta^4(x - x'). \tag{2.3}$$

Upon substituting Eq. (2.3) into Eq. (2.2) and integrating by parts, one obtains

$$i \frac{\partial}{\partial t} \Psi(x) = \left( -\frac{1}{2m} \boldsymbol{\nabla}^2 + V(x) \right) \Psi(x), \tag{2.4}$$

which is the well-known Schrödinger equation.

[3] This formulation can be easily extended to fields that depend on a larger set of parameters; for example, the scalar function $\Psi(t, \mathbf{x}_1, \mathbf{x}_2, ..., \mathbf{x}_n)$ is often used in many-body physics to denote the $n$-particle wave function.

[4] For the sake of clarity, here I refer to the $n$th component of the wave as the $n$th element of the column vector $\Psi(x)$. Thus, $\Psi^n(x)$ should not be confused with an amplitude corresponding to a particular wave mode.

[5] When parametric resonances are present, the kernel $\mathcal{D}(x, x')$ oscillates at harmonics of the wave frequency. In such cases, the general wave equation (2.2) will contain terms proportional to $\Psi^\dagger(x)$. Such terms usually break the conservation of the wave action (Dodin, 2014a).

In this thesis, I shall mainly be concerned with the dynamics of dissipationless waves (except in Chapter 9).[6] Hence, only Hermitian matrix kernels satisfying the property $\mathcal{D}(x, x') = \mathcal{D}^\dagger(x', x)$ are considered. Hence, an action functional $\mathcal{S}$ can be introduced for the wave equation:

$$\mathcal{S} \doteq \int \mathrm{d}^4x \, \mathrm{d}^4x' \, \Psi^\dagger(x) \mathcal{D}(x, x') \Psi(x'). \tag{2.5}$$

Here the action functional $\mathcal{S}$ is a bilinear functional of the fields $\Psi(x)$ and $\Psi^\dagger(x)$. Since only Hermitian matrix kernels are considered, the action $\mathcal{S}$ is guaranteed to be real, and the operation of finding the stationnary points of the action functional is well defined. Note also that $\Psi(x)$ and $\Psi^\dagger(x)$ can be treated as independent, which is equivalent to treating the real and imaginary parts of $\Psi(x)$ as independent (Dodin, 2014a).

The dynamics of nondissipative waves is described by the *principle of stationary action*

$$\delta \mathcal{S} = 0. \tag{2.6}$$

Varying the action with respect to $\Psi^\dagger$ leads to Eq. (2.2):

$$\delta \Psi^\dagger : \quad \int \mathrm{d}^4x' \, \mathcal{D}(x, x') \Psi(x') = 0. \tag{2.2 revisited}$$

Similarly, varying with respect to $\Psi$ gives the equation adjoint to Eq. (2.2) which is not necessary to discuss in further detail.

***Example.*** — Returning back to the example of the Schrödinger equation, one inserts Eq. (2.3) into Eq. (2.5) and integrates on the coordinate $x'$. One then obtains

$$\mathcal{S} = \int \mathrm{d}^4x \, \Psi^\dagger(x) \left( i \frac{\partial}{\partial t} + \frac{\boldsymbol{\nabla}^2}{2m} - V(x) \right) \Psi(x). \tag{2.7}$$

Certainly, varying the action with respect to $\Psi^\dagger$ leads to Eq. (2.4).

### 2.2.2 Abstract representation in the Hilbert space

It is convenient to relate the wave field $\Psi(x)$ to an abstract vector $|\, \Psi \,\rangle$ of the Hilbert space of states with inner product

$$\langle \Upsilon \mid \Psi \rangle \doteq \int \mathrm{d}^4x \, \Upsilon^\dagger(x) \Psi(x). \tag{2.8}$$

[6] See, e.g., Brizard (1994) and Dodin *et al.* (2017) for more information on the treatment of dissipative waves.

Here $|\,\Psi\,\rangle$ is called a "ket," and $\langle\,\Upsilon\,|$ is called a "bra." (This particular notation is known as the Dirac notation.) The function $\Psi(x)$ is related to the ket $|\,\Psi\,\rangle$ by[7]

$$\Psi(x) \doteq \langle\, x \,|\, \Psi \,\rangle\,, \tag{2.9}$$

where $|\,x\,\rangle \doteq |\,(t,\mathbf{x})\,\rangle$ are the eigenstates of the coordinate operator $\widehat{x}^\mu$ such that

$$\langle\, x \,|\, \widehat{x}^\mu \,|\, x' \,\rangle = x^\mu \,\langle\, x \,|\, x' \,\rangle = x^\mu \delta^4(x-x'). \tag{2.10}$$

On the left-hand side of Eq. (2.9), $\Psi(x)$ is a function mapping any point in spacetime to a complex $N$-component vector; on the right-hand side, $|\,\Psi\,\rangle = \int \mathrm{d}^4x'\,\Psi(x')\,|\,x'\,\rangle$ is a ket. Likewise, $\Psi^\dagger(x)$ is simply represented by $\Psi^\dagger(x) = (\langle\, x \,|\, \Psi\,\rangle)^\dagger = \langle\,\Psi\,|\,x\,\rangle$.

Now, let us consider a family of linear transformations of the form

$$\Phi(x) = \int \mathrm{d}^4x'\,\mathcal{T}(x,x')\Psi(x'), \tag{2.11}$$

where $\mathcal{T}(x,x')$ is a $N\times N$ matrix kernel. These linear transformations map the function $\Psi(x)$ into some family of image functions $\Phi(x)$ that belong in the same Hilbert space. In the abstract Hilbert representation, this transformation is written as follows:

$$\Phi(x) \doteq \langle\, x \,|\, \widehat{\mathcal{T}} \,|\, \Psi \,\rangle\,, \tag{2.12}$$

where $\widehat{\mathcal{T}}$ denotes the corresponding linear operator in the Hilbert space. Let us find a relation between the operator $\widehat{\mathcal{T}}$ and the kernel $\mathcal{T}(x,x')$. Since the eigenstates $|\,x\,\rangle$ form a complete orthogonal set, they satisfy the completeness relation

$$\int \mathrm{d}^4x\,|\,x\,\rangle\,\langle\,x\,| = \widehat{1}, \tag{2.13}$$

where $\widehat{1}$ is the identity operator. Inserting the latter into Eq. (2.12) leads to

$$\begin{aligned}\Phi(x) &= \langle\,x\,|\,\widehat{\mathcal{T}}\left(\int \mathrm{d}^4x'\,|\,x'\,\rangle\,\langle\,x'\,|\right)|\,\Psi\,\rangle \\ &= \int \mathrm{d}^4x'\,\langle\,x\,|\,\widehat{\mathcal{T}}\,|\,x'\,\rangle\,\langle\,x'\,|\,\Psi\,\rangle \\ &= \int \mathrm{d}^4x'\,\langle\,x\,|\,\widehat{\mathcal{T}}\,|\,x'\,\rangle\,\Psi(x').\end{aligned} \tag{2.14}$$

[7]For an extensive introduction to the properties of basis kets and operators, see Cohen-Tannoudji *et al.* (1977).

When comparing Eqs. (2.11) and (2.14), it is clear that the kernel of the linear transformation $\mathcal{T}(x, x')$ and the abstract operator $\widehat{\mathcal{T}}$ are related by

$$\mathcal{T}(x, x') \doteq \langle\, x \mid \widehat{\mathcal{T}} \mid x' \,\rangle \,. \tag{2.15}$$

For instance, the momentum (wavevector) operator $\widehat{p}_\mu$ is defined as

$$\langle\, x \mid \widehat{p}_\mu \mid x' \,\rangle \doteq i \frac{\partial}{\partial x^\mu} \delta^4(x - x') \tag{2.16}$$

in the physical-space representation. Thus, the transformation of the operator $\widehat{p}_\mu$ on $|\, \Psi \,\rangle$ is

$$\begin{aligned} \langle\, x \mid \widehat{p}_\mu \mid \Psi \,\rangle &= \int \mathrm{d}^4 x' \, \langle\, x \mid \widehat{p}_\mu \mid x' \,\rangle \langle\, x' \mid \Psi \,\rangle \\ &= i \frac{\partial}{\partial x^\mu} \int \mathrm{d}^4 x' \, \delta^4(x - x') \Psi(x') \\ &= i \frac{\partial}{\partial x^\mu} \Psi(x). \end{aligned} \tag{2.17}$$

In terms of contravariant components, the momentum operator is represented as $\langle\, x \mid \widehat{p}^0 \mid \Psi \,\rangle = i\partial_t \Psi(x)$ and $\langle\, x \mid \widehat{\mathbf{p}} \mid \Psi \,\rangle = -i\boldsymbol{\nabla}\Psi(x)$.

Now, one can rewrite the action (2.5) in the abstract representation. Upon substituting Eqs. (2.9) and (2.15) into Eq. (2.5), one can write the action functional as follows:

$$\mathcal{S} = \langle\, \Psi \mid \widehat{\mathcal{D}} \mid \Psi \,\rangle \,, \tag{2.18}$$

where $\widehat{\mathcal{D}}$ is the Hermitian *dispersion operator* such that $\mathcal{D}(x, x') = \langle\, x \mid \widehat{\mathcal{D}} \mid x' \,\rangle$. The action can also be expressed as

$$\mathcal{S} = \mathrm{Tr} \int \mathrm{d}^4 x \, \langle\, x \mid \widehat{\mathcal{D}}\widehat{\mathcal{N}} \mid x \,\rangle \,, \tag{2.19}$$

where $\widehat{\mathcal{N}} \doteq |\, \Psi \,\rangle \langle\, \Psi \,|$ is the *spectral-tensor operator* whose components are given by $\widehat{\mathcal{N}}^m_n = |\, \Psi^m \,\rangle \langle\, \Psi_n \,|$. For the case of scalar waves, $\widehat{\mathcal{N}}$ is also known as the *von Neumann density operator*. Also, "Tr" denotes the matrix trace so Eq. (2.19) can also be written as $\mathcal{S} = \int \mathrm{d}^4 x \, \langle\, x \mid \widehat{\mathcal{D}}^m_n \widehat{\mathcal{N}}^n_m \mid x \,\rangle$. Treating $\langle\, \Psi \,|$ and $|\, \Psi \,\rangle$ as independent variables and varying the action (2.18) gives

$$\delta \langle\, \Psi \,| : \quad \widehat{\mathcal{D}} \,|\, \Psi \,\rangle = 0 \tag{2.20}$$

(plus the conjugate equation), which is the generalized abstract form of Eq. (2.2). Specifically, Eq. (2.2) is obtained by projecting Eq. (2.20) with $\langle\, x \,|$ and using the completeness relation (2.13).

***Example.*** — In the abstract representation, the action (2.7) is written as

$$\mathcal{S} = \langle\, \Psi \,|\, \left( \widehat{p}_0 - \frac{\widehat{\mathbf{p}}^2}{2m} - V(\widehat{x}) \right) |\, \Psi \,\rangle\,, \tag{2.21}$$

and the corresponding ELE is

$$\delta\,\langle\, \Psi \,|: \quad \widehat{p}_0 \,|\, \Psi \,\rangle = \left( \frac{\widehat{\mathbf{p}}^2}{2m} + V(\widehat{x}) \right) |\, \Psi \,\rangle\,. \tag{2.22}$$

An important action functional of a linear wave, whose dispersion operator $\widehat{\mathcal{D}}$ is linear in the temporal momentum operator $\widehat{p}_0$, is

$$\mathcal{S} = \langle\, \Psi \,|\, (\widehat{p}_0 \mathbb{I}_N - \widehat{\mathcal{H}}) \,|\, \Psi \,\rangle\,, \tag{2.23}$$

where $\widehat{\mathcal{H}} = \mathcal{H}(\widehat{t}, \widehat{\mathbf{x}}, \widehat{\mathbf{p}})$ is a Hermitian operator called the Hamiltonian. Due to the fact that $\widehat{\mathcal{H}}$ is Hermitian, Eq. (2.20) leads to

$$\frac{\mathrm{d}}{\mathrm{d}t} \mathrm{Tr} \int \mathrm{d}^3\mathbf{x}\, \langle\, x \,|\, \widehat{\mathcal{N}} \,|\, x \,\rangle = \frac{\mathrm{d}}{\mathrm{d}t} \int \mathrm{d}^3\mathbf{x}\, |\Psi(t, \mathbf{x})|^2 = 0, \tag{2.24}$$

which is known as the action conservation theorem. The quantity $N(t) \doteq \mathrm{Tr} \int \mathrm{d}^3\mathbf{x}\, \langle\, x \,|\, \widehat{\mathcal{N}} \,|\, x \,\rangle$ is called the wave action, and $|\Psi(t, \mathbf{x})|^2 = \Psi^\dagger(t, \mathbf{x})\Psi(t, \mathbf{x})$ is called the action density. Note that the term "wave action" should not be confused with the action functional $\mathcal{S}$.

The advantages in writing wave equations using the abstract representation are as follows. First, this notation allows one to easily track transformations done on the wave field. As an example, one could define a transformation $\widehat{\mathcal{T}}$ such that $|\, \Psi \,\rangle = \widehat{\mathcal{T}} \,|\, \psi \,\rangle$. Then, the wave action becomes $\mathcal{S} = \langle\, \psi \,|\, \widehat{\mathcal{D}}_{\text{eff}} \,|\, \psi \,\rangle$, where $\widehat{\mathcal{D}}_{\text{eff}} \doteq \widehat{\mathcal{T}}^\dagger \, \widehat{\mathcal{D}} \, \widehat{\mathcal{T}}$ is an effective dispersion operator and $\psi(x)$ is the new independent wave field. In this thesis, reduced wave models will be obtained by finding transformations $\widehat{\mathcal{T}}$ such that $\widehat{\mathcal{D}}_{\text{eff}}$ is "simpler" than the original dispersion operator $\widehat{\mathcal{D}}$. (This will be shown in more detail in the subsequent chapters.) A second advantage of using the abstract representation is that one can seamlessly change to other representations of the wave dynamics, such as the $x$-coordinate space and $p$-momentum space (not mentioned here) representations. Alternatively, one can also seamlessly change to the phase-space representation, in which wave fields are described using phase-space coordinates, similar to the Hamiltonian formulation of classical mechanics. Such formulation is introduced below.

## 2.3 Weyl symbol calculus

In 1931, Weyl introduced what is now known as the Weyl symbol calculus that provides an invertible mapping between operators in the abstract Hilbert space and functions living in phase space. Weyl's original work was mainly motivated to show the correspondence between classical mechanics and quantum mechanics (Weyl, 1931). However, the tools that he developed have proven to be very useful for the study of general wave dynamics.

In what follows, some basic notions of the Weyl symbol calculus will be introduced. For a more complete overview on the mathematical construction of the Weyl symbol calculus, the reader is invited to read Appendix B. Another recommended reference is the book by Tracy *et al.* (2014).

### 2.3.1 Weyl transform

As mentioned before, the Weyl transformation is an invertible mapping between abstract operators and functions of phase space; the latter are called *Weyl symbols*. Let $\widehat{\mathcal{A}}$ be an operator of the Hilbert space with inner product (2.8). The Weyl transform $\mathsf{W}[\widehat{\mathcal{A}}]$ of this operator is defined as

$$\mathsf{W}[\widehat{\mathcal{A}}](x,p) \doteq \int \mathrm{d}^4 s \, e^{ip\cdot s} \, \langle \, x + s/2 \, | \, \widehat{\mathcal{A}} \, | \, x - s/2 \, \rangle \, . \tag{2.25}$$

The image of the Weyl transform $A(x,p) \doteq \mathsf{W}[\widehat{\mathcal{A}}](x,p)$ is called the Weyl symbol of the operator $\widehat{\mathcal{A}}$, and it is a function of the eight-dimensional phase space. Since vector waves are treated, the Weyl symbol $A(x,p)$ can in fact be a matrix whose elements are given by $A^m_n(x,p) = \mathsf{W}[\widehat{\mathcal{A}}^m_n]$.[8,9]

***Example.*** — Before continuing further, let us review some examples of Weyl transforms:

- First, the Weyl transform of the identity operator is unity. Upon using the fact that $\langle \, x \, | \, \widehat{1} \, | \, x' \, \rangle = \delta^4(x - x')$, one finds

$$\begin{aligned} \mathsf{W}[\,\widehat{1}\,] &= \int \mathrm{d}^4 s \, e^{ip\cdot s} \, \langle \, x + s/2 \, | \, \widehat{1} \, | \, x - s/2 \, \rangle \\ &= \int \mathrm{d}^4 s \, e^{ip\cdot s} \delta^4(s) \\ &= 1. \end{aligned} \tag{2.26}$$

As a corollary, the Weyl symbol of any constant is the same constant.

[8]In this thesis, I shall not be concerned on the conditions so that the mapping and its inverse exist. For the purposes of this thesis, I shall assume that these are well-defined operations.

[9]Notably, the mapping from abstract operators to functions of phase space is not unique. For more information on this subject, see McDonald (1988) and Dodin (2014a).

- The Weyl transform of the position operator $\widehat{x}^\mu$ is given by

$$\begin{aligned}\mathsf{W}[\widehat{x}^\mu] &= \int \mathrm{d}^4 s\, e^{ip\cdot s} \langle\, x + s/2 \mid \widehat{x}^\mu \mid x - s/2\,\rangle \\ &= \int \mathrm{d}^4 s\, e^{ip\cdot s} (x^\mu + s^\mu/2)\delta^4(s) \\ &= x^\mu, \end{aligned} \tag{2.27}$$

where I substituted Eq. (2.10). Hence, the Weyl transform conveniently maps the position operator $\widehat{x}^\mu$ to the position coordinate $x^\mu$.

- The Weyl symbol of the momentum operator $\widehat{p}_\mu$ is[10]

$$\begin{aligned}\mathsf{W}[\widehat{p}_\mu] &= \int \mathrm{d}^4 s\, e^{ip\cdot s} \langle\, x + s/2 \mid \widehat{p}_\mu \mid x - s/2\,\rangle \\ &= i\int \mathrm{d}^4 s\, e^{ip\cdot s} \frac{\partial}{\partial s^\mu}\delta^4(s) \\ &= -i\int \mathrm{d}^4 s\, \delta^4(s) \frac{\partial}{\partial s^\mu}\, e^{ip\cdot s} \\ &= p_\mu, \end{aligned} \tag{2.29}$$

where I substituted Eq. (2.16). Hence, the Weyl transform conveniently maps the position operator $\widehat{p}_\mu$ to the momentum coordinate $p_\mu$.

- For operators $f(\widehat{x})$ and $g(\widehat{p})$ depending solely on the position or momentum operators (but not both), it can be shown that the corresponding Weyl symbols are

$$\mathsf{W}[f(\widehat{x})] = f(x), \qquad \mathsf{W}[g(\widehat{p})] = g(p). \tag{2.30}$$

[10]To calculate the Weyl symbol of the momentum operator, it is convenient to adopt the equivalent definition

$$\langle\, x \mid \widehat{p}_\mu \mid x'\,\rangle \doteq i \lim_{a\to 0} \frac{\partial}{\partial a^\mu}\delta^4(x - x' + a). \tag{2.28}$$

- However, if an operator depends on both $\widehat{x}$ and $\widehat{p}$, the Weyl transformation can be less intuitive. For example, the Weyl transform of the operator $\widehat{x}^\mu \widehat{p}_\nu$ is

$$
\begin{aligned}
\mathsf{W}[\widehat{x}^\mu \widehat{p}_\nu] &= \int \mathrm{d}^4 s\, e^{ip\cdot s}\, \langle\, x + s/2 \mid \widehat{x}^\mu \widehat{p}_\nu \mid x - s/2\,\rangle \\
&= \int \mathrm{d}^4 s\, e^{ip\cdot s} (x^\mu + s^\mu/2)\, \langle\, x + s/2 \mid \widehat{p}_\nu \mid x - s/2\,\rangle \\
&= i \int \mathrm{d}^4 s\, e^{ip\cdot s} \left( x^\mu + \frac{s^\mu}{2} \right) \frac{\partial}{\partial s^\nu}\, \delta^4(s) \\
&= -i \int \mathrm{d}^4 s\, \delta^4(s) \frac{\partial}{\partial s^\nu} \left[ \left( x^\mu + \frac{s^\mu}{2} \right) e^{ip\cdot s} \right] \\
&= -i \int \mathrm{d}^4 s\, \delta^4(s) \left[ \frac{1}{2} \delta^\mu_\nu + i \left( x^\mu + \frac{s^\mu}{2} \right) p_\nu \right] e^{ip\cdot s} \\
&= x^\mu p_\nu - \frac{i}{2} \delta^\mu_\nu .
\end{aligned}
\tag{2.31}
$$

As shown, the Weyl transform of $\widehat{x}^\mu \widehat{p}_\nu$ is not $x^\mu p_\nu$ but involves an additional imaginary term. This occurs because the position and momentum operators do not commute. Similarly, one can show that $\mathsf{W}[\widehat{p}_\nu \widehat{x}^\mu] = x^\mu p_\nu + i\delta^\mu_\nu/2$. In consequence, the Weyl transform of the commutator $[\widehat{x}^\mu, \widehat{p}_\nu]_-$ is $\mathsf{W}[\,[\widehat{x}^\mu, \widehat{p}_\nu]_-\,] = -i\delta^\mu_\nu$ agrees with the commutation relation $[\widehat{x}^\mu, \widehat{p}_\nu]_- = \widehat{x}^\mu \widehat{p}_\nu - \widehat{p}_\nu \widehat{x}^\mu = -i\delta^\mu_\nu$.

### 2.3.2 The Wigner tensor

A particularly important Weyl symbol is the one corresponding to the spectral-tensor operator $\widehat{\mathsf{N}}$. With a factor $(2\pi)^{-4}$ added for convenience, it is also known as the *Wigner tensor* (Wigner, 1932) and is given by

$$
W_\Psi(x,p) \doteq \frac{1}{(2\pi)^4} \int \mathrm{d}^4 s\, e^{ip\cdot s}\, \langle\, x + s/2 \mid \Psi\,\rangle \langle\, \Psi \mid x - s/2\,\rangle\, ;
\tag{2.32}
$$

or in terms of components,

$$
[W_\Psi]^m_n(x,p) \doteq \frac{1}{(2\pi)^4} \int \mathrm{d}^4 s\, e^{ip\cdot s}\, \Psi^m(x + s/2)\, \Psi^*_n(x - s/2).
\tag{2.33}
$$

Note that the Wigner tensor is Hermitian; i.e., $W(x,p) = W^\dagger(x,p)$. This can be easily shown by calculating the conjugate transpose of Eq. (2.32) and adopting the variable transformation $s \to -s$. In the particular case where $\Psi(x)$ is a one-component scalar wave, the Wigner tensor becomes a real scalar, which is also commonly known as the *Wigner function*.

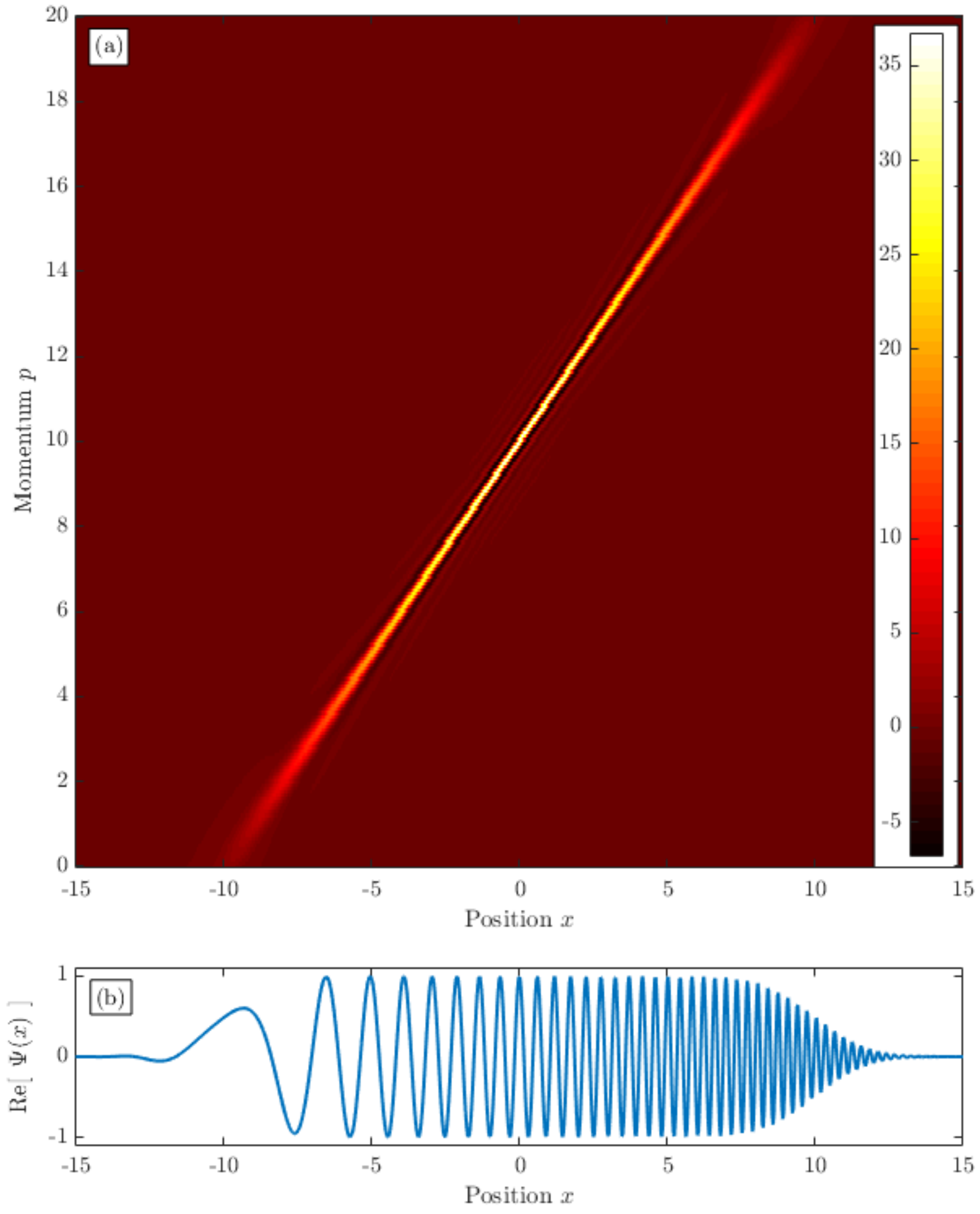

Figure 2.1: Wigner function of a chirped complex-valued scalar wave. (a) Corresponding Wigner function of the one-dimensional complex scalar wave. (b) The chirped scalar wave is generated by the function: $\Psi(x) = 0.5(2\pi)^3[\text{erf}(X+10) - \text{erf}(X-10)]\exp[-i\Theta(x)]$, where $\Theta(x) = 10x + 0.5x^2$. Here $x$ denotes a spatial coordinate, unlike in the main text, where $x$ denotes the spacetime coordinate.

***Example.*** — The Wigner tensor plays a fundamental role in the Weyl calculus because it is a natural tool for visualizing the system state in terms of phase-space variables. As an example, let us consider a one-dimensional scalar wave given by $\Psi(x) = (2\pi)^3(2\pi\sigma^2)^{-1/4}\exp\left[-(x-X)^2/(4\sigma^2)+iKx\right]$. (In this example, $x$ denotes a spatial coordinate, unlike in the main text, where $x$ denotes the spacetime coordinate.) Here the wave is spatially localized around $x = X$ with a constant momentum, or wavenumber, $K$. Upon introducing this expression into Eq. (2.32), one obtains

$$W_\Psi(x,p) = \frac{1}{\pi}\exp\left(-\frac{(x-X)^2}{2\sigma^2} - 2\sigma^2(p-K)^2\right). \tag{2.34}$$

As it can be seen, the Wigner function is proportional to a Gaussian distribution localized around the spatial position $X$ and momentum $K$. Although the wave function $\Psi(x)$ has a well-defined momentum $K$, the Wigner function shows a certain spread in the momentum coordinate. This is interpreted as the inability to measure simultaneously both the “position” and “momentum” of the wave with absolute certainty. In quantum mechanics, this issue is known as the Heisenberg uncertainty principle.

***Example.*** — Consider the chirped one-dimensional scalar wave shown in Fig. 2.1. Contrary to the Fourier transform, the Wigner function maps both the local spatial location and momentum of the wave. When the wave amplitude is increasing or decreasing, the uncertainty of the wave momentum is largest (as can be observed in the characteristic width in momentum of the Wigner function). During the interval where the wave amplitude is constant ($-5 \lesssim x \lesssim 5$), the uncertainty on the wave momentum is smallest.

As a measure of the system distribution in phase space, the Wigner function is commonly interpreted as a quasi-probability distribution. It is not a true probability distribution because in certain regions of phase space, the Wigner function can become negative (as shown in Fig. 2.1). This phenomenon is due to interference effects. However, $W_\Psi(x,p)$ becomes the standard positively defined probability distribution when smoothed over phase spacewith a Gaussian whose variance is greater than or equal to that of the minimum uncertainty of wave packets (Cartwright, 1976).

### 2.3.3 The Moyal product

Another valuable tool of the Weyl calculus is the Moyal product. Let $\widehat{\mathcal{C}}$ be an operator composed by the multiplication of two operators $\widehat{\mathcal{A}}$ and $\widehat{\mathcal{B}}$ such that $\widehat{\mathcal{C}} = \widehat{\mathcal{A}}\widehat{\mathcal{B}}$. Then, the corresponding Weyl symbol

$C(x,p) = \mathsf{W}[\widehat{\mathcal{C}}]$ satisfies (Moyal, 1949)

$$C(x,p) = A(x,p) \star B(x,p), \tag{2.35}$$

where $A(x,p)$ and $B(x,p)$ are the Weyl symbols corresponding to $\widehat{\mathcal{A}}$ and $\widehat{\mathcal{B}}$, respectively. Here '$\star$' refers to the *Moyal product*, which is given by

$$A(x,p) \star B(x,p) \doteq A(x,p) e^{i\widehat{\mathcal{L}}/2} B(x,p), \tag{2.36}$$

and $\widehat{\mathcal{L}}$ is the *Janus operator*

$$\widehat{\mathcal{L}} \doteq \overleftarrow{\partial_p} \cdot \overrightarrow{\partial_x} - \overleftarrow{\partial_x} \cdot \overrightarrow{\partial_p} = \{\,\cdot\,,\,\cdot\,\}. \tag{2.37}$$

The arrows indicate the direction in which the derivatives act, and $A\widehat{\mathcal{L}}B = \{A, B\}$ is the canonical Poisson bracket in the eight-dimensional phase space, namely,

$$\widehat{\mathcal{L}} = \frac{\overleftarrow{\partial}}{\partial p^0}\frac{\overrightarrow{\partial}}{\partial x^0} - \frac{\overleftarrow{\partial}}{\partial x^0}\frac{\overrightarrow{\partial}}{\partial p^0} + \frac{\overleftarrow{\partial}}{\partial \mathbf{x}} \cdot \frac{\overrightarrow{\partial}}{\partial \mathbf{p}} - \frac{\overleftarrow{\partial}}{\partial \mathbf{p}} \cdot \frac{\overrightarrow{\partial}}{\partial \mathbf{x}}. \tag{2.38}$$

***Example.*** — The Moyal product considerably facilitates the calculation of Weyl symbols. For example, the Weyl symbol of the operator $\widehat{x}^\mu \widehat{p}_\nu$ is easily calculated as follows:

$$\mathsf{W}[\widehat{x}^\mu \widehat{p}_\nu] = x^\mu \star p_\nu = x^\mu e^{i\widehat{\mathcal{L}}/2} p_\nu = x^\mu p_\nu + \frac{i}{2}\{x^\mu, p_\nu\} = x^\mu p_\nu - \frac{i}{2}\delta^\mu_\nu, \tag{2.39}$$

which is identical to the result obtained in Eq. (2.31).

Two additional properties of the Moyal product are worth mentioning. First, the Moyal product is associative. For arbitrary symbols $A(x,p)$, $B(x,p)$, and $C(x,p)$, then

$$A \star B \star C = (A \star B) \star C = A \star (B \star C). \tag{2.40}$$

The second property is that, when integrated over all phase space, the Moyal product between two symbols becomes an ordinary product between the symbols; i.e.,

$$\int \mathrm{d}^4x\, \mathrm{d}^4p\, A(x,p) \star B(x,p) = \int \mathrm{d}^4x\, \mathrm{d}^4p\, A(x,p) B(x,p). \tag{2.41}$$

A summary of these and other properties of the Weyl symbol calculus is given in Appendix A. Also, the mathematical formalism underlying the Weyl symbol calculus is presented in Appendix B.

## 2.4 Wave dynamics in the phase-space representation

After having introduced the Weyl symbol calculus, one may now write a phase-space variational principle for linear dissipationless waves. In the phase-space representation, the action $\mathcal{S}$ is written as follows (Kaufman *et al.*, 1987):

$$\mathcal{S} = \mathrm{Tr} \int \mathrm{d}^4x\, \mathrm{d}^4p\, D(x,p) W_\Psi(x,p), \tag{2.42}$$

where $D(x,p)$ is the Weyl symbol (2.25) corresponding to the dispersion operator $\widehat{\mathfrak{D}}$ and $W_\Psi(x,p)$ is the Wigner tensor (2.32) corresponding to the wave field $\Psi(x)$. It can be easily shown that Eq. (2.42) is equivalent to Eq. (2.18). Upon substituting Eqs. (2.25) and (2.32) into Eq. (2.42), one obtains

$$\begin{aligned}
\mathcal{S} &= \frac{1}{(2\pi)^4} \int \mathrm{d}^4x\, \mathrm{d}^4p\, \mathrm{d}^4s\, \mathrm{d}^4s'\, e^{ip\cdot s} e^{ip\cdot s'} \langle\, x + s/2 \mid \widehat{\mathfrak{D}}^m_n \mid x - s/2 \,\rangle \langle\, x + s'/2 \mid \Psi^n \,\rangle \langle\, \Psi_m \mid x - s'/2 \,\rangle \\
&= \int \mathrm{d}^4x\, \mathrm{d}^4s\, \mathrm{d}^4s'\, \delta^4(s+s') \langle\, \Psi_m \mid x - s'/2 \,\rangle \langle\, x + s/2 \mid \widehat{\mathfrak{D}}^m_n \mid x - s/2 \,\rangle \langle\, x + s'/2 \mid \Psi^n \,\rangle \\
&= \int \mathrm{d}^4x\, \mathrm{d}^4s\, \langle\, \Psi_m \mid x + s/2 \,\rangle \langle\, x + s/2 \mid \widehat{\mathfrak{D}}^m_n \mid x - s/2 \,\rangle \langle\, x - s/2 \mid \Psi^n \,\rangle \\
&= \int \mathrm{d}^4u\, \mathrm{d}^4v\, \langle\, \Psi_m \mid u \,\rangle \langle\, u \mid \widehat{\mathfrak{D}}^m_n \mid v \,\rangle \langle\, v \mid \Psi^n \,\rangle \\
&= \langle\, \Psi \mid \widehat{\mathfrak{D}} \mid \Psi \,\rangle ,
\end{aligned} \tag{2.43}$$

where the variable transformations $u = x + s/2$ and $v = x - s/2$ were used.

Since the Wigner tensor $W_\Psi(x,p)$ is bilinear in the wave field $\Psi(x)$, its variations cannot be completely arbitrary but rather satisfy certain constraints. Hence, if formulated in terms of $W_\Psi(x,p)$, the variational principle (2.42) becomes a constrained variational principle. Let the wave state $|\, \Psi \,\rangle$ be varied such that $\delta \,|\, \Psi \,\rangle = i\delta\widehat{\mathfrak{C}} \,|\, \Psi \,\rangle$, where $\delta\widehat{\mathfrak{C}}$ is an arbitrary $N \times N$ matrix operator. Then, varying the spectral-tensor operator $\widehat{\mathfrak{N}} \doteq |\, \Psi \,\rangle \langle\, \Psi \,|$ leads to $\delta\widehat{\mathfrak{N}} = i\delta\widehat{\mathfrak{C}} \,|\, \Psi \,\rangle \langle\, \Psi \,| - i \,|\, \Psi \,\rangle \langle\, \Psi \,|\, \delta\widehat{\mathfrak{C}}^\dagger = i(\delta\widehat{\mathfrak{C}})\widehat{\mathfrak{N}} - i\widehat{\mathfrak{N}}(\delta\widehat{\mathfrak{C}})^\dagger$ . The variation of the Wigner tensor is then calculated using the Weyl transform (2.25) and the Moyal product (2.35):

$$\delta W_\Psi(x,p) = i\delta C(x,p) \star W_\Psi(x,p) - iW_\Psi(x,p) \star \delta C^\dagger(x,p), \tag{2.44}$$

where $\delta C(x,p)$ is the corresponding Weyl symbol of the operator $\delta\widehat{\mathfrak{C}}$. Upon substituting Eq. (2.44) into the action (2.42), one obtains the variation of the action:

$$\delta\mathcal{S} = i\, \mathrm{Tr} \int \mathrm{d}^4x\, \mathrm{d}^4p\, [\, D\,(\delta C \star W_\Psi) - D\,(W_\Psi \star \delta C^\dagger)\, ]. \tag{2.45}$$

Next, one uses Eqs. (2.40) and (2.41) to eliminate some of the Moyal products. Also, since the trace is invariant under cyclic permutations, then

$$\begin{aligned}\delta\mathcal{S} &= i\,\mathrm{Tr}\int \mathrm{d}^4x\,\mathrm{d}^4p\,[\,D(\delta C\star W_\Psi) - (D\star W_\Psi)\delta C^\dagger\,]\\ &= i\,\mathrm{Tr}\int \mathrm{d}^4x\,\mathrm{d}^4p\,[\,(\delta C\star W_\Psi)D - \delta C^\dagger(D\star W_\Psi)\,]\\ &= i\,\mathrm{Tr}\int \mathrm{d}^4x\,\mathrm{d}^4p\,[\,\delta C\star W_\Psi\star D - \delta C^\dagger(D\star W_\Psi)\,]\\ &= i\,\mathrm{Tr}\int \mathrm{d}^4x\,\mathrm{d}^4p\,[\,\delta C(W_\Psi\star D) - \delta C^\dagger(D\star W_\Psi)\,]. \end{aligned} \tag{2.46}$$

The symbol $\delta C(x,p)$ can be decomposed into its Hermitian and anti-Hermitan parts so that $\delta C(x,p) = \delta C_H(x,p) + i\delta C_A(x,p)$. Here $\delta C_H = (\delta C + \delta C^\dagger)/2$ and $\delta C_A = (\delta C - \delta C^\dagger)/(2i)$ serve as independent generators of the variations. Then, $\delta\mathcal{S}$ becomes

$$\delta\mathcal{S} = \mathrm{Tr}\int \mathrm{d}^4x\,\mathrm{d}^4p\,[-i\delta C_H(D\star W_\Psi - W_\Psi\star D) - \delta C_A(D\star W_\Psi + W_\Psi\star D)]. \tag{2.47}$$

Since the action must be stationary ($\delta\mathcal{S}=0$) for arbitrary matrix symbols $\delta C_H(x,p)$ and $\delta C_A(x,p)$, the corresponding ELEs are given by

$$\delta C_H:\quad 0 = \{\!\{D, W_\Psi\}\!\} \doteq -i\,(D\star W_\Psi - W_\Psi\star D)\,, \tag{2.48a}$$

$$\delta C_A:\quad 0 = [[D, W_\Psi]] \doteq D\star W_\Psi + W_\Psi\star D, \tag{2.48b}$$

where $\{\!\{\cdot\,,\cdot\}\!\}$ and $[[\cdot\,,\cdot]]$ are the properly normalized anti-symmetric and symmetric Moyal products, respectively.[11] As it will become clear in the next Chapter, Eq. (2.48a) represents the eight-dimensional phase-space quantum kinetic equation, which describes the propagation of the wave action in phase space. Also, Eq. (2.48b) is related to the wave dispersion relation.

When summing Eqs. (2.48), one can write the ELEs as the single equation

$$D(x,p)\star W_\Psi(x,p) = 0. \tag{2.49}$$

This equation could have been derived more easily without using a variational principle. Indeed, upo using the outer product, one multiplies Eq. (2.20) from the right by $\langle\,\Psi\,|$ so that $\widehat{\mathfrak{D}}\,|\,\Psi\,\rangle\,\langle\,\Psi\,| = 0$. Then, using the Weyl transform and the Moyal product (2.36) leads to Eq. (2.49).

[11] In the literature, these brackets are also known as the *sine brakets* and *cosine brackets*, respectively.

***Example.*** — Let us consider the case of a scalar wave governed by the Schrödinger equation (2.4). After using Eqs. (2.29) and (2.30), one obtains the corresponding Weyl symbol of the dispersion operator in Eq. (2.21):

$$D(x,p) = p_0 - \frac{\mathbf{p}^2}{2m} - V(x). \tag{2.50}$$

Then, the first ELE is given by Eq. (2.48a) so that $\{\{p_0 - \mathbf{p}^2/(2m) - V(x), W_\Psi\}\} = 0$. Upon substituting property (A.12a) of the sine bracket and integrating over the temporal momentum $p_0$, one obtains

$$\frac{\partial}{\partial t} F_\Psi(t,\mathbf{x},\mathbf{p}) + \frac{\mathbf{p}}{m}\cdot\boldsymbol{\nabla} F_\Psi(t,\mathbf{x},\mathbf{p}) - 2V(t,\mathbf{x}) \sin\left(\frac{1}{2}\frac{\overleftarrow{\partial}}{\partial \mathbf{x}}\cdot\frac{\overrightarrow{\partial}}{\partial \mathbf{p}}\right) F_\Psi(t,\mathbf{x},\mathbf{p}) = 0, \tag{2.51}$$

where $F_\Psi(t,\mathbf{x},\mathbf{p}) \doteq \int \mathrm{d}p_0\, W_\Psi(x,p)$ is the time-dependent Wigner function in the six-dimensional phase space. Equation (2.51) is known as *Moyal's equation*, or the *Quantum Liouville equation*. Integrating Eq. (2.51) over the six-dimensional phase space gives

$$\frac{\mathrm{d}}{\mathrm{d}t}\int \mathrm{d}^3\mathbf{x}\,\mathrm{d}^3\mathbf{p}\, F_\Psi(t,\mathbf{x},\mathbf{p}) = 0, \tag{2.52}$$

which is the conservation of the wave action [Eq. (2.24)] written in phase-space variables.

## 2.5 Conclusions

This Chapter presented a general overview of variational formulations commonly used in wave dynamics. The variational formulations based on the physical-space, abstract, and phase-space representations were introduced. Also, a brief introduction to the Weyl symbol calculus was given. A final remark of this Chapter is the following. The advantage of writing the action functional $\mathcal{S}$ in the phase-space representation [Eq. (2.42)] is that it simplifies the asymptotic analysis of wave equations. As will be shown in this dissertation, the phase-space formulation is convenient for obtaining systematic derivations of reduced wave theories. Some of the basic methodologies that will be used in this dissertation are presented in Chapter 3, where I shall derive the geometrical-optics approximation of the wave dynamics from the phase-space representation of the action functional.

# Chapter 3

# Geometrical-optics approximation for scalar waves

This Chapter presents a brief introduction to the theory of geometrical optics (GO). The main assumptions underlying the GO approximation are first stated. Then, starting from the phase-space representation of the action functional, I derive the GO equations for linear scalar waves. For completeness, both eikonal waves and incoherent waves are treated. A brief comment is also given on how to extend GO in order to include diffraction effects.

## 3.1 Introduction

### 3.1.1 Motivation

In wave physics, there is only a small class of problems for which the wave dynamics can be exactly obtained. The simplest problem is the case of linear waves in homogeneous media. Here the traditional method for studying these waves consists of applying the well-known Fourier transform to the governing equations. The advantage of the Fourier transform is that it converts the original PDE problem into a simple algebraic problem. This method allows one to identify the wave dispersion relation and to obtain the corresponding natural oscillating wave modes.

However, for waves propagating in temporally and spatially varying media, the situation quickly becomes more complicated. In these cases, the Fourier transform is generally not very useful, as it converts the original PDE problem into a perhaps more complicated integral equation. In some ideal situations, one can still obtain exact solutions to some wave equations; for example, a plasma wave propagating in a parabolic density profile. However, for the vast majority of inhomogeneous linear wave equations, it is nearly impossible to obtain exact solutions. In order to describe waves in these non-ideal situations, approximate reduced wave models have been developed.

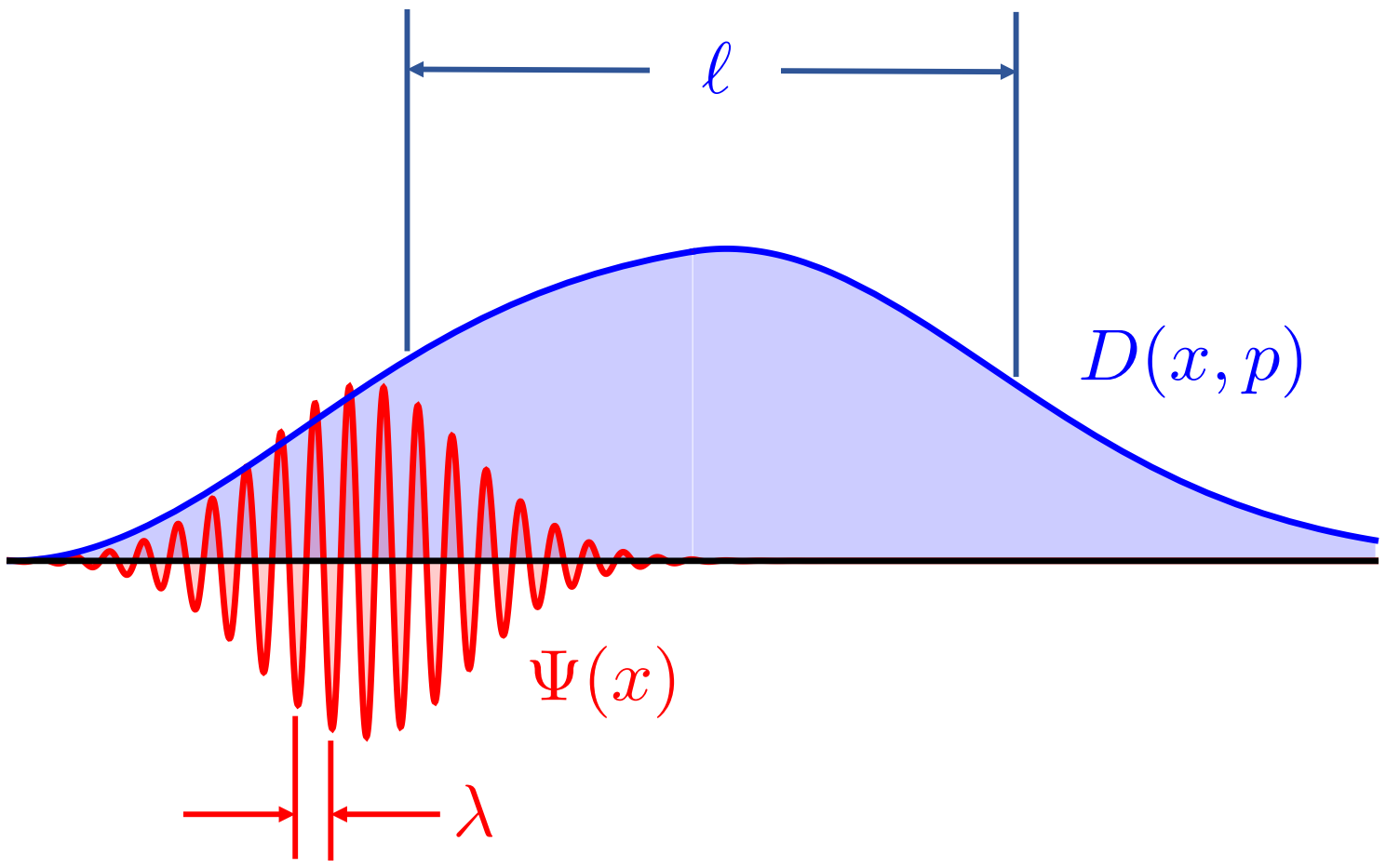

Figure 3.1: One-dimensional schematic of a wave (red) propagating in a medium (blue). GO describes the propagation of waves in the limit of small wavelengths; that is, when the characteristic length scale $\ell$ of the medium is much larger than the characteristic wavelength $\lambda$ of the wave.

Among the various reduced models for waves (Whitham, 2011; Tracy *et al.*, 2014), the GO approximation is perhaps the most well known.[1] GO describes the propagation of waves in the limit of small wavelengths; that is, when the medium varies slowly compared to the period and wavelength of the wave (see Fig. 3.1). This approximation has been widely used in many contexts ranging from quantum dynamics to electromagnetic, acoustic, and gravitational phenomena (Whitham, 1965a; Isaacson, 1968; Dodin and Fisch, 2012).

This Chapter presents a brief introduction to the theory of GO. The main goal is to obtain the GO approximation from the variational phase-space description of waves. In a broader perspective, the approach presented in this Chapter will provide a general overview of the basic concepts and methodologies that will be used throughout this dissertation.

### 3.1.2 Overview

This Chapter is organized as follows. In Sec. 3.2, the main assumptions underlying the GO approximation are presented. Then, starting from the phase-space representation of the action functional, I derive the GO equations for linear scalar waves propagating in a nondissipative medium. In Sec. 3.3, I discuss the dynamics of eikonal waves. In Sec. 3.4, I apply the GO approximation to describe incoherent waves. Finally, Sec. 3.5 presents the conclusions and final remarks.

[1] In mathematical physics, the GO approximation is also known as the JWKB, or WKB, approximation. The acronym stands for the names of the main developers of the theory: Jeffreys, Wentzel, Kramers, and Brillouin.

## 3.2 Basic equations

In this Chapter, I shall only consider scalar complex-valued waves $\Psi(x)$. (The case of multi-component waves will be treated in Chapter 4.) For scalar waves, the action in the phase-space representation [Eq. (2.42)] is given by

$$\mathcal{S} = \int \mathrm{d}^4x \, \mathrm{d}^4p \, D(x,p) W_\Psi(x,p), \tag{3.1}$$

where $D(x,p)$ is the Weyl symbol of the wave dispersion operator and $W_\Psi(x,p)$ is the Wigner function (2.32). For scalar waves, both $D(x,p)$ and $W_\Psi(x,p)$ are scalar functions of phase space.

We are going to consider waves propagating in a slowly varying medium. Specifically, I introduce the dimensionless parameter

$$\epsilon \doteq \max\left\{ \frac{1}{\omega T}, \frac{1}{|\mathbf{k}|\ell} \right\} \ll 1, \tag{3.2}$$

which is also known as the GO parameter. Here $T$ and $\ell$ are the characteristic temporal scale and length scale of the medium in which the wave propagates. Also, $\omega$ and $\mathbf{k}$ are the characteristic frequency and wavevector of the wave, respectively. The assumption in Eq. (3.2) considers that the characteristic temporal and spatial scales of the background medium are large with respect to the wave period and wavelength, respectively (see Fig. 3.1).

## 3.3 Eikonal waves

In this Section, I discuss coherent scalar waves, i.e., waves that have a well-defined frequency and wavevector. Let us express a scalar complex-valued wave in the polar (or Madelung) form

$$\Psi(x) = a(x) \, e^{i\theta(x)}, \tag{3.3}$$

where $a(x)$ and $\theta(x)$ are real functions. Let us assume that the phase $\theta(x)$ is fast compared to the slowly varying wave envelope $a(x)$ and to inhomogeneities in the background medium. Also,

$$\omega(x) \doteq -\partial_t \theta, \qquad \mathbf{k}(x) \doteq \boldsymbol{\nabla}\theta \tag{3.4}$$

are the local wave frequency and the wavevector, respectively. For convenience, let us introduce the four-wavevector

$$k_\mu(x) \doteq -\partial_\mu \theta(x) = (\omega, -\mathbf{k}) \tag{3.5}$$

or, in the contravariant representation, $k^\mu(x) = (\omega, \mathbf{k})$.

Upon inserting Eq. (3.3) into the Wigner function (2.32), one obtains

$$W_\Psi(x,p) = \frac{1}{(2\pi)^4} \int \mathrm{d}^4 s \, a(x+s/2) \, a(x-s/2) \, e^{ip\cdot s + i\theta(x+s/2) - i\theta(x-s/2)}. \tag{3.6}$$

I then use the ordering $|\partial a(x)| \sim |\partial k_\mu(x)| \sim \mathcal{O}(\epsilon)$.[2] Hence, one obtains

$$\begin{aligned} \theta(x+s/2) - \theta(x-s/2) &= s^\mu \partial_\mu \theta + \frac{\epsilon^2}{24} s^\mu s^\nu s^\alpha \partial^3_{\mu\nu\alpha}\theta + \cdots \\ &= -s\cdot k - \frac{\epsilon^2}{24} s^\nu s^\alpha \partial^2_{\mu\nu}(s\cdot k) + \cdots , \end{aligned} \tag{3.7}$$

where I substituted Eq. (3.5) and added the $\epsilon$ parameter to conveniently follow the ordering used. After inserting this result into Eq. (3.6) and asymptotically evaluating the integral, one obtains

$$\begin{aligned} W_\Psi(x,p) = & \frac{1}{(2\pi)^4} \int \mathrm{d}^4 s \left( a + \frac{\epsilon}{2} s\cdot\partial a + \frac{\epsilon^2}{8} s\cdot(\partial^2 a)\cdot s + \cdots \right) \left( a - \frac{\epsilon}{2} s\cdot\partial a + \frac{\epsilon^2}{8} s\cdot(\partial^2 a)\cdot s + \cdots \right) \\ & \times \exp\left( i(p-k)\cdot s - \frac{i\epsilon^2}{24} s\cdot\partial^2(s\cdot k)\cdot s + \cdots \right) \\ = & \frac{1}{(2\pi)^4} \int \mathrm{d}^4 s \left( a^2 - \frac{\epsilon^2}{4}(s\cdot\partial a)^2 + \frac{\epsilon^2}{4} a s\cdot(\partial^2 a)\cdot s - \frac{i\epsilon^2 a^2}{24} s\cdot\partial^2(s\cdot k)\cdot s \right) e^{i(p-k)\cdot s} + \mathcal{O}(\epsilon^3) \\ = & a^2(x)\delta^4(p-k) + \frac{\epsilon^2}{4}\left(\frac{\partial a}{\partial x^\mu}\right)\left(\frac{\partial a}{\partial x^\nu}\right)\frac{\partial^2\delta^4(p-k)}{\partial p_\mu \partial p_\nu} - \frac{\epsilon^2}{4} a \left(\frac{\partial^2 a}{\partial x^\mu \partial x^\nu}\right)\frac{\partial^2\delta^4(p-k)}{\partial p_\mu \partial p_\nu} \\ & + \frac{\epsilon^2 a^2}{24}\left(\frac{\partial^2 k_\mu}{\partial x^\nu \partial x^\alpha}\right)\frac{\partial^3\delta^4(p-k)}{\partial p_\mu \partial p_\nu \partial p_\alpha} + \mathcal{O}(\epsilon^3). \end{aligned} \tag{3.8}$$

In the following sections, I shall briefly discuss how different truncations of the Wigner function $W_\Psi(x,p)$ can lead to different physical models.

### 3.3.1 Geometrical optics approximation

In the limit of small $\epsilon$, the Wigner function can be approximated as

$$W_\Psi(x,p) = a^2(x)\delta^4(p-k) + \mathcal{O}(\epsilon^2). \tag{3.9}$$

[2] Variations of the four-wavevector $k_\mu(x)$ are related to inhomogeneities of the background medium in which the wave propagates. If the medium is homogeneous, then $\partial k_\mu = 0$. Similarly, since the width of the wave envelope is considered to be much larger than the wavelength, then $|\partial a| \sim \mathcal{O}(\epsilon)$.

This constitutes the so-called GO approximation. In phase space, the wave is located inside the support of $a^2(x)$ in spacetime,[3] and its four-frequency is $k_\mu(x) = -\partial_\mu \theta$. Note that in the case when $a^2(x)$ has compact support [in other words, when $a^2(x)$ is of finite extent], the GO approximation violates the Heisenberg uncertainty principle (see the discussion in Sec. 2.3.2). This occurs because the Wigner function in the GO approximation assigns a unique four-momentum $k_\mu(x)$ to the wave while still localizing to a certain degree the wave within the support of $a^2(x)$. For this reason, higher-order effects such as diffraction are lost in this approximation.

Upon inserting Eq. (3.9) into the phase-space action (3.1) and integrating along the momentum coordinate, one obtains the reduced GO wave action

$$\mathcal{S}_{\mathrm{GO}} = \int \mathrm{d}^4 x \, a^2(x) D(x, \underbrace{-\partial\theta}_{k(x)}), \tag{3.10}$$

where I dropped correction terms of $\mathcal{O}(\epsilon^2)$. Equation (3.10) is the general GO action for linear dissipationless scalar waves. In the following, I shall continue the discussion of GO by focusing on waves, whose dispersion symbol is given in the symplectic form $D(x,p) = p_0 - H(t, \mathbf{x}, \mathbf{p})$. Afterwards, I shall discuss waves with a more general dispersion symbol $D(x,p)$.

### 3.3.2 Scalar waves with $D(x,p) = p_0 - H(t, \mathbf{x}, \mathbf{p})$

#### Continuous wave model

Here I shall discuss the dynamics of scalar dissipationless waves, whose dispersion Weyl symbol is in the symplectic form; i.e.,

$$D(x,p) = p_0 - H(t, \mathbf{x}, \mathbf{p}), \tag{3.11}$$

where the symbol $H(t, \mathbf{x}, \mathbf{p})$ is called the Hamiltonian of the system and does not depend on the temporal momentum coordinate $p_0$. In this case, the GO action functional (3.10) is given by

$$\mathcal{S}_{\mathrm{GO}} = \int \mathrm{d}^4 x \, \mathcal{I} \, [\underbrace{\omega}_{-\partial_t \theta} - H(t, \mathbf{x}, \underbrace{\mathbf{k}}_{\boldsymbol{\nabla}\theta})], \tag{3.12}$$

where $\mathcal{I}(x) \doteq a^2(x)$ is Whitham's wave action. One may recognize Eq. (3.12) as the Hayes (1973) representation of the Lagrangian density of a GO wave .

[3]Suppose that $f : X \to \mathbb{R}$ is a real-valued function whose domain is an arbitrary set $X$. The support of $f$, written supp$(f)$, is the set of points in $X$ where $f$ is non-zero:

$$\mathrm{supp}(f) = \{x \in X | f(x) \neq 0\}.$$

Treating $\mathcal{I}(x)$ and $\theta(x)$ as independent variables yields the following ELEs:

$$\delta\theta : \quad \partial_t \mathcal{I} + \boldsymbol{\nabla} \cdot (\mathcal{I}\mathbf{v}) = 0, \tag{3.13a}$$

$$\delta\mathcal{I} : \quad \partial_t \theta + H(t, \mathbf{x}, \boldsymbol{\nabla}\theta) = 0, \tag{3.13b}$$

where

$$\mathbf{v}(t,\mathbf{x}) \doteq \left( \frac{\partial H(t,\mathbf{x},\mathbf{p})}{\partial \mathbf{p}} \right)_{\mathbf{p}=\mathbf{k}(t,\mathbf{x})} \tag{3.14}$$

is the wave group velocity. As usual, Eq. (3.13a) is a continuity equation known as the action conservation theorem (ACT). Also, Eq. (3.13b) is the Hamilton–Jacobi representation of the local dispersion relation $\omega = H(t, \mathbf{x}, \boldsymbol{\nabla}\theta)$. This equation determines a *dispersion manifold* for the wave dynamics (Tracy *et al.*, 2014). When calculating the gradient of Eq. (3.13b) and applying the chain rule, one obtains

$$\partial_t \boldsymbol{\nabla}\theta + \left( \frac{\partial H(t,\mathbf{x},\mathbf{p})}{\partial \mathbf{x}} \right)_{\mathbf{p}=\mathbf{k}} + \left( \frac{\partial H(t,\mathbf{x},\mathbf{p})}{\partial p_i} \right)_{\mathbf{p}=\mathbf{k}} \frac{\partial}{\partial \mathbf{x}} \frac{\partial \theta}{\partial x^i} = 0. \tag{3.15}$$

Using the consistency relation $\partial^2_{ij}\theta = \partial^2_{ji}\theta$, the equation can be written as follows:

$$\left( \frac{\partial}{\partial t} + \mathbf{v} \cdot \boldsymbol{\nabla} \right) \mathbf{k} = - \left( \frac{\partial H(t,\mathbf{x},\mathbf{p})}{\partial \mathbf{x}} \right)_{\mathbf{p}=\mathbf{k}}. \tag{3.16}$$

Hence, Eq. (3.16) represents a conservation equation of the wavevector $\mathbf{k}$; i.e., this is the wave momentum conservation equation. Note that the change in the wave momentum is determined by the spatial gradients of the Hamiltonian, i.e., by gradients of the underlying medium. Upon following the analogy with fluid mechanics, one denotes Eqs. (3.13a) and (3.16) as the "continuous fluid model" for wave dynamics.

> ***Example.*** — As in Chapter 2, let us consider a wave governed by the Schrödinger equation (2.4). The Weyl symbol of the dispersion operator is $D(x,p) = p_0 - \mathbf{p}^2/(2m) - V(x)$. Upon following the calculations above, one obtains the GO fluid equations
>
> $$\partial_t \mathcal{I} + \boldsymbol{\nabla} \cdot (\mathcal{I}\mathbf{v}) = 0, \tag{3.17a}$$
>
> $$m \left( \partial_t + \mathbf{v} \cdot \boldsymbol{\nabla} \right) \mathbf{v} = -\boldsymbol{\nabla} V, \tag{3.17b}$$
>
> where $\mathbf{v}(t,\mathbf{x}) = \boldsymbol{\nabla}\theta/m$ is the wave group velocity.

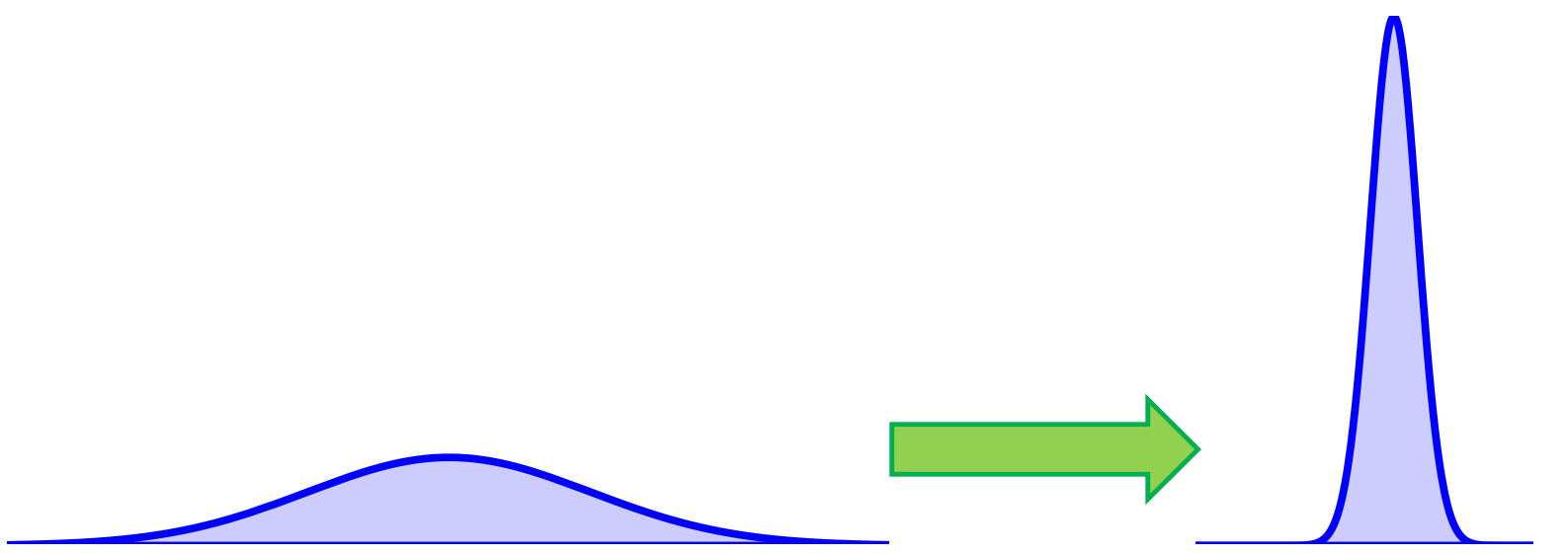

Figure 3.2: In the point-particle model, the wave envelope $\mathcal{I}(t, \mathbf{x})$ is localized around a domain nearby the spatial coordinate $\mathbf{X}(t)$.

### Point-particle model

The ray equations corresponding to the GO equations (3.13a) and (3.16) can be obtained as the point-particle limit of the wave envelope (see Fig. 3.2). In this limit, $\mathcal{I}(t, \mathbf{x})$ can be approximated with a Dirac delta function so that

$$\mathcal{I}(t, \mathbf{x}) = \mathcal{I}_0 \, \delta^3(\mathbf{x} - \mathbf{X}(t)). \tag{3.18}$$

Here $\mathcal{I}_0$ denotes the total action, which is conserved according to Eq. (3.13a). The value of $\mathcal{I}_0$ is not essential below, so I adopt $\mathcal{I}_0 = 1$ for brevity.

In this representation, the wave packet is located at the position $\mathbf{X}(t)$ in space, and the independent parameter is the physical time $t$. Inserting Eq. (3.18) into Eq. (3.12) leads to

$$\mathcal{S}_{\mathrm{GO}} = \int \mathrm{d}^4 x \, \delta^3(\mathbf{x} - \mathbf{X}(t)) \left[ -\partial_t \theta - H(t, \mathbf{x}, \boldsymbol{\nabla}\theta) \right]. \tag{3.19}$$

Calculating the first term in the action (3.19) gives the following (Ruiz and Dodin, 2015c):

$$\begin{aligned}
-\int \mathrm{d}^4 x \, \mathcal{I} \, \partial_t \theta &= -\int \mathrm{d}t \, \mathrm{d}^3 x \, \delta^3(\mathbf{x} - \mathbf{X}(t)) \, \partial_t \theta(t, \mathbf{x}) \\
&= \int \mathrm{d}t \, \mathrm{d}^3 x \, \theta(t, \mathbf{x}) \left[ \partial_t \delta^3(\mathbf{x} - \mathbf{X}(t)) \right] \\
&= -\int \mathrm{d}t \, \mathrm{d}^3 x \, \theta(t, \mathbf{x}) \left[ \dot{\mathbf{X}}(t) \cdot \boldsymbol{\nabla} \delta^3(\mathbf{x} - \mathbf{X}(t)) \right] \\
&= \int \mathrm{d}t \left( \dot{\mathbf{X}}(t) \cdot \int \mathrm{d}^3 x \, \boldsymbol{\nabla}\theta(t, \mathbf{x}) \, \delta^3(\mathbf{x} - \mathbf{X}(t)) \right) \\
&= \int \mathrm{d}t \, \mathbf{P}(t) \cdot \dot{\mathbf{X}}(t),
\end{aligned} \tag{3.20}$$

where $\mathbf{P}(t) \doteq \boldsymbol{\nabla}\theta(t, \mathbf{X}(t))$. Similarly, the second term in Eq. (3.19) gives

$$-\int \mathrm{d}^4 x\, \delta^3(\mathbf{x} - \mathbf{X}(t)) H(t, \mathbf{x}, \boldsymbol{\nabla}\theta) = -\int \mathrm{d}t\, H(t, \mathbf{X}(t), \mathbf{P}(t)). \tag{3.21}$$

Thus, the point-particle action is expressed as

$$\mathcal{S}_{\mathrm{GO}} = \int \mathrm{d}t\, [\, \mathbf{P}(t) \cdot \dot{\mathbf{X}}(t) - H(t, \mathbf{X}, \mathbf{P})\, ]. \tag{3.22}$$

One may recognize this action as the classical Hamiltonian variational principle, where $\mathbf{X}(t)$ and $\mathbf{P}(t)$ serve as the canonical coordinate and the canonical momentum, respectively. The first term in Eq. (3.22) represents the symplectic part of the phase-space action while the second term is the Hamiltonian part. For this reason, I refer to dispersion symbols in the form $D(x, p) = p_0 - H(t, \mathbf{x}, \mathbf{p})$ as being in the symplectic form.

Treating $\mathbf{X}(t)$ and $\mathbf{P}(t)$ as independent variables leads to ELEs matching Hamilton's equations in the canonical form:

$$\delta\mathbf{P}: \quad \frac{\mathrm{d}\mathbf{X}}{\mathrm{d}t} = \frac{\partial H}{\partial \mathbf{P}}, \tag{3.23a}$$

$$\delta\mathbf{X}: \quad \frac{\mathrm{d}\mathbf{P}}{\mathrm{d}t} = -\frac{\partial H}{\partial \mathbf{X}}. \tag{3.23b}$$

These equations serve as ray equations for Eqs. (3.13a) and (3.16).

***Example.*** — Let us consider the case of a scalar wave governed by the Schrödinger equation (2.4). The corresponding Weyl symbol of the dispersion operator is given by Eq. (2.50): $D(x, p) = p_0 - H(t, \mathbf{x}, \mathbf{p})$, where $H(t, \mathbf{x}, \mathbf{p}) = \mathbf{p}^2/(2m) - V(t, \mathbf{x})$. Upon following the discussion above, one obtains the corresponding GO ray equations

$$\delta\mathbf{P}: \quad \frac{\mathrm{d}\mathbf{X}}{\mathrm{d}t} = \frac{\mathbf{P}}{m}, \tag{3.24a}$$

$$\delta\mathbf{X}: \quad \frac{\mathrm{d}\mathbf{P}}{\mathrm{d}t} = -\boldsymbol{\nabla} V(t, \mathbf{x}). \tag{3.24b}$$

These equations coincide with Hamilton's equations for a particle in a potential field.

### 3.3.3 Scalar waves with general $D(x, p)$

#### Continuous wave model

Here I shall briefly discuss the case of waves whose dispersion operator is in the general form $D(x, p)$. (A more detailed discussion is presented in Chapter 4.) The GO action functional for a general scalar waves is

given by

$$\mathcal{S}_{\rm GO} = \int \mathrm{d}^4x \, a^2(x) D(x, \underbrace{-\partial\theta}_{k(x)}), \tag{3.10 revisited}$$

where I dropped the $\mathcal{O}(\epsilon^2)$ correction terms. Treating $a^2(x)$ and $\theta(x)$ as independent variables yields the following ELEs:

$$\delta\theta : \quad \partial_\mu \mathcal{J}^\mu = 0, \tag{3.25a}$$

$$\delta a^2 : \quad D\big(x, k(x)\big) = 0, \tag{3.25b}$$

where $\mathcal{J}^\mu(x) \doteq a^2(x) v^\mu(x)$ is the action flux density and

$$v^\mu(x) \doteq \left( \frac{\partial D(x,p)}{\partial p_\mu} \right)_{p=k(x)} \tag{3.26}$$

is the GO group four-velocity. These are the well-known GO equations. Similar to Eqs. (3.13), Eq. (3.25a) represents the action conservation theorem (ACT). In terms of components, Eq. (3.25a) can be written as

$$\partial_t \mathcal{I} + \boldsymbol{\nabla} \cdot (\mathcal{I} \mathbf{v}) = 0, \tag{3.27}$$

where $\mathcal{I}(x)$ serves as Whitham's wave action density (Whitham, 2011)[4]

$$\mathcal{I}(x) \doteq a^2(x) \left( \frac{\partial D(x,p)}{\partial p_0} \right)_{p=k(x)} \tag{3.28}$$

and the flow velocity $\mathbf{v}(x)$ is given by

$$\mathbf{v}(x) \doteq - \left( \frac{\partial D(x,p)}{\partial p_0} \right)^{-1}_{p=k(x)} \left( \frac{\partial D(x,p)}{\partial \mathbf{p}} \right)_{p=k(x)} . \tag{3.29}$$

Also, Eq. (3.25b) serves as a local dispersion relation. As before, upon calculating the spacetime derivative of Eq. (3.25b) and using the consistency relation, one finds the four-momentum conservation equation

$$\left( v^\nu \frac{\partial}{\partial x^\nu} \right) k_\mu = - \left( \frac{\partial D(x,p)}{\partial x^\mu} \right)_{p=k(x)} . \tag{3.30}$$

Hence, Eqs. (3.25a) and (3.30) constitute the continuous fluid model for wave dynamics.[5]

[4]The definition (3.28) of the wave action is consistent with that given in the previous section since $\partial D / \partial p_0 = 1$ for dispersion symbols in the symplectic form $D(x,p) = p_0 - H(t, \mathbf{x}, \mathbf{p})$.

[5]For an in-depth discussion of these equations, see, e.g., Dodin and Fisch (2012).

It is to be noted that, in principle, Eq. (3.25b) can be inverted in order to write the frequency $\omega(x)$ as some function that depends on the other phase-space coordinates:

$$\omega = H(t, \mathbf{x}, \mathbf{k}). \tag{3.31}$$

This determines a dispersion manifold (Tracy *et al.*, 2014). In general, there can be multiple dispersion branches. In that case, I assume that only one branch is excited while others have sufficiently different $\omega$ and thus can be neglected.[6]

Hence, using the phase-space approach, I have reproduced Whitham's equations for GO. These equations are valid for waves with arbitrary scalar dispersion symbols. For a more detailed discussion on the subject, the reader is encouraged to read the book by Tracy *et al.* (2014).

### Point-particle model

As in Sec. 3.3.2, it is also possible to obtain a variational point-particle model that generates the ray equations corresponding to Eqs. (3.25a) and (3.30). However, this will require an additional mathematical tool (the Feynman reparameterization) that is introduced in Chapter 4. For completeness, here I only state the main result; the covariant point-particle action is given by

$$\mathcal{S}_{\text{GO}} = \int \mathrm{d}\tau \, [\, P(\tau) \cdot \dot{X}(\tau) + D(X, P) \,], \tag{3.32}$$

where $X \cdot P = X^0 P^0 - \mathbf{X} \cdot \mathbf{P}$. Here $X^\mu(\tau) = (X^0, \mathbf{X})$ represents the wave four-position, and $P^\mu(\tau) = (P^0, \mathbf{P})$ serves as the wave canonical four-momentum. Both variables depend on $\tau$, which serves as an external variable that parameterizes the ray trajectory.

Treating $X^\mu(\tau)$ and $P_\mu(\tau)$ as independent leads to Hamilton's equations in covariant form:

$$\delta P_\mu : \quad \frac{\mathrm{d}X^\mu}{\mathrm{d}\tau} = -\frac{\partial D}{\partial P_\mu}, \tag{3.33a}$$

$$\delta X^\mu : \quad \frac{\mathrm{d}P_\mu}{\mathrm{d}\tau} = \frac{\partial D}{\partial X^\mu}. \tag{3.33b}$$

These are the well-known ray equations for GO [see, e.g., Tracy *et al.* (2014)], which are also supplemented by the dispersion relation

$$D(X, P) = 0. \tag{3.34}$$

[6]However, note that the effect of non-excited "passive" branches is not negligible when $\mathcal{O}(\epsilon)$ corrections are retained in the action functional $\mathcal{S}$, as discussed in Chapter 4.

In terms of components, Eqs. (3.33) can also be written as

$$\begin{aligned} \frac{\mathrm{d}X^0}{\mathrm{d}\tau} &= -\frac{\partial D}{\partial P_0}, & \frac{\mathrm{d}\mathbf{X}}{\mathrm{d}\tau} &= \frac{\partial D}{\partial \mathbf{P}}, \\ \frac{\mathrm{d}P_0}{\mathrm{d}\tau} &= \frac{\partial D}{\partial X^0}, & \frac{\mathrm{d}\mathbf{P}}{\mathrm{d}\tau} &= -\frac{\partial D}{\partial \mathbf{X}}. \end{aligned} \tag{3.35}$$

Note that the dispersion symbol $D(x,p)$ acts as a "Hamiltonian" in Eqs. (3.35).

***Example.*** — The Klein–Gordon equation[7]

$$(-2m)^{-1}[(i\partial_t - qV)^2 - (-i\boldsymbol{\nabla} - q\mathbf{A})^2 - m^2]\Psi(x) = 0 \tag{3.36}$$

describes a relativistic spinless particle interacting with an electromagnetic field, whose four-potential is given by $A^\mu(x) = (V, \mathbf{A})$. In this case, the corresponding Weyl symbol of the dispersion operator is $D(x,p) = (-2m)^{-1}(p_\mu - qA_\mu)(p^\mu - qA^\mu) - m^2$. After using Eqs. (3.33), one obtains the GO ray equations

$$\delta P_\mu : \quad \frac{\mathrm{d}X^\mu}{\mathrm{d}\tau} = \frac{\Pi^\mu}{m}, \tag{3.37a}$$

$$\delta X^\mu : \quad \frac{\mathrm{d}P_\mu}{\mathrm{d}\tau} = \frac{q}{m}\frac{\partial A_\nu}{\partial X^\mu}\Pi^\nu, \tag{3.37b}$$

where $\Pi_\mu(\tau) \doteq P_\mu(\tau) - qA_\mu\big(X(\tau)\big)$ is the kinetic four-momentum. From Eq. (3.26), one identifies $V^\mu(\tau) \doteq \Pi^\mu/m$ as the the particle four-velocity. Then, it is straightforward to show that Eqs. (3.37) lead to the well-known classical equations of relativistic mechanics:

$$\frac{\mathrm{d}X^\mu}{\mathrm{d}\tau} = V^\mu, \qquad m\frac{\mathrm{d}V^\mu}{\mathrm{d}\tau} = qF^{\mu\nu}V_\nu, \tag{3.38}$$

where $F_{\mu\nu}(x) \doteq \partial_\mu A_\nu(x) - \partial_\nu A_\mu(x)$ is the electromagnetic tensor. These equations are supplemented by Eq. (3.34), or more explicitly,

$$\Pi_\mu \Pi^\mu = m^2, \tag{3.39}$$

which is the mass-shell condition of classical mechanics.

[7] The $(-2m)^{-1}$ factor is inserted for convenience and does not alter the wave dynamics.

### 3.3.4 Higher-order effects: Diffraction

Here I show how the phase-space representation of the action allows one to systematically obtain higher-order effects that go beyond GO. In the case of scalar waves, diffraction effects are the next-order corrections to the GO model. Returning to Eq. (3.8), one inserts the asymptotically evaluated Wigner function into Eq. (3.1) and then integrates along the momentum coordinate. One then obtains the reduced action

$$\begin{aligned}\mathcal{S} = \int \mathrm{d}^4 x \bigg[ a^2 D(x, k(x)) &+ \frac{\epsilon^2}{4} \left( \frac{\partial a}{\partial x^\mu} \frac{\partial a}{\partial x^\nu} - a \frac{\partial^2 a}{\partial x^\mu \partial x^\nu} \right) \left( \frac{\partial^2 D(x,p)}{\partial p_\mu \partial p_\nu} \right)_{p=k(x)} \\ &- \frac{\epsilon^2 a^2}{24} \frac{\partial^2 k_\mu}{\partial x^\nu \partial x^\alpha} \left( \frac{\partial^3 D(x,p)}{\partial p_\mu \partial p_\nu \partial p_\alpha} \right)_{p=k(x)} \bigg] + \mathcal{O}(\epsilon^3). \end{aligned} \tag{3.40}$$

As before, Eq. (3.40) is parameterized by two functions: the wave amplitude $a(x)$ and the eikonal phase $\theta(x)$. Varying the action (3.40) with respect to the eikonal phase $\theta(x)$ gives a corrected action conservation theorem

$$\begin{aligned}\delta\theta: \quad \frac{\partial}{\partial x^\mu} \bigg\{ a^2 \left( \frac{\partial D(x,p)}{\partial p_\mu} \right)_{p=k(x)} &+ \frac{\epsilon^2}{4} \left( \frac{\partial a}{\partial x^\nu} \frac{\partial a}{\partial x^\alpha} - a \frac{\partial^2 a}{\partial x^\nu \partial x^\alpha} \right) \left( \frac{\partial^3 D(x,p)}{\partial p_\mu \partial p_\nu \partial p_\alpha} \right)_{p=k(x)} \\ - \frac{\epsilon^2}{24} \frac{\partial^2}{\partial x^\nu \partial x^\alpha} \left[ a^2 \left( \frac{\partial^3 D(x,p)}{\partial p_\mu \partial p_\nu \partial p_\alpha} \right)_{p=k(x)} \right] &- \frac{\epsilon^2}{24} a^2 \frac{\partial^2 k_\nu}{\partial x^\alpha \partial x^\beta} \left( \frac{\partial^4 D(x,p)}{\partial p_\mu \partial p_\nu \partial p_\alpha \partial p_\beta} \right)_{p=k(x)} \bigg\} = 0. \end{aligned} \tag{3.41}$$

As in the GO approximation, Eq. (3.41) is written in conservative form, which is not surprising since the action (3.40) only depends on derivatives of the eikonal phase. The first term inside the four-divergence is the action flux density introduced in Sec. 3.3.3. The next terms in Eq. (3.41) represent corrections of $\mathcal{O}(\epsilon^2)$. Varying the action with respect to the wave amplitude $a(x)$ leads to

$$\begin{aligned}\delta a: \quad a D(x, k(x)) = \frac{\epsilon^2}{4} \frac{\partial}{\partial x^\mu} \left[ \left( \frac{\partial^2 D(x,p)}{\partial p_\mu \partial p_\nu} \right)_{p=k(x)} \frac{\partial a}{\partial x^\nu} \right] &+ \frac{\epsilon^2}{8} \frac{\partial^2 a}{\partial x^\mu \partial x^\nu} \left( \frac{\partial^2 D(x,p)}{\partial p_\mu \partial p_\nu} \right)_{p=k(x)} \\ + \frac{\epsilon^2}{8} \frac{\partial^2}{\partial x^\mu \partial x^\nu} \left[ a \left( \frac{\partial^2 D(x,p)}{\partial p_\mu \partial p_\nu} \right)_{p=k(x)} \right] &+ \frac{\epsilon^2 a}{24} \frac{\partial^2 k_\mu}{\partial x^\nu \partial x^\alpha} \left( \frac{\partial^3 D(x,p)}{\partial p_\mu \partial p_\nu \partial p_\alpha} \right)_{p=k(x)} . \end{aligned} \tag{3.42}$$

Equation (3.42) serves as a corrected dispersion relation. The left-hand side of Eq. (3.42) is the previously obtained GO dispersion relation [Eq. (3.25b)]. The right-hand side contains $\mathcal{O}(\epsilon^2)$ corrections which are due to diffraction.

***Example.*** — Let us consider again the example of the Schödinger equation. The Weyl symbol of the dispersion operator is given by $D(x,p) = p_0 - \mathbf{p}^2/(2m) - V(x)$. One writes the wave function in the polar form $\Psi(x) = a(x)e^{i\theta(x)}$. The resulting action is given by Eq. (3.40), so one obtains

$$\mathcal{S} = \int \mathrm{d}^4x \left[ a^2 \left( \omega - \frac{\mathbf{k}^2}{2m} - V(x) \right) - \frac{1}{2m} \left( \boldsymbol{\nabla} a \right)^2 \right], \tag{3.43}$$

where I omitted writing the GO parameter $\epsilon$. [This example can be readily generalized to include the interaction with magnetic fields (Ruiz and Dodin, 2015c).] As a reminder, $\omega(x) \doteq -\partial_t \theta(x)$ and $\mathbf{k}(x) \doteq \boldsymbol{\nabla}\theta(x)$. Since the dispersion operator for the Schrödinger particle is only quadratic in momentum, the action (3.43) is exact. Upon using Eq. (3.41), one obtains the action conservation theorem

$$\delta\theta : \quad \partial_t \mathcal{I} + \boldsymbol{\nabla} \cdot (\mathcal{I}\mathbf{v}) = 0, \tag{3.44}$$

where $\mathcal{I}(x) = a^2(x)$ is the wave action density and $\mathbf{v}(x) = \mathbf{k}(x)/m$ is the group velocity. Likewise, upon using Eq. (3.42), one obtains

$$\delta a : \quad \omega - \frac{\mathbf{k}^2}{2m} - V(x) = -\frac{1}{2\sqrt{\mathcal{I}}} \boldsymbol{\nabla}^2 \sqrt{\mathcal{I}}, \tag{3.45}$$

The left-hand side of this equation represents the GO dispersion relation for the particle dynamics. The term on the right-hand side is the quantum Bohm potential. Taking the derivative of Eq. (3.45) gives the momentum conservation equation

$$m(\partial_t + \mathbf{v} \cdot \boldsymbol{\nabla})\mathbf{v} = -\boldsymbol{\nabla} V + \boldsymbol{\nabla} \left( \frac{1}{2\sqrt{\mathcal{I}}} \boldsymbol{\nabla}^2 \sqrt{\mathcal{I}} \right), \tag{3.46}$$

where the last term is the gradient of the well-known quantum potential, or Bohm potential. Equations (3.44) and (3.46) coincide with Madelung's hydrodynamical equations for quantum mechanics (Madelung, 1927). Hence, Eq. (3.43) can be identified as the Madelung action.

## 3.4 Incoherent waves

In this Section, I shall discuss the dynamics of incoherent waves. It will be shown that, within the GO approximation, incoherent waves are described by the wave kinetic equation (WKE). Let a scalar incoherent wave be governed by the phase-space action (3.1). As shown in Sec. 2.4, the constrained variation of the phase-space action (3.1) leads to

$$\delta\mathcal{S} = \int \mathrm{d}^4x \, \mathrm{d}^4p \, [ -i\delta C_H (D \star W_\Psi - W_\Psi \star D) - \delta C_A (D \star W_\Psi + W_\Psi \star D) ]. \tag{3.47}$$

Now, let us approximate $\delta\mathcal{S}$ as follows. Assuming that the waves propagate in a slowly varying medium, I adopt the following estimates:

$$\begin{aligned} &|\partial_t D| \sim T^{-1} D, && |\partial_{\mathbf{x}} D| \sim \ell^{-1} D, && |\partial_t W_\Psi| \sim T^{-1} W_\Psi, && |\partial_{\mathbf{x}} W_\Psi| \sim \ell^{-1} W_\Psi, \\ &|\partial_{p_0} D| \sim \omega^{-1} D, && |\partial_{\mathbf{p}} D| \sim |\mathbf{k}|^{-1} D, && |\partial_{p_0} W_\Psi| \sim \omega^{-1} W_\Psi, && |\partial_{\mathbf{p}} W_\Psi| \sim |\mathbf{k}|^{-1} W_\Psi, \end{aligned} \tag{3.48}$$

where $(T, \ell, \omega, \mathbf{k})$ are the characteristic parameters introduced in Sec. 3.2. These estimates assume that both the background medium and the Wigner function $W_\Psi(x,p)$ vary slowly in time and space.

With the estimates above, then

$$\frac{\partial D}{\partial p} \cdot \frac{\partial W_\Psi}{\partial x} \lesssim \epsilon D W_\Psi, \qquad \frac{\partial D}{\partial x} \cdot \frac{\partial W_\Psi}{\partial p} \lesssim \epsilon D W_\Psi. \tag{3.49}$$

Then, from the definition of the Moyal product [Eq. (A.5)], one uses the above ordering to approximate the Moyal product by

$$D(x,p) \star W_\Psi(x,p) = D(x,p) W_\Psi(x,p) + \frac{i\epsilon}{2} \{D(x,p), W_\Psi(x,p)\} + \mathcal{O}(\epsilon^2), \tag{3.50}$$

where $\{\cdot,\cdot\}$ is the canonical Poisson brackets in the eight-dimensional phase space [Eq. (A.8)] and the dimensionless $\epsilon$ parameter was added only to keep track of the GO ordering. Then, upon substituting Eq. (3.50) into Eq. (3.47), one obtains

$$\delta\mathcal{S}_{\mathrm{GO}} = \int \mathrm{d}^4 x\, \mathrm{d}^4 p \, [\, \delta C_H \{D, W_\Psi\} - 2\delta C_A (D W_\Psi)\, ], \tag{3.51}$$

where I dropped $\mathcal{O}(\epsilon^2)$ terms and used the skew-symmetry of the Poisson bracket. In this case, replacing the Moyal product by its lowest-order approximation constitutes the GO approximation. Since $\delta C_H$ and $\delta C_A$ are arbitrary, the corresponding ELEs are given by

$$\delta C_H : \quad \{D(x,p), W_\Psi(x,p)\} = 0, \tag{3.52a}$$

$$\delta C_A : \quad D(x,p) W_\Psi(x,p) = 0. \tag{3.52b}$$

As it will be shown below, Eq. (3.52a) reduces to the WKE (McDonald and Kaufman, 1985) and is equivalent to the action conservation theorem in Eq. (3.25a). Also, Eq. (3.52b) gives the wave dispersion relation (3.25b). Interestingly, while the ACT and the dispersion relations are usually considered to be different PDEs, in the phase-space formulation they both emerge as differential equations of the Moyal type.

I now follow the procedure proposed by McDonald and Kaufman (1985) to derive the wave kinetic equation. In order to satisfy Eq. (3.52b), either $D(x,p)$ or $W_\Psi(x,p)$ must be zero. However, in the GO approximation $W_\Psi(x,p)$ represents the probability density of finding wave quanta at a specific location in the eight-dimensional phase space; thus, it cannot be identically zero. We also know from the eikonal theory presented in Sec. 3.3.3 that $D(x,p)=0$ is the GO dispersion relation. Hence, in order to satisfy Eq. (3.52b), one chooses

$$W_\Psi(x,p) = \delta\big(D(x,p)\big) F_\Psi(t,\mathbf{x},\mathbf{p}). \tag{3.53}$$

This parameterization of $W_\Psi(x,p)$ satisfies Eq. (3.52b) in the distribution sense. Physically, Eq. (3.53) means that wave quanta is localized on the dispersion manifold defined by $D(x,p)=0$. It will be shown below that the function $F_\Psi(t,\mathbf{x},\mathbf{p})$ is actually the wave action density in the six-dimensional phase space.

As mentioned previously, the relation $D(x,p)=0$ can be, in principle, inverted so that one can express the temporal momentum as $p_0 = H(t,\mathbf{x},\mathbf{p})$. However, as explained in Sec. 3.3.3, there can be multiple solutions to $D(x,p)$. In the case of multiple modes, each mode could be described by a different wave–action density function. In the GO approximation, these modes are decoupled, but at higher orders, the resonant modes can interact (see Chapter 4). For simplicity, I only consider the case of a single excited wave mode.

Upon substituting Eq. (3.53) into Eq. (3.52a), one obtains

$$\{D,\delta(D)\}F_\Psi + \delta(D)\{D,F_\Psi\} = 0. \tag{3.54}$$

From the skew-symmetry of the Poisson bracket, the first term gives $\{D,\delta(D)\} = \{D,D\}\delta'(D) = 0$. Upon substituting Eq. (A.8) and noting that $F_\Psi(t,\mathbf{x},\mathbf{p})$ does not depend on the temporal momentum, one can explicitly write the second term as

$$\delta(D)\left(\frac{\partial D}{\partial p_0}\frac{\partial F_\Psi}{\partial t} + \frac{\partial D}{\partial \mathbf{x}}\cdot\frac{\partial F_\Psi}{\partial \mathbf{p}} - \frac{\partial D}{\partial \mathbf{p}}\cdot\frac{\partial F_\Psi}{\partial \mathbf{x}}\right) = 0. \tag{3.55}$$

One then integrates this equation along the temporal momentum $p_0$ and uses

$$\delta(D) = \left(\frac{D(x,p)}{\partial p_0}\right)^{-1}_{p=p_*} \delta\big(p_0 - H(t,\mathbf{x},\mathbf{p})\big) \tag{3.56}$$

where $p_*^\mu \doteq \big(H(t,\mathbf{x},\mathbf{p}),\mathbf{p}\big)$ is the four-momentum satisfying $D(x,p_*)=0$. Hence, one obtains

$$\frac{\partial F_\Psi}{\partial t} + \left(-\frac{\partial D/\partial \mathbf{p}}{\partial D/\partial p_0}\right)_{p=p_*}\cdot\frac{\partial F_\Psi}{\partial \mathbf{x}} - \left(-\frac{\partial D/\partial \mathbf{x}}{\partial D/\partial p_0}\right)_{p=p_*}\cdot\frac{\partial F_\Psi}{\partial \mathbf{p}} = 0, \tag{3.57}$$

Let us further simplify Eq. (3.57) as follows. Calculating the spatial gradient of $D(x, p_*) = 0$ and applying the chain rule gives

$$\left(\frac{\partial D}{\partial p_0}\right)_{p=p_*} \frac{\partial H}{\partial \mathbf{x}} + \left(\frac{\partial D}{\partial \mathbf{x}}\right)_{p=p_*} = 0. \tag{3.58}$$

Hence, one obtains the following relation:

$$-\left(\frac{\partial D/\partial \mathbf{x}}{\partial D/\partial p_0}\right)_{p=p_*} = \frac{\partial H}{\partial \mathbf{x}}. \tag{3.59}$$

Similarly, one can also show that

$$-\left(\frac{\partial D/\partial \mathbf{p}}{\partial D/\partial p_0}\right)_{p=p_*} = \frac{\partial H}{\partial \mathbf{p}}. \tag{3.60}$$

Inserting Eqs. (3.59) and (3.60) into Eq. (3.57) leads to

$$\frac{\partial F_\Psi}{\partial t} + \frac{\partial H}{\partial \mathbf{p}} \cdot \frac{\partial F_\Psi}{\partial \mathbf{x}} - \frac{\partial H}{\partial \mathbf{x}} \cdot \frac{\partial F_\Psi}{\partial \mathbf{p}} = 0. \tag{3.61}$$

Thus, the dynamics of the wave action density $F_\Psi(t, \mathbf{x}, \mathbf{p})$ is governed by the WKE (3.61). It is to be noted that further generalizations of the WKE to include the effects of small dissipation and random fluctuations in the medium are given in McDonald and Kaufman (1985) and McDonald (1991).

***Example.*** — Let us consider a wave bath composed of Langmuir waves in a plasma. For simplicity, let the dynamics of the Langmuir waves be governed by

$$\left[\partial_t^2 - 3v_T^2 \boldsymbol{\nabla}^2 + \omega_p^2(x)\right] \Psi(x) = 0, \tag{3.62}$$

where $v_t = (T/m)^{1/2}$ is the plasma thermal velocity (assumed to be constant here) and $\omega_p(x)$ is the plasma frequency. For the equation above, the dispersion operator is $\widehat{\mathfrak{D}} = \hat{p}_0^2 - 3v_T^2 \hat{\mathbf{p}}^2 - \hat{\omega}_p^2$, where $\hat{\omega}_p^2 \doteq \omega_p^2(\hat{x})$. Likewise, the dispersion Weyl symbol is $D(x, p) = p_0^2 - 3v_T^2 \mathbf{p}^2 - \omega_p^2(x)$. Upon solving for $D(x, p_*) = 0$, one considers waves with positive frequencies so that $p_*^0 = H(t, \mathbf{x}, \mathbf{p})$, where $H(t, \mathbf{x}, \mathbf{p}) = [3v_t^2 \mathbf{p}^2 + \omega_p^2(x)]^{1/2}$. Hence, upon following the discussion above, one derives the corresponding WKE describing a Langmuir-wave bath in a plasma:

$$\frac{\partial F_\Psi}{\partial t} + \frac{3v_T^2 \mathbf{p}}{H} \cdot \frac{\partial F_\Psi}{\partial \mathbf{x}} - \frac{\boldsymbol{\nabla}\omega_p^2}{2H} \cdot \frac{\partial F_\Psi}{\partial \mathbf{p}} = 0. \tag{3.63}$$

Since the WKE above is simply a transport equation in the six-dimensional phase space, it can be readily solved using a variety of methods, e.g., the method of characteristics.

Before concluding this Chapter, let us discuss why the WKE (3.61) describes the dynamics of incoherent waves. From the procedure shown above, it seems that the estimates provided in Eq. (3.48) are the key assumptions needed to obtain the WKE from the action functional. In the following, I shall demonstrate how a *wave bath* composed of incoherent waves can actually lead to a Wigner function that satisfies the estimates in Eq. (3.48).

Let us consider a wave function composed by a large number of individual eikonal waves so that $\Psi(x) = \sum_l a_l(x) \exp[i\theta_l(x)]$, where $\theta_l(x)$ is a fast real function compared to the real function $a_l(x)$. One then inserts $\Psi(x)$ into the Wigner function (3.6) and asymptotically evaluates the integral. This leads to

$$W_\Psi(x,p) = \sum_{l,m} a_l(x) a_m(x)\, \delta^4\left(p - \frac{k_l + k_m}{2}\right) \exp\left[i\theta_l(x) - i\theta_m(x)\right] + \mathcal{O}(\epsilon), \tag{3.64}$$

where $k_l \doteq -\partial\theta_l$ is the four-wavevector associated to the $l$th wave. For incoherent waves, the relative phases $\theta_l(x) - \theta_m(x)$ are considered random. Adding many fields with random phase factors causes the cross terms to effectively cancel out. Hence, for incoherent waves one obtains

$$W_\Psi(x,p) \simeq \sum_l a_l^2(x) \delta^4\left(p - k_l\right) + \mathcal{O}(\epsilon). \tag{3.65}$$

This approximation is known as the *random-phase approximation* (Nazarenko, 2011). Note that the Wigner function (3.65) does not satisfy the estimates in Eq. (3.48), as it is not smooth enough. For $W_\Psi(x,p)$ to be smooth, one can imagine that a large number of incoherent waves *densely packed* around a particular phase-space region could eventually lead to a Wigner function that varies slowly along the dispersion manifold (McDonald and Kaufman, 1985). Another possibility is to perform a *phase-space average* or *coarse graining* of $W_\Psi(x,p)$ in order to smooth out the fluctuations. For example, one can smooth the Wigner function with a Gaussian function, as was done by Cartwright (1976) to obtain a non-negative Wigner function. Thus, with these assumptions it seems that the WKE (3.61) can only describe the dynamics of a wave bath of incoherent waves.

## 3.5 Conclusions

In order to introduce some of the basic concepts and methodologies used in the thesis, this Chapter presented a brief discussion on the GO approximation. Starting from the variational phase-space formulation of wave dynamics, I derived the GO equations for linear scalar waves propagating in a nondissipative medium. Both eikonal waves and incoherent waves were discussed. In the case of eikonal waves, the resulting equations are the well-known Whitham's equations for wave dynamics. For incoherent waves, the governing equation is the wave kinetic equation.

As a final remark, it should be noted that within the GO approximation the ray trajectories of scalar waves are completely characterized by the wave position and momentum coordinates. However, this is no longer the case for multi-component (vector) waves. Vector waves have an additional degree of freedom, namely, the wave polarization. In the next Chapter, the theory of GO will be extended in order to include this extra degree of freedom. This extended model will include additional higher-order effects related to the wave polarization.

# Part II

# Extending geometrical optics to vector waves

# Chapter 4

# General theory of extended geometrical optics

Geometrical optics (GO) treats wave rays as classical particles, which are completely characterized by their coordinates and momenta. However, multi-component waves have another degree of freedom, namely, their polarization. The polarization degree of freedom manifests itself as an effective (classical) "wave spin" that can be assigned to rays and can affect the wave dynamics accordingly. A well-known manifestation of polarization dynamics is mode conversion, which is the linear exchange of wave action between different wave modes and can be interpreted as a precession of the wave spin. Another, less-known polarization effect is the polarization-driven bending of ray trajectories. This Chapter presents an extension and reformulation of GO as a first-principle Lagrangian theory whose action governs the aforementioned polarization phenomena simultaneously. Some of the results presented in this Chapter were published by Ruiz and Dodin (2017a).

## 4.1 Introduction

### 4.1.1 Motivation

As discussed in the previous Chapter, GO is an asymptotic theory with respect to a small parameter $\epsilon$ that is a ratio of the wave-relevant characteristic period (temporal or spatial) to the inhomogeneity scale of the underlying medium [see Eq. (3.2)]. Practical applications of GO are traditionally restricted to the lowest-order theory, where each wave is basically approximated with a local eigenmode of the underlying medium at each location in spacetime. Then, the wave dynamics are entirely determined by a single branch of the local dispersion relation. However, this approximation is not entirely accurate, even when diffraction is neglected (see Sec. 3.3.4). If a dispersion relation has more than one branch, i.e., a vector wave with more than one polarization at a given location, then the interaction between these branches can give rise to important polarization effects that are missed in the traditional lowest-order GO approximation.

One interesting manifestation of such polarization effects is the polarization-driven bending of ray trajectories. This effect is primarily known in two contexts. One is quantum mechanics, where polarization effects manifest as the Berry phase (Berry, 1984) and the associated Stern–Gerlach force experienced by vector particles, i.e., quantum particles with spin. Another one is optics, where a related effect has been known as the *Hall effect of light*; namely, even in an isotropic dielectric, rays propagate somewhat differently depending on their polarization.[1] But the same effect can also be anticipated for waves in plasmas, e.g., radiofrequency (RF) waves in tokamaks. In fact, since the GO parameter $\epsilon$ for RF waves in laboratory plasma is typically larger than that for quantum and optical waves, the polarization-driven bending of ray trajectories in this case can be more important and might need to be taken into account in practical ray tracing simulations. However, *ad hoc* theories of polarization effects available from optics are inapplicable to plasma waves, which have more complicated dispersion and thus require more fundamental approaches. Thus, a different theory is needed that would allow the calculation of the polarization-bending of the ray trajectories for plasma waves and, also more broadly, waves in general linear media.

Relevant work was done by Littlejohn and Flynn (1991) and Weigert and Littlejohn (1993), where a systematic procedure was proposed to asymptotically diagonalize the dispersion operator for linear vector waves. Polarization effects emerge as $\mathcal{O}(\epsilon)$ corrections to the GO dispersion relation. However, this approach excludes mode conversion, i.e., the linear exchange of wave action between different branches of the local dispersion relation. Since the group velocities of the different branches eventually separate, mode conversion is typically followed by ray splitting and was studied extensively in this context.[2] However, those works considered wave modes that are resonant in small, localized regions of phase space. Hence, the nonadiabatic dynamics was formulated as an asymptotic scattering problem between two wave modes, so the polarization-driven bending of ray trajectories was ignored.

In this Chapter, I propose a more general theory, which is a culmination of the research published by Ruiz and Dodin (2015b) and Ruiz and Dodin (2017a). It is shown that mode conversion and the polarization-driven bending of ray trajectories are two sides of the same coin and can be considered simultaneously within a unified variational theory. Specific applications of this general theory are given in the next two Chapters.

[1]For more information, see, for example, Bliokh *et al.* (2008), Bliokh *et al.* (2015), Onoda *et al.* (2004), Dooghin *et al.* (1992), and Liberman and Zel'dovich (1992).

[2]See, for example, Friedland *et al.* (1987), Friedland (1985), Tracy and Kaufman (1993), Flynn and Littlejohn (1994), Tracy *et al.* (2003), Tracy *et al.* (2007), and Richardson and Tracy (2008).

### 4.1.2 Outline

This Chapter is organized as follows. In Sec. 4.2, the variational formalism for describing vector waves is presented. In Sec. 4.3, a general procedure is proposed to block-diagonalize the wave dispersion operator. In Sec. 4.4, the action functional is reparameterized in order to facilitate a subsequent asymptotic analysis. In Sec. 4.5, the leading-order GO approximation is discussed. In Sec. 4.6, the more accurate model that includes polarization effects is discussed. Several aspects of the theory, such as mode conversion, precession of the wave spin, and gauge invariance, will also be discussed here. In Sec. 4.7, the main results are summarized.

## 4.2 Basic equations

### 4.2.1 Wave action principle

As in Sec. 2.2.1, let us describe the dynamics of any nondissipative linear wave by using the principle of stationnary action $\delta\mathcal{S} = 0$, where the real action functional $\mathcal{S}$ is bilinear in the wave field. The wave field is represented as a complex-valued field $\Psi(x)$. This field has an arbitrary (yet finite) number of components $\overline{N}$ and can be written explicitly as

$$\Psi(x) = \begin{pmatrix} \Psi^1(x) \\ \vdots \\ \Psi^{\overline{N}}(x) \end{pmatrix}, \tag{4.1}$$

where $\Psi^n(x)$ denotes the $n$th wave component.[3,4] In the absence of external sources and parametric resonances, the action functional is written as in Eq. (2.5) so

$$\mathcal{S} = \int \mathrm{d}^4x \, \mathrm{d}^4x' \, \Psi^\dagger(x)\mathcal{D}(x,x')\Psi(x'), \tag{4.2}$$

where $\mathcal{D}$ is a $\overline{N}\times\overline{N}$ Hermitian matrix kernel $[\mathcal{D}(x,x') = \mathcal{D}^\dagger(x',x)]$ that describes the underlying medium. Varying the action functional with respect to $\Psi^\dagger$ leads to the general wave equation (2.2),

$$\delta\Psi^\dagger : \quad 0 = \int \mathrm{d}^4x' \, \mathcal{D}(x,x')\Psi(x'). \tag{4.3}$$

Similarly, varying with respect to $\Psi$ gives the equation adjoint to Eq. (4.3), which I do not need to discuss in further detail.

[3]As in Chapter 2, here I refer to the $n$th component of the wave as the $n$th element of the column vector $\overline{\Psi}(x)$. Thus, $\overline{\Psi}^n(x)$ should not be confused with an amplitude corresponding to a particular wave mode.

[4]Note that I use an overline when referring to the $\overline{N}$-component wave $\overline{\Psi}(x)$. This additional notation is added in order to distinguish between resonant and nonresonant wave modes (to be defined later). The wave oscillations corresponding to the resonant wave modes will be denoted without using the overline. This will become clearer later.

Following Sec. 2.2.2, I represent the wave $\Psi(x)$ as an abstract vector $|\,\Psi\,\rangle$ in the Hilbert space of wave states with inner product given by Eq. (2.8). In the Dirac notation, the action (4.2) can be rewritten in the form of Eq. (2.18):

$$\mathcal{S} = \langle\,\Psi\,|\,\widehat{\mathcal{D}}\,|\,\Psi\,\rangle\,, \tag{4.4}$$

where $\widehat{\mathcal{D}}$ is the Hermitian *dispersion operator* such that $\mathcal{D}(x,x') = \langle\,x\,|\,\widehat{\mathcal{D}}\,|\,x'\,\rangle$.

### 4.2.2 Extended wave function

As in Dodin (2014a) and Ruiz and Dodin (2015b), reduced models of wave propagation are convenient to develop when the action is of the symplectic form; namely,

$$\mathcal{S}_{\text{symplectic}} \doteq \langle\,\Psi\,|\,(\widehat{p}_0 \mathbb{I}_{\overline{N}} - \widehat{\mathcal{H}})\,|\,\Psi\,\rangle\,, \tag{4.5}$$

where $\widehat{p}_0 = i\partial_t$ (in the physical-space representation) and "the wave Hamiltonian" $\widehat{\mathcal{H}} = \mathcal{H}(\widehat{t},\widehat{\mathbf{x}},\widehat{\mathbf{p}})$ is some Hermitian operator that is local in time, i.e., commutes with $\widehat{t}$.[5]

In order to cast the general action (4.4) into the symplectic form (4.5), let us perform the so-called Feynman reparameterization (Feynman, 1951; Aparicio *et al.*, 1995) that lifts the wave dynamics governed by Eq. (4.4) from $\mathbb{R}^4$ to $\mathbb{R}^5$. Specifically, I let the wave field depend on spacetime as well as on some parameter $\tau$ so that $\Psi(\tau,x) = \langle\,x\,|\,\Psi(\tau)\,\rangle$. Note that $|\,\Psi(\tau)\,\rangle$ belongs to the same Hilbert space defined in Eq. (2.8). Thus, the inner product remains the same; i.e.,

$$\langle\,\Upsilon(\tau')\,|\,\Psi(\tau)\,\rangle = \int \mathrm{d}^4x\,\Upsilon^\dagger(\tau',x)\Psi(\tau,x). \tag{4.6}$$

Let us then consider the following "extended" action

$$\mathcal{S}_{\mathrm{X}} \doteq \int \mathrm{d}\tau\; L, \tag{4.7}$$

where $L \doteq L_\tau + L_D$,

$$L_\tau \doteq -(i/2)\,[\,\langle\,\Psi(\tau)\,|\,\partial_\tau\Psi(\tau)\,\rangle - \text{c.c.}\,]\,, \tag{4.8a}$$

$$L_D \doteq \langle\,\Psi(\tau)\,|\,\widehat{\mathcal{D}}\,|\,\Psi(\tau)\,\rangle\,, \tag{4.8b}$$

[5]For extended discussions, see Dodin (2014a) and Bridges and Reich (2001).

and $\partial_\tau \Psi(\tau, x) = \langle\, x \mid \partial_\tau \Psi(\tau)\,\rangle$. Note that the Lagrangian $L$ is local in the parameter $\tau$; i.e., the abstract vectors are all evaluated at $\tau$. Thus, from henceforth all fields will be evaluated at $\tau$, and I shall avoid writing explicitly the dependence of $|\,\Psi\,\rangle$ on $\tau$. The corresponding ELE is

$$\delta \langle\, \Psi \,| : \quad i \frac{\partial}{\partial \tau} |\, \Psi \,\rangle = \widehat{\mathcal{D}} \,|\, \Psi \,\rangle \,. \tag{4.9}$$

Note that Eq. (4.9) can be interpreted as a vector Schrödinger equation in the extended variable space, where $\widehat{\mathcal{D}}$ acts as the Hamiltonian operator. The dynamics of the original system described by the action (4.4) is a special case of the dynamics governed by Eq. (4.9), which corresponds to a steady state with respect to the parameter $\tau$; i.e., $\partial_\tau \Psi = 0$. The advantage of the "extended" action (4.7) is that it is manifestly of the symplectic form, so one can proceed as follows.

## 4.3 Eigenmode representation

### 4.3.1 Variable transformation

Let us introduce an invertible, $\tau$-independent $\overline{N} \times \overline{N}$ transformation operator $\widehat{\mathcal{Q}}^{-1}$ that maps $|\,\Psi\,\rangle$ to some $\overline{N}$-component abstract vector $|\,\overline{\psi}\,\rangle$ yet to be defined. Hence, one has

$$|\, \Psi \,\rangle = \widehat{\mathcal{Q}} \,|\, \overline{\psi} \,\rangle \,. \tag{4.10}$$

Additionally, we choose $\widehat{\mathcal{Q}}$ to be unitary. Inserting Eq. (4.10) into Eqs. (4.8) leads to

$$L_\tau = -(i/2) \big( \langle\, \overline{\psi} \mid \partial_\tau \overline{\psi} \,\rangle - \mathrm{c.\,c.} \big) \,, \tag{4.11a}$$

$$L_D = \langle\, \overline{\psi} \mid \widehat{\mathcal{D}}_{\text{eff}} \mid \overline{\psi} \,\rangle \,, \tag{4.11b}$$

where $\widehat{\mathcal{D}}_{\text{eff}} \doteq \widehat{\mathcal{Q}}^\dagger \,\widehat{\mathcal{D}}\, \widehat{\mathcal{Q}}$ is the effective dispersion operator. In what follows, I seek to construct $\widehat{\mathcal{Q}}$ such that the operator $\widehat{\mathcal{D}}_{\text{eff}}$ is expressed in a block-diagonal form. Each block of $\widehat{\mathcal{D}}_{\text{eff}}$ will govern the dynamics of a corresponding set of resonant wave eigenmodes, which is yet to be specified. Resonant eigenmodes within a block will be allowed to interact with one another and exchange wave action between them. This will become clearer below.

### 4.3.2 Weyl representation

Following Eq. (2.42), let us write Eq. (4.11b) in the phase-space representation:

$$L_D = \mathrm{Tr} \int \mathrm{d}^4x\, \mathrm{d}^4p\, D_{\text{eff}}(x,p) W_{\overline{\psi}}(\tau,x,p). \tag{4.12}$$

The components of the Wigner tensor $W_{\overline{\psi}}(\tau,x,p)$ corresponding to $|\,\overline{\psi}\,\rangle$ are given by

$$[W_{\overline{\psi}}]^m_n(\tau,x,p) \doteq \frac{1}{(2\pi)^4} \int \mathrm{d}^4s\, e^{ip\cdot s} \,\langle\, x+s/2 \,|\, \overline{\psi}^m(\tau) \,\rangle \langle\, \overline{\psi}_n(\tau) \,|\, x-s/2 \,\rangle\,, \tag{4.13}$$

and $D_{\text{eff}}(x,p)$ is the Weyl symbol [Eq. (A.1)] corresponding to the operator $\widehat{\mathfrak{D}}_{\text{eff}}$. Upon using the Moyal product [Eq. (A.5)], one can write explicitly $D_{\text{eff}}(x,p)$ as

$$D_{\text{eff}}(x,p) = \big[Q^\dagger\big](x,p) \star D(x,p) \star Q(x,p), \tag{4.14}$$

where $D(x,p)$, $Q(x,p)$, and $\big[Q^\dagger\big](x,p)$ are the Weyl symbols corresponding to $\widehat{\mathfrak{D}}$, $\widehat{\mathfrak{Q}}$, and $\widehat{\mathfrak{Q}}^\dagger$, respectively. Also, in the Weyl representation the unitary condition $\widehat{\mathfrak{Q}}^\dagger\widehat{\mathfrak{Q}} = \widehat{\mathbb{I}}_{\overline{N}}$ is given by

$$\big[Q^\dagger\big](x,p) \star Q(x,p) = \mathbb{I}_{\overline{N}}, \tag{4.15}$$

which will be used below.

### 4.3.3 Eigenmode representation

Let us assume that $D_{\text{eff}}(x,p)$ and $Q(x,p)$ can be expanded in powers of the GO parameter (3.2). Hence, one has

$$D_{\text{eff}}(x,p) = \Lambda(x,p) + \epsilon U(x,p) + \mathcal{O}(\epsilon^2), \tag{4.16a}$$

$$Q(x,p) = Q_0(x,p) + \epsilon Q_1(x,p) + \mathcal{O}(\epsilon^2), \tag{4.16b}$$

where $(\Lambda, U, Q_0, Q_1)$ are $\overline{N}\times\overline{N}$ matrices of order unity. To the lowest order in $\epsilon$, the Moyal products in Eqs. (4.14) and (4.15) reduce to ordinary matrix products [see Eq. (3.50)], so

$$\Lambda(x,p) = \big[Q_0^\dagger\big](x,p) D(x,p) Q_0(x,p), \tag{4.17}$$

$$\big[Q_0^\dagger\big](x,p) Q_0(x,p) = \mathbb{I}_{\overline{N}}. \tag{4.18}$$

By properties of the Weyl transformation (Appendix A), the fact that $\widehat{\mathfrak{D}}$ is a Hermitian operator ensures that $D(x,p)$ is a Hermitian matrix. Hence, by the spectral theorem, $D(x,p)$ has $\overline{N}$ orthonormal eigenvectors $\mathbf{e}_q(x,p)$, which correspond to some $\overline{N}$ real eigenvalues $\lambda^{(q)}(x,p)$ (counted with algebraic multiplicity). Let us construct $Q_0(x,p)$ out of these eigenvectors so that

$$Q_0(x,p) = [\,\mathbf{e}_1(x,p),\,\dots,\,\mathbf{e}_{\overline{N}}(x,p)\,]\,, \tag{4.19}$$

where the individual $\mathbf{e}_q(x,p)$ form the columns of $Q_0(x,p)$. From Eq. (4.18), one then finds $\big[Q_0^\dagger\big](x,p) = Q_0^\dagger(x,p)$. Hence, the matrix $\Lambda(x,p)$ has the following diagonal form:

$$\Lambda(x,p) = \operatorname{diag}\big[\,\lambda^{(1)}(x,p),\,\dots,\,\lambda^{(\overline{N})}(x,p)\,\big]. \tag{4.20}$$

To the next order in $\epsilon$, Eq. (4.15) reads as follows:

$$Q_0^\dagger Q_1 + [Q_1^\dagger]Q_0 + \frac{i}{2}\{Q_0^\dagger, Q_0\} = 0. \tag{4.21}$$

Here I assumed that term involving the eight-dimensional Poisson bracket $\{Q_0^\dagger, Q_0\}$, which arises from the expansion of the Moyal star product [Eq. (A.8)], is of the first order in $\epsilon$. Following Littlejohn and Flynn (1991), let $Q_1 = Q_0(A+iG)$ and $[Q_1^\dagger] = Q_1^\dagger$, where $A(x,p)$ and $G(x,p)$ are $\overline{N}\times\overline{N}$ Hermitian matrices. Then, Eq. (4.21) gives

$$A(x,p) = -\frac{i}{4}\{Q_0^\dagger, Q_0\}. \tag{4.22}$$

In order to determine $G(x,p)$, let us write Eq. (4.14) to the first order in $\epsilon$. For convenience, I introduce the following bracket:

$$\{A,B\}_C \doteq \frac{\partial A}{\partial p_\mu} C \frac{\partial B}{\partial x^\mu} - \frac{\partial A}{\partial x^\mu} C \frac{\partial B}{\partial p_\mu}, \tag{4.23}$$

where $(A, B, C)$ are $\overline{N}$-dimensional matrices. Using the associative property of the Moyal product and noting that $DQ_0 = Q_0\Lambda$, one obtains

$$
\begin{aligned}
U(x,p) &= Q_1^\dagger D Q_0 + Q_0^\dagger D Q_1 + (i/2)\{Q_0^\dagger D, Q_0\} + (i/2)\{Q_0^\dagger, D\}Q_0 \\
&= (A - iG)Q_0^\dagger D Q_0 + Q_0^\dagger D Q_0 (A + iG) + (i/2)\{Q_0^\dagger D, Q_0\} + (i/2)\{Q_0^\dagger, D\}Q_0 \\
&= A\Lambda - iG\Lambda + \Lambda A + i\Lambda G + (i/2)\{Q_0^\dagger D, Q_0\} + (i/2)\{Q_0^\dagger, DQ_0\} - (i/2)\{Q_0^\dagger, Q_0\}_D \\
&= i\Lambda G - iG\Lambda - (i/4)\{Q_0^\dagger, Q_0\}\Lambda - (i/4)\Lambda\{Q_0^\dagger, Q_0\} + (i/2)\{\Lambda Q_0^\dagger, Q_0\} \\
&\quad + (i/2)\{Q_0^\dagger, Q_0\Lambda\} - (i/2)\{Q_0^\dagger, Q_0\}_D \\
&= i(\Lambda G - G\Lambda + \delta U),
\end{aligned}
\tag{4.24}
$$

where $\delta U(x,p)$ is a $\overline{N} \times \overline{N}$ matrix given by

$$
\begin{aligned}
\delta U(x,p) = (1/4)\{Q_0^\dagger, Q_0\}\Lambda + (1/4)\Lambda\{Q_0^\dagger, Q_0\} + (1/2)\{\Lambda, Q_0\}_{Q_0^\dagger} \\
+ (1/2)\{Q_0^\dagger, \Lambda\}_{Q_0} - (1/2)\{Q_0^\dagger, Q_0\}_D.
\end{aligned}
\tag{4.25}
$$

Since $[\Lambda G - G\Lambda]^m_n = G^m_n(\lambda^{(m)} - \lambda^{(n)})$ (no summation is assumed here over the repeating indices), one can diagonalize $U$ by adopting $G^m_n = \delta U^m_n/(\lambda^{(n)} - \lambda^{(m)})$ for $m \neq n$, as done in Littlejohn and Flynn (1991) and Weigert and Littlejohn (1993). However, this method is applicable only when $|\lambda^{(m)} - \lambda^{(n)}| \gtrsim \mathcal{O}(1)$; otherwise, when $|\lambda^{(m)} - \lambda^{(n)}| \simeq \mathcal{O}(\epsilon)$, $G^m_n \gtrsim \mathcal{O}(\epsilon^{-1})$ so $\epsilon Q_1 \gtrsim \mathcal{O}(1)$, which is in violation of the ordering assumed in Eq. (4.16b). Hence, instead of diagonalizing $U(x,p)$, I propose to only *block-diagonalize* $U(x,p)$ as follows. When $|\lambda^{(m)} - \lambda^{(n)}| \gtrsim \mathcal{O}(1)$, I choose the off-diagonal components of $G^m_n(x,p)$ so that $U^m_n(x,p) = 0$. (I call such modes *nonresonant*.) When $|\lambda^{(m)} - \lambda^{(n)}| \simeq \mathcal{O}(\epsilon)$, I let $G^m_n(x,p) = 0$. (I call such modes *resonant*.) By following this prescription and permuting the matrix rows, one can write $U(x,p)$ in the following form:

$$
U(x,p) = \begin{pmatrix}
[[U]]_1(x,p) & 0 & \dots & 0 \\
0 & [[U]]_2(x,p) & \dots & 0 \\
\vdots & \vdots & \ddots & \vdots \\
0 & 0 & \dots & [[U]]_J(x,p)
\end{pmatrix},
\tag{4.26}
$$

where $[[U]]_j(x,p)$ are $n_j \times n_j$ Hermitian matrices and $J$ is the total number of blocks, so $\sum_{j=1}^J n_j = \overline{N}$. Here $[[...]]$ denotes a matrix block. Note that, in the particular case where only nonresonant modes are present, $U(x,p)$ is diagonal, and one recovers the results obtained by Littlejohn and Flynn (1991) and Weigert and Littlejohn (1993).

Since the matrix $D_{\text{eff}} \approx \Lambda + \epsilon U$ is block-diagonal, the Lagrangian (4.12) is unaffected by the matrix elements $[W_{\overline{\Psi}}]^m_n$ with indices $(m, n)$ such that $U^m_n = 0$. Thus, without loss of generality, one can write

$$L_D = \sum_{j=1}^{J} \text{Tr} \int \mathrm{d}^4x \, \mathrm{d}^4p \, [[D_{\text{eff}}]]_j \, [[W]]_j + \mathcal{O}(\epsilon^2), \tag{4.27}$$

where $[[D_{\text{eff}}]]_j(x,p) \simeq [[\Lambda]]_j(x,p) + \epsilon[[U]]_j(x,p)$ and $[[W_{\overline{\psi}}]]_j(x,p)$ represent the $j$th matrix block of $\Lambda(x,p) + \epsilon U(x,p)$ and $W_{\overline{\psi}}(x,p)$, respectively. Hence, nonresonant eigenmodes are decoupled while resonant eigenmodes that belong to the same matrix block remain coupled.

## 4.4 Reduced action

### 4.4.1 Basic equations

Now that blocks of mutually nonresonant modes are decoupled, let us focus on the dynamics of modes within a single block of some size $N$. After projecting onto these modes and dropping the block index, one adopts the action $\mathcal{S} = \int \mathrm{d}\tau \, (L_\tau + L_D)$, where

$$L_\tau \doteq -(i/2) \int \mathrm{d}^4x \, \left[ \psi^\dagger (\partial_\tau \psi) - (\partial_\tau \psi^\dagger) \psi \right], \tag{4.28a}$$

$$L_D \doteq \text{Tr} \int \mathrm{d}^4x \, \mathrm{d}^4p \, ([[\Lambda]] + \epsilon[[U]]) \, W_\psi. \tag{4.28b}$$

Here $\psi(\tau, x)$ is a complex-valued function with $N$ components, and $W_\psi(\tau, x, p)$ is the $N \times N$ Wigner tensor, whose components are given by

$$[W_\psi]^m_n(\tau, x, p) = \frac{1}{(2\pi)^4} \int \mathrm{d}^4s \, e^{ip\cdot s} \, \langle \, x + s/2 \mid \psi^m(\tau) \, \rangle \, \langle \, \psi_n(\tau) \mid x - s/2 \, \rangle \, . \tag{4.29}$$

Since I consider the coupled dynamics of some $N$ resonant modes, $[[\Lambda]](x,p)$ is an $N \times N$ diagonal matrix whose components are the GO eigenvalues corresponding to the resonant modes, $[[\Lambda]](x,p) = \text{diag}[\lambda^{(1)}, ..., \lambda^{(N)}]$. [Note that the difference between the eigenvalues is $\mathcal{O}(\epsilon)$.] Similarly, only $N$ columns of $Q_0(x,p)$ actually contribute to $[[U]](x,p)$. For clarity, let us denote the resonant eigenmodes as $\mathbf{e}_q(x,p)$ with indices $q = 1, ..., N$. In order to calculate $[[U]](x,p)$, one can use Eq. (4.24). After block-diagonalizing $U(x,p)$ and introducing the (non-square) $\overline{N} \times N$ matrix

$$\Xi(x,p) = [\, \mathbf{e}_1(x,p), \, ... \, , \, \mathbf{e}_N(x,p) \,], \tag{4.30}$$

the corresponding matrix $[[U]](x,p)$ for the resonant modes is given by

$$[[U]](x,p) = \frac{i}{4}\{\Xi^\dagger,\Xi\}[[\Lambda]] + \frac{i}{4}[[\Lambda]]\{\Xi^\dagger,\Xi\} + \frac{i}{2}\{[[\Lambda]],\Xi\}_{\Xi^\dagger} + \frac{i}{2}\{\Xi^\dagger,[[\Lambda]]\}_{\Xi} - \frac{i}{2}\{\Xi^\dagger,\Xi\}_D, \tag{4.31}$$

which is a $N\times N$ Hermitian matrix. Note that $\Xi(x,p)$ only contains the eigenvectors of the resonant eigenmodes, contrary to $Q_0(x,p)$. Furthermore, it is convenient to split $[[D_{\text{eff}}]]$ as follows:

$$[[D_{\text{eff}}]](x,p) = \lambda(x,p)\mathbb{I}_N + \epsilon\,\mathcal{U}(x,p), \tag{4.32}$$

where $\lambda(x,p)$ is the average of the trace of $[[D_{\text{eff}}]]$,

$$\lambda(x,p) \doteq N^{-1}\text{Tr}\left[[[\Lambda]](x,p) + \epsilon\,[[U]](x,p)\right], \tag{4.33}$$

and $\epsilon\,\mathcal{U}(x,p)$ is the remaining traceless part,

$$\epsilon\,\mathcal{U}(x,p) \doteq [[\Lambda]](x,p) + \epsilon[[U]](x,p) - \lambda(x,p)\mathbb{I}_N. \tag{4.34}$$

In the special case when all the GO eigenvalues within the block are degenerate (that is, when $\lambda^{(q)}(x,p)$ are identical), and when $[[U]](x,p)$ is traceless, then $\Lambda(x,p) = \lambda(x,p)\mathbb{I}_N$ and $\mathcal{U}(x,p) = [[U]](x,p)$. I call such modes *degenerate*. Then, $[[U]](x,p)$ in Eq. (4.31) can be written as follows:

$$\begin{aligned}
\mathcal{U}(x,p) &= \frac{i}{4}\{\Xi^\dagger,\Xi\}\lambda + \frac{i}{4}\lambda\{\Xi^\dagger,\Xi\} + \frac{i}{2}\{\lambda,\Xi\}_{\Xi^\dagger} + \frac{i}{2}\{\Xi^\dagger,\lambda\}_{\Xi} - \frac{i}{2}\{\Xi^\dagger,\Xi\}_D \\
&= -\frac{1}{2i}\Xi^\dagger\{\lambda,\Xi\} - \frac{1}{2i}\{\Xi^\dagger,\lambda\}\Xi + \frac{1}{2i}\{\Xi^\dagger,\Xi\}_D - \frac{1}{2i}\{\Xi^\dagger,\Xi\}_\lambda \\
&= -\left[\Xi^\dagger\{\lambda,\Xi\}\right]_A + \left[(\partial_p\Xi^\dagger)(D-\lambda\mathbb{I}_N)(\partial_x\Xi)\right]_A,
\end{aligned} \tag{4.35}$$

where I substituted the bracket introduced in Eq. (4.23) and the subscript "$A$" means "anti-Hermitian part;" namely, for any given matrix $M$, one has $M_A \doteq (M-M^\dagger)/(2i)$. The expression in Eq. (4.35) can also be written more explicitly in terms of components as follows:

$$\mathcal{U}(x,p) = \left(-\frac{\partial\lambda}{\partial p_\mu}\right)\left(\Xi^\dagger\frac{\partial\Xi}{\partial x^\mu}\right)_A + \left(\frac{\partial\lambda}{\partial x^\mu}\right)\left(\Xi^\dagger\frac{\partial\Xi}{\partial p_\mu}\right)_A + \left(\frac{\partial\Xi^\dagger}{\partial p_\mu}(D-\lambda\mathbb{I}_N)\frac{\partial\Xi}{\partial x^\mu}\right)_A. \tag{4.36}$$

As we shall see, the matrix $\mathcal{U}(x,p)$ will couple the different wave polarizations, allowing them to exchange wave action. Hence, I denote $\mathcal{U}(x,p)$ as the *polarization-coupling matrix*.

### 4.4.2 Parameterization of the action

In the following, I shall reparameterize the action functional in order to facilitate the subsequent asymptotic analysis of the wave dynamics. Let us adopt the following parameterization of the wave:

$$\psi(\tau, x) = a(\tau, x)\, z(\tau, x)\, e^{i\theta(\tau,x)}. \tag{4.37}$$

Here $\theta(\tau, x)$ is a real variable that serves as the rapid phase common for all $N$ modes. (Remember that all modes within the block of interest are approximately resonant to each other.) Also, $a(\tau, x)$ is a real function representing the wave envelope, and $z(\tau, x)$ is a $N$-dimensional complex unit vector ($z^\dagger z = 1$), whose components describe the amount of wave action (or loosely speaking, wave quanta) in the corresponding modes.[6] Both $a(\tau, x)$ and $z(\tau, x)$ are considered to be slow compared to the fast phase $\theta(\tau, x)$.

After substituting the ansatz (4.37) into Eq. (4.28a), one obtains[7]

$$L_\tau = \int \mathrm{d}^4x\, a^2 \left[\partial_\tau\theta - (i\epsilon/2)(z^\dagger\partial_\tau z - \text{c.\,c.})\right]. \tag{4.38}$$

Now, let us calculate the Wigner tensor (4.29). Upon substituting Eq. (4.37) into Eq. (4.29), one obtains the following expression for the Wigner tensor:

$$\begin{aligned}
[W_\psi]^m_n(\tau, x, p) =& \frac{1}{(2\pi)^4}\int \mathrm{d}^4s\, a(\tau, x+s/2) z^m(\tau, x+s/2) e^{i\theta(\tau,x+s/2)} \\
& \qquad \times a(\tau, x-s/2) z^*_n(\tau, x-s/2) e^{-i\theta(\tau,x-s/2)} e^{ip\cdot s} \\
=& \frac{1}{(2\pi)^4}\int \mathrm{d}^4s\, e^{i(p-k)\cdot s}\Bigg[a^2(\tau,x) z^m(\tau,x) z^*_n(\tau,x) \\
& \qquad + \epsilon\frac{s^\mu}{2}\left(\frac{\partial(a\,z^m)}{\partial x^\mu} a z^*_n - a z^m \frac{\partial(a\,z^*_n)}{\partial x^\mu}\right)\Bigg] + \mathcal{O}(\epsilon^2) \\
=& \; a^2(\tau,x)\, z^m(\tau,x) z^*_n(\tau,x)\, \delta^4(p-k) \\
& - \frac{i\epsilon}{2}\frac{\partial[\delta^4(p-k)]}{\partial p_\mu}\left(\frac{\partial(a\,z^m)}{\partial x^\mu} a z^*_n - a z^m \frac{\partial(a\,z^*_n)}{\partial x^\mu}\right) + \mathcal{O}(\epsilon^2),
\end{aligned} \tag{4.39}$$

[6] Since I reparameterize the $N$-dimensional complex vector $\psi(\tau, x)$ by the $N$-dimensional complex vector $z(\tau, x)$ plus two independent real functions $\theta(\tau, x)$ and $a(\tau, x)$, not all components of $z(\tau, x)$ are truly independent (see Sec. 4.6).

[7] Here I formally introduce $\epsilon$ to denote that $z$ is a slowly-varying quantity; however, this ordering parameter will be removed later.

where $k_\mu(\tau, x) \doteq -\partial_\mu \theta(\tau, x)$ is the four-wavevector introduced in Eq. (3.5). Upon inserting Eq. (4.39) into Eq. (4.28b) and integrating over momentum, one obtains

$$\begin{aligned} L_D =& \mathrm{Tr} \int \mathrm{d}^4 x \, \mathrm{d}^4 p \; (\lambda[[W_\psi]] + \epsilon \mathcal{U}[[W_\psi]]) \\ =& \int \mathrm{d}^4 x \, a^2 \left[ \lambda(x, k) z^\dagger z + \epsilon z^\dagger \mathcal{U}(x, k) z \right] \\ & - \frac{i\epsilon}{2} \int \mathrm{d}^4 x \, \mathrm{d}^4 p \, \lambda \frac{\partial [\delta^4(p - k)]}{\partial p_\mu} \left( \frac{\partial (a \, z^m)}{\partial x^\mu} a z_m^* - \mathrm{c.\,c.} \right) + \mathcal{O}(\epsilon^2) \\ =& \int \mathrm{d}^4 x \, a^2 \left[ \lambda(x, k) + \epsilon z^\dagger \mathcal{U}(x, k) z \right] - \frac{i\epsilon}{2} \int \mathrm{d}^4 x \, a^2 \, v^\mu(\tau, x) \left( z^\dagger \frac{\partial z}{\partial x^\mu} - \mathrm{c.\,c.} \right) + \mathcal{O}(\epsilon^2), \end{aligned} \tag{4.40}$$

where we integrated by parts and used $z^\dagger z = 1$. Here

$$v^\mu(\tau, x) \doteq - \left( \frac{\partial \lambda(x, p)}{\partial p_\mu} \right)_{p = k(x)} \tag{4.41}$$

is the zeroth-order (in $\epsilon$) group velocity of the wave. Upon summing Eqs. (4.38) and (4.40), one obtains the action $\mathcal{S}_{\mathrm{XGO}} = \int \mathrm{d}\tau \, L_{\mathrm{XGO}} + \mathcal{O}(\epsilon^2)$, where the Lagrangian is given by

$$L_{\mathrm{XGO}} = \int \mathrm{d}^4 x \, a^2 \left[ \partial_\tau \theta + \lambda(x, k) - \frac{i\epsilon}{2} \left( z^\dagger \frac{\mathrm{d}z}{\mathrm{d}\tau} - \frac{\mathrm{d}z^\dagger}{\mathrm{d}\tau} z \right) + \epsilon z^\dagger \mathcal{U}(x, k) z \right]. \tag{4.42}$$

Here I introduced the convective derivative

$$\frac{\mathrm{d}}{\mathrm{d}\tau} \doteq \frac{\partial}{\partial \tau} + v^\mu(\tau, x) \frac{\partial}{\partial x^\mu}. \tag{4.43}$$

Equation (4.42), along with the definitions in Eqs. (4.30)–(4.32), (4.41), and (4.43), is the main result of this Chapter. The first two terms of the integrand in Eq. (4.42) represent the lowest-order GO Lagrangian. The next two terms are $\mathcal{O}(\epsilon)$ and introduce polarization effects. Since the action (4.42) captures polarization effects, which are not included in GO, this new theory is called *Extended Geometrical Optics* (XGO). It is important to note that diffraction effects are $\mathcal{O}(\epsilon^2)$ (see Sec. 3.3.4) and thus are safe to neglect in this first-order theory. In the remainder of this Chapter, I shall discuss the theoretical consequences of this formalism; e.g., I shall derive the corresponding ELEs. Then, in the next two Chapters, I shall apply the theory to two concrete problems.

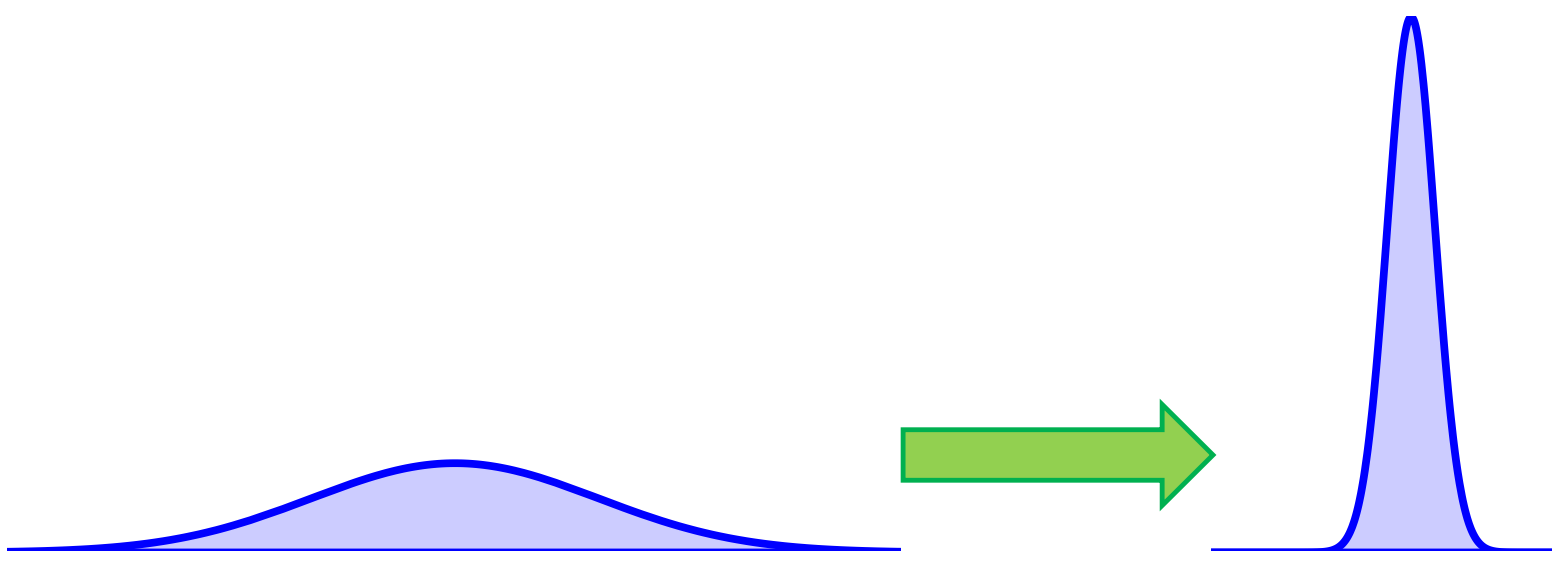

Figure 4.1: In the point-particle model, the wave envelope $\mathcal{I}(\tau, x)$ is localized near some spacetime location $X^\mu(\tau)$.

## 4.5 Geometrical optics revisited

In this Section, I recover the GO approximation. In particular, I derive the point-particle model presented in Sec. 3.3.3.

### 4.5.1 Continuous wave model

To lowest order in $\epsilon$, the Lagrangian (4.42) can be approximated simply with

$$L_{\text{GO}} \doteq \int \mathrm{d}^4x \; a^2 \left[ \partial_\tau \theta + \lambda(x, -\partial\theta) \right]. \tag{4.44}$$

This Lagrangian is parameterized by just two functions, the rapid phase $\theta(\tau, x)$ and the wave amplitude $a(\tau, x)$. Thus, upon varying the action $\mathcal{S}_{\text{GO}} = \int \mathrm{d}\tau \, L_{\text{GO}}$, one obtains the following ELEs:

$$\delta\theta : \quad \partial_\tau a^2 + \partial_\mu (a^2 v^\mu) = 0, \tag{4.45a}$$

$$\delta a^2 : \quad \partial_\tau \theta + \lambda(x, k) = 0. \tag{4.45b}$$

As mentioned in Sec. 4.2, the dynamics of the physical wave propagating in spacetime is obtained by adopting $\partial_\tau \Psi = 0$. Thus, letting $\partial_\tau a^2 = 0$ and $\partial_\tau \theta = 0$ leads to Eqs. (3.25), where the scalar dispersion symbol $D(x, p)$ is replaced by the resonant eigenvalue $\lambda(x, p)$. Thus, one identifies $v^\mu(\tau, x)$ in Eq. (4.41) as the GO four-group-velocity. Similarly as in Sec. 3.3.3, one can then obtain the corresponding fluid equations.

### 4.5.2 Point-particle model

The ray equations corresponding to the previous field equations can be obtained as the point-particle limit of the wave envelope (see Fig. 4.1). In this limit, $a^2(\tau, x)$ can be approximated with a Dirac delta

function in spacetime:

$$a^2(\tau, x) = a_0^2\, \delta^4\big(x - X(\tau)\big). \tag{4.46}$$

Here $a_0^2$ is conserved according to Eq. (4.45a). The value of $a_0^2$ is not essential below so I adopt $a_0^2 = 1$ for brevity.[8]

In this representation, the wave packet is located at the position $X(\tau)$ in spacetime, and the independent parameter is $\tau$.[9] When one inserts Eq. (4.46) into Eq. (4.44), the first term in the action gives the following:

$$\begin{aligned}
\int \mathrm{d}\tau\, \mathrm{d}^4x\, a^2\, \partial_\tau \theta &= \int \mathrm{d}\tau\, \mathrm{d}^4x\, \delta^4\big(x - X(\tau)\big)\, \partial_\tau \theta(\tau, x) \\
&= - \int \mathrm{d}\tau\, \mathrm{d}^4x\, \theta(\tau, x) [\partial_\tau \delta^4\big(x - X(\tau)\big)] \\
&= \int \mathrm{d}\tau\, \mathrm{d}^4x\, \theta(\tau, x) [\dot{X}^\mu(\tau) \partial_\mu \delta^4\big(x - X(\tau)\big)] \\
&= - \int \mathrm{d}\tau\, \mathrm{d}^4x\ \partial_\mu \theta(\tau, x) \dot{X}^\mu(\tau) \delta^4\big(x - X(\tau)\big) \\
&= \int \mathrm{d}\tau\, P_\mu(\tau) \dot{X}^\mu(\tau),
\end{aligned} \tag{4.47}$$

where $P_\mu(\tau) \doteq -\partial_\mu \theta\big(\tau, X(\tau)\big)$. Similarly, the second term in Eq. (4.44) gives

$$\int \mathrm{d}^4x\, \delta^4\big(x - X(\tau)\big) \lambda(x, -\partial\theta) = \lambda\big(X(\tau), P(\tau)\big). \tag{4.48}$$

Thus, the point-particle action is expressed as

$$\mathcal{S}_{\mathrm{GO}} = \int \mathrm{d}\tau\, [\, P(\tau) \cdot \dot{X}(\tau) + \lambda(X, P)\, ]. \tag{4.49}$$

This is a covariant phase-space action, where $X^\mu(\tau) = (X^0, \mathbf{X})$ and $P^\mu(\tau) = (P^0, \mathbf{P})$ serve as canonical position and momentum coordinates, respectively. Also, $\lambda(X, P)$ serves as the ray Hamiltonian.[10] Treating $X(\tau)$ and $P(\tau)$ as independent variables leads to ELEs matching Hamilton's covariant equations:

$$\delta P_\mu : \quad \frac{\mathrm{d}X^\mu}{\mathrm{d}\tau} = -\frac{\partial \lambda}{\partial P_\mu}, \tag{4.50a}$$

$$\delta X^\mu : \quad \frac{\mathrm{d}P_\mu}{\mathrm{d}\tau} = \frac{\partial \lambda}{\partial X^\mu}. \tag{4.50b}$$

[8] As will be shown later, this choice corresponds to considering a single wave quantum, such as an electron, provided that one works with units $\hbar = 1$.

[9] This means that, at a given $\tau$, the wave packet is located at the spatial point $\mathbf{X}(\tau)$ at time $t(\tau)$.

[10] The signs in Eq. (4.49) might seem unusual, but they can be changed by adopting a different signature for the Minkowski metric.

Since the Hamiltonian part $\lambda(X,P)$ does not depend explicitly on $\tau$, then $\mathrm{d}\lambda(X,P)/\mathrm{d}\tau = 0$ along the ray trajectories. Thus, the ray dynamics lies on the dispersion manifold

$$\lambda(X,P) = 0. \tag{4.51}$$

As a reminder, $\lambda(x,p)$ is defined in Eq. (4.33) as the average eigenvalue of the resonant block. Because the action (4.49) is only accurate to $\mathcal{O}(\epsilon^0)$ and the resonant eigenvalues only differ by $\mathcal{O}(\epsilon)$, one can approximate $\lambda(x,p) \simeq \lambda^{(n)}(x,p)$, where $\lambda^{(n)}$ is any particular resonant eigenvalue.

## 4.6 Extended geometrical optics

### 4.6.1 Continuous wave model

To obtain the continuous wave model, I shall reparameterize $z(\tau,x)$ since not all components of the field $z(\tau,x)$ are independent (see Sec. 4.4.2). As in Sec. 4.5.1, I adopt $(\theta,a)$ as independent variables. Since $\psi(\tau,x)$ is a $N$-component complex-valued wave, only $2N-2$ functions can be introduced to reparameterize $z(\tau,x)$. These $2N-2$ functions must also satisfy the normalization condition $z^\dagger(\tau,x)z(\tau,x) = 1$. Following Ruiz and Dodin (2015b), $z(\tau,x)$ is parameterized with the $(N-1)$ spherical angles $\zeta^r(\tau,x)$ on the $N$-dimensional sphere and some $(N-1)$ relative phases $\vartheta^q(\tau,x)$ of individual $z^q(\tau,x)$. With these considerations, the components of $z(\tau,x)$ can be parameterized as

$$z^q(\zeta,\vartheta) = \Phi^q(\zeta^1,\dots,\zeta^{N-1}) \times \begin{cases} e^{-i\vartheta^q} & q<N, \\ 1 & q=N. \end{cases} \tag{4.52}$$

Here $\Phi^q(\zeta^1,\dots,\zeta^{N-1})$ are the well-known real functions parameterizing the location of a point on a unit sphere,

$$\Phi^1 \doteq \cos(\zeta^1), \tag{4.53a}$$

$$\Phi^2 \doteq \sin(\zeta^1)\cos(\zeta^2), \tag{4.53b}$$

$$\Phi^3 \doteq \sin(\zeta^1)\sin(\zeta^2)\cos(\zeta^3), \tag{4.53c}$$

$$\vdots$$

$$\Phi^{N-1} \doteq \sin(\zeta^1)\dots\sin(\zeta^{N-2})\cos(\zeta^{N-1}), \tag{4.53d}$$

$$\Phi^N \doteq \sin(\zeta^1)\dots\sin(\zeta^{N-2})\sin(\zeta^{N-1}), \tag{4.53e}$$

so that $\sum_{q=1}^{N}(\Phi^q)^2 = 1$. Upon substituting Eqs. (4.52) and (4.53) into the Lagrangian (4.42), one obtains

$$L = \int \mathrm{d}^4x \, a^2 \left( \partial_\tau \theta + \lambda(x,k) + \epsilon \langle \mathcal{U} \rangle (x,k,\zeta,\vartheta) + \epsilon \sum_{q=1}^{N-1} (\Phi^q)^2 \frac{\mathrm{d}\vartheta^q}{\mathrm{d}\tau} \right), \tag{4.54}$$

where $\langle \mathcal{U} \rangle (x,k,\zeta,\vartheta) \doteq z^\dagger(\zeta,\vartheta)\mathcal{U}(x,k)\, z(\zeta,\vartheta)$. Note that the variable $\zeta(\tau,x)$ denotes the whole set of $(N-1)$ variables $\zeta^r(\tau,x)$, and similarly for $\vartheta(\tau,x)$.

The independent variables in this formulation are $(\theta, a^2, \zeta, \vartheta)$. This leads to the following ELEs. Varying the action with respect to $\theta$ leads to the action conservation theorem

$$\delta\theta : \quad \partial_\tau a^2 + \partial_\mu (a^2 V^\mu) = 0, \tag{4.55}$$

which is the corrected continuity equation for $a^2(x)$. Note that the corresponding flow velocity is $V^\mu \doteq v^\mu + \epsilon u^\mu + \epsilon w^\mu$, where $v^\mu(\tau,x)$ is the GO group velocity (4.41) and

$$u^\mu(\tau,x) \doteq - \left( \frac{\partial}{\partial p_\mu} \langle \mathcal{U} \rangle (x,k,\zeta,\vartheta) \right)_{p=k(x)}, \tag{4.56a}$$

$$w^\mu(\tau,x) \doteq \sum_{q=1}^{N-1} (\Phi^q)^2 \left( \frac{\partial \lambda(x,p)}{\partial p_\mu \partial p_\nu} \right)_{p=k(x)} \frac{\partial \vartheta^q}{\partial x^\nu}. \tag{4.56b}$$

Notably, one can also recast the latter formula as $w^\mu = M^{-1} \cdot \kappa$, where $M(x,k)$ is understood as the mass tensor of a wave quantum and $\kappa_\mu$ is the wave vector of $z(\tau,x)$; namely,

$$[M^{-1}]^{\mu\nu}(\tau,x) \doteq \left( \frac{\partial \lambda(x,p)}{\partial p_\mu \partial p_\nu} \right)_{p=k(x)}, \tag{4.57a}$$

$$\kappa_\mu(\tau,x) \doteq -(i/2) \left[ z^\dagger (\partial_\mu z) - (\partial_\mu z^\dagger) z \right] = -i z^\dagger \partial_\mu z. \tag{4.57b}$$

Varying the action with respect to $a^2(x)$ leads to a generalization of the Hamilton–Jacobi equation to vector waves:

$$\delta a^2 : \quad \partial_\tau \theta + \lambda(x,k) + \epsilon \langle \mathcal{U} \rangle (x,k,\zeta,\vartheta) + \epsilon \sum_{q=1}^{N-1} (\Phi^q)^2 \frac{\mathrm{d}\vartheta^q}{\mathrm{d}\tau} = 0. \tag{4.58}$$

Another set of $(N-1)$ ELEs is obtained by varying the action with respect to $\zeta^r$:

$$\delta\zeta^r : \quad \frac{\partial}{\partial \zeta^r} \langle \mathcal{U} \rangle (x,k,\zeta,\vartheta) + \sum_{q=1}^{N-1} \frac{\partial (\Phi^q)^2}{\partial \zeta^r} \frac{\mathrm{d}\vartheta^q}{\mathrm{d}\tau} = 0. \tag{4.59}$$

Finally, varying the action with respect to $\vartheta^q$ leads to another set of $(N-1)$ ELEs:

$$\delta\vartheta^q: \quad \frac{\partial}{\partial\tau}[a^2(\Phi^q)^2] + \frac{\partial}{\partial x^\mu}[v^\mu a^2(\Phi^q)^2] = a^2\,\frac{\partial}{\partial\vartheta^q}\langle\mathcal{U}\rangle(x,k,\zeta,\vartheta). \tag{4.60}$$

Each of these equations describes the evolution of the action of an individual ($q$th) mode, $a^2\,(\Phi^q)^2$. By using Eq. (4.55), one can rewrite Eq. (4.60) also as

$$a^2\left(\frac{\partial}{\partial\tau} + V^\mu\frac{\partial}{\partial x^\mu}\right)(\Phi^q)^2 + \frac{\partial}{\partial x^\mu}[a^2(\Phi^q)^2(v^\mu - V^\mu)] = a^2\,\frac{\partial}{\partial\vartheta^q}\langle\mathcal{U}\rangle(x,k,\zeta,\vartheta). \tag{4.61}$$

In the case of a localized wave packet, averaging over the packet volume eliminates the divergence term and predicts the advection of $(\Phi^q)^2$ at the four-velocity $V^\mu(\tau,x)$ with an averaged source term. Moreover, taking the spacetime derivative of Eq. (4.58) leads to the four-momentum conservation equation. As in Sec. 3.3.3, a straightforward calculation leads to

$$\left(\frac{\partial}{\partial\tau} + V^\nu\frac{\partial}{\partial x^\nu}\right)(k_\mu - \epsilon\kappa_\mu) = \frac{\partial\lambda}{\partial x^\mu} + \epsilon\frac{\partial}{\partial x^\mu}\langle\mathcal{U}\rangle(x,k,\zeta,\vartheta) + \mathcal{O}(\epsilon^2). \tag{4.62}$$

As in Sec. 4.5.1, the dynamics of the physical wave propagating in spacetime is obtained by setting $\partial_\tau = 0$ in Eqs. (4.55)–(4.62). Combined together, these $2N$ equations can be viewed as a generalization of the classical spin–fluid equations that were earlier derived for a Pauli particle ($N=2$) (Ruiz and Dodin, 2015c). The generalization consists of the fact that the new equations apply to general waves, both quantum and classical. (This will become more apparent in Sec. 4.6.4.)

### 4.6.2 Covariant point-particle model

The ray equations with polarization effects included can be obtained as a point-particle limit of the Lagrangian (4.42). As in Sec. 4.5.2, the wave envelope is localized to a single point in spacetime [Eq. (4.46)]. Following the discussion surrounding Eqs. (4.52) and (4.53), one also needs to constrain $Z(\tau) \doteq Z\big(\tau, X(\tau)\big)$ so that it is normalized $[Z^\dagger(\tau)Z(\tau) = 1]$, and also the $N$th component of $Z(\tau)$ must be real $[Z^*_N(\tau) = Z^N(\tau)]$. These constraints can be implemented into the action principle by introducing two Lagrange multipliers $(\mu, i\nu)$ so that the resulting point-particle action becomes

$$\mathcal{S}_{\text{XGO}} = \int \mathrm{d}\tau\left[P\cdot\dot{X} - \frac{i}{2}\left(Z^\dagger\dot{Z} - \dot{Z}^\dagger Z\right) + \lambda(X,P) + Z^\dagger\mathcal{U}(X,P)Z + \mu(Z^\dagger Z - 1) + i\nu(Z^*_N - Z^N)\right], \tag{4.63}$$

where the GO ordering parameter $\epsilon$ is dropped. The corresponding ELEs are

$$\delta Z_q^*: \quad i\dot{Z}^q = (\mathcal{U}Z)^q + \mu Z^q + i\delta_q^N \nu, \tag{4.64a}$$

$$\delta Z^q: \quad -i\dot{Z}_q^* = (Z^\dagger \mathcal{U})_q + \mu Z_q^* - i\delta_N^q \nu, \tag{4.64b}$$

$$\delta\mu: \quad Z^\dagger Z - 1 = 0, \tag{4.64c}$$

$$\delta\nu: \quad Z_N^* - Z^N = 0, \tag{4.64d}$$

$$\delta P_\mu: \quad \dot{X}^\mu = -\frac{\partial}{\partial P_\mu}(\lambda + Z^\dagger \mathcal{U} Z), \tag{4.64e}$$

$$\delta X^\mu: \quad \dot{P}_\mu = \frac{\partial}{\partial X^\mu}(\lambda + Z^\dagger \mathcal{U} Z), \tag{4.64f}$$

where $\delta_q^p$ is the Kronecker symbol. The complex conjugate of Eq. (4.64a) must be equal to Eq. (4.64b). Hence, it is seen that both $\mu$ and $\nu$ are real. Combining these two equations leads to

$$\frac{\mathrm{d}}{\mathrm{d}\tau}\left(Z^\dagger Z\right) = 2\nu Z^N. \tag{4.65}$$

However, from Eq. (4.64c), it follows that $\mathrm{d}(Z^\dagger Z)/\mathrm{d}\tau = 0$. Since $Z^N(\tau)$ cannot remain zero identically, one concludes that $\nu = 0$, so the entire vector $Z(\tau)$ satisfies

$$i\dot{Z} = \mathcal{U}Z + \mu Z. \tag{4.66}$$

Notice that the system of ELEs is not closed because there is no independent equation for $\mu$. The issue can be evaded by performing the transformation $Z(\tau) \to Z(\tau)\exp[-i\int^\tau \mathrm{d}\tau'\,\mu(\tau')]$. With this transformation, the equation above becomes

$$i\dot{Z} = \mathcal{U}Z, \tag{4.67}$$

which is the expected polarization-precession equation for $Z(\tau)$. In contrast to the original ELEs, Eqs. (4.64e), (4.64f), and (4.67) *do* form a closed system. It is also seen now that the effect of polarization on the wave ray, or point-particle, dynamics is caused by gradients in phase-space of the polarization-coupling term $Z^\dagger \mathcal{U} Z$. This is similar to the effect of the Stern–Gerlach Hamiltonian on the dynamics of a Pauli particle (Ruiz and Dodin, 2015c).

The resulting equations for the variables $(X, P, Z^\dagger, Z)$ can also be assigned the action

$$\mathcal{S}_{\mathrm{XGO}} = \int \mathrm{d}\tau \left[ P\cdot\dot{X} - \frac{i}{2}\left(Z^\dagger \dot{Z} - \dot{Z}^\dagger Z\right) + \lambda(X,P) + Z^\dagger \mathcal{U}(X,P) Z \right], \tag{4.68}$$

assuming that the initial conditions are restricted to

$$Z^\dagger(\tau)Z(\tau) = 1. \tag{4.69}$$

Hence, the independent variables are $(X, P, Z, Z^\dagger)$, and the corresponding ELEs are

$$\delta P_\mu : \quad \frac{\mathrm{d}X^\mu}{\mathrm{d}\tau} = -\frac{\partial \lambda}{\partial P_\mu} - Z^\dagger \frac{\partial \mathcal{U}}{\partial P_\mu} Z, \tag{4.70a}$$

$$\delta X^\mu : \quad \frac{\mathrm{d}P_\mu}{\mathrm{d}\tau} = \frac{\partial \lambda}{\partial X^\mu} + Z^\dagger \frac{\partial \mathcal{U}}{\partial X^\mu} Z, \tag{4.70b}$$

$$\delta Z^\dagger : \quad \frac{\mathrm{d}Z}{\mathrm{d}\tau} = -i\mathcal{U}Z, \tag{4.70c}$$

$$\delta Z : \quad \frac{\mathrm{d}Z^\dagger}{\mathrm{d}\tau} = iZ^\dagger \mathcal{U}. \tag{4.70d}$$

Together with Eqs. (4.30)–(4.34), Eqs. (4.70) form a complete set of equations. The first terms on the right-hand side of Eqs. (4.70a) and (4.70b) describe the ray dynamics in the GO limit. The second terms describe the coupling to the mode polarization. Note that this generalized polarization-coupling Hamiltonian can contribute to the expression for the particle velocity $\dot{X}^\mu$, so the latter is not necessarily equal to the GO group velocity $-\partial\lambda/\partial P_\mu$. Finally, Eqs. (4.70c) and (4.70d) describe the wave-polarization dynamics, such as the polarization precession.

As in Sec. 4.5.2, the Hamiltonian part of Eq. (4.68) is constant along the ray trajectories. As before, the ray dynamics lies on the dispersion manifold defined by setting the Hamiltonian part to zero; i.e.,

$$\lambda(X, P) + Z^\dagger \mathcal{U}(X, P) Z = 0. \tag{4.71}$$

As a reminder, $\lambda(x, p)$ and $\mathcal{U}$ are defined in Eqs. (4.33) and (4.34), respectively.

### 4.6.3 Non-covariant point-particle model

In certain cases, it is possible to obtain an explicit expression for the wave frequency $P_0$ from the lowest-order GO dispersion relation $\lambda(X, P) = 0$. In other words, one can find the wave frequency as a function of the wave position and momentum:

$$P_0 = H_0(t, \mathbf{X}, \mathbf{P}), \tag{4.72}$$

which is the lowest-order GO dispersion manifold (Tracy *et al.*, 2014). When this is the case, it is possible to simplify the covariant point-particle Lagrangian (4.68) in order to obtain a non-covariant point-particle Lagrangian that depends on a smaller number of independent variables ($\mathbf{X}$, $\mathbf{P}$, $Z$, and $Z^\dagger$ only). Let us define the four-momentum $P_*^\mu(t, \mathbf{X}, \mathbf{P}) \doteq (H_0(t, \mathbf{X}, \mathbf{P}),\ \mathbf{P})$ so that $\lambda(X, P_*) = 0$. Then, after Taylor expanding the XGO dispersion relation, one obtains

$$\left( \frac{\partial \lambda(X,P)}{\partial P_0} \right)_{P=P_*} [P_0 - H_0(t, \mathbf{X}, \mathbf{P})] + Z^\dagger \mathcal{U}(X, P_*) Z = 0, \tag{4.73}$$

where I kept only the lowest-order approximation for $Z^\dagger \mathcal{U}(X,P) Z$ since it is already $\mathcal{O}(\epsilon)$. After solving for $P_0$, one obtains the corrected wave frequency $P_0 = H_{\text{XGO}}(t, \mathbf{X}, \mathbf{P}, Z, Z^\dagger)$, where

$$H_{\text{XGO}}(t, \mathbf{X}, \mathbf{P}, Z, Z^\dagger) \doteq H_0(t, \mathbf{X}, \mathbf{P}) - \left( \frac{\partial \lambda(X,P)}{\partial P_0} \right)^{-1}_{P=P_*} Z^\dagger \mathcal{U}(X, P_*) Z \tag{4.74}$$

is the XGO Hamiltonian. Thus, the polarization-coupling term causes a shift in the wave frequency. Similarly, upon Taylor expanding the action (4.68) to the lowest order, one obtains[11]

$$\mathcal{S}_{\text{XGO}} = \int \mathrm{d}\tau \left[ P \cdot \dot{X} - \frac{i}{2} \left( Z^\dagger \dot{Z} - \text{c.c.} \right) + \left( \frac{\partial \lambda(X,P)}{\partial P_0} \right)_{P=P_*} [P_0 - H_0(t, \mathbf{X}, \mathbf{P})] + Z^\dagger \mathcal{U}(X, P_*) Z \right]. \tag{4.75}$$

Writing the action above explicitly in terms of components leads to

$$\begin{aligned} \mathcal{S}_{\text{XGO}} = \int \mathrm{d}\tau \, P_0 \left[ \dot{X}^0 + \left( \frac{\partial \lambda(X,P)}{\partial P_0} \right)_{P=P_*} \right] \\ - \int \mathrm{d}\tau \left[ \mathbf{P} \cdot \dot{\mathbf{X}} + \frac{i}{2} \left( Z^\dagger \dot{Z} - \text{c.c.} \right) + \left( \frac{\partial \lambda(X,P)}{\partial P_0} \right)_{P=P_*} H_0(t, \mathbf{X}, \mathbf{P}) - Z^\dagger \mathcal{U}(X, P_*) Z \right]. \end{aligned} \tag{4.76}$$

When the action is written in this form, the wave frequency $P_0$ acts as a Lagrange multiplier. Varying the action with respect to $P_0$ leads to $\dot{X}^0 = - \left( \partial \lambda / \partial p_0 \right)_{p=P_*}$. Thus, the $\tau$ parameter and the physical time $t(\tau)$ are related by

$$\mathrm{d}\tau = - \left( \frac{\partial \lambda(X,P)}{\partial P_0} \right)^{-1}_{P=P_*} \mathrm{d}t. \tag{4.77}$$

Up to a negative sign multiplying the action integral, substituting Eq. (4.77) into Eq. (4.76) leads to the non-covariant XGO action

$$\mathcal{S}_{\text{XGO}} = \int \mathrm{d}t \left[ \mathbf{P} \cdot \dot{\mathbf{X}} + \frac{i}{2} \left( Z^\dagger \dot{Z} - \dot{Z}^\dagger Z \right) - H_{\text{XGO}}(t, \mathbf{X}, \mathbf{P}, Z, Z^\dagger) \right]. \tag{4.78}$$

[11]Although the action was only expanded to the first order in $\epsilon$, no accuracy is lost since the action $\mathcal{S}_{\text{XGO}}$ is already $\mathcal{O}(\epsilon)$-accurate only.

(Here the dots denote derivatives with respect to the physical time $t$.) The independent variables of this action are $\mathbf{X}(t)$, $\mathbf{P}(t)$, $Z(t)$, and $Z^\dagger(t)$ only. Varying the action leads to

$$\delta\mathbf{P}: \quad \frac{\mathrm{d}\mathbf{X}}{\mathrm{d}t} = \frac{\partial}{\partial \mathbf{P}} H_{\text{XGO}}, \tag{4.79a}$$

$$\delta\mathbf{X}: \quad \frac{\mathrm{d}\mathbf{P}}{\mathrm{d}t} = -\frac{\partial}{\partial \mathbf{X}} H_{\text{XGO}}, \tag{4.79b}$$

$$\delta Z^\dagger: \quad \frac{\mathrm{d}Z}{\mathrm{d}t} = i\left(\frac{\partial \lambda(X,P)}{\partial P_0}\right)^{-1}_{P=P_*} \mathcal{U} Z, \tag{4.79c}$$

$$\delta Z: \quad \frac{\mathrm{d}Z^\dagger}{\mathrm{d}t} = -i\left(\frac{\partial \lambda(X,P)}{\partial P_0}\right)^{-1}_{P=P_*} Z^\dagger \mathcal{U}. \tag{4.79d}$$

As in the covariant point-particle XGO model, these equations self-consistently describe the coupling between the wave point-particle ray trajectories and the wave polarization dynamics.

### 4.6.4 Precession of the wave spin

Let us also describe the rotation of $Z(\tau)$ as follows. Since $\mathcal{U}(X,P)$ is a traceless Hermitian $N \times N$ matrix, it can be decomposed into a linear combination of $N^2 - 1$ generators $\mathcal{T}_u$ of SU($N$), which are traceless Hermitian matrices, with some real coefficients $-W^u$ (Anisovich *et al.*, 2004):

$$\mathcal{U} = -\sum_{u=1}^{N^2-1} \mathcal{T}_u W^u \equiv -\boldsymbol{\mathcal{T}} \cdot \mathbf{W}. \tag{4.80}$$

Then, I introduce the $(N^2-1)$-dimensional vector

$$\mathbf{S}(\tau) \doteq Z^\dagger(\tau) \boldsymbol{\mathcal{T}} Z(\tau) \tag{4.81}$$

so that $Z^\dagger \mathcal{U} Z = -\mathbf{S}\cdot\mathbf{W}$. The components of $\mathbf{S}(\tau)$ satisfy the following equation:

$$\begin{aligned}
\mathrm{d}_\tau S_w &= Z^\dagger \mathcal{T}_w (\mathrm{d}_\tau Z) + (\mathrm{d}_\tau Z^\dagger)\mathcal{T}_w Z \\
&= i Z^\dagger \mathcal{U}\mathcal{T}_w Z - i Z^\dagger \mathcal{T}_w \mathcal{U} Z \\
&= i Z^\dagger [\mathcal{U}, \mathcal{T}_w] Z \\
&= -i\, Z^\dagger [\mathcal{T}_u, \mathcal{T}_w] Z W^u \\
&= f_{uwv} (Z^\dagger \mathcal{T}^v Z) W^u \\
&= f_{wvu} S^v W^u.
\end{aligned} \tag{4.82}$$

Here $f_{abc}$ are structure constants, which are defined via $[\mathcal{T}_a, \mathcal{T}_b] = i f_{abc} \mathcal{T}^c$ and are antisymmetric in all indices (Anisovich *et al.*, 2004).

As an example, let us consider the case when only two waves are resonant. Then, $N^2 - 1 = 3$, and one can choose $\mathcal{T}^v$ to be the three Pauli matrices divided by two (so $|\mathbf{S}|^2 = 1/2$), and $f_{wuv}$ is the Levi–Civita symbol, so $f_{wvu} S^v W^u = (\mathbf{S} \times \mathbf{W})_w$. For a Dirac electron, which is a special case, such $\mathbf{S}$ is recognized as the spin vector undergoing the precession equation, $\mathrm{d}_\tau \mathbf{S} = \mathbf{S} \times \mathbf{W}$ (see Chapter 5). In optics, this is an equation for the Stokes vector that characterizes the polarization of transverse EM waves propagating in dielectrics (Bliokh *et al.*, 2008; Kravtsov *et al.*, 2007; Ruiz and Dodin, 2015a) and in weakly magnetized plasmas (see Chapter 6).

Hence, it is convenient to extend this quantum terminology also to $N$ resonant waves. I shall call the corresponding $(N^2 - 1)$-dimensional vector $\mathbf{S}$ a generalized *wave-spin vector* and express $f_{wvu} S^v W^u$ symbolically as $(\mathbf{S} * \mathbf{W})_w$, where "$*$" can be viewed as a generalized vector product. Notably, upon using the concept of the spin vector $\mathbf{S}$, one can rewrite Eqs. (4.70c) and (4.70d) as

$$\frac{\mathrm{d}}{\mathrm{d}\tau} \mathbf{S} = \mathbf{S} * \mathbf{W}, \tag{4.83}$$

which is understood as a generalized precession equation.

In the particular case when $\mathbf{S}$ is conserved (I call such waves *pure states*), Eqs. (4.70a), (4.70b), and (4.83) form a closed set of equations, and $\lambda - \mathbf{S} \cdot \mathbf{W}$ serves as an effective scalar Hamiltonian. The dynamics of $Z$ and $Z^\dagger$ do not need to be resolved in this case, so one can rewrite $S_{\text{XGO}}$ as a functional of $(X, P)$ alone:

$$S_{\text{XGO}} = \int \mathrm{d}\tau \, [\, P \cdot \dot{X} + \lambda(X, P) - \mathbf{S} \cdot \mathbf{W}(X, P) \,]. \tag{4.84}$$

A more general case is when $\mathbf{S}$ is close to some eigenvector $\mathbf{w}$ of $\mathbf{W}$ that corresponds to some nondegenerate eigenvalue $\Omega_w$. If $\Omega_w$ is large enough, then $\mathbf{S}(\tau)$ will remain close to $\mathbf{w}(\tau)$ and will only experience small-amplitude oscillations. These oscillations can be understood as a generalized *zitterbewegung* effect and they are transient, i.e., vanish when $\dot{\mathbf{W}}$ becomes zero. In this regime, no mode conversion occurs at $\tau \to \infty$. In contrast, if $\Omega_w$ is not large enough, the change of $\mathbf{S}$ governed by Eq. (4.83) is not necessarily negligible. This corresponds to mode conversion and causes ray splitting at $\tau \to \infty$.[12] This is discussed below.

[12] See, for example, Friedland (1985), Tracy and Kaufman (1993), Flynn and Littlejohn (1994), Tracy *et al.* (2003), Tracy *et al.* (2007), and Richardson and Tracy (2008).

### 4.6.5 Mode conversion as a form of spin precession

Equation (4.70c) [and thus Eq. (4.83)] can also describe mode conversion as it is understood by Friedland *et al.* (1987). This is shown as follows. Let us consider the resonant interaction between two modes as an example; then, $\mathcal{U}(x,p)$ is a $2\times 2$ matrix. From Eq. (4.34), $\mathcal{U}(x,p)$ is Hermitian and traceless and can be represented as

$$\mathcal{U}(x,p) = \begin{pmatrix} \Delta\lambda/2 & U_{12} \\ U_{12}^* & -\Delta\lambda/2 \end{pmatrix}, \tag{4.85}$$

where $\Delta\lambda(x,p) \doteq [\lambda^{(1)} - \lambda^{(2)}]/2 + (U_{11} - U_{22})/2$ and the coefficient $U_{12}$ determines the mode coupling. Suppose that, absent coupling ($U_{12} = 0$), the dispersion curves of two modes cross at some point $(X_\star, P_\star)$. Suppose also that $\Delta\lambda(\tau) = \Delta\lambda\big(X(\tau), P(\tau)\big)$ changes along the ray trajectory approximately linearly in $\tau$. Then, $\Delta\lambda \approx \alpha\tau$, where $\alpha$ is some constant coefficient and one choses the origin on the time axis such that $\Delta\lambda(\tau = 0) = 0$ for simplicity. Similarly, $U_{12}(\tau) \doteq U_{12}\big(X(\tau), P(\tau)\big) \simeq \beta + \gamma\tau$, where $\beta$ and $\gamma$ are some constants. Assuming $\beta$ is sufficiently large, I neglect the term $\gamma\tau$, for it only causes a correction to the dominant effect. Thus, near the mode-conversion region, Eq. (4.70c) is approximately written as

$$i\frac{\mathrm{d}}{\mathrm{d}\tau}\begin{pmatrix} Z_1 \\ Z_2 \end{pmatrix} = \begin{pmatrix} \alpha\tau/2 & \beta \\ \beta^* & -\alpha\tau/2 \end{pmatrix}\begin{pmatrix} Z_1 \\ Z_2 \end{pmatrix}. \tag{4.86}$$

Equation (4.86) is the well-known equation for mode conversion that was studied by Zener (1932). After eliminating $Z_2$, one obtains the governing equation for $Z_1$:

$$\ddot{Z}_1(\tau) + \left(|\beta|^2 + i\frac{\alpha}{2} + \frac{\alpha^2\tau^2}{4}\right) Z_1(\tau) = 0. \tag{4.87}$$

Upon introducing $w \doteq \tau\sqrt{\alpha}\, e^{i\pi/4}$ and $n \doteq -i|\beta|^2/\alpha$, one writes Eq. (4.87) as the Weber equation

$$Z_1''(w) + \left(n + \frac{1}{2} - \frac{w^2}{4}\right) Z_1(w) = 0, \tag{4.88}$$

whose solutions are the parabolic cylinder functions $D_n(w)$. In Flynn and Littlejohn (1994) and Zener (1932), the matrix connecting the wave amplitude entering and exiting the resonance is obtained by analyzing the asymptotics of $D_n(w)$. Specifically,

$$\begin{pmatrix} Z_{1,\mathrm{out}} \\ Z_{2,\mathrm{out}} \end{pmatrix} = \begin{pmatrix} \mathcal{T} & -\mathcal{C}^* \\ \mathcal{C} & \mathcal{T} \end{pmatrix}\begin{pmatrix} Z_{1,\mathrm{in}} \\ Z_{2,\mathrm{in}} \end{pmatrix}, \tag{4.89}$$

where

$$\mathcal{T} = \exp(-\pi|\eta|^2), \qquad \mathcal{C} = -\frac{\sqrt{2\pi\eta}}{\eta\Gamma(-i|\eta|^2)}, \tag{4.90}$$

$\Gamma$ is the Gamma function, and $\eta \doteq \beta/\sqrt{\alpha}$. The transmission and conversion coefficients for the wave quanta are, correspondingly,

$$|\mathcal{T}|^2 = \exp(-2\pi|\beta|^2/|\alpha|), \tag{4.91}$$

$$|\mathcal{C}|^2 = 1 - |\mathcal{T}|^2. \tag{4.92}$$

[Also see Tracy and Kaufman (1993) for a somewhat different approach leading to the same answer.]

This calculation shows that mode conversion, in the way as commonly described in literature, is nothing but a manifestation of the wave-spin precession described by Eqs. (4.70c) and (4.83). Note that the present point-particle model cannot capture ray-splitting because it introduces only one ray for the whole field. However, this theory does predict the transfer of wave quanta, which is a prerequisite for ray-splitting.[13]

### 4.6.6 XGO as a gauge-invariant theory?

This section presents a discussion regarding gauge invariance of the XGO theory. For simplicity, I consider a $N$-component vector wave with degenerate eigenvalues $\lambda^{(q)}(x,p)$ and traceless spin-coupling Hamiltonian $\mathcal{U}(x,p)$. In the point-particle limit, the XGO action is given by Eq. (4.68):

$$\mathcal{S}_{\text{XGO}} = \int \mathrm{d}\tau \left[ P \cdot \dot{X} - \frac{i}{2}\left(Z^\dagger \dot{Z} - \dot{Z}^\dagger Z\right) + \lambda(X,P) + Z^\dagger \mathcal{U}(X,P) Z \right], \tag{4.68 revisited}$$

where $\lambda(x,p)$ is the degenerate eigenvalue of the resonant modes. The spin-coupling Hamiltonian $\mathcal{U}(X,P) = \mathcal{U}_{\text{B}}(X,P) + \mathcal{U}_{\text{NN}}(X,P)$ is given by Eq. (4.35), where

$$\mathcal{U}_{\text{B}}(x,p) = -\left[\Xi^\dagger\{\lambda,\Xi\}\right]_A, \tag{4.93a}$$

$$\mathcal{U}_{\text{NN}}(X,P) = \left[(\partial_p\Xi^\dagger)(D - \lambda\mathbb{I}_N)(\partial_x\Xi)\right]_A. \tag{4.93b}$$

When constructing the matrix $\Xi(x,p)$ in Eq. (4.30), the resonant eigenvectors $\mathbf{e}_q(x,p)$ are only determined up to a phase. Moreover, there is also a certain degree of arbitrariness when choosing the order in which these eigenvectors are located inside $\Xi(x,p)$. Given these facts, it is important to determine if the XGO action (4.68) remains invariant under changes of the phases of the eigenvectors $\mathbf{e}_q(x,p)$ or interchanges of the eigenvectors in the columns of the matrix $\Xi(x,p)$.

[13]For a complete analysis of ray-splitting mode conversion, please refer to Tracy *et al.* (2014), Tracy *et al.* (2003), and Tracy *et al.* (2007).

This can be investigated as follows. Suppose one changes the eigenvector basis by applying some U($N$) transformation. Specifically, let

$$\Xi(x,p) \to \Xi(x,p)M(x,p), \qquad Z \to M^\dagger(X,P)Z(\tau), \tag{4.94}$$

where $M$ is a unitary $N \times N$ complex matrix ($M^\dagger M = \mathbb{I}_N = MM^\dagger$).[14] If $M$ is constant, substituting Eq. (4.94) into Eq. (4.68) leaves the XGO action invariant, so the transformation (4.94) has no effect on the ray dynamics.

More generally, let us consider an arbitrary (yet differentiable) unitary transformation matrix $M(x,p)$. Being an element of a continuous Lie group of U($N$) matrices, $M(x,p)$ can be written as

$$M(x,p) = \exp\left[i\phi(x,p)J\right], \tag{4.95}$$

where $\phi(x,p)$ is a differentiable scalar function and $J$ is a constant $N \times N$ Hermitian matrix. [Since $\Xi(x,p)$ is required to be slow, the function $\phi(x,p)$ is also considered to be slowly varying.] After substituting Eq. (4.95) into Eq. (4.35) and noting that $D(x,p)\Xi(x,p) = \lambda(x,p)\Xi(x,p)$, one obtains

$$\begin{aligned}
\mathcal{U}(x,p) \to & -[M^\dagger\Xi^\dagger\{\lambda,\Xi M\}]_A + \left[\partial_p\,(M^\dagger\Xi^\dagger)(D-\lambda\mathbb{I}_{\overline{N}})\partial_x(\Xi M)\right]_A \\
&= M^\dagger\mathcal{U}M - [M^\dagger\Xi^\dagger\Xi\{\lambda,M\}]_A + \left[M^\dagger(\partial_p\,\Xi^\dagger)(D-\lambda\mathbb{I}_{\overline{N}})\Xi(\partial_x M)\right]_A \\
&\quad + \left[(\partial_p\,M^\dagger)\Xi^\dagger(D-\lambda\mathbb{I}_{\overline{N}})(\partial_x\Xi)M\right]_A + \left[(\partial_p\,M^\dagger)\Xi^\dagger(D-\lambda\mathbb{I}_{\overline{N}})\Xi(\partial_x M)\right]_A \\
&= M^\dagger\mathcal{U}M - [M^\dagger\{\lambda,M\}]_A \\
&= M^\dagger\mathcal{U}M - [M^\dagger\{\lambda,e^{i\phi J}\}]_A \\
&= M^\dagger\mathcal{U}M - [M^\dagger\{\lambda,i\phi\}JM]_A \\
&= M^\dagger\mathcal{U}M - M^\dagger\{\lambda,\phi\}JM \\
&= M^\dagger\mathcal{U}M - \{\lambda,\phi\}J,
\end{aligned} \tag{4.96}$$

where the derivatives of $M(x,p)$ are given by $M' = i\phi' JM$ and I used the fact that $J$ commutes with $M(x,p)$. As shown in Eq. (4.96), the spin-coupling term $\mathcal{U}_{\rm B}(x,p)$ leads to an additional gauge-dependent term. The term $\mathcal{U}_{\rm B}(x,p)$ is, in fact, closely related to the Berry phase (Berry, 1984); for this reason, I assign the subscript "B" to this term. Also note that the $\mathcal{U}_{\rm NN}(x,p)$ does not lead to additional new terms.

[14] U($N$) transformations are considered so that the unitary condition $\Xi^\dagger(x,p)\Xi(x,p) = \mathbb{I}_N$ is satisfied.

Substituting Eqs. (4.94) and (4.96) into Eq. (4.68) leads to

$$\mathcal{S}_{\text{XGO}} = \int \mathrm{d}\tau \left[ P \cdot \dot{X} - \frac{i}{2}\left( Z^\dagger \dot{Z} - \text{c.c.}\right) + \lambda(X,P) + Z^\dagger \mathcal{U}(X,P) Z - \dot{\phi} Z^\dagger J Z - \{\lambda, \phi\} Z^\dagger J Z \right], \tag{4.97}$$

where $\dot{\phi}(X,P) = \dot{X} \cdot \partial_X \phi + \dot{P} \cdot \partial_P \phi$ is the total $\tau$-derivative of $\phi(X,P)$. As shown, the action (4.97) is different from the original XGO action (4.68). Thus, under the continuous $\mathrm{U}(N)$ transformation (4.94), the wave dynamics will depend on the choice of $M(x,p)$. In other words, the dynamics in the current variables $(X, P, Z^\dagger, Z)$ are gauge dependent, which is something one might find problematic.

To fix this problem, one must find a new set of variables so that the XGO action remains invariant under continuous $\mathrm{U}(N)$ transformations. Fortunately, these variables do exist. Following Littlejohn and Flynn (1991), I introduce a new set of canonical variables $(X', P')$ given by[15]

$$X^\mu = X'^\mu - Z^\dagger \left( \Xi^\dagger \frac{\partial \Xi}{\partial P'_\mu} \right)_A Z, \tag{4.98a}$$

$$P^\mu = P'_\mu + Z^\dagger \left( \Xi^\dagger \frac{\partial \Xi}{\partial X'^\mu} \right)_A Z. \tag{4.98b}$$

Note that $|X^\mu - X'^\mu|$ is of the order of the wavelength, so both $X^\mu$ and $X'^\mu$ are equally suitable as ray coordinates. One then substitutes Eqs. (4.98) into Eq. (4.68). The symplectic part of the action becomes

$$\begin{aligned} P \cdot \dot{X} &= \left[ P'_\mu + Z^\dagger \left( \Xi^\dagger \frac{\partial \Xi}{\partial X'^\mu} \right)_A Z \right] \frac{\mathrm{d}}{\mathrm{d}\tau} \left[ X'^\mu - Z^\dagger \left( \Xi^\dagger \frac{\partial \Xi}{\partial P'_\mu} \right)_A Z \right] \\ &\simeq P'_\mu \dot{X}'^\mu + \dot{X}'^\mu Z^\dagger \left( \Xi^\dagger \frac{\partial \Xi}{\partial X'^\mu} \right)_A Z - P'_\mu \frac{\mathrm{d}}{\mathrm{d}\tau} \left[ Z^\dagger \left( \Xi^\dagger \frac{\partial \Xi}{\partial P'_\mu} \right)_A Z \right] \\ &= P' \cdot \dot{X}' + Z^\dagger \left[ \dot{X}' \cdot \left( \Xi^\dagger \frac{\partial \Xi}{\partial X'} \right)_A + \dot{P}' \cdot \left( \Xi^\dagger \frac{\partial \Xi}{\partial P'} \right)_A \right] Z, \\ &= P' \cdot \dot{X}' + Z^\dagger \left( \Xi^\dagger \frac{\mathrm{d}\Xi}{\mathrm{d}\tau} \right)_A Z, \end{aligned} \tag{4.99}$$

where I dropped a second-order term as well as a total-derivative term that is not important once inserted into the action. Note that the second term in Eq. (4.99) involves a total $\tau$ derivative. Similarly, $\lambda(X,P)$ can be approximated as follows:

$$\lambda(X,P) \simeq \lambda(X', P') + Z^\dagger \left[ \Xi^\dagger \{ \lambda, \Xi \} \right]_A Z = \lambda(X', P') - Z^\dagger \mathcal{U}_{\mathrm{B}}(X', P') Z. \tag{4.100}$$

[15] Littlejohn and Flynn (1991) treat the case of a single resonant eigenmode, so only continuous $\mathrm{U}(1)$ transformations are considered. Here, I generalize the main ideas to $N$ resonant eigenmodes and continuous $\mathrm{U}(N)$ transformations.

Hence, in these new variables the action (4.68) is given by

$$\mathcal{S}_{\rm XGO} = \int \mathrm{d}\tau \left[ P' \cdot \dot{X}' + Z^\dagger \left( \Xi^\dagger \frac{\mathrm{d}\Xi}{\mathrm{d}\tau} \right)_A Z - \frac{i}{2} \left( Z^\dagger \dot{Z} - \dot{Z}^\dagger Z \right) + \lambda(X', P') + Z^\dagger \mathcal{U}_{\rm NN}(X', P') Z \right], \tag{4.101}$$

where I approximated $\mathcal{U}(X, P) \simeq \mathcal{U}(X', P')$ since it is already $\mathcal{O}(\epsilon)$. As a reminder, the $\mathcal{U}_{\rm NN}$ term is given in Eq. (4.93b). It is a straightforward to show that the transformation (4.94) leaves the action (4.101) invariant. Hence, the dynamics in the independent variables $(X', P', Z^\dagger, Z)$ are intrinsically gauge invariant and represent physical reality.

Note that the new phase-space variables $(X', P')$ are generally noncanonical. This occurs because of the additional second term appearing in the symplectic part of the action [see Eq. (4.99)]. When varying the action (4.101), the corresponding ELEs are

$$\delta P'_\mu : \quad \frac{\mathrm{d}X'^\mu}{\mathrm{d}\tau} = -\frac{\partial}{\partial P'_\mu} \left[ \lambda + \left( \Xi^\dagger \frac{\mathrm{d}\Xi}{\mathrm{d}\tau} \right)_A + \mathcal{U}_{\rm NN} \right] + \frac{\mathrm{d}}{\mathrm{d}\tau} \left[ Z^\dagger \left( \Xi^\dagger \frac{\partial \Xi}{\partial P'_\mu} \right)_A Z \right], \tag{4.102a}$$

$$\delta X'^\mu : \quad \frac{\mathrm{d}P'_\mu}{\mathrm{d}\tau} = \frac{\partial}{\partial X'^\mu} \left[ \lambda + \left( \Xi^\dagger \frac{\mathrm{d}\Xi}{\mathrm{d}\tau} \right)_A + \mathcal{U}_{\rm NN} \right] - \frac{\mathrm{d}}{\mathrm{d}\tau} \left[ Z^\dagger \left( \Xi^\dagger \frac{\partial \Xi}{\partial X'^\mu} \right)_A Z \right], \tag{4.102b}$$

$$\delta Z^\dagger : \quad \frac{\mathrm{d}Z}{\mathrm{d}\tau} = -i \left[ \left( \Xi^\dagger \frac{\mathrm{d}\Xi}{\mathrm{d}\tau} \right)_A + \mathcal{U}_{\rm NN} \right] Z, \tag{4.102c}$$

$$\delta Z : \quad \frac{\mathrm{d}Z^\dagger}{\mathrm{d}\tau} = i Z^\dagger \left[ \left( \Xi^\dagger \frac{\mathrm{d}\Xi}{\mathrm{d}\tau} \right)_A + \mathcal{U}_{\rm NN} \right]. \tag{4.102d}$$

These are the corresponding dynamical equations for the new variables. When compared to the original ELEs (4.70), Eqs. (4.102) are significantly more complicated.

Within the accuracy of the theory, Eqs. (4.70) and (4.102) are mathematically equivalent. One may wonder though which variables are more convenient for practical applications. The answer is that this depends on the specific applications. For example, the noncanonical variables $(X', P')$ are more convenient for obtaining the correct Bohr–Sommerfield quantization rule for multi-component vector waves (Littlejohn and Flynn, 1991). However, the governing equations for $(X, P)$ are simpler, so the canonical coordinates may be preferable for numerical simulations, such as ray tracing.

## 4.7 Conclusions

Even when diffraction is neglected, the well-known equations of geometrical optics (GO) are not entirely accurate. Traditional GO treats wave rays as classical particles, which are completely described by their position and momentum coordinates. However, vector waves have another degree of freedom, namely, their polarization. Polarization dynamics are manifested in two forms: (i) mode conversion, which is the transfer of wave quanta between resonant eigenmodes and can be understood as the precession of the wave spin; and (ii) polarization-driven bending of ray trajectories, which refers to deviations of the GO ray trajectories arising from first-order corrections to the GO dispersion relation. These effects are easily understood by drawing parallels with quantum mechanics, where similar effects are known as spin rotation and spin–orbital coupling.

In this Chapter, I described a first-principle variational formulation that captures both types of polarization-related effects simultaneously. General linear nondissipative waves were considered. Using the Feynman reparameterization and the Weyl calculus, I obtained a reduced Lagrangian model for such general waves. In contrast with the traditional GO action functional, which is $\mathcal{O}(\epsilon^0)$-accurate in the GO parameter $\epsilon$, the XGO action is $\mathcal{O}(\epsilon)$-accurate. In this procedure, polarization effects are contained in the $\mathcal{O}(\epsilon)$ corrections in the XGO action. In the next two Chapters, examples of concrete applications of the XGO theory are presented.

From the theoretical standpoint, the work presented in this Chapter could be extended in several directions. First, XGO considers a wave propagating in a flat spacetime Minkowski metric. It would be interesting to investigate polarization effects on waves propagating in curved spacetime manifolds (or as a special case, curvilinear coordinates). In this regard, concepts of differential geometry or the basic machinery proposed by Dodin (2014a) could be useful to tackle this problem. This theory could be useful for studying waves in media where general-relativity effects are important, for example, electromagnetic waves propagating in the vicinity of black holes.

A second avenue of future research could be extended the present theory in order to describe dissipative waves. This problem could perhaps be solved by using the recent technique proposed by Dodin *et al.* (2017), where dissipation is included into variational principles for linear waves. If such formulation is developed, this could perhaps be useful to obtain a reduced classical model for the spin-1/2 electron with radiation-damping effects included.

A third project of interest could be to further expand the XGO theory in order to include wave diffraction. Roughly speaking, this theory theory could be obtained by calculating the second-order corrections in $\epsilon$ of the wave action function. As discussed in Chapter 2, diffraction effects are expected to appear in the second-order terms of the action functional. This theory could be applied to describe optical vortex beams propagating through inhomogeneous dielectric media (Bliokh, 2006). It is expected that the intrinsic angular momentum of such beams, as well as the wave polarization, will cause a deviation of the GO ray trajectories. The proposed theory could also be useful in order to include diffraction effects in ray tracing codes for radio-frequency waves in tokamak plasmas.

# Chapter 5

# Point-particle Lagrangian model for the Dirac particle

In this Chapter, I apply the XGO theory to study the dynamics of quantum particles. Specifically, I obtain the first-ever point-particle Lagrangian model for the relativistic spin-1/2 particle. The starting point of the theory is the first-principle Dirac action that describes relativistic spin-1/2 quantum particles. By applying the procedure in Chapter 4, I obtain a point-particle phase-space Lagrangian, whose Hamiltonian captures spin dynamics such as the spin precession and the Stern–Gerlach (SG) force. The model is then compared with previously obtained classical theories for spinning particles. The results of this Chapter are published by Ruiz and Dodin (2015b).

## 5.1 Introduction

### 5.1.1 Motivation

Spinning particles can couple to external EM fields via their charge and their spin. In addition to the Lorentz force, spinning particles can show additional dynamics that are related to spin, such as the spin precession and the SG force (Jackson, 1999). In recent years, there has been a growing interest in different areas of physics to study particle-spin effects; these include condensed matter (Dyakonov and Perel, 1971; Hirsch, 1999), astrophysics (Mahajan *et al.*, 2015), and quantum plasma physics (Marklund and Brodin, 2007; Brodin *et al.*, 2008a,b). Particularly, in particle-accelerator physics, spin effects are essential for obtaining spin-polarized particle beams in storage rings (Niinikoski and Rossmanith, 1987; Conte *et al.*, 1995; Hoffstaetter, 2009).

For describing relativistic spin-1/2 particles, the model of excellence is the Dirac equation (Dirac, 1928). The key element of this quantum-mechanical formalism is a complex-valued, four-component wave function that simultaneously describes particle and anti-particle states (Thaller, 1992). The Dirac equation plays an important role in quantum electrodynamics, which is perhaps the most rigorously experimentally tested theory of physics (Odom *et al.*, 2006). Still, ODE-based classical reductions of this equation (point-particle models) have been attracting substantial interest due to their simplicity.[1]

In this Chapter, I apply the XGO formulation to obtain a classical model for the relativistic spin-1/2 particle. The main result is given in Eqs. (5.32)–(5.35). This variational formulation captures both the particle spin precession and the relativistic SG force. By considering the results of this Chapter as a specific application of the XGO theory, one sees that the particle spin-precession is simply the polarization precession for vector waves. Likewise, the SG force on spin-1/2 particles is nothing but the polarization-driven bending of ray trajectories in the XGO theory.

When one compares this model to previously developed classical models, one finds that the point-particle Lagrangian model obtained here agrees with the Frenkel model (Frenkel, 1926) in the limit of low energies but differs in the relativistic-energy limit.[2] Since the Frenkel model is based on physical considerations only and was not derived from a first-principle model such as the Dirac equation, it is not necessary to further discuss possible causes for the discrepancies.

As an application of XGO theory, the formulation is related to the many existing semiclassical GO descriptions of the Dirac particle [for instance, Pauli (1932), Rubinow and Keller (1963), Bolte and Keppeler (1998), and Spohn (2000)], but it is still different in several aspects. First, previous semiclassical calculations obtain the correct Bargmann–Michel–Telegdi (BMT) spin-precession equation (Bargmann *et al.*, 1959) but do not capture the SG force in the equations of motion for the particle orbit. This discrepancy occurs because these studies consider the SG force as a higher-order effect when compared to the spin precession. However, based on the analysis in Sec. 5.3, this argument is erroneous because both the spin precession and the SG force originate from the same spin-coupling term in the Hamiltonian (5.33). Second, this theory is Lagrangian; hence, it introduces a classical theory unambiguously [see also Ruiz and Dodin (2015c)] and keeps the equations manifestly conservative.

Since the model presented here is manifestly Lagrangian, the theory is reminiscent of that by Barut and Zanghi (1984), Barut and Pavsic (1987), and Barut and Cruz (1993). Both theories are deduced from the Dirac model, use the concept of the Feynmann parameterization (Sec. 4.2.2), and apply the point-particle limit (Sec. 4.6.2). However, in contrast to the present theory, Barut *et al.*'s theory is formulated in terms of bispinors, as opposed to spinors.[3] This might be an unnecessary complication because a general bispinor describes the superposition of particle and antiparticle states (Barut and Zanghi, 1984). It seems that Barut *et al.* apply the point-particle limit *before* the projection to the particle states is made; in other words, Barut *et al.* locate wave quanta on both leading-order dispersion manifolds (5.15). For this reason, Barut *et al.*'s model includes the *Zitterbewegung* effect (Barut and Cruz, 1994), which is simply mode conversion between the particle and anti-particle states. It must be noted that the procedure of Barut *et al.* may be valid for describing particles in extremely intense EM fields, where pair creation occurs rather frequently. However, for slowly-varying EM fields, this procedure is not valid because particles and antiparticles respond differently to external fields and do not travel as a whole. The present theory avoids these issues altogether by being explicitly restricted to pure (i.e., particle *or* antiparticle) states.

[1] For more information, see Gaioli and Garcia Alvarez (1998), Ternov and Bordovitsyn (1980), and Rohrlich (1973) for reviews.

[2] For the interested reader, the Frenkel model is reviewed in Wen *et al.* (2016) using modern notation.

### 5.1.2 Overview

This Chapter is organized as follows. In Sec. 5.2, a brief introduction to the Dirac formalism for the relativistic spin-1/2 is presented. In Sec. 5.3, the Dirac dispersion operator is block-diagonalized using the general procedure given in Chapter 4. In Sec. 5.4, effects due to the anomalous magnetic moment are added to the theory. In Sec. 5.5, the point-particle Lagrangian model is derived, and the corresponding equations of motion are discussed. In Sec. 5.6, the developed model is compared to yet another model, which originates from the Foldy–Wouthuysen transformation. In Sec. 5.7, the main results are summarized and future work is discussed. Some auxiliary calculations are presented in Appendix C.1 of the thesis.

## 5.2 Basic model

### 5.2.1 The Dirac action

As for any quantum particle or nondissipative wave, the dynamics of a Dirac particle is governed by the principle of stationary action $\delta\mathcal{S} = 0$. In the Dirac notation, the action functional for the Dirac particle is (Peskin and Schroeder, 1995)

$$\mathcal{S} = \langle\, \bar{\Psi} \mid (\widehat{\not{p}} - q\widehat{\not{A}} - m\mathbb{I}_4) \mid \Psi \,\rangle\,, \tag{5.1}$$

where $\Psi(x) \doteq \langle\, x \mid \Psi \,\rangle$ is a complex-valued four-component particle wave function (also called a Dirac spinor), $\langle\, \bar{\Psi} \mid x \,\rangle = \bar{\Psi}(x) \doteq \Psi^\dagger(x)\gamma^0$ is the "Dirac conjugate" of $\Psi(x)$, and the particle mass and charge are denoted by $m$ and $q$, respectively. The Feynman slash notation is used: $\not{a} \doteq a_\mu \gamma^\mu$, where $\gamma^\mu = (\gamma^0, \boldsymbol{\gamma})$ are the

[3] In the theory of Barut *et al.*, the complex valued $Z(\tau)$ variables are composed of four components, whereas in the present theory, $Z(\tau)$ is two dimensional.

Dirac matrices (defined below). As in Sec. 2.2.2, $\widehat{p}_\mu$ is the four-momentum operator, and $\widehat{A}^\mu = A^\mu(\widehat{x})$ is the four-potential operator describing the interaction of the particle with the surrounding EM field. The components of the four-potential are given by $A^\mu = (V, \mathbf{A})$, where $V(x)$ is the electrostatic potential and $\mathbf{A} = \mathbf{A}(x)$ is the vector potential. The Dirac matrices $\gamma^\mu$ are $4 \times 4$ matrices satisfying the condition

$$\gamma^\mu \gamma^\nu + \gamma^\nu \gamma^\mu = 2 g^{\mu\nu} \mathbb{I}_4, \tag{5.2}$$

where $g^{\mu\nu} = \text{diag}[+1, -1, -1, -1]$ is the Minkowski metric tensor. Hence,

$$\not{a}\not{b} + \not{b}\not{a} = 2(a \cdot b)\mathbb{I}_4, \qquad \not{a}\not{a} = a^2 \mathbb{I}_4, \tag{5.3}$$

for any pair of four-vectors $a^\mu$ and $b^\mu$. There are different representations of the Dirac matrices that satisfy Eq. (5.2). In this thesis, the standard representation will be used; namely,

$$\gamma^0 = \begin{pmatrix} \mathbb{I}_2 & 0 \\ 0 & -\mathbb{I}_2 \end{pmatrix}, \qquad \boldsymbol{\gamma} = \begin{pmatrix} 0 & \boldsymbol{\sigma} \\ -\boldsymbol{\sigma} & 0 \end{pmatrix}, \tag{5.4}$$

where $\boldsymbol{\sigma} \doteq (\sigma_x, \sigma_y, \sigma_z)$ are the $2 \times 2$ Pauli matrices. These matrices satisfy

$$(\gamma^\mu)^\dagger = \gamma^0 \gamma^\mu \gamma^0. \tag{5.5}$$

Upon substituting the Dirac conjugate $\langle \bar{\Psi} | = \langle \Psi | \gamma^0$, one can write the action (5.1) as

$$\mathcal{S} = \langle \Psi | \widehat{\mathfrak{D}} | \Psi \rangle, \tag{5.6}$$

where $\widehat{\mathfrak{D}} \doteq \widehat{p}_0 \mathbb{I}_4 - \widehat{\mathcal{H}}$ is the dispersion operator,

$$\widehat{\mathcal{H}} \doteq \boldsymbol{\alpha} \cdot (\widehat{\mathbf{p}} - q\widehat{\mathbf{A}}) + m\beta + q\widehat{V}\, \mathbb{I}_4 \tag{5.7}$$

is the Hamiltonian operator, $\boldsymbol{\alpha} \doteq \gamma^0 \boldsymbol{\gamma}$ are $4 \times 4$ Hermitian matrices, and $\beta \doteq \gamma^0$. As in Sec. 2.2.2, varying the action (5.6) with respect to $\langle \Psi |$ leads to

$$\widehat{p}_0 | \Psi \rangle = [\, \boldsymbol{\alpha} \cdot (\widehat{\mathbf{p}} - q\widehat{\mathbf{A}}) + m\beta + q\widehat{V}\, \mathbb{I}_4 \,] | \Psi \rangle\,. \tag{5.8}$$

Projecting this equation onto the position eigenstates leads to the well-known Dirac equation

$$i\partial_t \Psi(x) = \{\, \boldsymbol{\alpha} \cdot [-i\boldsymbol{\nabla} - q\mathbf{A}(x)] + m\beta + qV(x)\,\mathbb{I}_4 \,\}\, \Psi(x). \tag{5.9}$$

In the literature, the action (5.1) is more commonly used than the action (5.6), as it is manifestly Lorentz-covariant. However, the action (5.6) has the advantage that its dispersion operator is written in the symplectic form which is convenient for determining particle-spin effects. In what follows, the XGO theory presented in Chapter 4 will be applied to determine the point-particle Lagrangian for the relativistic spin-1/2 particle.

### 5.2.2 Extended action principle

Having obtained the dispersion operator $\widehat{\mathcal{D}} \doteq \widehat{p}_0 \mathbb{I}_4 - \widehat{\mathcal{H}}$ that governs the particle dynamics, let us construct the principle of stationary action from which the XGO equations can be obtained. As in Sec. 4.2, the action in the extended space is given by

$$\mathcal{S}_{\mathrm{X}} \doteq \int \mathrm{d}\tau \; L, \tag{5.10}$$

where $L \doteq L_\tau + L_D$ serves as the Lagrangian and

$$L_\tau \doteq -(i/2)\, [\langle\, \Psi(\tau) \mid \partial_\tau \Psi(\tau) \,\rangle - \mathrm{c.\,c.}]\,, \tag{5.11a}$$

$$L_D \doteq \langle\, \Psi(\tau) \mid \widehat{\mathcal{D}} \mid \Psi(\tau) \,\rangle\,. \tag{5.11b}$$

As a reminder, here $\Psi(\tau, x)$ depends not only on spacetime but also on some parameter $\tau$, so $\Psi(\tau, x) = \langle\, x \mid \Psi(\tau) \,\rangle$ and $\partial_\tau \Psi(\tau, x) = \langle\, x \mid \partial_\tau \Psi(\tau) \,\rangle$. Also, the abstract vector state $|\, \Psi(\tau) \,\rangle$ belongs to the same Hilbert space with inner product defined by Eq. (4.6).

## 5.3 Block-diagonalizing the dispersion operator

As in Secs. 4.3 and 4.4, let us block-diagonalize the dispersion operator. Upon using the Weyl transformation (A.1) along with the Moyal product (A.5), one obtains the Weyl symbol of the dispersion operator:

$$D(x, p) \doteq p_0 \mathbb{I}_4 - H(t, \mathbf{x}, \mathbf{p}), \tag{5.12}$$

where the Weyl symbol of the Hamiltonian operator $\widehat{\mathcal{H}}$ is given by

$$H(t,\mathbf{x},\mathbf{p}) \doteq \boldsymbol{\alpha}\cdot\boldsymbol{\pi} + m\beta + qV(t,\mathbf{x})\,\mathbb{I}_4 \tag{5.13}$$

and $\boldsymbol{\pi}(t,\mathbf{x},\mathbf{p}) \doteq \mathbf{p} - q\mathbf{A}(t,\mathbf{x})$ is the particle kinetic momentum.

As in Sec. 4.3.3, let us now identify the eigenvalues and eigenmodes of the dispersion symbol $D(x,p)$. Since $D(x,p)$ is a $4\times 4$ Hermitian matrix, it has four real eigenvalues which are given by

$$\lambda^{(1,2)}(x,p) = p_0 - \sqrt{\boldsymbol{\pi}^2 + m^2} - qV, \tag{5.14a}$$

$$\lambda^{(3,4)}(x,p) = p_0 + \sqrt{\boldsymbol{\pi}^2 + m^2} - qV. \tag{5.14b}$$

One sees that the Weyl symbol $D(x,p)$ has two doubly degenerate eigenvalues. The eigenvalues $\lambda^{(1,2)}(x,p)$ correspond to the particles states, and the eigenvalues $\lambda^{(3,4)}(x,p)$ correspond to the antiparticle states. Upon setting the eigenvalues to zero, one obtains the lowest-order GO frequencies:

$$p^0_{\pm}(t,\mathbf{x},\mathbf{p}) = \pm\sqrt{\boldsymbol{\pi}^2 + m^2} + qV, \tag{5.15}$$

As shown in Fig. 5.1, the difference between the two frequencies is equal to or greater than twice the particle mass; i.e., $p^0_+ - p^0_- = 2\varepsilon \geq 2m$. Physically, this is understood as the minimal energy needed for particle/anti-particle pair creation. In the following, I shall consider EM fields that are weak enough such that pair creation can be neglected. In this regime, the particle and antiparticle eigenmodes can be decoupled (Foldy and Wouthuysen, 1950; Chen and Chiou, 2010, 2014), and the theory presented in Chapter 4 can be applied.

Before we continue, let us first identify the GO parameter for this problem. From Eq. (5.15), the relevant frequency scale for the particle dynamics is $\omega = 2mc^2/\hbar$, which is the Compton frequency.[4] Similarly, the wavevector associated to the particle dynamics is given by the de Broglie relation, $\boldsymbol{\pi} = \hbar\mathbf{k}$. Hence, the asymptotic analysis presented in Sec. 4.3 can be applied as long as the characteristic timescales $T$ and length scales $\ell$ of the four-potential field $A^\mu(x)$ satisfy the following condition:

$$\epsilon = \max\left\{\frac{\hbar}{2mc^2 T}, \frac{\hbar}{|\boldsymbol{\pi}|\ell}\right\} \ll 1. \tag{5.16}$$

In what follows, I assume that this condition is satisfied.

Let us consider only the dynamics of the particle states. From Sec. 4.4, the matrix $\Lambda(x,p)$ is given by $\Lambda(x,p) = (\pi_0 - \varepsilon)\mathbb{I}_2$, where $\pi_0(t,\mathbf{x},p_0) \doteq p_0 - qV(t,\mathbf{x})$ and $\varepsilon(t,\mathbf{x},\mathbf{p}) \doteq (\boldsymbol{\pi}^2 + m^2)^{1/2}$. For the two particle states, the corresponding matrix $\Xi(t, \mathbf{x}, \mathbf{p})$ [see Eq. (4.30)] is a $4 \times 2$ matrix given by

$$\Xi(t, \mathbf{x}, \mathbf{p}) \doteq \sqrt{\frac{m+\varepsilon}{2\varepsilon}} \begin{pmatrix} \mathbb{I}_2 \\ \frac{\boldsymbol{\pi}\cdot\boldsymbol{\sigma}}{m+\varepsilon} \end{pmatrix}. \tag{5.17}$$

[4] For this discussion, I restored the $c$ and $\hbar$ constants for simplicity.

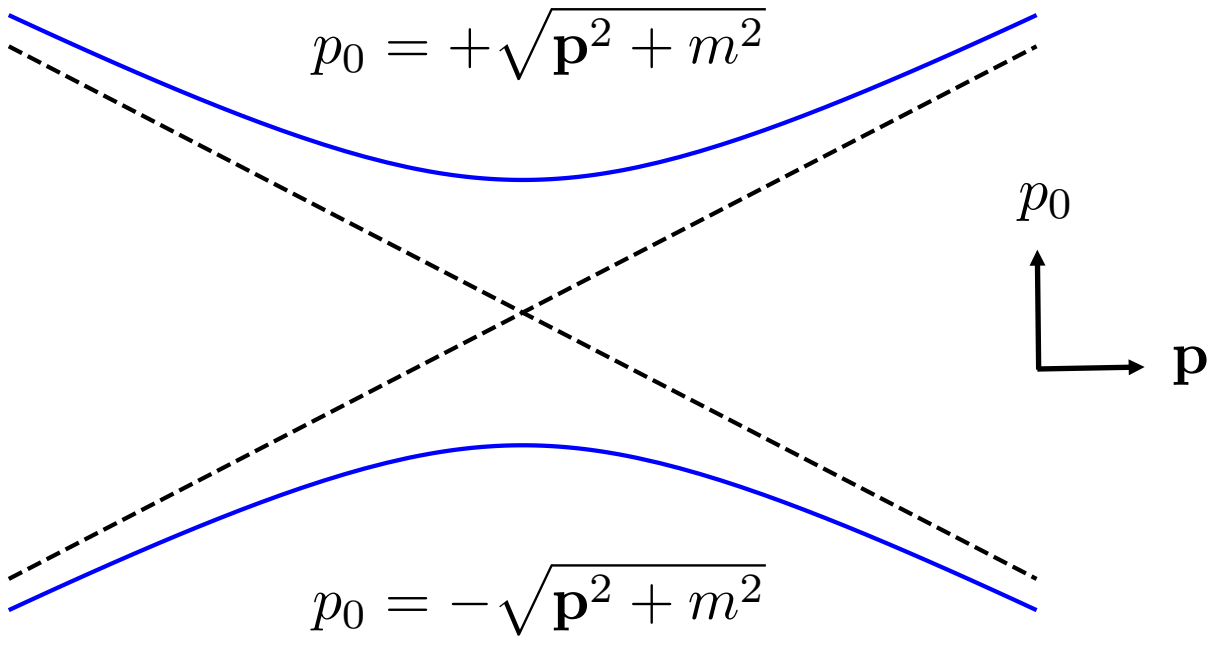

Figure 5.1: Diagram showing the energy values corresponding to the Dirac particle in vacuum.

It is easy to verify that $\Xi(x, \mathbf{p})$ diagonalizes the dispersion Weyl symbol:

$$\begin{aligned}
\Xi^\dagger(t, \mathbf{x}, \mathbf{p}) D(x, p) \Xi(t, \mathbf{x}, \mathbf{p}) &= \frac{m+\varepsilon}{2\varepsilon} \begin{pmatrix} \mathbb{I}_2 & \frac{\boldsymbol{\sigma}\cdot\boldsymbol{\pi}}{m+\varepsilon} \end{pmatrix} \begin{pmatrix} p_0 - m - qV & -\boldsymbol{\sigma}\cdot\boldsymbol{\pi} \\ -\boldsymbol{\sigma}\cdot\boldsymbol{\pi} & p_0 + m - qV \end{pmatrix} \begin{pmatrix} \mathbb{I}_2 \\ \frac{\boldsymbol{\sigma}\cdot\boldsymbol{\pi}}{m+\varepsilon} \end{pmatrix} \\
&= \frac{m+\varepsilon}{2\varepsilon} \begin{pmatrix} \mathbb{I}_2 & \frac{\boldsymbol{\sigma}\cdot\boldsymbol{\pi}}{m+\varepsilon} \end{pmatrix} \begin{pmatrix} \left(p_0 - m - qV - \frac{\boldsymbol{\pi}^2}{m+\varepsilon}\right) \mathbb{I}_2 \\ -\boldsymbol{\sigma}\cdot\boldsymbol{\pi} + \frac{\boldsymbol{\sigma}\cdot\boldsymbol{\pi}}{m+\varepsilon}(p_0 + m - qV) \end{pmatrix} \\
&= \frac{m+\varepsilon}{2\varepsilon} \left( p_0 - m - qV - \frac{2\boldsymbol{\pi}^2}{m+\varepsilon} + \frac{\boldsymbol{\pi}^2}{(m+\varepsilon)^2}(p_0 + m - qV) \right) \mathbb{I}_2 \\
&= \frac{1}{2\varepsilon(m+\varepsilon)} \left[ (p_0 - m - qV)(m+\varepsilon)^2 + \boldsymbol{\pi}^2 (p_0 - m - qV - 2\varepsilon) \right] \mathbb{I}_2 \\
&= \frac{1}{m+\varepsilon} \left[ (p_0 - m - qV)(m+\varepsilon) - \boldsymbol{\pi}^2 \right] \mathbb{I}_2 \\
&= (p_0 - \varepsilon - qV) \mathbb{I}_2 \\
&= \Lambda(x, p).
\end{aligned} \tag{5.18}$$

Similarly, it can also be shown that $\Xi(t, \mathbf{x}, \mathbf{p})$ is normalized so that $\Xi^\dagger(t, \mathbf{x}, \mathbf{p}) \Xi(t, \mathbf{x}, \mathbf{p}) = \mathbb{I}_2$.

Having obtained $\Xi(t, \mathbf{x}, \mathbf{p})$, I now proceed to calculate the spin-coupling term $\mathcal{U}(x, p)$. Since the eigenvalues of the particle states are degenerate, one may use the expression for the spin-coupling term given by

Eq. (4.35):

$$\mathcal{U}(x,p) = \left(-\frac{\partial \lambda}{\partial p_\mu}\right)\left(\Xi^\dagger \frac{\partial \Xi}{\partial x^\mu}\right)_A + \left(\frac{\partial \lambda}{\partial x^\mu}\right)\left(\Xi^\dagger \frac{\partial \Xi}{\partial p_\mu}\right)_A + \left(\frac{\partial \Xi^\dagger}{\partial p_\mu}(D - \lambda \mathbb{I}_4)\frac{\partial \Xi}{\partial x^\mu}\right)_A, \tag{5.19}$$

where $\lambda(x,p) \doteq \pi_0(t,\mathbf{x},p_0) - \varepsilon(t,\mathbf{x},\mathbf{p})$. A straightforward calculation of the spin-coupling matrix $\mathcal{U}(x,p)$ leads to (Appendix C.1)

$$\mathcal{U}(t,\mathbf{x},\mathbf{p}) = \frac{1}{2\varepsilon}\left(\mathbf{B} - \frac{\boldsymbol{\pi}\times\mathbf{E}}{m+\varepsilon}\right)\cdot\boldsymbol{\sigma}. \tag{5.20}$$

Thus, the $2\times 2$ effective dispersion symbol (4.32) that governs the dynamics of the particle states is

$$[[D_{\text{eff}}]](x,p) = (p_0 - \varepsilon - qV)\,\mathbb{I}_2 + \frac{1}{2\varepsilon}\left(\mathbf{B} - \frac{\boldsymbol{\pi}\times\mathbf{E}}{m+\varepsilon}\right)\cdot\boldsymbol{\sigma}. \tag{5.21}$$

The first term describes the dynamics of the relativistic spinless particle, and the second term provides a correction due to the particle spin. As will be shown below, this term describes spin effects such as the SG force and the spin precession. It is worth noting that the spin-coupling term $\mathcal{U}(t,\mathbf{x},\mathbf{p})$ depends explicitly on the electric and magnetic fields. Hence, $\mathcal{U}(t,\mathbf{x},\mathbf{p})$ scales as the inverse of the characteristic time and length scales of the four-potential $A^\mu(x)$. In other words, $\mathcal{U}(t,\mathbf{x},\mathbf{p})$ is $\mathcal{O}(\epsilon)$ as expected.

## 5.4 Anomalous magnetic moment

Before analyzing the dynamics generated by $[[D_{\text{eff}}]](x,p)$, I shall first include a correction to the action (5.1) due to the particle's anomalous magnetic moment. This correction is included by adding to the action (5.1) the Pauli term (Gaioli and Garcia Alvarez, 1998; Salamin, 1993)

$$\mathcal{S}_{\text{anom}} \doteq -\langle \bar{\Psi} \mid \frac{q}{4m}\left(\frac{g}{2}-1\right)\sigma^{\mu\nu}F_{\mu\nu}(\widehat{x}) \mid \Psi \rangle, \tag{5.22}$$

where $g$ is a scalar constant known as the $g$ factor. The $g$ factor characterizes the magnetic moment of a spin-1/2 particle; e.g., $g \simeq 2.002319$ for the electron. Also, $\sigma^{\mu\nu} \doteq (i/2)[\gamma^\mu,\gamma^\nu]$ is the relativistic spin operator, and $F_{\mu\nu}(x) \doteq \partial_\mu A_\nu(x) - \partial_\nu A_\mu(x)$ is the EM field tensor. A brief calculation of the term $\sigma^{\mu\nu}F_{\mu\nu}$ leads to

$$\sigma^{\mu\nu}F_{\mu\nu} = 2\begin{pmatrix} -\boldsymbol{\sigma}\cdot\mathbf{B} & i\boldsymbol{\sigma}\cdot\mathbf{E} \\ i\boldsymbol{\sigma}\cdot\mathbf{E} & -\boldsymbol{\sigma}\cdot\mathbf{B}\end{pmatrix}. \tag{5.23}$$

As in the previous section, I introduce a transformation such that $|\,\Psi\,\rangle = \widehat{\Xi}\,|\,\psi\,\rangle$, where $\widehat{\Xi}$ is the operator corresponding to the Weyl symbol $\Xi(x,\mathbf{p})$. Then, $\mathcal{S}_{\text{anom}}$ becomes

$$\mathcal{S}_{\text{anom}} = -\,\langle\,\psi\,|\,\widehat{\Xi}^{\dagger}\frac{q}{4m}\left(\frac{g}{2}-1\right)\beta\sigma^{\mu\nu}F_{\mu\nu}(\widehat{x})\widehat{\Xi}\,|\,\psi\,\rangle\,. \tag{5.24}$$

Upon rewriting $\mathcal{S}_{\text{anom}}$ in the phase-space representation, one obtains

$$\mathcal{S}_{\text{anom}} = \operatorname{Tr}\int \mathrm{d}^4x\,\mathrm{d}^4p\,D_{\text{anom}}(t,\mathbf{x},\mathbf{p})W_{\psi}(x,p), \tag{5.25}$$

where the Weyl symbol $D_{\text{anom}}(t,\mathbf{x},\mathbf{p})$ is obtained using the Moyal product (A.5):

$$D_{\text{anom}}(t,\mathbf{x},\mathbf{p}) \doteq -\Xi^{\dagger}(t,\mathbf{x},\mathbf{p})\star\frac{q}{4m}\left(\frac{g}{2}-1\right)\beta\sigma^{\mu\nu}F_{\mu\nu}(x)\star\Xi(t,\mathbf{x},\mathbf{p}). \tag{5.26}$$

Note that the EM tensor is given by $F_{\mu\nu}(x) \doteq \partial_{\mu}A_{\nu}(x) - \partial_{\nu}A_{\mu}(x)$. Since I only consider smoothly varying EM fields that satisfy the condition (5.16), $F_{\mu\nu}(x)$ is a matrix of $\mathcal{O}(\epsilon)$ like the spin-coupling Hamiltonian (5.20). Thus, to the leading order in $\epsilon$ the Moyal products can be simplified and replaced by simple matrix products, so $D_{\text{anom}}(t,\mathbf{x},\mathbf{p})$ can be approximated as

$$D_{\text{anom}}(t,\mathbf{x},\mathbf{p}) = -\frac{q}{4m}\left(\frac{g}{2}-1\right)\Xi^{\dagger}(t,\mathbf{x},\mathbf{p})\beta\sigma^{\mu\nu}F_{\mu\nu}(x)\Xi(t,\mathbf{x},\mathbf{p}) + \mathcal{O}(\epsilon^2). \tag{5.27}$$

A simple calculation leads to (Appendix C.1)

$$D_{\text{anom}}(t,\mathbf{x},\mathbf{p}) = \frac{q}{2m}\left(\frac{g}{2}-1\right)\left(\mathbf{B} - \frac{\boldsymbol{\pi}\times\mathbf{E}}{\varepsilon} - \frac{(\boldsymbol{\pi}\cdot\mathbf{B})\boldsymbol{\pi}}{\varepsilon(m+\varepsilon)}\right)\cdot\boldsymbol{\sigma} + \mathcal{O}(\epsilon^2). \tag{5.28}$$

Now, I add the anomalous correction $D_{\text{anom}}(t,\mathbf{x},\mathbf{p})$ to the spin-coupling Hamiltonian $\mathcal{U}(t,\mathbf{x},\mathbf{p})$ in Eq. (5.20). The total spin-coupling term $\mathcal{U}_{\text{tot}}(t,\mathbf{x},\mathbf{p}) \doteq \mathcal{U}(t,\mathbf{x},\mathbf{p}) + D_{\text{anom}}(t,\mathbf{x},\mathbf{p})$ is given by

$$\mathcal{U}_{\text{tot}}(t,\mathbf{x},\mathbf{p}) = \frac{q}{2m}\left[\left(\frac{g}{2}-1+\frac{m}{\varepsilon}\right)\mathbf{B} - \left(\frac{g}{2}-\frac{\varepsilon}{1+\varepsilon}\right)\frac{\boldsymbol{\pi}\times\mathbf{E}}{\varepsilon} - \left(\frac{g}{2}-1\right)\frac{(\boldsymbol{\pi}\cdot\mathbf{B})\boldsymbol{\pi}}{\varepsilon(m+\varepsilon)}\right]\cdot\boldsymbol{\sigma}. \tag{5.29}$$

In what follows, I shall discuss the dynamics described by $[[D_{\text{tot}}]](x,p) \doteq \lambda(x,p) + \mathcal{U}_{\text{tot}}(t,\mathbf{x},\mathbf{p})$.

## 5.5 XGO action in the point-particle limit

For the sake of simplicity, I shall only consider the point-particle dynamics. As in Sec. 4.6, one substitutes $\lambda(x,p)$ and $\mathcal{U}_{\text{tot}}(t,\mathbf{x},\mathbf{p})$ into the general point-particle XGO action (4.68), and one obtains

$$\mathcal{S} = \int \mathrm{d}\tau \left[ P \cdot \dot{X} - \frac{i}{2}\left( Z^\dagger \dot{Z} - \dot{Z}^\dagger Z \right) + \lambda(X,P) + Z^\dagger \mathcal{U}_{\text{tot}}(X^0, \mathbf{X}, \mathbf{P}) Z \right], \tag{5.30}$$

where $X^\mu(\tau) = (X^0, \mathbf{X})$ is the particle four-position, $P^\mu(\tau) = (P_0, \mathbf{P})$ is the particle canonical four-momentum, and $Z(\tau)$ is a complex-valued two-component function that describes the particle spin state. It is normalized such that $Z^\dagger(\tau)Z(\tau) = 1$. For completeness, $\lambda(x,p) = p_0 - \varepsilon - qV$, and $\mathcal{U}_{\text{tot}}(t,\mathbf{x},\mathbf{p})$ is the total spin-coupling matrix given by Eq. (5.29). The action (5.30) is complemented by the corrected dispersion relation

$$\lambda(X,P) + Z^\dagger \mathcal{U}_{\text{tot}}(X^0, \mathbf{X}, \mathbf{P}) Z = 0. \tag{5.31}$$

In the covariant XGO action (5.30), the independent dynamical variables are $(X^\mu, P_\mu, Z, Z^\dagger)$. Let us follow the procedure in Sec. 4.6.3 in order to eliminate the temporal momentum variable $P_0(\tau)$ from the action. Since the dispersion operator is in the symplectic form, a brief calculation leads to the noncovariant point-particle XGO action:[5]

$$\mathcal{S} = \int \mathrm{d}t \left[ \mathbf{P} \cdot \dot{\mathbf{X}} + \frac{i\hbar}{2}\left( Z^\dagger \dot{Z} - \dot{Z}^\dagger Z \right) - H(t, \mathbf{X}, \mathbf{P}, Z, Z^\dagger) \right], \tag{5.32}$$

where, in this case, the dots represent derivatives with respect to the physical time $t$. The Hamiltonian $H(t, \mathbf{X}, \mathbf{P}, Z, Z^\dagger)$ is given by

$$H(t, \mathbf{X}, \mathbf{P}, Z, Z^\dagger) \doteq \gamma(t, \mathbf{X}, \mathbf{P}) mc^2 + qV(t, \mathbf{X}) - \frac{\hbar}{2} Z^\dagger \left[ \boldsymbol{\Omega}_{\text{BMT}}(t, \mathbf{X}, \mathbf{P}) \cdot \boldsymbol{\sigma} \right] Z \tag{5.33}$$

and $\boldsymbol{\Omega}_{\text{BMT}}(t, \mathbf{X}, \mathbf{P})$ is the Bargmann–Michel–Teledgi spin-precession frequency (Bargmann *et al.*, 1959)

$$\boldsymbol{\Omega}_{\text{BMT}}(t, \mathbf{X}, \mathbf{P}) \doteq \frac{q}{mc} \left[ \left( \frac{g}{2} - 1 + \frac{1}{\gamma} \right) \mathbf{B} - \left( \frac{g}{2} - \frac{\gamma}{1+\gamma} \right) \frac{\mathbf{v}_0 \times \mathbf{E}}{c} - \left( \frac{g}{2} - 1 \right) \frac{\gamma}{1+\gamma} \frac{(\mathbf{v}_0 \cdot \mathbf{B}) \mathbf{v}_0}{c^2} \right]. \tag{5.34}$$

Here $\mathbf{v}_0 \doteq \boldsymbol{\Pi}/(\gamma m)$ is the lowest-order GO velocity,

$$\gamma(t, \mathbf{X}, \mathbf{P}) \doteq \sqrt{1 + \left( \frac{\boldsymbol{\Pi}}{mc} \right)^2}, \tag{5.35}$$

[5]For simplicity, from henceforth and until the end of this Chapter, I restore the constants $c$ and $\hbar$.

is the relativistic Lorentz factor, and $\mathbf{\Pi}(t) \doteq \mathbf{P}(t) - q\mathbf{A}(t, \mathbf{X}(t))/c$ is the particle kinetic momentum.

In the action Eq. (5.32), the independent dynamical variables are $(\mathbf{X}, \mathbf{P}, Z, Z^\dagger)$; they depend on the physical time $t$. As before, $Z(t)$ is normalized so that $Z^\dagger(t)Z(t) = 1$. Note that $Z = (1, 0)^T$ represents the particle spin-up state, and $Z = (0, 1)^T$ corresponds to the spin-down state. As shown in Eq. (5.33), the first two terms are the well-known classical relativistic Hamiltonian for the spinless particle. The last term in Eq. (5.33) governs the spin dynamics.

The ELEs corresponding to the action (5.32) are given by

$$\delta\mathbf{P} : \quad \frac{\mathrm{d}}{\mathrm{d}t}\mathbf{X} = \mathbf{v}_0 - \frac{\partial}{\partial\mathbf{P}}(\mathbf{S}\cdot\mathbf{\Omega}_{\mathrm{BMT}}), \tag{5.36a}$$

$$\delta\mathbf{X} : \quad \frac{\mathrm{d}}{\mathrm{d}t}\mathbf{P} = -\frac{\partial}{\partial\mathbf{X}}(\gamma mc^2 + qV) + \frac{\partial}{\partial\mathbf{X}}(\mathbf{S}\cdot\mathbf{\Omega}_{\mathrm{BMT}}), \tag{5.36b}$$

$$\delta Z^\dagger : \quad \frac{\mathrm{d}}{\mathrm{d}t}Z = \frac{i}{2}\mathbf{\Omega}_{\mathrm{BMT}}\cdot\boldsymbol{\sigma}Z, \tag{5.36c}$$

$$\delta Z : \quad \frac{\mathrm{d}}{\mathrm{d}t}Z^\dagger = -\frac{i}{2}Z^\dagger\mathbf{\Omega}_{\mathrm{BMT}}\cdot\boldsymbol{\sigma}, \tag{5.36d}$$

where the particle spin vector is

$$\mathbf{S}(t) \doteq \frac{\hbar}{2}Z^\dagger(t)\boldsymbol{\sigma}Z(t), \tag{5.37}$$

and $|\mathbf{S}(t)| = \hbar/2$. Equations (5.33)–(5.36d) form a complete set of equations. The first terms on the right-hand side of Eqs. (5.36a) and (5.36b) describe the dynamics of a relativistic spinless particle. The second terms describe the spin-orbit coupling. Interestingly, when one includes the spin correction into the theory, the particle velocity $\dot{\mathbf{X}}(t)$ and kinetic momentum $\mathbf{\Pi}(t)$ are not necessarily colinear. Also, the second term on the right-hand side of Eq. (5.36b) gives the relativistic SG force. Also, Eqs. (5.36c) and (5.36d) lead to the spin-precession equation. Upon taking the derivative of $\mathbf{S}(t)$, one obtains

$$\begin{aligned}
\frac{\mathrm{d}}{\mathrm{d}t}\mathbf{S} &= \frac{\hbar}{2}\dot{Z}^\dagger\boldsymbol{\sigma}Z + \frac{\hbar}{2}Z^\dagger\boldsymbol{\sigma}\dot{Z} \\
&= -\frac{i\hbar}{4}Z^\dagger\left(\mathbf{\Omega}_{\mathrm{BMT}}\cdot\boldsymbol{\sigma}\right)\boldsymbol{\sigma}Z + \frac{i\hbar}{4}Z^\dagger\boldsymbol{\sigma}\left(\mathbf{\Omega}_{\mathrm{BMT}}\cdot\boldsymbol{\sigma}\right)Z \\
&= \frac{i\hbar}{4}Z^\dagger[\boldsymbol{\sigma}, \mathbf{\Omega}_{\mathrm{BMT}}\cdot\boldsymbol{\sigma}]Z \\
&= \frac{\hbar}{2}Z^\dagger\left(\boldsymbol{\sigma}\times\mathbf{\Omega}_{\mathrm{BMT}}\right)Z \\
&= \mathbf{S}\times\mathbf{\Omega}_{\mathrm{BMT}},
\end{aligned} \tag{5.38}$$

where I used the anti-commutation property of the Pauli matrices: $[\sigma_a, \sigma_b] = 2i\varepsilon_{abc}\sigma_c$. As shown, the spin vector $\mathbf{S}(t)$ precesses at the BMT spin-precession frequency $\Omega_{\mathrm{BMT}}$.

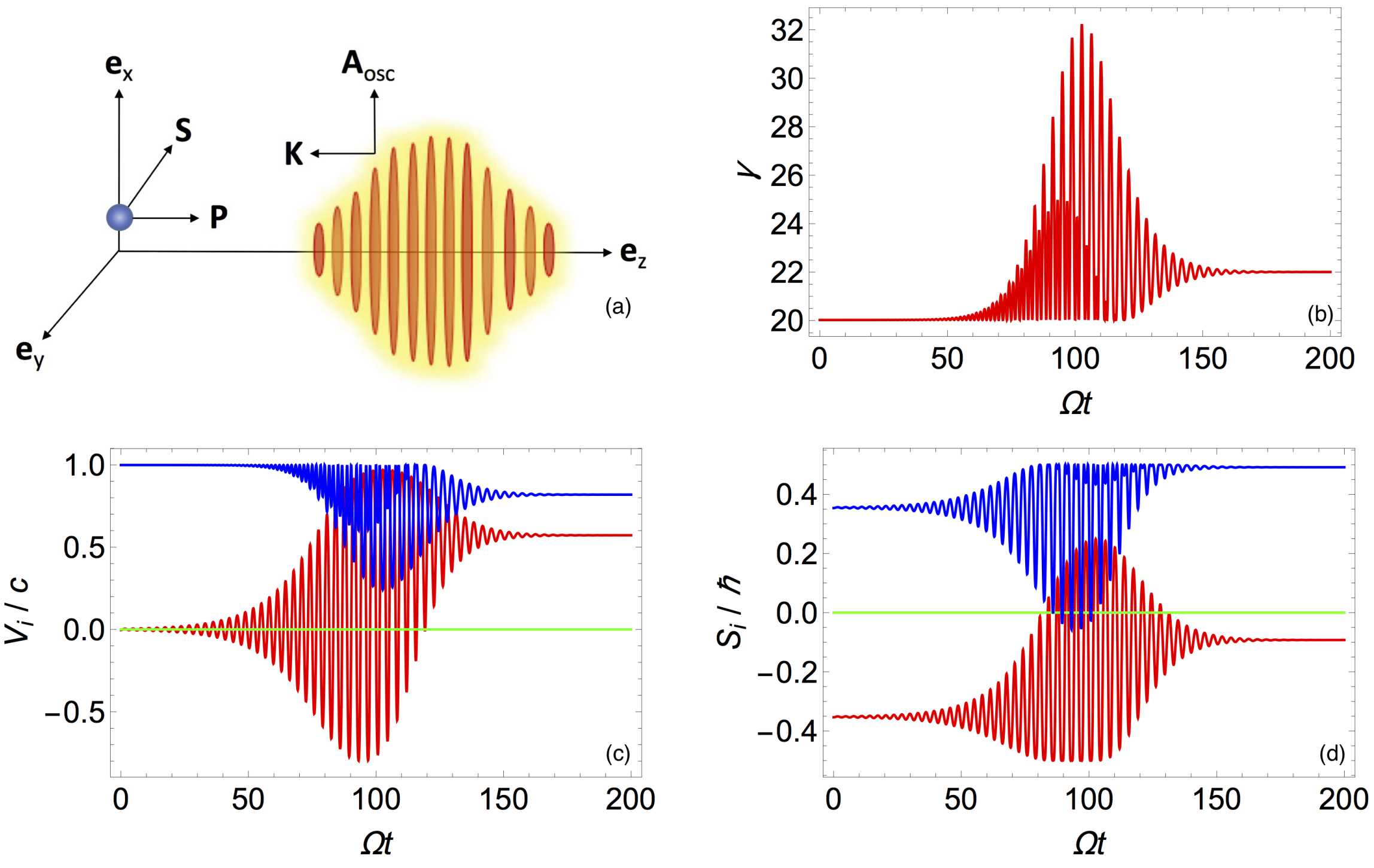

Figure 5.2: Dynamics of a relativistic spin-1/2 electron interacting with a relativistically intense laser pulse (numerical simulation). (a) Schematic of the interaction; yellow and red is the laser field, blue is the particle; arrows denote the direction of the laser wave vector $\mathbf{k}$, the oscillating vector potential $\mathbf{A}$, the particle canonical momentum $\mathbf{P}$, and the particle spin $\mathbf{S}$. The unit vectors along the reference axes are denoted by $\mathbf{e}_i$. Subfigures (b)–(d) show the components of the particle Lorentz factor $\gamma$, velocity $\dot{\mathbf{X}}$, and spin $\mathbf{S}$. The red, green, and blue lines correspond to projections on the $x$, $y$, and $z$ axes, respectively. Here an electron is initially traveling along the $z$ axis and then collides with a counter-propagating laser pulse. The initial position of the particle is $\mathbf{X}_0 = (\ell/2)\mathbf{e}_x$, the initial momentum is $\mathbf{P}_0/(mc) = 20\mathbf{e}_z$, and the normalized initial spin vector is $\mathbf{S}_0/\hbar = 0.14\mathbf{e}_x + 0.33\mathbf{e}_y + 0.35\mathbf{e}_z$. The envelope of the vector potential of the laser pulse is $q\mathbf{A}_{\text{osc}}/(mc^2) = 30\ \text{sech}\left[(z - 5\ell + ct)/\ell\right] \exp\left[-(x^2+y^2)/\ell^2\right] \mathbf{e}_x$, where $\ell = 20|\mathbf{k}|^{-1}$. These parameters correspond to a maximum intensity $I_{\max} \simeq 1.23 \times 10^{21}\ \text{W/cm}^2$ for a $1\,\mu$m laser.

As an example, Eqs. (5.36) were solved using MATHEMATICA's numerical solver `NDSOLVE`. Figure 5.2 shows the dynamics of a relativistic spin-1/2 electron colliding with a counter-propagating relativistically strong ($a_0 \gg 1$) laser pulse. The simulation parameters are given in the caption of Fig. 5.2, and a schematic of the interaction is presented in Fig. 5.2(a). Figure 5.2(b) shows the evolution of the particle Lorentz factor. Figures 5.2(c) and 5.2(d) show the temporal evolution of the particle velocity and spin. Since the particle collides with the laser pulse away from the $\mathbf{e}_z$ axis, the particle is accelerated transversely.

## 5.6 Discussion

The point-particle Lagrangian model [Eqs. (5.32)–(5.35)] agrees with the findings of many studies based on the Foldy–Wouthuysen (FW) transformation.[6] In some sense, it is surprising that both approaches obtain equivalent equations of motion since the corresponding procedures are entirely different. The FW transformation is a unitary transformation that asymptotically block-diagonalizes the Dirac Hamiltonian by using a perturbative procedure that involves the small parameter $\epsilon_{\rm FW} \doteq E/(2mc^2) \ll 1$, where $E$ is the characteristic energy of the particle. In Chen and Chiou (2014), it was shown that up to $\mathcal{O}(\epsilon_{\rm FW}^8)$, the Hamiltonian obtained using the FW transform agrees with the Hamiltonian (5.33). In contrast, the model obtained in this Chapter was derived by expanding the Dirac action to the first order in the GO parameter $\epsilon$ [Eq. (5.16)].[7] Another somewhat technical difference between the two approaches is that the theories based on the FW transform are Hamiltonian. In contrast, the present theory is constructed straightforwardly by reduction of the Dirac action functional (5.1). There is no need to postulate additional symmetries, Poisson brackets, or any other correspondence between quantum and classical dynamics except the GO limit.

As a final note, Wen *et al.* (2016) recently compared the Frenkel model and the FW model against numerical simulations of the quantum Dirac equation. The case studied was that of a relativistic electron interacting with a strong EM vacuum pulse. The numerical simulations showed that the FW model was more accurate in describing the dynamics described by the Dirac equation. The results of this Chapter can be considered as an explanation of this fact. Here I show that, unlike the Frenkel model, the FW model is derivable from generic first principles based on GO without using any additional postulates. Hence, it is more fundamental and robust.

## 5.7 Conclusions

The XGO theory developed in Chapter 4 was used to derive a classical model for the relativistic spin-1/2 particle. The resulting formulation includes the well-known spin effects, such as the spin precession and the SG force. This model is equivalent to other previously obtained formulations based on the FW transform.

Overall, the advantages of this new theory are as follows: (i) This theory is the first point-particle model that is derived from generic first principles based on GO. (ii) The theory is derived in a variational form. Therefore, the resulting equations are manifestly conservative. Due to the variational structure, this single-particle model can be easily implemented in kinetic models for studying relativistic spin-1/2 plasmas (Burby, 2017). (iii) This new theory is manifestly symplectic. Hence, symplectic integrators could be used to accurately solve the resulting equations of motion (Qin *et al.*, 2015; Hairer *et al.*, 2006; McLachlan and Quispel, 2006). (iv) Since a variational approach is used, the theory is naturally suited to serve as a stepping stone to develop further reduced models for relativistic spin-1/2 particles.

[6] See, for example, Silenko (2008), Chen and Chiou (2010), Chen and Chiou (2014), and Bliokh (2005).

[7] This is a noticeable advantage of the method shown here because approaches based on the FW transform often rely on expansions to the eight order in $\epsilon_{\rm FW}$ or even higher.

This work can be expanded in at least three directions. First, to ensure the stability and high quality of spin-polarized beams in storage rings, it is necessary to accurately describe the beam dynamics over a large ($\sim 10^{10}$) number of turns (Hoffstaetter, 2009). Since the derived model is symplectic, variational integrators with good conservation properties could be useful to more accurately simulate the dynamics of these beams. Second, the action (5.32) is relativistically invariant, but it is not explicitly Lorentz invariant. From a theoretical standpoint, an interesting project would be to develop a classical manifestly Lorentz-invariant action for the Dirac particle. Third, being a conservative model, this formulation does not include radiation-reaction effects. Given the recent developments on Lagrangian models for dissipative waves (Dodin *et al.*, 2017), it would be interesting to investigate where radiation-reaction effects can be incorporated into the theory.

# Chapter 6

# Polarization effects on waves in magnetized plasma

In this Chapter, I apply the XGO theory to study polarization effects on classical waves. Specifically, I study radio-frequency (RF) waves propagating in magnetized plasma. Analytical results are presented for the case of waves propagating in weakly magnetized plasma. It is shown that polarization effects are manifested as a precession of the wave polarization and as a polarization-driven bending of ray trajectories. Numerical simulations are presented for the case of waves in strongly magnetized plasma. Some of the results in this Chapter were already published by Ruiz and Dodin (2017a). The remaining part is presented here for the first time.

## 6.1 Introduction

### 6.1.1 Motivation

As mentioned in Chapter 1, waves play an essential role in many aspects of plasma dynamics. In the specific context of magnetic confinement fusion, RF waves are important for manipulating and diagnosing plasmas. For this purpose, various reduced models based on geometrical optics (GO) have been extensively used to calculate RF wave trajectories in tokamak plasmas (Bernstein, 1975). However, as discussed in Chapter 4, GO is an approximate reduced wave model that does not capture polarization effects.

In this Chapter, I apply the developed XGO model in Chapter 4 to study polarization effects on RF waves in magnetized plasma. For transverse EM waves propagating in weakly magnetized plasma, I analytically show that polarization effects are manifested as a precession of the wave polarization and as an additional polarization-driven bending of the ray trajectories. For RF waves propagating in tokamak plasmas, the numerical simulations presented here suggest that polarization effects are negligible for waves at or above the electron-cyclotron-frequency regime. However, there is reason to speculate that polarization effects can remain important for waves with lower frequencies (lower-hybrid waves or ion-cyclotron waves). This issue remains a matter of future research.

### 6.1.2 Overview

This Chapter is organized as follows. Section 6.2 introduces the basic model used for describing linear waves in magnetized plasma. The governing equations are then cast into the variational XGO formalism. Section 6.3 discusses polarization effects on waves propagating in weakly magnetized plasma. The analysis presented in this section is mainly analytical and provides a concrete example of how the XGO theory is applied. Section 6.4 considers the case of waves in strongly magnetized plasma. Since the strongly-magnetized-plasma case is algebraically more complicated, the analysis presented in this section mainly relies on numerical simulations. Finally, Sec. 6.5 summarizes the main results obtained.

## 6.2 Basic model

### 6.2.1 Governing equations

Let us consider waves propagating in cold magnetized plasma, and let us assume that the background plasma is stationnary and non-drifting. The linearized equations of motion are (Stix, 1992)

$$\partial_t \mathbf{v}_s = \frac{q_s}{m_s}\mathbf{E} + \frac{q_s}{m_s c}\mathbf{v}_s \times \mathbf{B}_0, \tag{6.1a}$$

$$\partial_t \mathbf{E} = -\sum_s 4\pi q_s n_{s0} \mathbf{v}_s + c\boldsymbol{\nabla} \times \mathbf{B}, \tag{6.1b}$$

$$\partial_t \mathbf{B} = -c\boldsymbol{\nabla} \times \mathbf{E}, \tag{6.1c}$$

where $n_{s0}(\mathbf{x})$ is the unperturbed number density and $\mathbf{v}_s(t,\mathbf{x})$ is the flow velocity of the particles of species $s$. The constants $m_s$ and $q_s$ denote the mass and charge of the particles of species $s$, respectively, and $c$ is the speed of light. Also, $\mathbf{E}(t,\mathbf{x})$ and $\mathbf{B}(t,\mathbf{x})$ denote the wave electric field and magnetic fields, respectively. The background magnetic field is given by $\mathbf{B}_0(\mathbf{x})$. I introduce a rescaled velocity field $\overline{\mathbf{v}}_s(t,\mathbf{x}) \doteq$

$\mathbf{v}_s(t,\mathbf{x})[4\pi n_{s0}(\mathbf{x})m_s]^{1/2}$ so Eqs. (6.1) can be rewritten as

$$\partial_t \overline{\mathbf{v}}_s = \omega_{ps}\mathbf{E} + \overline{\mathbf{v}}_s \times \boldsymbol{\Omega}_s, \tag{6.2a}$$

$$\partial_t \mathbf{E} = -\sum_s \omega_{ps}\overline{\mathbf{v}}_s + c\boldsymbol{\nabla}\times\mathbf{B}, \tag{6.2b}$$

$$\partial_t \mathbf{B} = -c\boldsymbol{\nabla}\times\mathbf{E}, \tag{6.2c}$$

where $\omega_{ps}(\mathbf{x}) \doteq \operatorname{sgn}(q_s)[4\pi q_s^2 n_{s0}(\mathbf{x})/m_s]^{1/2}$ is the plasma frequency and $\boldsymbol{\Omega}_s(\mathbf{x}) \doteq q_s\mathbf{B}_0(\mathbf{x})/(m_s c)$ is the gyrofrequency associated with the particles of species $s$.

As in Sec. 2.2.2, I rewrite Eqs. (6.2) using the abstract Dirac notation. Let $|\,\mathbf{v}_s\,\rangle$ be a three-component state vector representing the velocity field such that $\mathbf{v}(t,\mathbf{x}) = \langle\,(t,\mathbf{x})\,|\,\mathbf{v}_s\,\rangle$. Likewise, $|\,\mathbf{E}\,\rangle$ and $|\,\mathbf{B}\,\rangle$ are the state vectors corresponding to $\mathbf{E}(t,\mathbf{x})$ and $\mathbf{B}(t,\mathbf{x})$, respectively. Then, Eqs. (6.2) are written as follows:

$$\widehat{p}_0\,|\,\overline{\mathbf{v}}_s\,\rangle = i\widehat{\omega}_{ps}\,|\,\mathbf{E}\,\rangle - (\boldsymbol{\alpha}\cdot\widehat{\boldsymbol{\Omega}}_s)\,|\,\overline{\mathbf{v}}_s\,\rangle\,, \tag{6.3a}$$

$$\widehat{p}_0\,|\,\mathbf{E}\,\rangle = -i\sum_s \widehat{\omega}_{ps}\,|\,\overline{\mathbf{v}}_s\,\rangle + ic(\boldsymbol{\alpha}\cdot\widehat{\mathbf{p}})\,|\,\mathbf{B}\,\rangle\,, \tag{6.3b}$$

$$\widehat{p}_0\,|\,\mathbf{B}\,\rangle = -ic(\boldsymbol{\alpha}\cdot\widehat{\mathbf{p}})\,|\,\mathbf{E}\,\rangle\,, \tag{6.3c}$$

where $\widehat{\omega}_{ps} \doteq \omega_{ps}(\widehat{\mathbf{x}})$ and $\widehat{\boldsymbol{\Omega}}_s \doteq \boldsymbol{\Omega}_s(\widehat{\mathbf{x}})$ are now considered as operators. (As a reminder, $\widehat{p}_0 = i\partial_t$ and $\widehat{\mathbf{p}} = -i\boldsymbol{\nabla}$ are the components of the four-momentum operator in the physical-space representation.) Also, $\boldsymbol{\alpha} \doteq \left(\alpha^1, \alpha^2, \alpha^3\right)$ are $3\times 3$ Hermitian matrices[1]

$$\alpha^1 \doteq \begin{pmatrix} 0 & 0 & 0 \\ 0 & 0 & -i \\ 0 & i & 0 \end{pmatrix}, \qquad \alpha^2 \doteq \begin{pmatrix} 0 & 0 & i \\ 0 & 0 & 0 \\ -i & 0 & 0 \end{pmatrix}, \qquad \alpha^3 \doteq \begin{pmatrix} 0 & -i & 0 \\ i & 0 & 0 \\ 0 & 0 & 0 \end{pmatrix}. \tag{6.4a}$$

These matrices are useful for representing the vector product between three-component vectors. Namely, for any three-component column vectors $\mathbf{A}$ and $\mathbf{B}$, one can easily verify

$$(\boldsymbol{\alpha}\cdot\mathbf{A})\,\mathbf{B} = (\alpha^1 A^1 + \alpha^2 A^2 + \alpha^3 A^3)\,\mathbf{B} = i\mathbf{A}\times\mathbf{B}, \tag{6.5a}$$

$$\mathbf{A}^{\mathrm{T}}\alpha^j\mathbf{B} = -i(\mathbf{A}\times\mathbf{B})^j, \tag{6.5b}$$

where the superscript T denotes the matrix transpose.

[1] Note that the $\alpha$ matrices are related to the Gell-Mann matrices, which serve as the infinitesimal generators of the special unitary group SU(3).

### 6.2.2 Dispersion operator

The next step is to construct a dispersion operator for the electric field state $|\,\mathbf{E}\,\rangle$. Starting from Eq. (6.3a), one solves for the velocity state $|\,\overline{\mathbf{v}}_s\,\rangle$ in terms of the electric field $|\,\mathbf{E}\,\rangle$. One formally obtains $|\,\overline{\mathbf{v}}_s\,\rangle = i(\widehat{p}_0\mathbb{I}_3 + \boldsymbol{\alpha}\cdot\widehat{\boldsymbol{\Omega}}_s)^{-1}\widehat{\omega}_{ps}\,|\,\mathbf{E}\,\rangle$. After inverting the operator, one obtains

$$|\,\overline{\mathbf{v}}_s\,\rangle = i\left(\frac{1}{\widehat{p}_0}\mathbb{I}_3 - \frac{\boldsymbol{\alpha}\cdot\widehat{\boldsymbol{\Omega}}_s}{\widehat{p}_0^2 - \widehat{\Omega}_s^2} + \frac{(\boldsymbol{\alpha}\cdot\widehat{\boldsymbol{\Omega}}_s)^2}{\widehat{p}_0(\widehat{p}_0^2 - \widehat{\Omega}_s^2)}\right)\widehat{\omega}_{ps}\,|\,\mathbf{E}\,\rangle\,, \tag{6.6}$$

where $\widehat{\Omega}_s \doteq q_s|\mathbf{B}(\widehat{\mathbf{x}})|/(m_s c)$. This result can be easily verified. Upon letting $\mathbf{A}$ be any arbitrary three-component column vector, one has

$$\begin{aligned}
(p_0\mathbb{I}_3 + \boldsymbol{\alpha}\cdot\boldsymbol{\Omega}_s)^{-1}(p_0\mathbb{I}_3 + \boldsymbol{\alpha}\cdot\boldsymbol{\Omega}_s)\mathbf{A} &= \left(\frac{1}{p_0}\mathbb{I}_3 - \frac{\boldsymbol{\alpha}\cdot\boldsymbol{\Omega}_s}{p_0^2 - \Omega_s^2} + \frac{(\boldsymbol{\alpha}\cdot\boldsymbol{\Omega}_s)^2}{p_0(p_0^2 - \Omega_s^2)}\right)(p_0\mathbb{I}_3 + \boldsymbol{\alpha}\cdot\boldsymbol{\Omega}_s)\mathbf{A} \\
&= \mathbf{A} + \frac{1}{p_0(p_0^2 - \Omega_s^2)}\left[(\boldsymbol{\alpha}\cdot\boldsymbol{\Omega}_s)^3 - \Omega_s^2(\boldsymbol{\alpha}\cdot\boldsymbol{\Omega}_s)\right]\mathbf{A} \\
&= \mathbf{A} - \frac{i\{\boldsymbol{\Omega}_s\times[\boldsymbol{\Omega}_s\times(\boldsymbol{\Omega}_s\times\mathbf{A})] + \Omega_s^2(\boldsymbol{\Omega}_s\times\mathbf{A})\}}{p_0(p_0^2 - \Omega_s^2)} \\
&= \mathbf{A},
\end{aligned} \tag{6.7}$$

where I used Eq. (6.5a) in the third line. The same proof applies to Eq. (6.6) since the operators commute.[2] Similarly from Eq. (6.3c), the magnetic field is $|\,\mathbf{B}\,\rangle = -ic\widehat{p}_0^{-1}(\boldsymbol{\alpha}\cdot\widehat{\mathbf{p}})\,|\,\mathbf{E}\,\rangle$. Upon substituting these results into Eq. (6.3b), one finds that the state $|\,\mathbf{E}\,\rangle$ satisfies

$$\widehat{\mathcal{D}}\,|\,\mathbf{E}\,\rangle = 0, \tag{6.8}$$

where $\widehat{\mathcal{D}} = \mathcal{D}(\widehat{\mathbf{x}}, \widehat{p}_0, \widehat{\mathbf{p}})$ is the wave dispersion operator, defined as

$$\mathcal{D}(\widehat{\mathbf{x}}, \widehat{p}_0, \widehat{\mathbf{p}}) \doteq -\widehat{p}_0^2\mathbb{I}_3 + c^2(\boldsymbol{\alpha}\cdot\widehat{\mathbf{p}})^2 + \sum_s \widehat{\omega}_{ps}^2\left(\mathbb{I}_3 - \frac{\widehat{p}_0(\boldsymbol{\alpha}\cdot\widehat{\boldsymbol{\Omega}}_s)}{\widehat{p}_0^2 - \widehat{\Omega}_s^2} + \frac{(\boldsymbol{\alpha}\cdot\widehat{\boldsymbol{\Omega}}_s)^2}{\widehat{p}_0^2 - \widehat{\Omega}_s^2}\right). \tag{6.9}$$

Since $\widehat{\omega}_{ps}$ and $\widehat{\boldsymbol{\Omega}}_s$ are time independent, $\widehat{p}_0$ commutes with $\widehat{\omega}_{ps}$ and $\widehat{\boldsymbol{\Omega}}_s$. It is clear that $\widehat{\mathcal{D}}$ is Hermitian, so one can proceed with the construction of the variational principle.

[2]Note that, if the background plasma is time dependent, then the result in Eq. (6.6) is no longer valid.

### 6.2.3 Extended action principle

After obtaining the dispersion operator (6.9), let us construct a principle of stationary action from which the XGO equations can be obtained. As in Sec. 4.2, I propose the following action:

$$\mathcal{S}_{\mathrm{X}} \doteq \int \mathrm{d}\tau \; L, \tag{6.10}$$

where $L \doteq L_\tau + L_D$ serves as the Lagrangian and

$$L_\tau \doteq -(i/2)\left[\langle\, \mathbf{E}(\tau) \mid \partial_\tau \mathbf{E}(\tau)\,\rangle - \mathrm{c.\,c.}\right], \tag{6.11a}$$

$$L_D \doteq \langle\, \mathbf{E}(\tau) \mid \widehat{\mathfrak{D}} \mid \mathbf{E}(\tau)\,\rangle\,. \tag{6.11b}$$

As a reminder, here the electric field $\mathbf{E}$ not only depends on spacetime but also on some parameter $\tau$ so that $\mathbf{E}(\tau, x) = \langle\, x \mid \mathbf{E}(\tau)\,\rangle$ and $\partial_\tau \mathbf{E}(\tau, x) = \langle\, x \mid \partial_\tau \mathbf{E}(\tau)\,\rangle$. Also, the abstract vector state $|\,\mathbf{E}(\tau)\,\rangle$ belongs to the Hilbert space with inner product (4.6).

## 6.3 RF waves in weakly magnetized plasma

### 6.3.1 Block-diagonalizing the dispersion operator

As in Secs. 4.3 and 4.4, let us block-diagonalize the dispersion operator. Upon using the Weyl transformation (A.1) along with the Moyal product (A.5), one obtains the Weyl symbol $D(\mathbf{x}, p_0, \mathbf{p})$ corresponding to the operator $\widehat{\mathfrak{D}}$:

$$D(\mathbf{x}, p_0, \mathbf{p}) = -p_0^2 \mathbb{I}_3 + c^2(\boldsymbol{\alpha}\cdot\mathbf{p})^2 + \sum_s \omega_{ps}^2 \left( \mathbb{I}_3 - \frac{p_0(\boldsymbol{\alpha}\cdot\boldsymbol{\Omega}_s)}{p_0^2 - \Omega_s^2} + \frac{(\boldsymbol{\alpha}\cdot\boldsymbol{\Omega}_s)^2}{p_0^2 - \Omega_s^2} \right) \tag{6.12}$$

In this section, I consider waves propagating in a weakly magnetized plasma. (The strongly magnetized case is discussed in the next section.) I assume that the typical wave frequency is much larger than the particle gyrofrequency so that $p_0 \gg |\Omega_s|$. Upon expanding the dispersion symbol (6.12) in powers of $\Omega_s/p_0$, one obtains

$$D(\mathbf{x}, p_0, \mathbf{p}) \simeq D_0(\mathbf{x}, p_0, \mathbf{p}) + D_1(\mathbf{x}, p_0) + \mathcal{O}(\Omega_s^2/p_0^2), \tag{6.13}$$

where

$$D_0(\mathbf{x}, p_0, \mathbf{p}) \doteq -p_0^2 + c^2(\boldsymbol{\alpha} \cdot \mathbf{p})^2 + \omega_p^2(\mathbf{x}), \tag{6.14a}$$

$$D_1(\mathbf{x}, p_0) \doteq -\sum_s \omega_{ps}^2(\mathbf{x})[\boldsymbol{\alpha} \cdot \boldsymbol{\Omega}_s(\mathbf{x})]/p_0. \tag{6.14b}$$

Here $\omega_p^2(\mathbf{x}) \doteq \sum_s \omega_{ps}^2(\mathbf{x})$ is the square of the plasma frequency. In what follows, I assume that $D_1(\mathbf{x}, p_0) \sim \mathcal{O}(\Omega/p_0)$ is comparable in magnitude to the GO parameter (3.2). Hence, $D_1(\mathbf{x}, p_0)$ is considered as a perturbation only.

As in Sec. 4.3.3, let us proceed by identifying the eigenvalues and eigenmodes of the dispersion symbol $D_0(\mathbf{x}, p_0, \mathbf{p})$. The corresponding eigenvalues are given by

$$\lambda^{(1)}(\mathbf{x}, p_0, \mathbf{p}) = -p \cdot p + \omega_p^2(\mathbf{x}), \tag{6.15a}$$

$$\lambda^{(2)}(\mathbf{x}, p_0, \mathbf{p}) = -p \cdot p + \omega_p^2(\mathbf{x}), \tag{6.15b}$$

$$\lambda^{(3)}(\mathbf{x}, p_0) = -p_0^2 + \omega_p^2(\mathbf{x}), \tag{6.15c}$$

where $p \cdot p = p_0^2 - \mathbf{p}^2$. (For convenience, from henceforth I omit writing the speed of light constant $c$.) The eigenvalues $\lambda^{(1,2)}(\mathbf{x}, p_0, \mathbf{p})$ correspond to the dispersion relations of two transverse EM waves, and $\lambda^{(3)}(\mathbf{x}, p_0)$ corresponds to longitudinal Langmuir oscillations. The matrix $Q_0(\mathbf{p})$ defined in Eq. (4.19) is given by

$$Q_0(\mathbf{p}) = [\,\mathbf{e}_1(\mathbf{p}), \quad \mathbf{e}_2(\mathbf{p}), \quad \mathbf{e}_{\mathbf{p}}(\mathbf{p})\,], \tag{6.16}$$

where $\mathbf{e}_1(\mathbf{p})$ and $\mathbf{e}_2(\mathbf{p})$ are any two orthonormal vectors in the plane normal to $\mathbf{e}_{\mathbf{p}}(\mathbf{p}) \doteq \mathbf{p}/|\mathbf{p}|$. A right-hand convention is adopted such that $\mathbf{e}_1(\mathbf{p}) \times \mathbf{e}_2(\mathbf{p}) = \mathbf{e}_{\mathbf{p}}(\mathbf{p})$. One can easily verify that these vectors are indeed eigenvectors of $D_0(\mathbf{x}, p_0, \mathbf{p})$. For example, one obtains for $\mathbf{e}_1(\mathbf{p})$:

$$\begin{aligned}
D_0(\mathbf{x}, p_0, \mathbf{p})\, \mathbf{e}_1(\mathbf{p}) &= [-p_0^2 + (\boldsymbol{\alpha} \cdot \mathbf{p})(\boldsymbol{\alpha} \cdot \mathbf{p}) + \omega_p^2]\, \mathbf{e}_1 \\
&= (-p_0^2 + \omega_p^2)\, \mathbf{e}_1 - \mathbf{p} \times (\mathbf{p} \times \mathbf{e}_1) \\
&= (-p_0^2 + \omega_p^2)\, \mathbf{e}_1 - |\mathbf{p}|\, (\mathbf{p} \times \mathbf{e}_2) \\
&= \lambda^{(1)}\, \mathbf{e}_1,
\end{aligned} \tag{6.17}$$

where I substituted Eq. (6.5a). A similar calculation follows for the other two eigenvectors.

Now, let us analyze the dynamics of the transverse EM waves. Following Sec. 4.4, the diagonal matrix of eigenvalues is $\Lambda(\mathbf{x}, p_0, \mathbf{p}) \doteq \lambda(\mathbf{x}, p_0, \mathbf{p})\mathbb{I}_2$, where $\lambda(\mathbf{x}, p_0, \mathbf{p}) \doteq -p \cdot p + \omega_p^2(\mathbf{x})$. Also, one has the $3 \times 2$ matrix $\Xi(\mathbf{p}) \doteq [\mathbf{e}_1(\mathbf{p}), \mathbf{e}_2(\mathbf{p})]$. Since $\Xi(\mathbf{p})$ depends only on the spatial momentum coordinate, the polarization-coupling term $\mathcal{U}(\mathbf{x}, \mathbf{p})$ in Eq. (4.36) is given by

$$\begin{aligned}
\mathcal{U}(\mathbf{x}, \mathbf{p}) &= \frac{\partial \lambda}{\partial x^\mu} \left( \Xi^\dagger \frac{\partial \Xi}{\partial p_\mu} \right)_A \\
&= -\frac{\partial \lambda}{\partial \mathbf{x}} \cdot \left( \Xi^\dagger \frac{\partial \Xi}{\partial \mathbf{p}} \right)_A \\
&= -\frac{\partial \omega_p^2}{\partial \mathbf{x}} \cdot \left[ \begin{pmatrix} \mathbf{e}^1 \\ \mathbf{e}^2 \end{pmatrix} \frac{\partial}{\partial \mathbf{p}} \begin{pmatrix} \mathbf{e}_1 & \mathbf{e}_2 \end{pmatrix} \right]_A \\
&= -\frac{\partial \omega_p^2}{\partial \mathbf{x}} \cdot \begin{pmatrix} \mathbf{e}^1 \partial_\mathbf{p} \mathbf{e}_1 & \mathbf{e}^1 \partial_\mathbf{p} \mathbf{e}_2 \\ \mathbf{e}^2 \partial_\mathbf{p} \mathbf{e}_1 & \mathbf{e}^2 \partial_\mathbf{p} \mathbf{e}_2 \end{pmatrix}_A \\
&= -\frac{1}{i} \frac{\partial \omega_p^2}{\partial \mathbf{x}} \cdot \begin{pmatrix} 0 & \mathbf{e}^1 \partial_\mathbf{p} \mathbf{e}_2 \\ -\mathbf{e}^1 \partial_\mathbf{p} \mathbf{e}_2 & 0 \end{pmatrix},
\end{aligned} \tag{6.18}$$

where $\mathbf{e}^q$ is the dual to $\mathbf{e}_q$, so $\mathbf{e}^q \mathbf{e}_r = \delta^q_r$. (Specifically, $\mathbf{e}^q$ is a row vector, whose elements are complex conjugates of those of $\mathbf{e}_q$.) One can also write Eq. (6.18) as

$$\mathcal{U}(\mathbf{x}, \mathbf{p}) = -\sigma_y (\boldsymbol{\nabla} \omega_p^2) \cdot \mathbf{F}(\mathbf{p}), \tag{6.19}$$

where

$$\sigma_y = \begin{pmatrix} 0 & -i \\ i & 0 \end{pmatrix} \tag{6.20}$$

is the $y$ component of the Pauli matrices and $\mathbf{F}(\mathbf{p})$ is a vector with components given by $\mathbf{F}(\mathbf{p}) \doteq \mathbf{e}^1 \partial_\mathbf{p} \mathbf{e}_2$.[3] For an explicit expression for $\mathbf{F}(\mathbf{p})$, one must choose a basis for the vectors $\mathbf{e}_1(\mathbf{p})$ and $\mathbf{e}_2(\mathbf{p})$. (As shown in Sec. 6.3.4, changing the basis does not affect the variational principle as long as it is expressed using the gauge-invariant noncanonical coordinates.) For example, let us adopt

$$\mathbf{e}_1(\mathbf{p}) \doteq \begin{pmatrix} \frac{p_x p_z}{|\mathbf{p}|\sqrt{p_x^2+p_y^2}} \\ \frac{p_y p_z}{|\mathbf{p}|\sqrt{p_x^2+p_y^2}} \\ -\frac{\sqrt{p_x^2+p_y^2}}{p} \end{pmatrix}, \qquad \mathbf{e}_2(\mathbf{p}) \doteq \begin{pmatrix} -\frac{p_y}{\sqrt{p_x^2+p_y^2}} \\ \frac{p_x}{\sqrt{p_x^2+p_y^2}} \\ 0 \end{pmatrix}. \tag{6.21}$$

[3]It is interesting to note that a vector similar to $\mathbf{F}$ appears in the theory of guiding center motion in plasma physics. Such vector is called the gyrogauge-transformation vector. For more information, see, for example (Littlejohn, 1983).

Then, $\mathbf{F}(\mathbf{p}) = (\mathbf{p}_\perp \times \mathbf{p})/(|\mathbf{p}||\mathbf{p}_\perp|^2)$, where $\mathbf{p}_\perp \doteq (p_x, p_y, 0)^T$; or, more explicitly,

$$\mathbf{F}(\mathbf{p}) = \frac{p_z}{|\mathbf{p}||\mathbf{p}_\perp|^2} \begin{pmatrix} p_y \\ -p_x \\ 0 \end{pmatrix}. \tag{6.22}$$

Returning to the perturbation caused by the background magnetic field, let us calculate the projection of the eigenmodes on the matrix $D_1(\mathbf{x}, p_0)$:

$$\begin{aligned}
\Xi^\dagger(\mathbf{p}) D_1(\mathbf{x}, p_0) \Xi(\mathbf{p}) &= -\sum_s \frac{\omega_{ps}^2}{p_0} \begin{pmatrix} \mathbf{e}^1 \\ \mathbf{e}^2 \end{pmatrix} (\boldsymbol{\alpha} \cdot \boldsymbol{\Omega}_s) \begin{pmatrix} \mathbf{e}_1 & \mathbf{e}_2 \end{pmatrix} \\
&= -\sum_s \frac{\omega_{ps}^2}{p_0} \begin{pmatrix} \mathbf{e}^1 (\boldsymbol{\alpha} \cdot \boldsymbol{\Omega}_s) \mathbf{e}_1 & \mathbf{e}^1 (\boldsymbol{\alpha} \cdot \boldsymbol{\Omega}_s) \mathbf{e}_2 \\ \mathbf{e}^2 (\boldsymbol{\alpha} \cdot \boldsymbol{\Omega}_s) \mathbf{e}_1 & \mathbf{e}^2 (\boldsymbol{\alpha} \cdot \boldsymbol{\Omega}_s) \mathbf{e}_2 \end{pmatrix} \\
&= -i \sum_s \frac{\omega_{ps}^2}{p_0} \begin{pmatrix} \mathbf{e}_1 \cdot (\boldsymbol{\Omega}_s \times \mathbf{e}_1) & \mathbf{e}_1 \cdot (\boldsymbol{\Omega}_s \times \mathbf{e}_2) \\ \mathbf{e}_2 \cdot (\boldsymbol{\Omega}_s \times \mathbf{e}_1) & \mathbf{e}_2 \cdot (\boldsymbol{\Omega}_s \times \mathbf{e}_2) \end{pmatrix} \\
&= -\sum_s \frac{\omega_{ps}^2}{p_0} (\mathbf{e}_\mathbf{p} \cdot \boldsymbol{\Omega}_s) \sigma_y,
\end{aligned} \tag{6.23}$$

where I used Eq. (6.5b) so that $\mathbf{e}^q (\boldsymbol{\alpha} \cdot \boldsymbol{\Omega}_s) \mathbf{e}_r = i \mathbf{e}_q \cdot (\boldsymbol{\Omega}_s \times \mathbf{e}_r) = i \boldsymbol{\Omega}_s \cdot (\mathbf{e}_r \times \mathbf{e}_q)$.

As a side note, repeating the calculations for Langmuir waves leads to $\mathcal{U} = 0$ and $D_1 = 0$. Hence, there is no coupling between the Langmuir mode polarization and the ray curvature or magnetic field. Physically, this means that Langmuir wave rays behave as spinless particles with no polarization effects. In what follows, I discuss more specifically how polarization effects can affect the dynamics of transverse EM waves.

### 6.3.2 XGO action in the point-particle limit

For simplicity, let us discuss the wave ray dynamics. Following Sec. 4.6, let us substitute $\lambda(\mathbf{x}, p_0, \mathbf{p}) = -p \cdot p + \omega_p^2(\mathbf{x})$, Eq. (6.19), and Eq. (6.23) into Eq. (4.68). One then obtains the point-particle action in canonical coordinates:

$$\begin{aligned}
\mathcal{S}_{\text{XGO}} = \int \mathrm{d}\tau \bigg[ & P \cdot \dot{X} - \frac{i}{2} \left( Z^\dagger \dot{Z} - \dot{Z}^\dagger Z \right) - P \cdot P + \omega_p^2(\mathbf{X}) \\
& - \left( (\boldsymbol{\nabla} \omega_p^2) \cdot \mathbf{F} + \sum_s \frac{\omega_{ps}^2}{P_0} (\mathbf{e}_\mathbf{p} \cdot \boldsymbol{\Omega}_s) \right) Z^\dagger \sigma_y Z \bigg],
\end{aligned} \tag{6.24}$$

where $X^\mu(\tau) = (X^0, \mathbf{X})$ is the wave four-position, $P^\mu(\tau) = (P^0, \mathbf{P})$ is the wave four-momentum, and $Z(\tau)$ is a complex-valued vector with two components that describe the degree of polarization along the vectors $\mathbf{e}_1(\mathbf{P})$ and $\mathbf{e}_2(\mathbf{P})$. It is normalized such that $Z^\dagger(\tau)Z(\tau) = 1$.

In the action (6.24), the two polarization modes are coupled through the matrix $\sigma_y$. One can decouple the two modes by using the basis of circularly polarized modes. Let us introduce the transformation $Z(\tau) = \mathcal{R}\,\Gamma(\tau)$, where $\mathcal{R}$ is a U(2) matrix whose columns are the normalized eigenvectors of $\sigma_y$,

$$\mathcal{R} \doteq \frac{1}{\sqrt{2}} \begin{pmatrix} 1 & 1 \\ i & -i \end{pmatrix} \tag{6.25}$$

and $\Gamma(\tau)$ is a new vector with components denoted as

$$\Gamma(\tau) \doteq \begin{pmatrix} \Gamma_+ \\ \Gamma_- \end{pmatrix}. \tag{6.26}$$

Inserting $Z(\tau) = \mathcal{R}\,\Gamma(\tau)$ into the action (6.24) leads to

$$\begin{aligned} \mathcal{S}_{\text{XGO}} = \int \mathrm{d}\tau \bigg[ & P \cdot \dot{X} - \frac{i}{2}\left(\Gamma^\dagger \dot{\Gamma} - \dot{\Gamma}^\dagger \Gamma\right) - P \cdot P + \omega_p^2(\mathbf{X}) \\ & - \left( (\boldsymbol{\nabla}\omega_p^2) \cdot \mathbf{F} + \sum_s \frac{\omega_{ps}^2}{P_0} (\mathbf{e}_\mathbf{p} \cdot \boldsymbol{\Omega}_s) \right) \Gamma^\dagger \sigma_z \Gamma \bigg], \end{aligned} \tag{6.27}$$

where $\sigma_z$ is another Pauli matrix,

$$\sigma_z = \begin{pmatrix} 1 & 0 \\ 0 & -1 \end{pmatrix}. \tag{6.28}$$

In this description, the components $\Gamma_\pm(\tau)$ represent the wave action belonging to the right-hand and left-hand circularly polarized modes, respectively (as defined from the point of view of the source). Also, $\Gamma(\tau)$ is normalized such that $\Gamma^\dagger(\tau)\Gamma(\tau) = 1$.

The action (6.27) can be further simplified by adopting the noncovariant description and eliminating the frequency variables $P_0(\tau)$ from the action. Following Sec. 4.6.3, one obtains the frequency from the corrected XGO dispersion relation [see Eq. (4.71)]

$$P \cdot P = \omega_p^2(\mathbf{X}) - \left( (\boldsymbol{\nabla}\omega_p^2) \cdot \mathbf{F} + \sum_s \frac{\omega_{ps}^2}{P_0} (\mathbf{e}_\mathbf{p} \cdot \boldsymbol{\Omega}_s) \right) \Gamma^\dagger \sigma_z \Gamma. \tag{6.29}$$

Since the polarization coupling is small, the wave frequency $P_0$ can be approximated by

$$P_0 \simeq H_0(\mathbf{X}, \mathbf{P}) - \Sigma(\mathbf{X}, \mathbf{P})\Gamma^\dagger \sigma_z \Gamma, \tag{6.30}$$

where $H_0(\mathbf{x}, \mathbf{p}) \doteq [\mathbf{p}^2 + \omega_p^2(\mathbf{x})]^{1/2}$ is the GO wave frequency and the spin-coupling term is

$$\Sigma(\mathbf{x}, \mathbf{p}) \doteq \boldsymbol{\nabla} H_0(\mathbf{x}, \mathbf{p}) \cdot \mathbf{F}(\mathbf{p}) + \frac{1}{2H_0^2(\mathbf{x}, \mathbf{p})} \sum_s \omega_{ps}^2(\mathbf{x})[\mathbf{e}_\mathbf{p}(\mathbf{p}) \cdot \boldsymbol{\Omega}_s(\mathbf{x})]. \tag{6.31}$$

Then, following the procedure in Sec. 4.6.3, one obtains the non-covariant XGO point-particle action

$$\mathcal{S}_{\text{XGO}} = \int \mathrm{d}t \left[ \mathbf{P} \cdot \dot{\mathbf{X}} + \frac{i}{2} \left( \Gamma^\dagger \dot{\Gamma} - \dot{\Gamma}^\dagger \Gamma \right) - H_0(\mathbf{X}, \mathbf{P}) + \Sigma(\mathbf{X}, \mathbf{P}) \Gamma^\dagger \sigma_z \Gamma \right], \tag{6.32}$$

where in this case the dots represent derivatives with respect to the physical time $t$.

Upon treating $\mathbf{X}(t)$, $\mathbf{P}(t)$, $\Gamma(t)$, and $\Gamma^\dagger(t)$ as independent, one obtains the following ELEs:

$$\delta \mathbf{P}: \quad \frac{\mathrm{d}\mathbf{X}}{\mathrm{d}t} = \frac{\partial H_0}{\partial \mathbf{P}} - \frac{\partial \Sigma}{\partial \mathbf{P}} \Gamma \sigma_z \Gamma, \tag{6.33a}$$

$$\delta \mathbf{X}: \quad \frac{\mathrm{d}\mathbf{P}}{\mathrm{d}t} = -\frac{\partial H_0}{\partial \mathbf{X}} + \frac{\partial \Sigma}{\partial \mathbf{X}} \Gamma \sigma_z \Gamma, \tag{6.33b}$$

$$\delta \Gamma^\dagger: \quad \frac{\mathrm{d}\Gamma}{\mathrm{d}t} = i \Sigma \sigma_z \Gamma, \tag{6.33c}$$

$$\delta \Gamma: \quad \frac{\mathrm{d}\Gamma^\dagger}{\mathrm{d}t} = -i \Gamma \Sigma \sigma_z. \tag{6.33d}$$

Equations (6.33) form a complete set of equations. The first terms on the right-hand side of Eqs. (6.33a) and (6.33b) describe the ray dynamics in the GO limit. The second terms describe the coupling of the mode polarization with the ray curvature and magnetic field.

To better understand the equations for the wave polarization, let us rewrite Eq. (6.33c) as an equation in the basis of linearly polarized modes:[4]

$$\dot{Z} = \mathcal{R}\dot{\Gamma} = i\Sigma \mathcal{R} \sigma_z \Gamma = i\Sigma(\mathcal{R}\sigma_z \mathcal{R}^{-1}) Z = i \Sigma \sigma_y Z. \tag{6.34}$$

Since $\Sigma$ is a scalar and $\sigma_y$ is constant, this equation can be readily integrated, yielding[5]

$$Z(t) = \exp[i\Theta(t)\sigma_y] Z_0 = [\, \mathbb{I}_2 \cos\Theta(t) + i\sigma_y \sin\Theta(t) \,] \, Z_0, \tag{6.35}$$

[4]This equation could also be obtained if the ray equations were derived directly from the action (6.24).
[5]Here I used the well-known Euler formula for Pauli matrices, $e^{ia(\mathbf{n}\cdot\boldsymbol{\sigma})} = \mathbb{I}_2 \cos a + i(\mathbf{n}\cdot\boldsymbol{\sigma}) \sin a$.

where $\Theta(t) \doteq \int_{t_0}^{t} \mathrm{d}t' \, \Sigma(\mathbf{X}(t'), \mathbf{P}(t'))$ is the polarization precession angle and $Z_0 \doteq Z(t_0)$. This result can also be expressed explicitly as

$$Z(t) = \begin{pmatrix} \cos\Theta & -\sin\Theta \\ \sin\Theta & \cos\Theta \end{pmatrix} Z_0. \tag{6.36}$$

Hence, the polarization of the EM field rotates at the rate $\Sigma(t)$ in the reference frame defined by the basis vectors $(\mathbf{e}_1, \mathbf{e}_2)$. The first term of $\Sigma$ [see Eq. (6.31)] is identified as the rate of change of the wave Berry phase (Berry, 1984), and the second term in Eq. (6.31) is identified as the rate of change due to Faraday rotation. In optics, the rotation of the polarization plane caused by the Berry phase is also known as the *Rytov rotation* (Bliokh *et al.*, 2007; Tomita and Chiao, 1986).

### 6.3.3 Dynamics of pure states

If a ray initially corresponds to a strictly circular polarization such that $\sigma_z \Gamma_0 = \pm\Gamma_0$, then it will remain in the given polarization during later times. This is an example of a pure state, and the action (6.27) can be simplified to $\mathcal{S}_{\text{XGO}} = \int \mathrm{d}\tau \, L_\pm$, where the Lagrangian is given by

$$L_\pm = \mathbf{P} \cdot \dot{\mathbf{X}} - H_0(\mathbf{X}, \mathbf{P}) \pm \Sigma(\mathbf{X}, \mathbf{P}). \tag{6.37}$$

Here $L_\pm$ governs the propagation of right-hand and left-hand polarization modes, respectively. In this model, only the variables $\mathbf{X}(t)$ and $\mathbf{P}(t)$ are independent. The corresponding ELEs are

$$\delta\mathbf{P}: \quad \frac{\mathrm{d}\mathbf{X}}{\mathrm{d}t} = \frac{\partial H_0}{\partial \mathbf{P}} \mp \frac{\partial \Sigma}{\partial \mathbf{P}}, \tag{6.38a}$$

$$\delta\mathbf{X}: \quad \frac{\mathrm{d}\mathbf{P}}{\mathrm{d}t} = -\frac{\partial H_0}{\partial \mathbf{X}} \pm \frac{\partial \Sigma}{\partial \mathbf{X}}. \tag{6.38b}$$

Hence, it is clear that left-hand and right-hand polarized waves will follow different ray trajectories.

To illustrate the polarization-driven divergence of the ray trajectories, Eqs. (6.38) were implemented in a ray tracing code. The equations were solved using MATLAB's numerical solver `ODE45`. The calculated ray trajectories are compared with those determined by the lowest-order GO action

$$\mathcal{S}_{\text{GO}} = \int \mathrm{d}t \left[ \mathbf{P} \cdot \dot{\mathbf{X}} - H_0(\mathbf{X}, \mathbf{P}) \right], \tag{6.39}$$

which does not account for polarization effects. The corresponding ELEs are identical to Eqs. (6.38) when setting the polarization-coupling term to zero:[6]

$$\delta \mathbf{P}: \quad \frac{\mathrm{d}\mathbf{X}}{\mathrm{d}t} = \frac{\partial H_0}{\partial \mathbf{P}}, \tag{6.40a}$$

$$\delta \mathbf{X}: \quad \frac{\mathrm{d}\mathbf{P}}{\mathrm{d}t} = -\frac{\partial H_0}{\partial \mathbf{X}}. \tag{6.40b}$$

The following procedure was used to compare the ray trajectories. First, the initial conditions $(\mathbf{X}_0, \mathbf{P}_0)$ for the ray trajectories are chosen, and the GO ray equations (6.40) are solved. In order to correctly compare the GO and XGO ray trajectories, the GO and XGO frequencies must be the same. (In experiments, the wave generator determines the wave frequency; thus, the GO and XGO frequencies must be equal.) However, as seen from Eq. (6.30), the XGO frequency will have a small correction due to the polarization-coupling term. Hence, the initial conditions for the XGO ray trajectories must be modified. To determine the new initial conditions, note that in experiments the position and direction of the antenna can only fix the initial position of the wave packet and the direction of its wavevector. Hence, for the XGO ray trajectories the initial wavevector is modified to $\mathbf{P}_0 \to a\mathbf{P}_0$. Here $a$ is a real scalar constant and is determined by equalizing the GO and XGO wave frequencies:

$$H_0(\mathbf{X}_0, \mathbf{P}_0) = H_0(\mathbf{X}, a\mathbf{P}_0) \mp \Sigma(\mathbf{X}, a\mathbf{P}_0). \tag{6.41}$$

(Since the polarization-coupling term is small, the constant $a$ should be close to unity.) Once the new initial conditions are obtained, the XGO ray equations (6.38) are solved. Then, the GO and XGO ray trajectories are compared.

As shown in Figure 6.1, the XGO ray trajectories for a right-polarized and left-polarized waves differ noticeably from the GO ray trajectory. The divergence along the $x$ axis is driven by the polarization coupling. In optics, this effect is known as the Hall effect of light and has been observed for EM waves propagating in isotropic non-birefringent dielectrics (Bliokh *et al.*, 2008).

### 6.3.4 Noncanonical representation and the Berry connection

As discussed in Sec. 4.6.6, the variables $(\mathbf{X}, \mathbf{P})$ are not gauge invariant with respect to continuous $\mathrm{U}(N)$ transformations on the matrix $\Xi(\mathbf{p})$.[7] Here I obtain the alternative, noncanonical representation of the ray Lagrangian (6.37) that is invariant and that explicitly shows the so-called Berry connection. Following the

[6] For simplicity, I assume that no background magnetic fields are present.

[7] In the case of pure states, a continuous transformation on $\Xi(\mathbf{p})$ leads to adding $\partial_{\mathbf{P}'}\chi(\mathbf{P})$ to $\mathbf{F}(\mathbf{P})$, where $\chi(\mathbf{P})$ is an arbitrary scalar function.

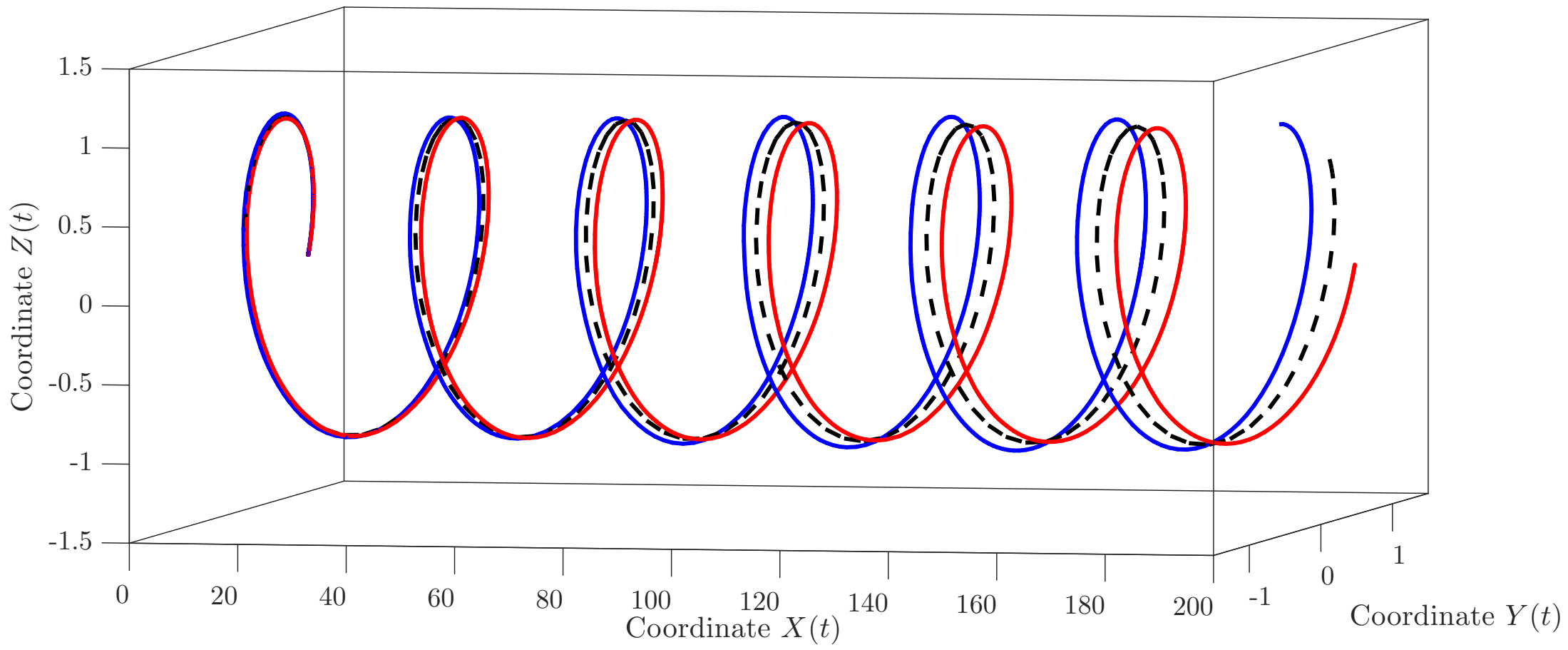

Figure 6.1: Comparison between ray trajectories calculated using the equations of traditional GO [Eqs. (6.40), dashed line] and extended GO [Eqs. (6.38)]. The blue and red lines represent the ray trajectories for the right-hand and left-hand polarized rays, respectively. For simplicity, non-magnetized plasma is considered, so the Faraday effect is absent. The plasma frequency is given by $\omega_p^2(\mathbf{x}) = y^2 + z^2$. The initial location of the ray trajectories is $\mathbf{X}_0 = (0, 1, 0)$, and the initial momentum is $\mathbf{P}_0 = (5, 0, 1)$. [The units are arbitrary ($c = 1$) since the figure is a general illustration only.] For this simulation, the GO parameter is roughly $\epsilon \sim 1/|\mathbf{P}_0| \sim 0.2$. Due to the radial gradient in the plasma frequency, the wave rays follow helical trajectories along the $x$ axis.

procedure given in Sec. 4.6.6, I introduce the variable transformation

$$\mathbf{X}(t) = \mathbf{X}'(t) \pm \mathbf{F}(\mathbf{P}'(t)), \qquad \mathbf{P}(t) = \mathbf{P}'(t). \tag{6.42}$$

The new variables $(\mathbf{X}', \mathbf{P}')$ will be shown to be gauge independent. Upon substituting the variable transformation and Eq. (6.31) into Eq. (6.37), one obtains

$$L_\pm = \mathbf{P}' \cdot \dot{\mathbf{X}}' \mp \dot{\mathbf{P}}' \cdot \mathbf{F}(\mathbf{P}') - H_0(\mathbf{X}', \mathbf{P}') \pm \sum_s \frac{\omega_{ps}^2(\mathbf{X}')}{2H_0^2(\mathbf{X}', \mathbf{P}')} [\mathbf{e}_{\mathbf{P}'}(\mathbf{P}') \cdot \boldsymbol{\Omega}(\mathbf{X}')], \tag{6.43}$$

where I dropped a perfect time derivative. Here I assumed that $\omega_p^2(\mathbf{x})$ is smooth, and I neglected terms of $\mathcal{O}(\epsilon^2)$. I also approximated $\mathbf{X} \simeq \mathbf{X}'$ in the Faraday rotation term since it is already $\mathcal{O}(\epsilon)$.

The term $\dot{\mathbf{P}}' \cdot \mathbf{F}(\mathbf{P}')$ in Eq. (6.43) is known as the Berry connection term (Bliokh *et al.*, 2015). Note that adding $\partial_{\mathbf{P}'} \chi(\mathbf{P}')$ to $\mathbf{F}(\mathbf{P}')$, where $\chi(\mathbf{P}')$ is an arbitrary scalar function, changes $L_\pm$ by a perfect derivative and does not affect the equations of motion. Hence, these variables are gauge invariant. The ELEs corresponding to the Lagrangian (6.43) are given by

$$\frac{\mathrm{d}\mathbf{X}'}{\mathrm{d}t} = \frac{\partial H_0}{\partial \mathbf{P}'} \pm \dot{\mathbf{P}}' \times (\boldsymbol{\nabla}_{\mathbf{P}'} \times \mathbf{F}) \mp \frac{\partial}{\partial \mathbf{P}'} \left( \sum_s \frac{\omega_{ps}^2(\mathbf{X}')}{2H_0^2(\mathbf{X}', \mathbf{P}')} [\mathbf{e}_{\mathbf{P}'}(\mathbf{P}') \cdot \boldsymbol{\Omega}(\mathbf{X}')] \right), \tag{6.44a}$$

$$\frac{\mathrm{d}\mathbf{P}'}{\mathrm{d}t} = -\frac{\partial H_0}{\partial \mathbf{X}'} \pm \frac{\partial}{\partial \mathbf{X}'} \left( \sum_s \frac{\omega_{ps}^2(\mathbf{X}')}{2H_0^2(\mathbf{X}', \mathbf{P}')} [\mathbf{e}_{\mathbf{P}'}(\mathbf{P}') \cdot \boldsymbol{\Omega}(\mathbf{X}')] \right). \tag{6.44b}$$

These equations are equivalent to Eqs. (6.38) within the accuracy of the theory. After substituting Eq. (6.22), one can also write Eq. (6.44a) as

$$\frac{\mathrm{d}\mathbf{X}'}{\mathrm{d}t} = \frac{\partial H_0}{\partial \mathbf{P}'} \pm \frac{\dot{\mathbf{P}}' \times \mathbf{P}'}{|\mathbf{P}'|^3} \mp \frac{\partial}{\partial \mathbf{P}'} \left( \sum_s \frac{\omega_{ps}^2(\mathbf{X}')}{2H_0^2(\mathbf{X}', \mathbf{P}')} [\mathbf{e}_{\mathbf{P}'}(\mathbf{P}') \cdot \boldsymbol{\Omega}(\mathbf{X}')] \right). \tag{6.45}$$

With the use of the noncanonical coordinates $(\mathbf{X}', \mathbf{P}')$, the equations of motion no longer depend on the specific choice of $\mathbf{F}(\mathbf{P}')$; i.e., they are invariant with respect to the choice (6.21) of vectors $\mathbf{e}_1(\mathbf{P}')$ and $\mathbf{e}_2(\mathbf{P}')$.

## 6.4 RF waves in strongly magnetized plasma

In this section, I discuss polarization effects on waves propagating in strongly magnetized plasma. Specifically, I focus on the polarization-driven bending of ray trajectories. To do this, one ideally should repeat the analysis shown in the previous section; however, the eigenvectors $\mathbf{e}(\mathbf{x}, p_0, \mathbf{p})$ for waves in magnetized plasmas are much more complicated. Thus, the polarization-coupling term $\mathcal{U}(\mathbf{x}, p_0, \mathbf{p})$, which involves phase-space derivatives of $\mathbf{e}(\mathbf{x}, p_0, \mathbf{p})$, becomes too cumbersome to calculate analytically. For this reason, the results of this section will mainly rely on numerical simulations.

### 6.4.1 Schrödinger formulation of the wave dynamics

In order to simplify the numerical simulations, I shall adopt the noncovariant XGO model described in Sec. 4.6.3, as it involves a smaller number of independent variables. Moreover, I shall use an alternative formulation of the wave dynamics, which will help us avoid solving the XGO dispersion relation for waves in magnetized plasmas. For simplicity, in this section I consider a two-component plasma composed by electrons and a single species of ions.

Let us first cast Eqs. (6.3) into a vector Schrödinger equation. Let us define the following twelve-component vector composed of the electron velocity, ion velocity, electric, and magnetic fields:[8]

$$\Psi(t, \mathbf{x}) = \langle\, (t, \mathbf{x}) \mid \Psi \,\rangle \doteq \begin{pmatrix} \overline{\mathbf{v}}_e \\ \overline{\mathbf{v}}_i \\ \mathbf{E} \\ \mathbf{B} \end{pmatrix}. \tag{6.46}$$

[8]For the general case of an $n$-component plasma, $\Psi(t, \mathbf{x})$ would be a $(3n + 6)$-component vector.

From Eqs. (6.3), it can be easily shown that $|\,\Psi\,\rangle$ satisfies a vector Schrödinger equation

$$\widehat{p}_0\,|\,\Psi\,\rangle = \widehat{\mathcal{H}}\,|\,\Psi\,\rangle\,, \tag{6.47}$$

where the Hamiltonian operator $\widehat{\mathcal{H}} = \mathcal{H}(\widehat{\mathbf{x}}, \widehat{\mathbf{p}})$ is a nine-dimensional matrix given by

$$\mathcal{H}(\widehat{\mathbf{x}}, \widehat{\mathbf{p}}) = \begin{pmatrix} -(\boldsymbol{\alpha}\cdot\widehat{\boldsymbol{\Omega}}_e) & 0 & i\widehat{\omega}_{pe}\,\mathbb{I}_3 & 0 \\ 0 & -(\boldsymbol{\alpha}\cdot\widehat{\boldsymbol{\Omega}}_i) & i\widehat{\omega}_{pi}\,\mathbb{I}_3 & 0 \\ -i\widehat{\omega}_{pe}\,\mathbb{I}_3 & -i\widehat{\omega}_{pi}\,\mathbb{I}_3 & 0 & ic(\boldsymbol{\alpha}\cdot\widehat{\mathbf{p}}) \\ 0 & 0 & -ic(\boldsymbol{\alpha}\cdot\widehat{\mathbf{p}}) & 0 \end{pmatrix}. \tag{6.48}$$

The Hamiltonian operator $\widehat{\mathcal{H}}$ is Hermitian. Hence, a variational principle can be constructed where the action is $\mathcal{S} = \langle\,\Psi\,|\,(\widehat{p}_0\mathbb{I}_{12} - \widehat{\mathcal{H}})\,|\,\Psi\,\rangle$.

### 6.4.2 Point-particle XGO action

The dispersion operator corresponding to Eq. (6.47) is in the symplectic form: $\widehat{\mathcal{D}} = \widehat{p}_0\mathbb{I}_{12} - \widehat{\mathcal{H}}$. The corresponding Weyl symbol is $D(\mathbf{x}, p_0, \mathbf{p}) = p_0\mathbb{I}_{12} - \mathcal{H}(\mathbf{x}, \mathbf{p})$. Since $D(\mathbf{x}, p_0, \mathbf{p})$ is diagonal on the wave frequency $p_0$, the corresponding twelve-component eigenvectors diagonalize the Hamiltonian $\mathcal{H}(\mathbf{x}, \mathbf{p})$. Hence, they depend on the spatial position and momentum coordinates only, so $\mathbf{e} = \mathbf{e}(\mathbf{x}, \mathbf{p})$. Upon denoting $H_0(\mathbf{x}, \mathbf{p})$ as the eigenvalue of the Hamiltonian matrix $\mathcal{H}(\mathbf{x}, \mathbf{p})$, one has

$$\mathcal{H}(\mathbf{x}, \mathbf{p})\,\mathbf{e}(\mathbf{x}, \mathbf{p}) = H_0(\mathbf{x}, \mathbf{p})\,\mathbf{e}(\mathbf{x}, \mathbf{p}). \tag{6.49}$$

Hence, the eigenvalue of $D(\mathbf{x}, p_0, \mathbf{p})$ is given by $\lambda(\mathbf{x}, p_0, \mathbf{p}) \doteq p_0 - H_0(\mathbf{x}, \mathbf{p})$.

In the following, I focus on the polarization-driven bending of ray trajectories and assume that the eigenmodes are degenerate and decoupled. Thus, only a single resonant eigenmode is considered, and I ignore other polarization effects such as mode conversion.[9] (In other words, I limit the discussion to the dynamics of pure states.) Since only a single resonant eigenmode is considered, the matrix $\Xi(\mathbf{x}, \mathbf{p})$ is a twelve-component column vector given by $\Xi(\mathbf{x}, \mathbf{p}) = \mathbf{e}(\mathbf{x}, \mathbf{p})$. After following the procedure given in Sec. 4.6, one obtains the point-particle XGO action for a single eigenmode:

$$\mathcal{S}_{\text{XGO}} = \int \mathrm{d}\tau \left[ P\cdot\dot{X} - \frac{i}{2}\left(Z^\dagger\dot{Z} - \dot{Z}^\dagger Z\right) + P_0 - H_0(\mathbf{X}, \mathbf{P}) + Z^\dagger\mathcal{U}(\mathbf{X}, \mathbf{P})Z \right], \tag{6.50}$$

[9]For more information on mode conversion in tokamak plasmas, see Jaun *et al.* (1998, 2007).

where the polarization-coupling term is

$$\mathcal{U}(\mathbf{x},\mathbf{p}) = -\left(\frac{\partial H_0}{\partial \mathbf{p}}\right)\left(\mathbf{e}^\dagger \frac{\partial \mathbf{e}}{\partial \mathbf{x}}\right)_A + \left(\frac{\partial H_0}{\partial \mathbf{x}}\right)\left(\mathbf{e}^\dagger \frac{\partial \mathbf{e}}{\partial \mathbf{p}}\right)_A - \left(\frac{\partial \mathbf{e}^\dagger}{\partial \mathbf{p}}(\mathcal{H} - H_0\mathbb{I}_{12})\frac{\partial \mathbf{e}}{\partial \mathbf{x}}\right)_A. \tag{6.51}$$

Here $\mathcal{U}(\mathbf{x},\mathbf{p})$ is a scalar so one can use the normalization condition ($Z^\dagger Z = 1$) to eliminate the state vector $Z$ from the spin-coupling Hamiltonian. Then, the dynamics of the state vector $Z$ becomes trivial so one can drop the second term of the action (6.50). This leads to

$$\mathcal{S}_{\text{XGO}} = \int \mathrm{d}\tau \, [\, P \cdot \dot{X} + P_0 - H_0(\mathbf{X},\mathbf{P}) + \mathcal{U}(\mathbf{X},\mathbf{P})\,]. \tag{6.52}$$

The advantage of using the Schrödinger formulation of the wave dynamics is that the wave frequency $P_0$ can be obtained explicitly. From Eq. (4.71), the wave frequency $P_0$ can be obtained from the action (6.52) so that

$$P_0(\mathbf{X},\mathbf{P}) = H_0(\mathbf{X},\mathbf{P}) - \mathcal{U}(\mathbf{X},\mathbf{P}). \tag{6.53}$$

As shown, $H_0(\mathbf{x},\mathbf{p})$ represents the lowest-order GO frequency. Any numerical algorithm that solves the algebraic eigenvalue problem can be used to solve Eq. (6.49) and obtain the corresponding GO frequencies $H_0(\mathbf{x},\mathbf{p})$ and eigenvectors $\mathbf{e}(\mathbf{x},\mathbf{p})$. The mode eigenvector $\mathbf{e}(\mathbf{x},\mathbf{p})$ is then substituted into the polarization-coupling term (6.51) to calculate the first-order correction to the wave frequency.

Following the procedure given in Sec. 4.6.3, one eliminates the frequency variable $P_0$ from the action (6.52). One then obtains the noncovariant XGO action[10]

$$\mathcal{S}_{\text{XGO}} = \int \mathrm{d}t \, [\, \mathbf{P} \cdot \dot{\mathbf{X}} - H_0(\mathbf{X},\mathbf{P}) + \mathcal{U}(\mathbf{X},\mathbf{P})\,], \tag{6.54}$$

where in this case the dots represents derivatives with respect to the physical time $t$. In this formulation, the independent variables are the wave packet position $\mathbf{X}(t)$ and the wave packet momentum $\mathbf{P}(t)$. The corresponding ELEs are given by

$$\frac{\mathrm{d}\mathbf{X}}{\mathrm{d}t} = \frac{\partial}{\partial \mathbf{P}}\,(H_0 - \mathcal{U})\,, \tag{6.55a}$$

$$\frac{\mathrm{d}\mathbf{P}}{\mathrm{d}t} = -\frac{\partial}{\partial \mathbf{X}}\,(H_0 - \mathcal{U})\,. \tag{6.55b}$$

[10]Since the eigenmodes are assumed to be degenerate and decoupled, the theory presented here is identical to that proposed by Littlejohn and Flynn (1991).

### 6.4.3 Numerical simulations

To illustrate the polarization-driven divergence of the ray trajectories, Eqs. (6.55) were implemented into a ray tracing code. The equations were solved using MATLAB's numerical solver `ODE45`. For completeness, the calculated ray trajectories were compared with those determined by the lowest-order GO action (6.39) with the appropriate Hamiltonian $H_0(\mathbf{X}, \mathbf{P})$. The corresponding ELEs are identical to Eqs. (6.40). The same procedure described in Sec. 6.3.3 was used to compare the GO and XGO ray trajectories.

Let us consider RF waves propagating in a tokamak hydrogen plasma with the following density and magnetic field profiles:

$$n_{s0}(\mathbf{x}) = n_{\text{core}} \left\{ \exp[-10(R-1)^2 - z^2] + .00001 \right\}, \tag{6.56}$$

$$\mathbf{B}(\mathbf{x}) = \frac{B_{\text{core}}}{R} \mathbf{e}_\psi, \tag{6.57}$$

where $R \doteq \sqrt{x^2 + y^2}$ and $\mathbf{e}_\psi$ is the toroidal direction. The major radius of the tokamak is considered to be 1 m. The plasma density in the tokamak core is $n_{\text{core}} = 10^{13}$ cm$^{-3}$, and the toroidal magnetic field in the core is $B_{\text{core}} = 0.5$ T. These parameters correspond to a plasma frequency $\omega_{p,\text{core}} \simeq 28$ GHz and electron gyrofrequency $\Omega_{e,\text{core}} \simeq 14$ GHz.

Let us first consider the case of an electron cyclotron wave launched from the low-magnetic-field side of the plasma at a frequency $\omega \simeq 54.7$ GHz. As seen in Figs. 6.2 and 6.3, the calculated GO and XGO ray trajectories are identical and no polarization effects can be observed. This can be explained as follows. At the given wave frequency $\omega \simeq 54.7$ GHz, the initial wavelength of the wave is approximately $\lambda \simeq 6$ mm in the plasma. Given that the characteristic scale length of the plasma is approximately $\ell = 1$ m, the initial GO parameter is approximately $\epsilon = \lambda/(2\pi\ell) \sim (10)^{-3}$. Since the GO parameter $\epsilon$ is small, polarization effects are weak and the resulting GO and XGO trajectories are nearly identical.

As a second example, let us consider the case of a RF wave launched from the low-magnetic-field side of the plasma at a lower frequency $\omega \simeq 0.41$ GHz. As seen in Figs. 6.4, 6.5, and 6.6, the GO and XGO ray trajectories clearly diverge due to polarization effects. For the given frequency, the initial wavelength of the wave is approximately $\lambda \simeq 0.17$ m in the plasma. This corresponds to the GO parameter $\epsilon = \lambda/(2\pi\ell) \sim 0.03$. In this case, the larger GO parameter leads to a stronger polarization-driven bending of the ray trajectories.

These findings show that polarization effects could become important for RF waves in the low-frequency range. For example, RF waves used in Lower-Hybrid Current Drive have typical frequencies in the range of 1–5 GHz (Rax, 2011), which correspond to wavelengths of 6–30 cm in vacuum. In this regime, the GO parameter could reach values of $\epsilon \simeq 0.05$, where polarization effects are expected to be important. Hence, a more careful analysis of polarization effects on RF wave trajectories is needed. Future research will be oriented to further analyze the implications of these effects.

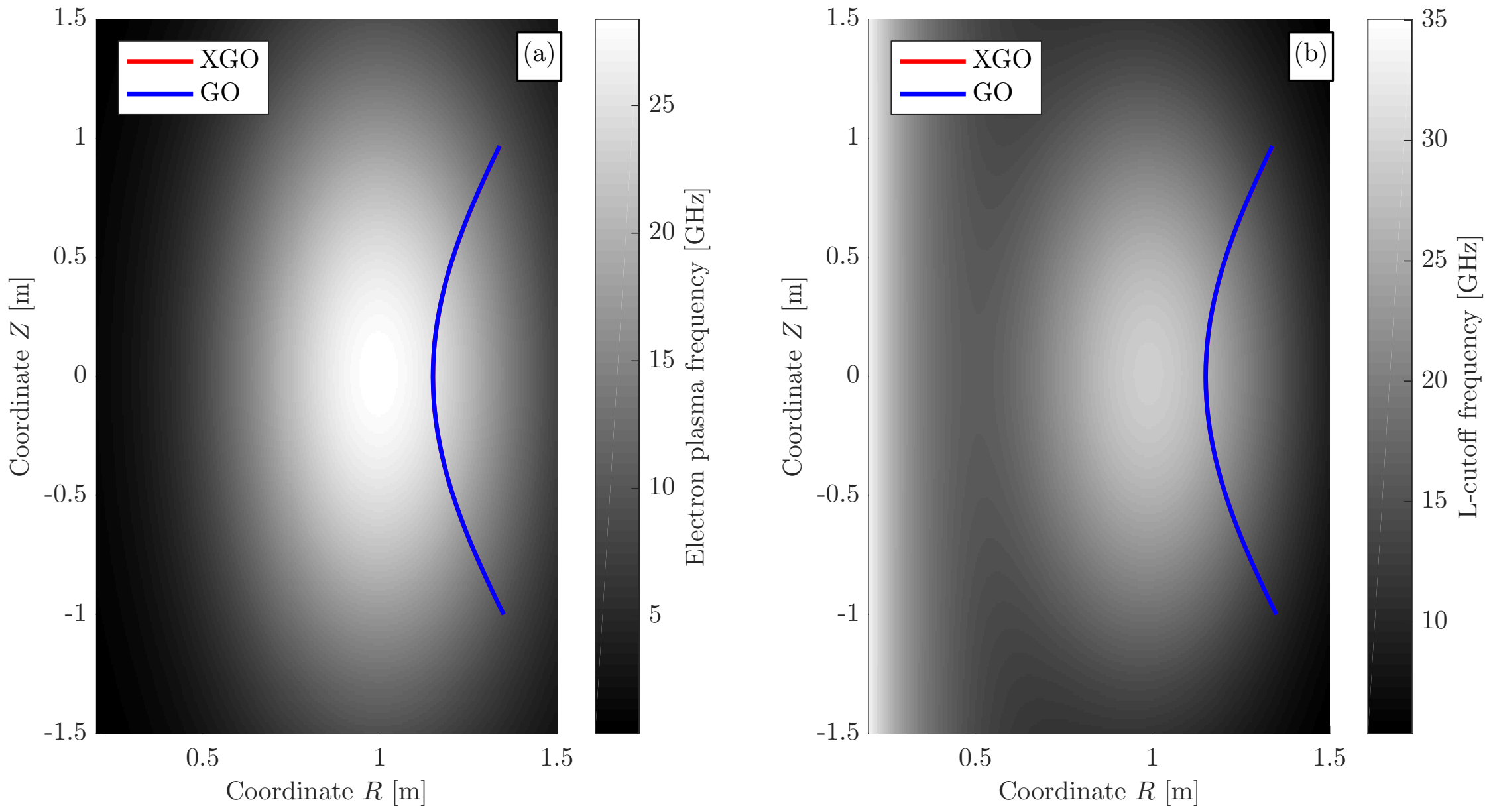

Figure 6.2: Comparison between ray trajectories calculated using the traditional GO [Eqs. (6.40)] and the XGO [Eqs. (6.55)] models. The red curve denotes the XGO ray trajectory with polarization effects included, while the blue curve is the GO ray trajectory. Due to the small GO parameter, the ray trajectories are identical. In the background of subfigure (a), the profile of the electron plasma frequency is shown; in subfigure (b), the profile of the electron L-cutoff frequency $\omega_{\mathrm{L}}(\mathbf{x}) \doteq [\omega_{pe}^2(\mathbf{x}) + \Omega_e^2(\mathbf{x})/4]^{1/2} - \Omega_e(\mathbf{x})/2$ is given.

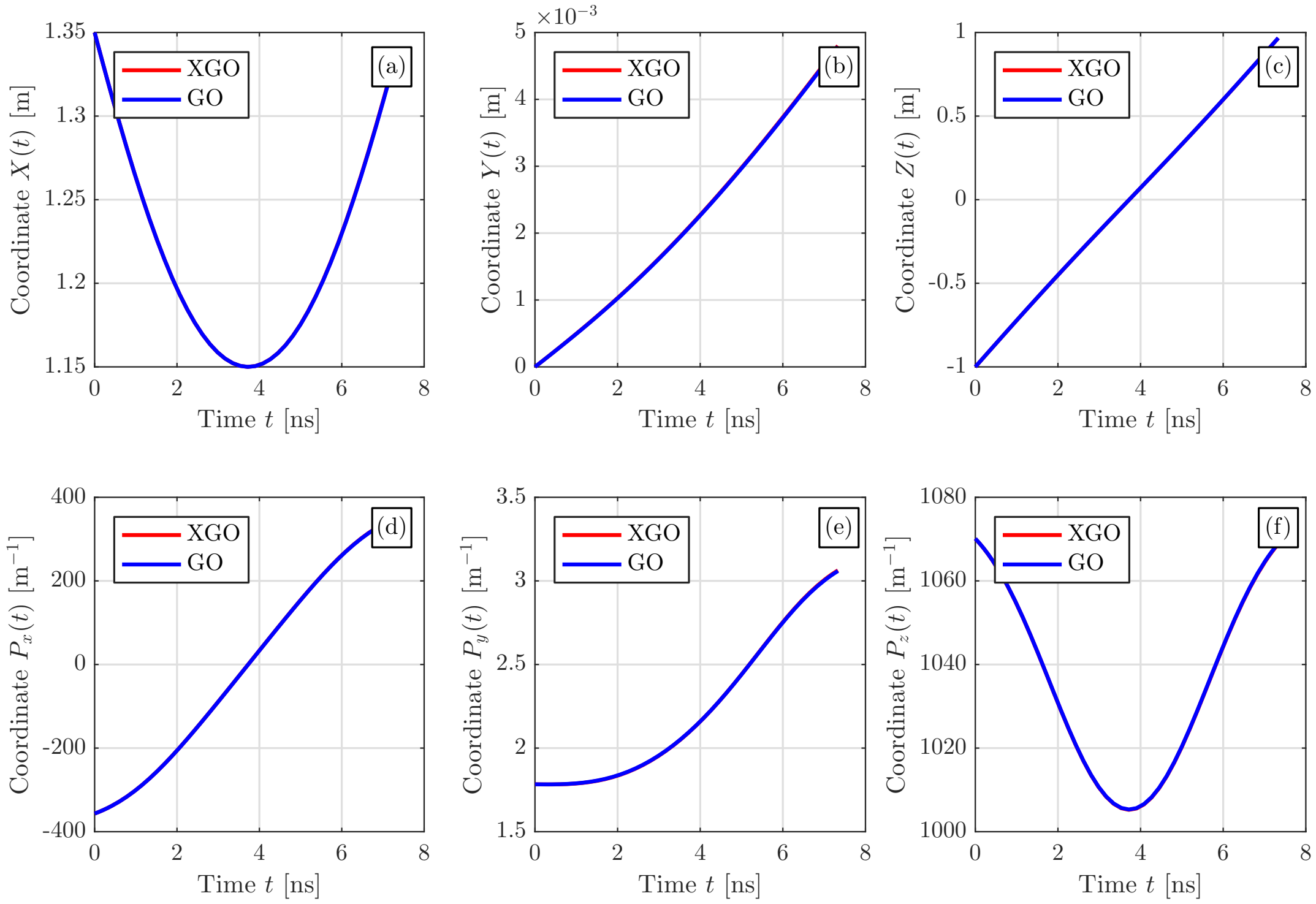

Figure 6.3: Evolution of the GO and XGO phase-space coordinates of a electron cyclotron wave propagating in a tokamak plasma. Subfigures (a)–(c) show the position coordinates of the wave packet, and subfigures (d)–(f) show the momentum coordinates of the wave packet.

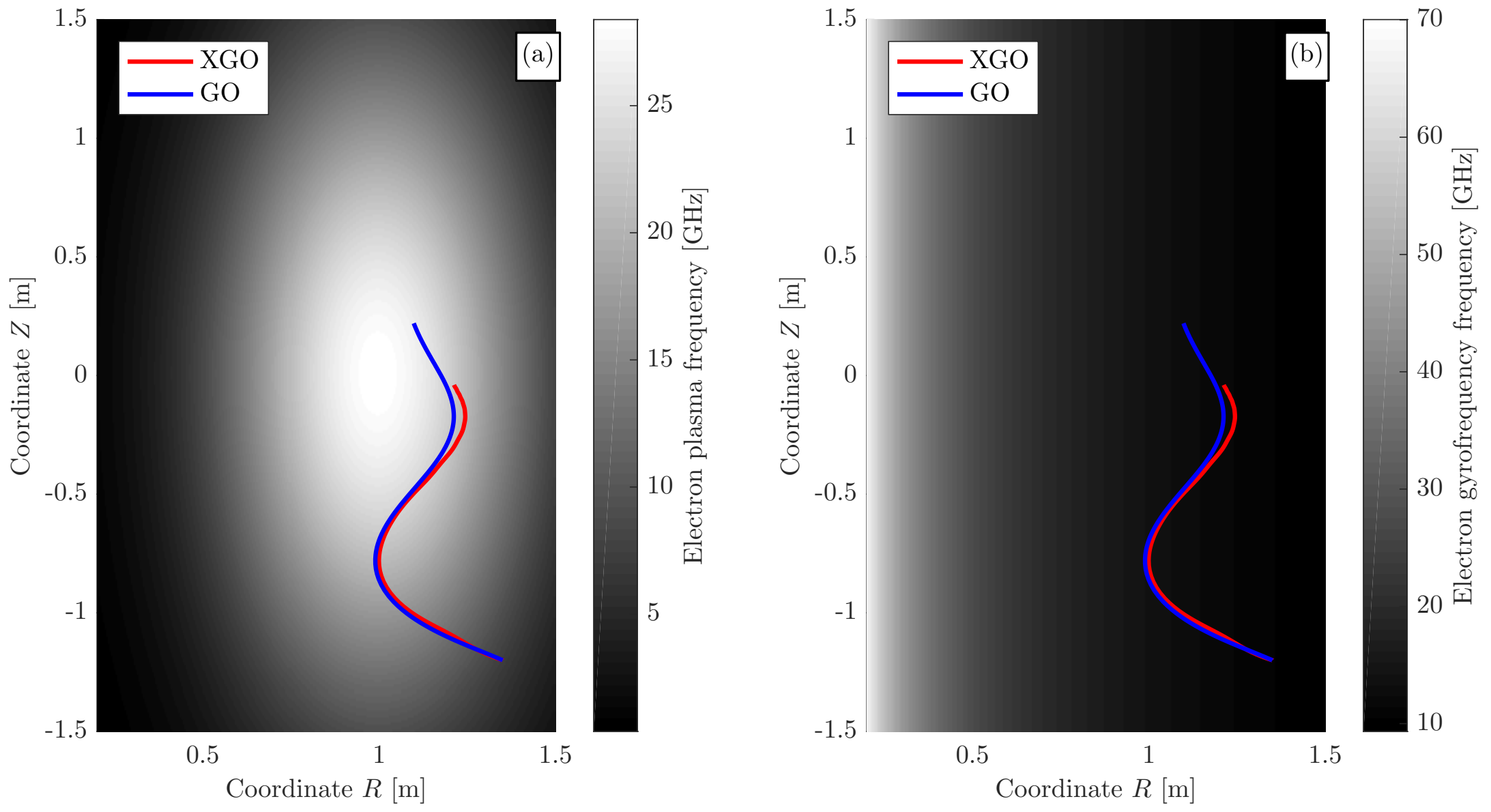

Figure 6.4: Comparison between ray trajectories of a lower-hybrid wave calculated using the traditional GO [Eqs. (6.40)] and the XGO [Eqs. (6.55)] models. The red curve denotes the XGO ray trajectory with polarization effects included, while the blue curve is the GO ray trajectory. In the background of subfigure (a), the profile of the electron plasma frequency is shown; in subfigure (b), the profile of the electron gyrofrequency is given.

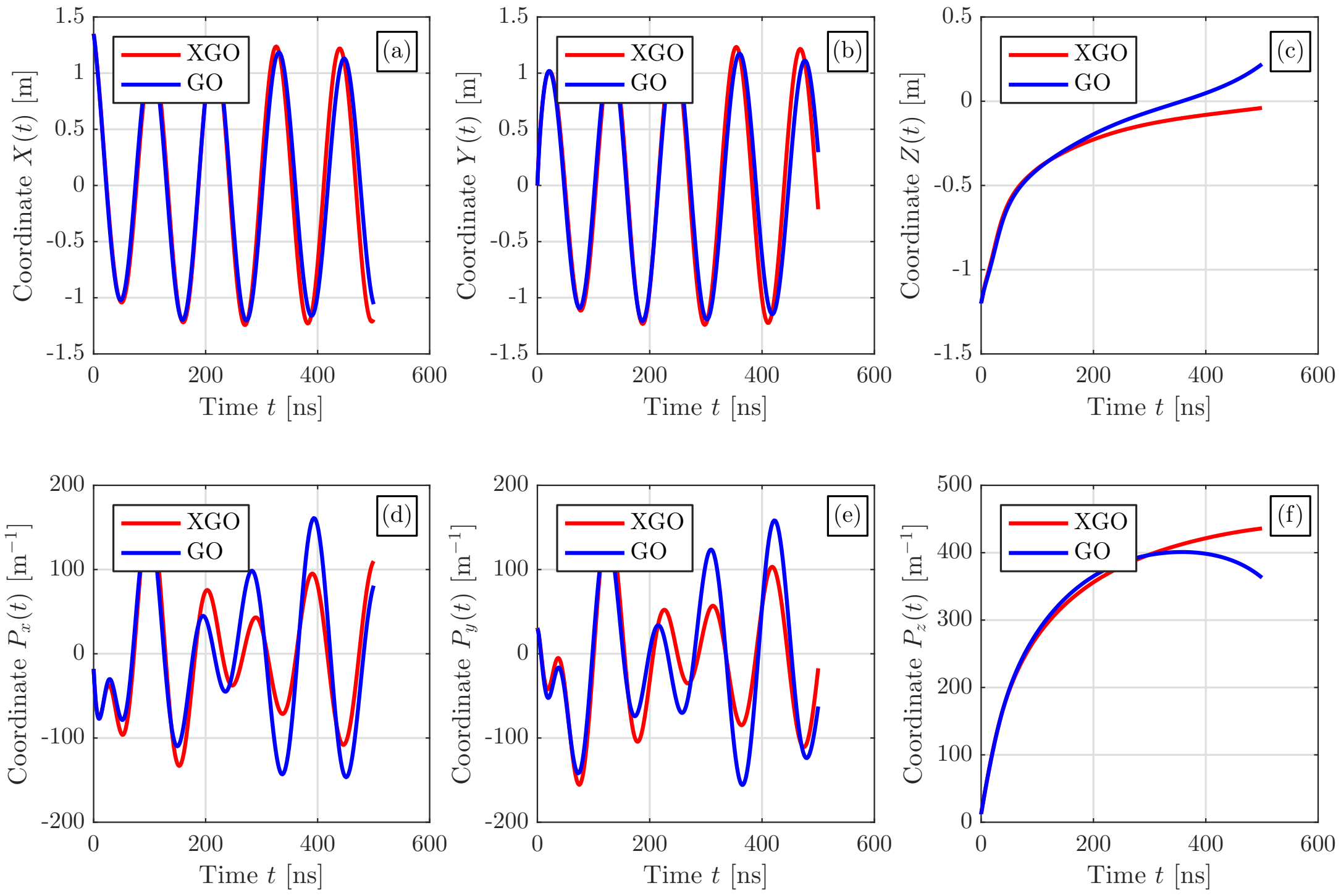

Figure 6.5: Evolution of the GO and XGO phase-space coordinates of a lower-hybrid wave propagating in a tokamak plasma. Subfigures (a)–(c) show the position coordinates of the wave packet, and subfigures (d)–(f) show the momentum coordinates of the wave packet.

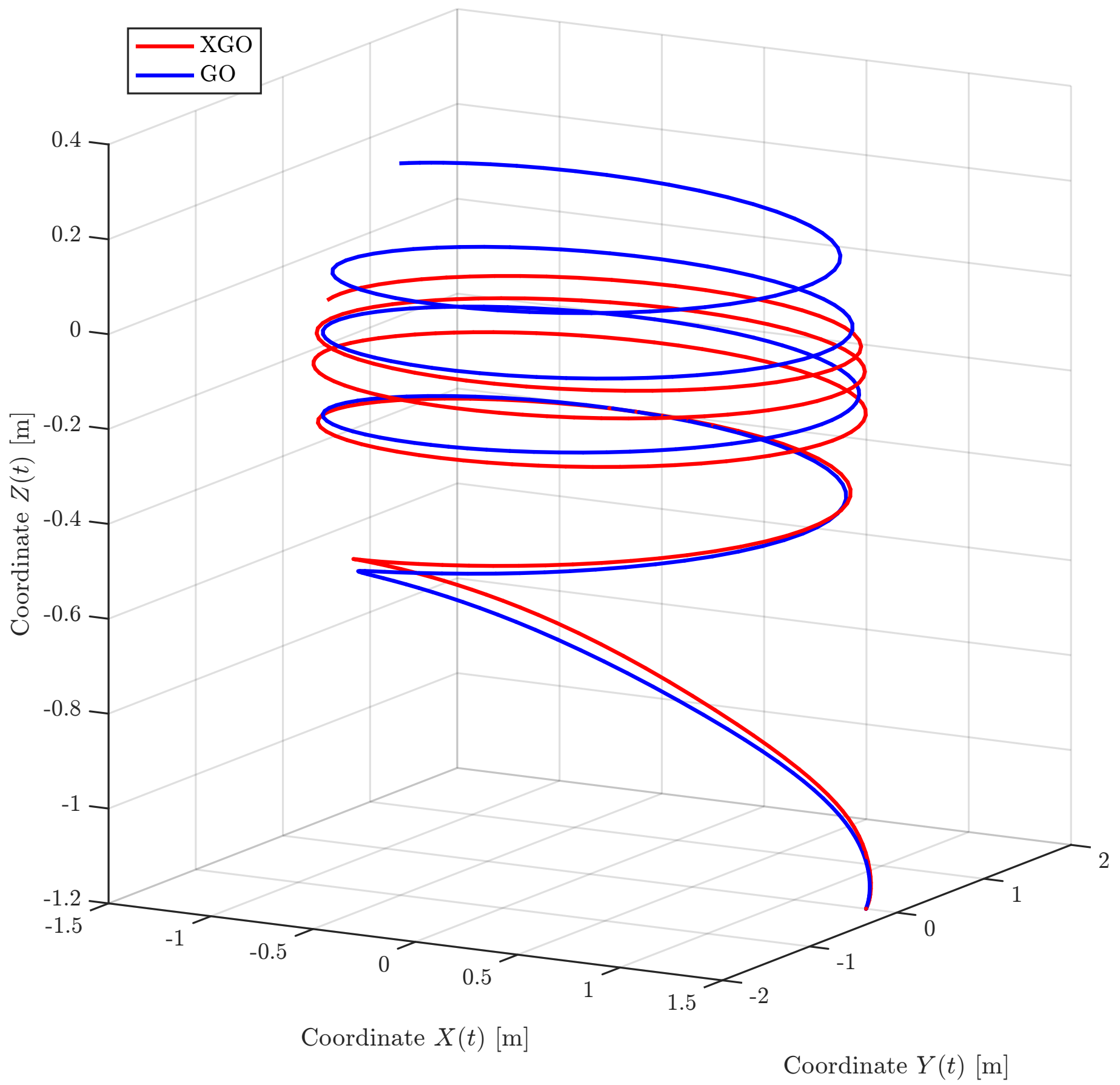

Figure 6.6: Three-dimensional parametric plot of the GO and XGO ray trajectories of a lower-hybrid wave propagating in a tokamak plasma.

## 6.5 Conclusions

In this Chapter, I discussed the polarization effects on waves propagating in magnetized plasma. For waves in weakly magnetized plasma, I analytically calculated how polarization effects cause a precession of the wave polarization and also polarization-driven bending of the GO ray trajectories. Regarding waves in strongly magnetized plasma, I presented a formalism that is useful to analyze the polarization-driven bending of ray trajectories. It is speculated that polarization effects will remain important for waves with lower frequencies; e.g., waves used for ion-cyclotron heating or for lower-hybrid current drive.

From the practical standpoint, this work can be extended in several directions. First, it is important to fully understand how polarization effects can affect the ray trajectories of low-frequency waves in tokamaks. If a more careful analysis shows that polarization effects need to be considered, corrections due to polarization effects will need to be included in ray tracing codes. Second, in this Chapter I did not mention another polarization-related phenomenon which is localized mode conversion (Friedland, 1985). The only existing ray tracing code that captures mode conversion is the `RAYCON` code (stands for "ray conversion") (Jaun *et al.*, 2007), but it does not capture the polarization-driven bending of ray trajectories. It could be useful for future RF applications to develop a general ray tracing code that could capture both of these phenomena simultaneously. Doing so is beyond the scope of the present thesis.

# Part III

# Nonlinear wave–wave interactions

# Chapter 7

# Ponderomotive dynamics of waves

The machinery developed in the previous Chapters can also be naturally applied to another class of problems, which is as follows. Similarly to how charged particles experience time-averaged ponderomotive forces in high-frequency (HF) fields, linear waves also experience time-averaged refraction in modulated media. As found in this research, this refraction, or *ponderomotive effect on waves*, can be described within a general variational theory. Using the Weyl symbol calculus, this theory can accommodate waves with temporal and spatial periods comparable to those of the modulation (provided that parametric resonances are avoided). The same theory also shows that any wave is, in fact, a polarizable object that contributes to the linear dielectric tensor of the ambient medium. In this Chapter, I present this theory and also calculate some ponderomotive Hamiltonians of quantum particles and photons within a number of models. I also explain a fundamental connection between these results and the well-known electrostatic dielectric tensor of quantum plasmas. The results presented in this Chapter were published by Ruiz and Dodin (2017b).

## 7.1 Introduction

### 7.1.1 Motivation

A non-uniform HF electromagnetic (EM) field can produce a time-averaged force, known as the ponderomotive force, on any particle that is charged or, more generally, has nonzero polarizability.[1] This effect has permitted a number of applications ranging from atomic cooling to particle acceleration (Ashkin, 1970; Malka and Miquel, 1996), but many other interesting opportunities remain. In particular, similar manipulations can be practiced on waves too. Dodin and Fisch (2014) showed that any wave propagating through a temporally and (or) spatially modulated medium generally experiences time-averaged refraction determined by the modulation intensity.[2] This *ponderomotive effect on waves* subsumes the ponderomotive dynamics of particles as a special case because, in quantum mechanics, particles can be represented as waves. However, Dodin and Fisch (2014) assume that the wave period (both temporal and spatial) is much smaller than the modulation period. In other words, this theory is only applicable within the GO regime of validity. This approximation limits the applicability of the theory. One may wonder, then, whether this limitation can be relaxed (without specifying the type of waves being considered) and whether new interesting physics remain to be discovered.

[1] See, for example, the works by Boot and Harvie (1957), Gaponov and Miller (1958), Dewar (1972), Cary and Kaufman (1977), and Grebogi *et al.* (1979).

In this Chapter, I answer these questions positively by proposing a general theory of the ponderomotive effect on waves. In contrast with the formulation proposed by Dodin and Fisch (2014), this theory can describe waves with temporal and spatial periods comparable to those of the modulation (provided that parametric resonances are avoided). Hence, this theory is not limited to the GO approximation. Here I explicitly derive the effective dispersion symbol (7.27) that governs the time-averaged dynamics of a wave in a quasiperiodically modulated medium. This result is later used to obtain the wave *ponderomotive Hamiltonian* (7.41). This formulation can be understood as a generalization of the oscillation-center (OC) theory, which is known from classical plasma physics (Dewar, 1973; Cary, 1981; Brizard, 2009), to any linear waves and quantum particles in particular. The theory also shows that *any wave is, in fact, a polarizable object* that contributes to the linear dielectric tensor of the ambient medium. As an illustration, ponderomotive energies of quantum particles and photons are calculated within a number of models and compared with simulations. In particular, I find that quantum effects can change the sign of the ponderomotive force. I also explain a fundamental connection between these results and the commonly known expression for the electrostatic dielectric function of quantum plasmas. In particular, I calculate the contribution of photons to the dielectric function of nonmagnetized plasmas.

It is to be noted that effective Hamiltonians for temporally driven systems have been studied before in condensed matter physics.[3] However, these studies are mainly focused on systems described by the Schrödinger equation and use the modulation period as the small parameter. In contrast, here I discuss more general waves and expand in the modulation amplitude rather than period. In this way, one can calculate the ponderomotive effect on waves using the Weyl calculus, which provides a direct connection with classical physics and the aforementioned OC theory in particular.

---

[2]In specific contexts, particularly for EM waves that have local dispersion, a similar phenomenon is known as the instantaneous Kerr effect due to cross-phase modulation (Mendonça, 2001; Alfano, 2006; Agrawal, 2007).

[3]See, e.g., Grozdanov and Raković (1988), Gilary *et al.* (2003), Mikami *et al.* (2016), Goldman *et al.* (2015), Eckardt and Anisimovas (2015), Bukov *et al.* (2015), Goldman and Dalibard (2014), and Novičenko *et al.* (2017).

### 7.1.2 Overview

This work is organized as follows. In Sec. 7.2, I present the variational formalism and the main assumptions used throughout the work. In Sec. 7.3, I derive a general expression for the effective wave action. In Sec. 7.4, I present a theory of ponderomotive dynamics for eikonal waves. In Sec. 7.5, I apply the theory to specific examples. In Sec. 7.6, I show the fundamental connection between the ponderomotive energy that is derived here and the commonly known dielectric tensor of quantum plasma. In Sec. 7.7, I summarize the main results. Some auxiliary calculations are included in Appendix C.2 of the thesis.

## 7.2 Physical model

### 7.2.1 Wave action principle

As discussed in Sec. 2.2.1, let us represent a wave field, either quantum or classical, as a scalar complex function $\Psi(x)$. The dynamics of any nondissipative linear wave can be described by the principle of stationary action $\delta\mathcal{S} = 0$. In the absence of external sources and parametric resonances, the action functional $\mathcal{S}$ is written as in Eq. (2.5) so

$$\mathcal{S} \doteq \int \mathrm{d}^4x\,\mathrm{d}^4x'\,\Psi^\dagger(x)\mathcal{D}(x,x')\Psi(x'), \tag{7.1}$$

where $\mathcal{D}$ is a Hermitian $[\mathcal{D}(x,x') = \mathcal{D}^*(x',x)]$ scalar kernel that describes the underlying medium. Varying $\mathcal{S}$ with respect to $\Psi^\dagger$ leads to the general wave equation

$$\delta\Psi^\dagger: \quad 0 = \int \mathrm{d}^4x'\,\mathcal{D}(x,x')\Psi(x'). \tag{7.2}$$

Similarly, varying with respect to $\Psi$ gives the equation adjoint to Eq. (7.2).

As in Sec. 2.2.2, I consider the wave $\Psi(x)$ as an abstract vector $|\,\Psi\,\rangle$ in the Hilbert space of wave states with inner product (2.8). In the Dirac notation, the action (7.1) is written as in Eq. (2.18):

$$\mathcal{S} = \langle\,\Psi\,|\,\widehat{\mathcal{D}}\,|\,\Psi\,\rangle\,, \tag{7.3}$$

where $\widehat{\mathcal{D}}$ is the Hermitian dispersion operator such that $\mathcal{D}(x,x') = \langle\,x\,|\,\widehat{\mathcal{D}}\,|\,x'\,\rangle$.

### 7.2.2 Problem outline

Below, I consider the propagation of a wave $|\,\Psi\,\rangle$, called the *probe wave* (PW), in a medium whose parameters are modulated by some other wave, which I call the *modulating wave* (MW). Note that the PW $|\,\Psi\,\rangle$ can represent a classical wave (e.g., a EM wave) or a quantum particle (e.g., a Schrödinger particle). Due to the HF modulations, $\mathcal{D}(x, x')$ is a rapidly oscillating function. The goal is to derive a reduced action for Eq. (7.3) that describes ponderomotive dynamics of the PW.

Let us assume that $\widehat{\mathcal{D}}$ can be decomposed as

$$\widehat{\mathcal{D}} = \widehat{\mathcal{D}}_0 + \widehat{\mathcal{D}}_{\text{osc}}, \tag{7.4}$$

where $\widehat{\mathcal{D}}_0$ represents the effect of the unperturbed background medium and $\widehat{\mathcal{D}}_{\text{osc}}$ represents a weak perturbation caused by the MW. Additionally, let us assume

$$\widehat{\mathcal{D}}_{\text{osc}} = \sum_{n=1}^{\infty} \sigma^n \widehat{\mathcal{D}}_n, \tag{7.5}$$

where $\sigma \ll 1$ is some linear measure of the MW amplitude[4] and $\widehat{\mathcal{D}}_n$ are Hermitian. Finally, let us require the MW (but not necessarily the PW) to satisfy the standard assumptions of geometrical optics (GO). This means that the MW frequency $\Omega$ and wavevector $\mathbf{K}$ must be large compared to the inverse temporal and spatial scales at which the envelope evolves. In a homogeneous medium, those scales would be simply the MW envelope duration $T_{\text{mw}}$ and the MW envelope length $\ell_{\text{mw}}$. More generally, one also has the scales $T_{\text{bg}}$ and $\ell_{\text{bg}}$ that characterize the background temporal and spatial inhomogeneities, correspondingly. Thus, the applicability of the theory relies on the smallness of the following parameter:

$$\epsilon_{\text{mw}} \doteq \max\left\{\frac{1}{\Omega T}, \frac{1}{|\mathbf{K}|\ell}\right\} \ll 1, \tag{7.6}$$

where $T \doteq \min\{T_{\text{bg}}, T_{\text{mw}}\}$ and $\ell \doteq \min\{\ell_{\text{bg}}, \ell_{\text{mw}}\}$. A more rigorous definition of the GO regime that also covers waves near natural resonances is somewhat subtle, so it is not discussed here.[5]

## 7.3 General theory

The oscillating terms in the dispersion operator will be eliminated by introducing an appropriate variable transformation on the PW. Specifically, let $|\,\Psi\,\rangle = \widehat{\mathcal{U}}\,|\,\psi\,\rangle$, where $\widehat{\mathcal{U}}$ is a unitary operator such that $\widehat{\mathcal{U}}^\dagger\,\widehat{\mathcal{U}} = \widehat{1} = \widehat{\mathcal{U}}\,\widehat{\mathcal{U}}^\dagger$. One requires that $\widehat{\mathcal{U}}$ be unitary so that the norm of the wave is conserved, $\langle\,\Psi\,|\,\Psi\,\rangle = \langle\,\psi\,|\,\psi\,\rangle$. Then, the action (7.3) transforms to

$$\mathcal{S} = \langle\,\psi\,|\,\widehat{\mathcal{D}}_{\text{eff}}\,|\,\psi\,\rangle\,, \tag{7.7}$$

[4] I shall incorporate the factor $\sigma$ in the definition of $\widehat{\mathcal{D}}_n$ eventually, but for now, it is convenient to have it written explicitly.
[5] For further details, see, for example, Tracy *et al.* (2014).

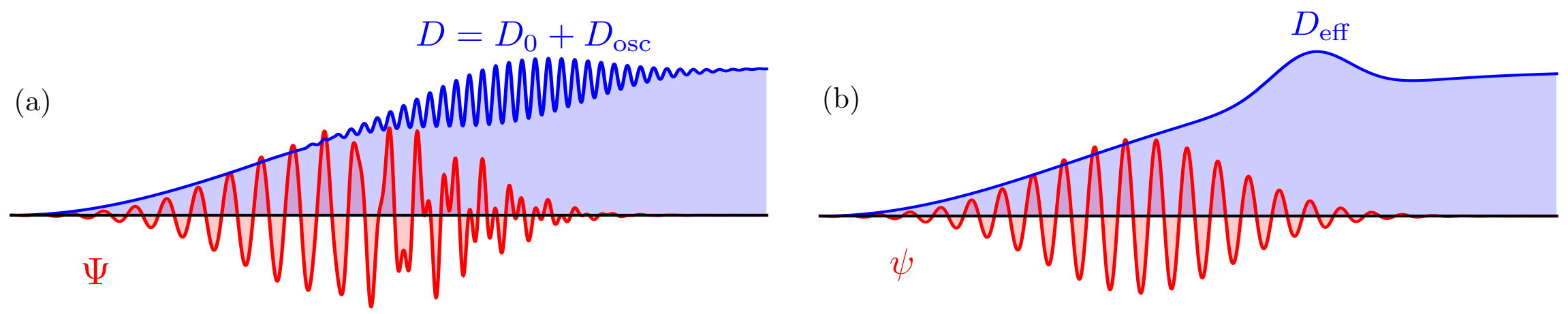

Figure 7.1: One-dimensional schematic of a PW (red) propagating in a medium with a given dispersion operator (blue) affected by some MW. (a) Dynamics in the original variables. (b) Dynamics in the OC representation, in which the oscillations at the MW phase and its harmonics are removed.

where $\widehat{\mathcal{D}}_{\text{eff}}$ is the effective dispersion operator

$$\widehat{\mathcal{D}}_{\text{eff}} \doteq \widehat{\mathcal{U}}^\dagger \, \widehat{\mathcal{D}} \, \widehat{\mathcal{U}}. \tag{7.8}$$

In the following, I search for a transformation $\widehat{\mathcal{U}}$ such that, unlike $\widehat{\mathcal{D}}$, the operator $\widehat{\mathcal{D}}_{\text{eff}}$ contains no dependence on the MW phase; i.e., it contains no HF modulations. The corresponding state $|\,\psi\,\rangle$ is then understood as the OC state of the PW in the modulated medium. A schematic of the transformation is shown in Fig. 7.1.

### 7.3.1 Near-identity unitary transformation

In light of the ordering shown in Eq. (7.4), I search for a unitary operator $\widehat{\mathcal{U}}$ using the standard perturbation approach based on Lie transforms (Cary, 1981; Cary and Littlejohn, 1983). Let $\widehat{\mathcal{U}}$ be given in the following form:

$$\widehat{\mathcal{U}} = \lim_{n\to\infty} \widehat{\mathcal{U}}_1 \, \widehat{\mathcal{U}}_2 \, \widehat{\mathcal{U}}_3 \cdots \widehat{\mathcal{U}}_n. \tag{7.9}$$

In order to satisfy the unitary condition, the operators $\widehat{\mathcal{U}}_n$ can be represented as follows:

$$\widehat{\mathcal{U}}_n = \exp(i\sigma^n \widehat{\mathcal{G}}_n), \tag{7.10}$$

where $\widehat{\mathcal{G}}_n$ is a scalar Hermitian operator called the *generator* of the unitary transformation $\widehat{\mathcal{U}}_n$. Similarly, I decompose $\widehat{\mathcal{D}}_{\text{eff}}$ as a power series in $\sigma$ so that

$$\widehat{\mathcal{D}}_{\text{eff}} = \sum_{n=0}^{\infty} \sigma^n \widehat{\mathcal{D}}_{\text{eff},n}, \tag{7.11}$$

where $\widehat{\mathcal{D}}_{\text{eff},n}$ is Hermitian. One then substitutes Eqs. (7.4), (7.5), (7.9), and (7.11) into Eq. (7.8). Upon using the Baker–Hausdorff lemma[6] and collecting terms by equal powers in $\sigma$, one obtains the following set of equations (Gramespacher and Weigert, 1996):

$$\widehat{\mathcal{D}}_{\text{eff},0} = \widehat{\mathcal{D}}_0, \tag{7.13a}$$

$$\widehat{\mathcal{D}}_{\text{eff},1} = \widehat{\mathcal{D}}_1 + i[\widehat{\mathcal{D}}_0, \widehat{\mathcal{G}}_1]_-, \tag{7.13b}$$

$$\widehat{\mathcal{D}}_{\text{eff},2} = \widehat{\mathcal{D}}_2 + i[\widehat{\mathcal{D}}_0, \widehat{\mathcal{G}}_2]_- + \widehat{\mathcal{C}}_2, \tag{7.13c}$$

and so on. The operator $\widehat{\mathcal{C}}_2$ is given by

$$\widehat{\mathcal{C}}_2 \doteq i[\widehat{\mathcal{D}}_1, \widehat{\mathcal{G}}_1]_- - \frac{1}{2}[[\widehat{\mathcal{D}}_0, \widehat{\mathcal{G}}_1]_-, \widehat{\mathcal{G}}_1]_-. \tag{7.14}$$

Here the brackets $[\cdot,\cdot]_-$ denote commutators of the operators; i.e., $[\widehat{\mathcal{A}}, \widehat{\mathcal{B}}]_- \doteq \widehat{\mathcal{A}}\widehat{\mathcal{B}} - \widehat{\mathcal{B}}\widehat{\mathcal{A}}$. If needed, this procedure can be iterated to higher orders in $\sigma$.

### 7.3.2 Phase-space representation

In order to obtain the effective dispersion operator $\widehat{\mathcal{D}}_{\text{eff}}$ and the transformation generators $\widehat{\mathcal{G}}_n$, let us consider the hierarchy equations (7.13) in the Weyl representation. Calculating the Weyl transformation [Eq. (A.1)] of Eqs. (7.13) leads to

$$D_{\text{eff},0} = D_0, \tag{7.15a}$$

$$D_{\text{eff},1} = D_1 - \{\{D_0, G_1\}\}, \tag{7.15b}$$

$$D_{\text{eff},2} = D_2 - \{\{D_0, G_2\}\} + C_2, \tag{7.15c}$$

where $D_n(x,p)$, $D_{\text{eff},n}(x,p)$ and $G_n(x,p)$ are the Weyl symbols of the operators $\widehat{\mathcal{D}}$, $\widehat{\mathcal{D}}_{\text{eff},n}$ and $\widehat{\mathcal{G}}_n$, respectively. Here the bracket $\{\{\cdot\,,\,\cdot\}\}$ denote the the eight-dimensional phase-space Moyal brackets [Eq. (A.10)]. Also, $C_2(x,p)$ is given by

$$C_2(x,p) = -\{\{D_1, G_1\}\} + \frac{1}{2}\{\{\,\{\{D_0, G_1\}\}, G_1\,\}\}. \tag{7.16}$$

[6] The Baker–Hausdorff lemma states that, given two Hermitian operators $\widehat{\mathcal{A}}$ and $\widehat{\mathcal{B}}$, then

$$e^{-i\lambda\widehat{\mathcal{A}}}\widehat{\mathcal{B}}e^{i\lambda\widehat{\mathcal{A}}} = \widehat{\mathcal{B}} + i\lambda[\widehat{\mathcal{B}},\widehat{\mathcal{A}}]_- + \frac{(i\lambda)^2}{2!}[[\widehat{\mathcal{B}},\widehat{\mathcal{A}}]_-,\widehat{\mathcal{A}}]_- + \cdots + \frac{(i\lambda)^n}{n!}[\cdots[[\widehat{\mathcal{B}},\underbrace{\widehat{\mathcal{A}}]_-,\widehat{\mathcal{A}}]_-,\cdots,\widehat{\mathcal{A}}]_-}_{n\text{ times}} + \cdots, \tag{7.12}$$

where $\lambda$ denotes a real parameter and the brackets denote the commutators of the operators.

It is to be noted that $D_n(x,p)$, $D_{\text{eff},n}(x,p)$, and $G_n(x,p)$ are real functions of the eight-dimensional phase space because the corresponding operators are Hermitian (see Appendix A).

The goal is to find $G_n(x,p)$ so that $D_{\text{eff},n}(x,p)$ contains no HF modulations. Hence, $D_{\text{eff},n}(x,p)$ is chosen such that

$$D_{\text{eff},0} = D_0, \tag{7.17a}$$

$$D_{\text{eff},1} = \langle\!\langle D_1 \rangle\!\rangle, \tag{7.17b}$$

$$D_{\text{eff},2} = \langle\!\langle D_2 \rangle\!\rangle + \langle\!\langle C_2 \rangle\!\rangle, \tag{7.17c}$$

where "$\langle\!\langle \ldots \rangle\!\rangle$" is a time average over a period of the modulational wave. In other words, $\langle\!\langle \ldots \rangle\!\rangle$ averages out the fast oscillations. After subtracting Eqs. (7.17) from Eqs. (7.15), one finds that the symbols $G_n(x,p)$ satisfy the following equations:

$$\{\!\{ D_0, G_1 \}\!\} = D_1 - \langle\!\langle D_1 \rangle\!\rangle, \tag{7.18a}$$

$$\{\!\{ D_0, G_2 \}\!\} = D_2 - \langle\!\langle D_2 \rangle\!\rangle + C_2 - \langle\!\langle C_2 \rangle\!\rangle. \tag{7.18b}$$

As usual, this procedure can be iterated to higher orders in $\sigma$. However, to obtain the lowest-order ponderomotive effect on waves, one only needs to calculate $\widehat{\mathfrak{D}}_{\text{eff}}(x,p)$ up to $\mathcal{O}(\sigma^2)$. Below, I demonstrate how to solve Eqs. (7.18) for $G_1(x,p)$ and $G_2(x,p)$.

### 7.3.3 $\mathfrak{D}_{\text{eff}}$ within the leading-order approximation

For $n = 0$, Eq. (7.17a) gives $D_{\text{eff},0}(x,p) = D_0(x,p)$. As expected, $D_{\text{eff},0}(x,p)$ describes the PW dynamics in the absence of the MW. Now, let us analyze the equations to the first order in $\sigma$. Since $D_1(x,p)$ is a linear measure of the MW field, I adopt

$$D_1(x,p) = \text{Re}\big[\mathcal{D}_1(x,p) e^{i\Theta(x)}\big], \tag{7.19}$$

where the real function $\Theta(x)$ is the MW phase and $\mathcal{D}_1(x,p)$ is the Weyl symbol characterizing the slowly varying MW envelope.[7,8] The gradients of the phase,

$$\Omega(x) \doteq -\partial_t \Theta, \qquad \mathbf{K}(x) \doteq \boldsymbol{\nabla}\Theta, \tag{7.20}$$

[7]One can also easily generalize further calculations to the case of multiple incoherent GO MWs.

[8]By a slowly varying function $f(x,p)$ in spacetime, I mean an infinitely differentiable function that may be written as $f(\epsilon x, p)$ so that its $n$th spacetime derivatives are $\mathcal{O}(\epsilon)$.

determine the MW local frequency and wavevector, respectively. For convenience, I introduce the MW four-wavevector $K_\mu(x) \doteq -\partial_\mu \Theta = (\Omega, -\mathbf{K})$, which is considered to be a slow function. [In other words, derivatives of the four-wavevector are considered to be $\mathcal{O}(\epsilon_{\text{mw}})$.] Accordingly, the contravariant representation of the MW four-wavevector is $K^\mu(x) = (\Omega, \mathbf{K})$.

Since $D_1(x,p)$ is quasiperiodic,[9] on has $\langle\!\langle D_1 \rangle\!\rangle(x,p) = 0$. From Eq. (7.17b), one concludes that $D_{\text{eff},1}(x,p) = 0$. In consequence, Eq. (7.18a) is written as

$$\{\!\{D_0, G_1\}\!\} = D_1 = \text{Re}\big[\mathcal{D}_1(x,p) e^{i\Theta(x)}\big]. \tag{7.21}$$

In order to find a solution to this equation, let us search for $G_1(x,p)$ in the polar representation:

$$G_1(x,p) = \text{Re}\big[\mathcal{G}_1(x,p) e^{i\Theta(x)}\big], \tag{7.22}$$

where $\mathcal{G}_1(x,p)$ serves as a slowly varying envelope for $G_1(x,p)$. Upon substituting Eq. (7.22) into Eq. (7.21) and equating terms with the same phase, one obtains (Appendix C.2)

$$\begin{aligned}
\mathcal{D}_1(x,p) e^{i\Theta(x)} &= \{\!\{D_0(x,p), \mathcal{G}_1(x,p) e^{i\Theta(x)}\}\!\} \\
&= \mathcal{G}_1(x,p) \{\!\{D_0(x,p), e^{i\Theta(x)}\}\!\} + \mathcal{O}(\epsilon_{\text{mw}}) \\
&= -i\mathcal{G}_1 \big[D_0(x,p) \star e^{i\Theta(x)} - e^{i\Theta(x)} \star D_0(x,p)\big] + \mathcal{O}(\epsilon_{\text{mw}}) \\
&= -i \left[D_0(x, p + K/2) - D_0(x, p - K/2)\right] \mathcal{G}_1(x,p) e^{i\Theta(x)} + \mathcal{O}(\epsilon_{\text{mw}}),
\end{aligned} \tag{7.23}$$

where "$\star$" is the Moyal product (A.5) and $\mathcal{G}_1(x,p)$ is pulled out of the Moyal bracket because it is a slowly varying function. After solving for $\mathcal{G}_1(x,p)$, one obtains

$$\mathcal{G}_1(x,p) = \frac{i\mathcal{D}_1(x,p)}{D_0(x, p+K/2) - D_0(x, p-K/2)} + \mathcal{O}(\epsilon_{\text{mw}}). \tag{7.24}$$

Now let us calculate the second-order effective dispersion symbol $D_{\text{eff},2}(x,p)$. Upon substituting Eq. (7.21) into (7.16), one obtains $C_2(x,p) = -(1/2)\{\!\{D_1, G_1\}\!\}$. After substituting $D_1(x,p)$ and $G_1(x,p)$, one finds the following expression for the Weyl symbol $C_2(x,p)$ (Appendix C.2):

$$C_2(x,p) = -\frac{1}{4} \sum_{n=\pm 1} \frac{|\mathcal{D}_1(x, p + nK/2)|^2}{D_0(x, p+nK) - D_0(x,p)} + \text{Re}\big[\mathcal{C}_2(x,p) e^{i2\Theta(x)}\big] + \mathcal{O}(\epsilon_{\text{mw}}), \tag{7.25}$$

[9] By a quasiperiodic function $f(x,p)$ in spacetime, I mean one that may be written as $g(\epsilon x, p, \Theta)$, where $g(a,p,\Theta)$ is $2\pi$-periodic for any constant $a$ and where the phase $\Theta(x)$ is a function such that its derivatives are nonzero and slowly varying.

where $\mathcal{C}_2(x,p)$ is a slowly varying function whose explicit expression will not be needed for the present purposes. From Eqs. (7.17c) and (7.18b), one obtains $D_{\text{eff},2} = \langle\!\langle D_2 \rangle\!\rangle + \langle\!\langle C_2 \rangle\!\rangle$. Thus, the symbol $G_2(x,p)$ satisfies

$$\{\!\{D_0, G_2\}\!\} = D_2 - \langle\!\langle D_2 \rangle\!\rangle + \text{Re}\big(\mathcal{C}_2 e^{i2\Theta}\big). \tag{7.26}$$

The procedure in Eqs. (7.22)–(7.24) is then repeated to obtain the expression for $G_2(x,p)$ that satisfies Eq. (7.26). Finally, after collecting the results of this section, one obtains the effective dispersion symbol

$$D_{\text{eff}}(x,p) = D_0(x,p) + \sigma^2 \langle\!\langle D_2(x,p) \rangle\!\rangle - \frac{\sigma^2}{4} \sum_{n=\pm 1} \frac{|\mathcal{D}_1(x, p + nK/2)|^2}{D_0(x, p+nK) - D_0(x,p)} + \mathcal{O}(\epsilon_{\text{mw}}, \sigma^4). \tag{7.27}$$

The leading-order correction to $D_{\text{eff}}(x,p)$, which scales as $\epsilon_{\text{mw}}^0$, can be only of the fourth power of $\sigma$. This occurs because the third and other odd powers of the MW field have zero average and thus cannot contribute to the effective dispersion symbol $D_{\text{eff}}(x,p)$ that governs the $\Theta$-averaged motion.

The Weyl symbol $D_{\text{eff}}(x,p)$ in Eq. (7.27) constitutes one of the main results of this Chapter. It determines the asymptotic form of the effective dispersion operator that governs the dynamics of the PW averaged over the MW oscillations at small enough GO parameter $\epsilon_{\text{mw}}$ [Eq. (7.6)] and small enough MW amplitude $\sigma$. The actual operator $\widehat{\mathcal{D}}_{\text{eff}}$ can be obtained from the symbol (7.27) using the inverse Weyl transform (A.2). Alternatively, its coordinate representation $\mathcal{D}_{\text{eff}}(x,x')$ can be obtained using Eq. (A.3).

## 7.4 Ponderomotive dynamics

After calculating the effective dispersion operator $\widehat{\mathcal{D}}_{\text{eff}}$, one can obtain the corresponding ELE from the action (7.8):

$$\widehat{\mathcal{D}}_{\text{eff}} \,|\, \psi \,\rangle = 0. \tag{7.28}$$

Alternatively, one can apply the variational approach and study the action (7.7) in the *phase-space representation*. As in Sec. 2.4, the effective action is written as

$$\mathcal{S} = \int \mathrm{d}^4x \, \mathrm{d}^4p \, D_{\text{eff}}(x,p) \, W_\psi(x,p), \tag{7.29}$$

where $W_\psi(x,p)$ is the Wigner function (2.32) corresponding to the OC state $|\,\psi\,\rangle$; namely,

$$W_\psi(x,p) \doteq \frac{1}{(2\pi)^4} \int \mathrm{d}^4s \, e^{ip\cdot s} \, \langle\, x + s/2 \,|\, \psi \,\rangle \, \langle\, \psi \,|\, x - s/2 \,\rangle \,. \tag{7.30}$$

As already seen from the previous Chapters, the variational approach is particularly convenient for deriving approximate models of wave dynamics. In the following, I shall focus on the OC dynamics of PWs in the eikonal approximation.

### 7.4.1 Eikonal approximation

As in Sec. 3.3, let us consider the complex function $\psi(x) \doteq \langle\, x \mid \psi \,\rangle$ in the polar representation

$$\langle\, x \mid \psi \,\rangle = \psi(x) = a(x)\, e^{i\theta(x)}, \tag{7.31}$$

where $a(x)$ and $\theta(x)$ are real functions. The phase $\theta$ is considered fast compared to the slowly varying function $a(x)$. I also assume

$$\epsilon_{\rm pw} \doteq \max\left\{\frac{1}{\omega T}, \frac{1}{|\mathbf{k}|\ell}\right\} \ll 1, \tag{7.32}$$

where

$$\omega(x) \doteq -\partial_t \theta, \quad \mathbf{k}(x) \doteq \boldsymbol{\nabla}\theta \tag{7.33}$$

are the local PW frequency and the wavevector, respectively. In other words, I consider that the characteristic scale lengths of the inhomogeneities of the background medium and of the MW envelope are large with respect to the wavelength of the PW. For simplicity, I shall combine the small parameters (7.6) and (7.32) into a single parameter

$$\epsilon \doteq \max\{\epsilon_{\rm mw}, \epsilon_{\rm pw}\} \ll 1. \tag{7.34}$$

In particular, note that in order to apply the standard GO approximation to the original problem (see Chapter 3), the PW parameters must satisfy $\Omega/\omega \ll 1$ and $|\mathbf{K}|/|\mathbf{k}| \ll 1$; e.g., the PW wavelength must be smaller than the MW wavelength. However, after the transformation, the MW oscillations are eliminated (see Fig. 7.1), so the PW parameters must satisfy the less restrictive condition (7.32).

Since $\psi(x)$ is assumed to be quasi-monochromatic, I approximate the Wigner function (7.30) with the lowest-order term in $\epsilon$ [see Eq. (3.8) for more details]:

$$W_\psi(x,p) = a^2(x)\delta^4(p-k) + \mathcal{O}(\epsilon), \tag{7.35}$$

where $k_\mu(x) \doteq -\partial_\mu \theta$ is the four-wavevector [Eq. (3.5)] of the PW. Substituting Eq. (7.35) into Eq. (7.29) leads to the action functional

$$\mathcal{S} = \int \mathrm{d}^4 x\, a^2(x) D_{\rm eff}(x,k). \tag{7.36}$$

The action functional (7.36) has the form of Whitham's action (see Sec. 3.3.3), where

$$\mathcal{I}(x) \doteq a^2(x) \left( \frac{D_{\text{eff}}(x,p)}{\partial p_0} \right)_{p=k(x)} \tag{7.37}$$

serves as the wave action density (Whitham, 2011). [From now on, $k(x) = -\partial\theta$.] As discussed in Sec. 3.3.3, one can treat $a^2(x)$ and $\theta(x)$ as independent variables. The corresponding ELEs are

$$\delta\theta: \quad \partial_t \mathcal{I} + \boldsymbol{\nabla} \cdot (\mathcal{I}\mathbf{v}) = 0, \tag{7.38a}$$

$$\delta a^2: \quad D_{\text{eff}}(x,k) = 0, \tag{7.38b}$$

where the flow velocity $\mathbf{v}(x)$ is given by

$$\mathbf{v}(x) \doteq -\left( \frac{\partial D_{\text{eff}}(x,p)}{\partial p_0} \right)^{-1}_{p=k(x)} \left( \frac{\partial D_{\text{eff}}(x,p)}{\partial \mathbf{p}} \right)_{p=k(x)}. \tag{7.39}$$

As discussed in Sec. 3.3.3, Eq. (7.38a) represents the action conservation theorem, and Eq. (7.38b) is the local wave dispersion relation. Calculating the gradient of Eq. (7.38b) leads to the four-momentum conservation equation.

### 7.4.2 Hayes representation

Equation (7.38b) can be, in principle, inverted in order to express the PW frequency $\omega(x)$ as some function that depends on the other phase-space coordinates:

$$\omega = H_{\text{eff}}(t, \mathbf{x}, \mathbf{k}), \tag{7.40}$$

where $H_{\text{eff}}(t, \mathbf{x}, \mathbf{k})$ is the *wave ponderomotive Hamiltonian*.[10] The function $H_{\text{eff}}(t, \mathbf{x}, \mathbf{p})$ can be written as

$$H_{\text{eff}}(t, \mathbf{x}, \mathbf{p}) \doteq H_0(t, \mathbf{x}, \mathbf{p}) + \sigma^2 \Phi(t, \mathbf{x}, \mathbf{p}), \tag{7.41}$$

where higher powers of $\sigma$ are neglected, as in the previous section. (Henceforth, the small parameter $\sigma$ will be omitted for clarity.) Here $H_0(t, \mathbf{x}, \mathbf{p})$ is the *unperturbed frequency* of the PW, so it satisfies $D_0(x, p_*) = 0$, where

$$p_*^\mu(t, \mathbf{x}, \mathbf{p}) \doteq (H_0(t, \mathbf{x}, \mathbf{p}),\ \mathbf{p}) \tag{7.42}$$

[10] In general, there can be multiple dispersion branches. Here I assume that only one branch is excited while others have sufficiently different $\omega$ and thus can be neglected. For corrections caused by the presence of multiple branches, see Chapter 4.

is the unperturbed PW four-wavevector. The function $\Phi(t, \mathbf{x}, \mathbf{p})$ can be understood as the PW *ponderomotive frequency shift*. When multiplied by $\hbar$, $\Phi$ is also understood as the ponderomotive energy or ponderomotive potential; that said, one may want to restrict usage of the term "potential" to cases when $\Phi$ is independent of the momentum coordinate $\mathbf{p}$.

After using Eqs. (7.27) and (7.38b) together with the Taylor expansion

$$D_{\text{eff}}(x,p) \approx D_{\text{eff}}(x,p_*) + \left( \frac{\partial D_{\text{eff}}(x,p)}{\partial p_0} \right)_{p=p_*} [p_0 - H_0(t,\mathbf{x},\mathbf{p})], \tag{7.43}$$

one obtains an explicit expression for the ponderomotive frequency shift

$$\Phi(t,\mathbf{x},\mathbf{p}) = - \left( \frac{\partial D_{\text{eff}}(x,p)}{\partial p_0} \right)^{-1}_{p=p_*} \left( \langle\langle D_2(x,p) \rangle\rangle - \frac{1}{4} \sum_{n=\pm 1} \frac{|\mathcal{D}_1(x,p+nK/2)|^2}{D_0(x,p+nK) - D_0(x,p)} \right)_{p=p_*}. \tag{7.44}$$

Hence, the action (7.36) can be written in the Hayes (1973) form; namely,

$$\mathcal{S} \simeq - \int \mathrm{d}^4 x \, \mathcal{I} \left[ \partial_t \theta + H_{\text{eff}}(t,\mathbf{x},\boldsymbol{\nabla}\theta) \right]. \tag{7.45}$$

In this case, the corresponding ELEs are

$$\delta\theta: \quad \partial_t \mathcal{I} + \boldsymbol{\nabla} \cdot (\mathcal{I}\mathbf{u}) = 0, \tag{7.46a}$$

$$\delta\mathcal{I}: \quad \partial_t \theta + H_{\text{eff}}(t,\mathbf{x},\boldsymbol{\nabla}\theta) = 0, \tag{7.46b}$$

where $\mathbf{u}(t,\mathbf{x})$ is the effective PW group velocity,

$$\mathbf{u}(t,\mathbf{x}) \doteq \left( \frac{\partial H_{\text{eff}}(t,\mathbf{x},\mathbf{p})}{\partial \mathbf{p}} \right)_{\mathbf{p}=\mathbf{k}(t,\mathbf{x})}. \tag{7.47}$$

Equation (7.46a) serves as a ponderomotive continuity equation for the wave action. Also, Eq. (7.46b) is a Hamilton–Jacobi equation representing the local ponderomotive wave dispersion.

As a side note, when $|K| \ll |p_*|$ the effective Hamiltonian can be approximated to

$$H_{\text{eff}}(t,\mathbf{x},\mathbf{p}) \simeq H_0(t,\mathbf{x},\mathbf{p}) - \left[ \frac{\langle\langle D_2(x,p) \rangle\rangle}{\partial_{p_0} D_0(x,p)} - \frac{\sigma^2}{4\partial_{p_0} D_0(x,p)} K \cdot \frac{\partial}{\partial p} \left( \frac{|\mathcal{D}_1(x,p)|^2}{K \cdot \partial_p D_0(x,p)} \right) \right]_{p=p_*}, \tag{7.48}$$

where $K \cdot \partial_p \doteq K_\mu \partial_{p_\mu} = \Omega \partial_{p_0} + \mathbf{K} \cdot \partial_{\mathbf{p}}$. When $D(x,p)$ in Eq. (7.4) is of the Hayes (1973) form $[D(x,p) = p_0 - H(t,\mathbf{x},\mathbf{p})]$, Eq. (7.48) recovers the same expression for $H_{\text{eff}}(t,\mathbf{x},\mathbf{p})$ that was previously reported by Dodin and Fisch (2014).

### 7.4.3 Point-particle model and ray equations

From Eq. (7.45), one can obtain the ponderomotive action in the point-particle limit. Following Sec. 3.3.2, one readily finds

$$\mathcal{S} = \int \mathrm{d}t \, [\, \mathbf{P} \cdot \dot{\mathbf{X}} - H_{\text{eff}}(t, \mathbf{X}, \mathbf{P}) \,], \tag{7.49}$$

where $\mathbf{X}(t)$ is the wave position coordinate and $\mathbf{P}(t)$ is the wave canonical momentum. The corresponding ELEs are given by

$$\delta \mathbf{P} : \quad \dot{\mathbf{X}} = \partial_{\mathbf{P}} H_{\text{eff}}(t, \mathbf{X}, \mathbf{P}), \tag{7.50a}$$

$$\delta \mathbf{X} : \quad \dot{\mathbf{P}} = -\partial_{\mathbf{X}} H_{\text{eff}}(t, \mathbf{X}, \mathbf{P}). \tag{7.50b}$$

Equations (7.50) describe the ponderomotive dynamics of PW rays. These equations include the time-averaged refraction of a PW caused by the MW oscillations. In the present theory, $\Psi(x)$ can represent both classical waves and quantum particles. Hence, the ponderomotive dynamics of charged particles is subsumed here as a special case. Also note that, since $H_{\text{eff}}$ is generally not separable into a kinetic energy and a potential energy, the dynamics governed by Eqs. (7.50) may be quite complicated and perhaps counter-intuitive (Dodin and Fisch, 2008a,b).

### 7.4.4 Obtaining the original wave function $\Psi(x)$

The advantage of introducing the transformation $|\, \Psi \,\rangle = \widehat{\mathfrak{U}} \,|\, \psi \,\rangle$ is that the resulting dynamics for the OC wave $\psi(x)$ can be easily solved using the GO approximation. Supposing that the solution for the eikonal wave $\psi(x)$ is known, let us determine the original wave function $\Psi(x)$. To accomplish this, one may use the following expression:

$$\Psi(x) = \langle\, x \,|\, \widehat{\mathfrak{U}} \,|\, \psi \,\rangle = \int \mathrm{d}^4 y \, \langle\, x \,|\, \widehat{\mathfrak{U}} \,|\, y \,\rangle \langle\, y \,|\, \psi \,\rangle \,, \tag{7.51}$$

where I inserted the identity operator $\int \mathrm{d}^4 y \,|\, y \,\rangle \langle\, y \,| = \widehat{1}$. One then approximates the unitary operator $\widehat{\mathfrak{U}}$ with its lowest-order approximation so that $\widehat{\mathfrak{U}} \simeq \widehat{1} + i\sigma \widehat{\mathfrak{G}}_1$. Upon using Eq. (A.3) for the coordinate representation of $\widehat{\mathfrak{G}}_1$, one writes $\langle\, x \,|\, \widehat{\mathfrak{G}}_1 \,|\, y \,\rangle$ in terms of the Weyl symbol $G_1(x,p)$ so that

$$\Psi(x) = \psi(x) + \frac{i\sigma}{(2\pi)^4} \int \mathrm{d}^4 y \, \mathrm{d}^4 p \, e^{-ip\cdot(x-y)} \, G_1\left( \frac{x+y}{2}, p \right) \psi(y) + \mathcal{O}(\epsilon, \sigma^2), \tag{7.52}$$

where $G_1(x,p)$ is the Weyl symbol corresponding to the generator $\widehat{\mathfrak{G}}_1$. One then substitutes the general expression for $G_1(x,p)$ [Eqs. (7.22) and (7.24)] and writes $\psi(x)$ in the eikonal form (7.31). Upon asymptotically evaluating the integral, one obtains (Appendix C.2)

$$\Psi(x) = a(x)e^{i\theta(x)} \left(1 + \frac{i\sigma}{2}\big[\mathcal{G}_1(x, k + K/2)e^{i\Theta(x)} + \mathcal{G}_1^*(x, k - K/2)e^{-i\Theta(x)}\big]\right) + \mathcal{O}(\sigma^2, \epsilon), \tag{7.53}$$

where $k_\mu(x) = -\partial_\mu\theta$ and $K_\mu(x) = -\partial_\mu\Theta$. As shown in Eq. (7.53), the presence of the MW causes additional $\mathcal{O}(\sigma)$ oscillations on the wave $\Psi(x)$. Hence, the probe wave $\Psi(x)$ oscillates with phases $\theta(x) \pm \Theta(x)$. For completeness, the norm $|\Psi(x)|^2$ is given by

$$|\Psi(x)|^2 = a^2(x) \left(1 - \sigma \mathrm{Im}\big[\mathcal{G}_1(x, k + K/2)e^{i\Theta(x)} - \mathcal{G}_1(x, k - K/2)e^{i\Theta(x)}\big]\right) + \mathcal{O}(\sigma^2, \epsilon). \tag{7.54}$$

This concludes the presentation of the general theory of the ponderomotive effect on waves. In the following Sections, I shall present some applications of the theory. I shall also discuss how this theory of ponderomotive effects is related to the theory of dispersion of waves.

## 7.5 Discussion and examples

### 7.5.1 Example 1: Schrödinger particle in an electrostatic field

Let us consider a nonrelativistic particle interacting with a modulated electrostatic potential. The particle dynamics can be described using the Schrödinger equation

$$i\partial_t\Psi = \left[-\boldsymbol{\nabla}^2/(2m) + qV(x)\right]\Psi, \tag{7.55}$$

where $m$ and $q$ are the particle mass and charge, the electrostatic potential $V(x) = \mathrm{Re}\big[\mathcal{V}(x)e^{i\Theta(x)}\big]$ is assumed to be small, $\Theta(x)$ is a fast real phase, and $\mathcal{V}(x)$ is a complex function describing the slowly varying potential envelope. In this case, the dispersion operator is

$$\widehat{\mathcal{D}} \doteq \hat{p}_0 - \widehat{\mathbf{p}}^2/(2m) - qV(\widehat{x}). \tag{7.56}$$

The corresponding Weyl symbols are

$$D_0(p) = p_0 - \mathbf{p}^2/(2m), \qquad D_{\mathrm{osc}}(x) = -qV(x). \tag{7.57}$$

The symbol $D_{\text{eff}}$ is calculated using Eq. (7.27). Note that $D_1 = -\text{Re}\big(q\mathcal{V}e^{i\Theta}\big)$ so $\mathcal{D}_1 = -q\mathcal{V}$ and $D_n = 0$ for $n \geq 2$. Upon substituting into Eq. (7.27), one obtains

$$\begin{aligned} D_{\text{eff}}(x,p) &= D_0(p) - \frac{1}{4}\sum_{n=\pm 1}\frac{|\mathcal{D}_1(x)|^2}{D_0(p+nK) - D_0(p)} \\ &= p_0 - \frac{\mathbf{p}^2}{2m} - \sum_{n=\pm 1}\frac{q^2|\mathcal{V}(x)|^2/4}{[p_0 + n\Omega - (\mathbf{p}+n\mathbf{K})^2/(2m)] - [p_0 - \mathbf{p}^2/(2m)]} \\ &= p_0 - \frac{\mathbf{p}^2}{2m} - \frac{q^2|\mathbf{K}\mathcal{V}|^2/m}{4(\Omega - \mathbf{p}\cdot\mathbf{K}/m)^2 - (\mathbf{K}^2/m)^2}. \end{aligned} \tag{7.58}$$

Inserting Eq. (7.58) into Eq. (7.36) leads to the action in the Hayes form (7.45), where the effective Hamiltonian is

$$H_{\text{eff}}(t,\mathbf{x},\mathbf{p}) = \frac{\mathbf{p}^2}{2m} + \frac{q^2|\mathbf{K}\mathcal{V}|^2/m}{4(\Omega - \mathbf{p}\cdot\mathbf{K}/m)^2 - (\mathbf{K}^2/m)^2}. \tag{7.59}$$

In the fluid description of the particle wave packet, the corresponding ELEs are given by Eqs. (7.46). The corresponding ray equations are obtained from the point-particle Lagrangian (7.49). When one introduces the missing $\hbar$ factors, the effective Hamiltonian becomes

$$H_{\text{eff}}(t,\mathbf{X},\mathbf{P}) = \frac{\mathbf{P}^2}{2m} + \frac{q^2|\mathbf{K}\mathcal{V}|^2/m}{4(\Omega - \mathbf{P}\cdot\mathbf{K}/m)^2 - (\hbar\mathbf{K}^2/m)^2}. \tag{7.60}$$

In contrast with the classical ponderomotive Hamiltonian (Dewar, 1972; Cary and Kaufman, 1977; Dodin, 2014b)

$$H_{\text{eff,cl}}(t,\mathbf{X},\mathbf{P}) = \frac{\mathbf{P}^2}{2m} + \frac{q^2|\mathbf{K}\mathcal{V}|^2}{4m(\Omega - \mathbf{P}\cdot\mathbf{K}/m)^2}, \tag{7.61}$$

which is recovered from Eq. (7.60) at small enough $\mathbf{K}$, Eq. (7.60) predicts that the ponderomotive force can be attractive. This is seen from the fact that, when the second term in the denominator in Eq. (7.60) dominates, the ponderomotive energy becomes an effective negative potential (i.e., does not depend on $\mathbf{P}$). The effect, which is similar to that reported in Gilary *et al.* (2003) and Rahav *et al.* (2003), does not have a classical analogue.

The attractive regime of the ponderomotive energy was confirmed with numerical full-wave simulations using MATLAB. The results are presented in Fig. 7.2. As shown, the wave packet is attracted to the region of higher potential and oscillates within it. Note that the ray trajectories generated by $H_{\text{eff}}$ accurately match the motion of the wave packet's center. The approximate solution of $|\Psi(x)|^2$ using Eq. (7.54) is shown in Fig. 7.3. Here I approximated the solution of the GO wave envelope by $\mathcal{I}(t,\mathbf{x}) \simeq |\Psi_0\big(\mathbf{x}-\mathbf{X}(t)\big)|^2$ so that the wave envelope is simply transported by the ray trajectory. As shown, the differences between the numerical solution (Fig. 7.2) and the approximate solution (Fig. 7.3) are minimal. As a final note, Fig. 7.3(b) shows the approximate wave envelope $|\Psi(x)|^2$ at $t = 1000$. As can be seen, the local maxima of the wave function are located in the regions near the minima of the potential. This agrees with the interpretation that the particle has a higher probability of being located in the regions of weaker potential.

Also note that at $\Omega = 0$ the ponderomotive energy is resonant at $2\mathbf{K}\cdot\mathbf{P} = \pm\hbar\mathbf{K}^2$. This relation can be written as $\lambda_{\mathrm{dB}} = 2d\cos\zeta$, where $\lambda_{\mathrm{dB}}$ is the particle de Broglie wavelength, $d \doteq 2\pi/|\mathbf{K}|$ is the characteristic length of the lattice, and $\zeta$ is the angle between the $\mathbf{K}$ and $\mathbf{P}$ vectors. One may recognize this as the first harmonic of the Bragg resonance. Hence, this wave theory presents Bragg scattering as a variation of the ponderomotive effect. One can also identify a parallel between Eq. (7.60) and the linear susceptibility of quantum plasma (Lifshitz and Pitaevskii, 1981; Haas, 2011; Daligault, 2014). This will be further discussed in the next Section.

### 7.5.2 Example 2: Relativistic spinless particle in an oscillating EM field

In this section, I calculate the ponderomotive Hamiltonian of a relativistic spinless particle interacting with a slowly varying background EM field and a high-frequency EM modulation.[11] The particle dynamics can be described using the Klein–Gordon equation

$$\left[(i\partial_t - qV)^2 - (-i\boldsymbol{\nabla} - q\mathbf{A})^2 - m^2\right]\Psi = 0, \tag{7.62}$$

where $V(x)$ and $\mathbf{A}(x)$ are the scalar and vector potentials, respectively. Let $A^\mu(x) \doteq (V, \mathbf{A})$ be the associated four-potential, which can be written as

$$A^\mu(x) \doteq A^\mu_{\mathrm{bg}}(x) + A^\mu_{\mathrm{osc}}(x). \tag{7.63}$$

Here $A^\mu_{\mathrm{bg}}(x)$ is the four-potential describing the background EM field, and $A^\mu_{\mathrm{osc}}(x) \doteq \mathrm{Re}\big[\mathcal{A}^\mu(x)e^{i\Theta(x)}\big]$ is the four-potential of the modulated EM wave with small amplitude. As before, $\Theta(x)$ is a fast real phase, and $\mathcal{A}^\mu(x)$ is a slowly varying function. The corresponding dispersion operator of the Klein–Gordon particle is $\widehat{\mathfrak{D}} = \widehat{\mathfrak{D}}_0 + \widehat{\mathfrak{D}}_{\mathrm{osc}}$, where

$$\widehat{\mathfrak{D}}_0 = [\widehat{p}^\mu - qA^\mu_{\mathrm{bg}}(\widehat{x})][\widehat{p}_\mu - qA_{\mathrm{bg},\mu}(\widehat{x})] - m^2, \tag{7.64a}$$

$$\widehat{\mathfrak{D}}_{\mathrm{osc}} = -\left\{qA^\mu_{\mathrm{osc}}(\widehat{x})[\widehat{p}_\mu - qA_{\mathrm{bg},\mu}(\widehat{x})] + \mathrm{h.\,c.}\right\} + q^2A^\mu_{\mathrm{osc}}(\widehat{x})A_{\mathrm{osc},\mu}(\widehat{x}). \tag{7.64b}$$

[11] A related calculation is presented in Chapter 8 that describes the fully relativistic ponderomotive force on the Dirac electron with spin forces included.

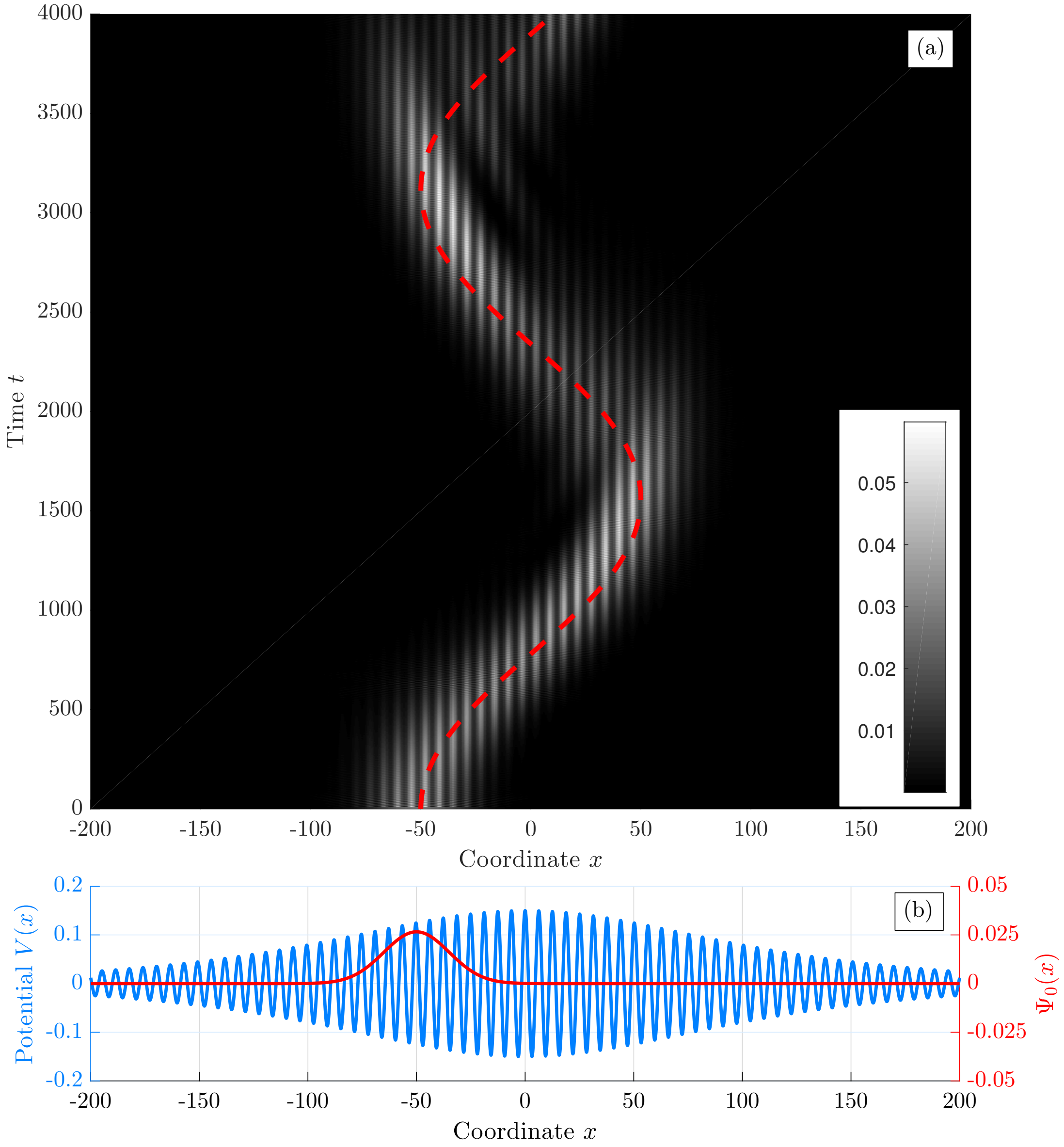

Figure 7.2: (a) Comparison of the simulation results obtained by numerically integrating the full-wave Eq. (7.55) (solid fill) and the ray trajectory calculated using Eqs. (7.50) with $H_{\text{eff}}$ taken from Eq. (7.60) (dashed). The initial wave packet is $\Psi_0(x) = (2\pi\eta^2)^{-1/4} \exp\left[-(x-\mu)^2/(4\eta^2)\right]$, where $\mu = -50$ and $\eta = 15$, and it is normalized such that $\int \mathrm{d}x \, |\Psi_0(x)|^2 = 1$. (The simulation is one-dimensional, and $x$ denotes the spatial coordinate, unlike in the main text, where $x$ denotes the spacetime coordinate.) The initial conditions for the ray trajectory are $X(0) = -50$ and $P(0) = 0$. (b) In blue, MW profile of the form $V(x) = 0.15 \, \operatorname{sech}(x/80) \cos(x)$, and in red, initial envelope of the PW. Natural units are used such that $m = 1$, $q = 1$, $\hbar = 1$, and $K = 1$. At later times (not shown), diffraction effects become important, so the eikonal theory becomes inapplicable.

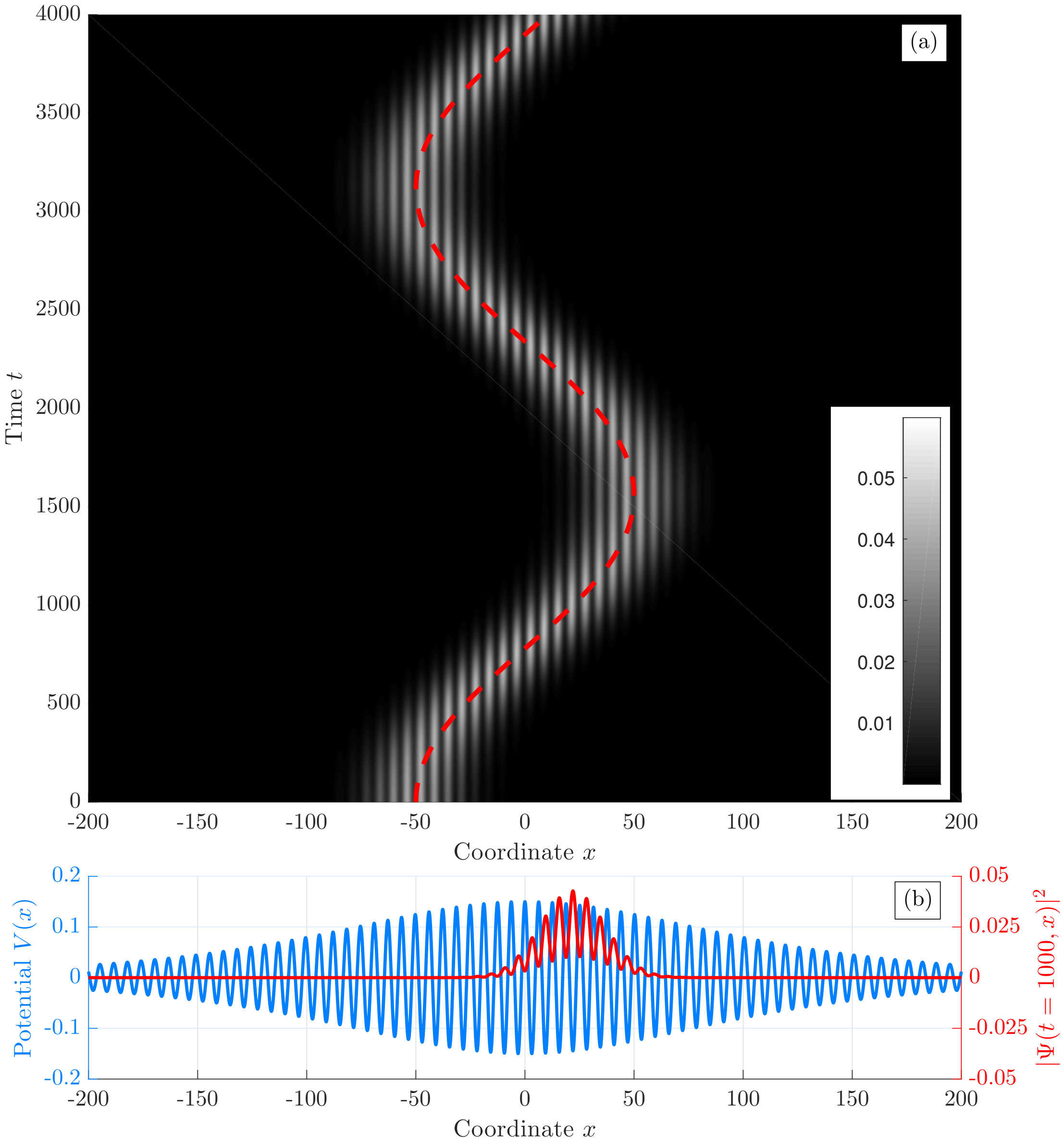

Figure 7.3: (a) Approximate solution for $|\Psi(t,x)|^2$ obtained by using Eq. (7.54) (solid fill) and the ray trajectory calculated using Eqs. (7.50) with $H_{\text{eff}}$ taken from Eq. (7.60) (dashed). (b) In blue, MW profile of the form $V(x) = 0.15\,\text{sech}(x/80)\cos(x)$, and in red, envelope of the PW at $t = 1000$. The initial conditions and parameters used here are identical to those described in Fig. 7.2.

The corresponding Weyl symbols are (Appendix A)

$$D_0(x,p) = \pi^2 - m^2, \tag{7.65a}$$

$$D_1(x,p) = -\,2q\pi \cdot A_{\text{osc}}(x), \tag{7.65b}$$

$$D_2(x,p) = q^2 A_{\text{osc}}(x) \cdot A_{\text{osc}}(x), \tag{7.65c}$$

where $\pi^\mu(x,p) \doteq p^\mu - qA^\mu_{\text{bg}}(x)$ is the kinetic four-momentum. Upon substituting Eqs. (7.65) into Eq. (7.27), one obtains

$$D_{\text{eff}}(x,p) = \pi^2 - m^2 + \frac{q^2|\mathcal{A}|^2}{2} - \sum_{n=\pm1} \frac{q^2|\mathcal{A}\cdot(\pi + nK/2)|^2}{2n\pi\cdot K + n^2 K\cdot K}, \tag{7.66}$$

where $|\mathcal{A}|^2 = \mathcal{A}\cdot\mathcal{A}^* = |\mathcal{V}|^2 - |\boldsymbol{\mathcal{A}}|^2$.

As in Sec. 7.4, one can derive the effective Hamiltonian for a point particle. After introducing the missing $c$ and $\hbar$ factors, one obtains

$$H_{\text{eff}}(t,\mathbf{X},\mathbf{P}) = \gamma mc^2 + qV_{\text{bg}} - \frac{q^2|\mathcal{A}|^2}{4\gamma mc^2} + \frac{1}{2\gamma mc^2}\sum_{n=\pm1}\frac{q^2|\mathcal{A}\cdot(\Pi_* + n\hbar K/2)|^2}{2n\Pi_*\cdot\hbar K + n^2\hbar^2 K\cdot K}, \tag{7.67}$$

where

$$\gamma(t,\mathbf{X},\mathbf{P}) \doteq \sqrt{1 + \left(\frac{\mathbf{P}}{mc} - \frac{q\mathbf{A}_{\text{bg}}}{mc^2}\right)^2} \tag{7.68}$$

is the unperturbed Lorentz factor, $\Pi^\mu_* \doteq (\gamma mc, \mathbf{P} - q\mathbf{A}_{\text{bg}}/c)$ is the unperturbed kinetic four-momentum, $\mathcal{A}^\mu(t,\mathbf{X}) = (\mathcal{V},\boldsymbol{\mathcal{A}})$ is the modulated four-potential, and $K^\mu(t,\mathbf{X}) = (\Omega/c,\mathbf{K})$ is the MW four-wavevector. All quantities are evaluated at the particle position $\mathbf{X}(t)$.

Several interesting limits can be studied with the effective Hamiltonian (7.67). In the Lorentz gauge where $\partial_\mu A^\mu_{\text{osc}} = 0$, then $K\cdot\mathcal{A} = \mathcal{O}(\epsilon)$. Hence, $H_{\text{eff}}$ becomes

$$H_{\text{eff}}(t,\mathbf{X},\mathbf{P}) = \gamma mc^2 + qV_{\text{bg}} - \frac{q^2|\mathcal{A}|^2}{4\gamma mc^2} - \left(\frac{q^2|\mathcal{A}\cdot\Pi_*|^2}{\gamma mc^2}\right)\frac{K\cdot K}{4(\Pi_*\cdot K)^2 - (\hbar K\cdot K)^2}. \tag{7.69}$$

Let us analyze the terms appearing in Eq. (7.69). For example, in the case of a vacuum wave $K\cdot K = (\Omega/c)^2 - \mathbf{K}^2 = 0$, so the last term vanishes. The remaining terms can be understood as the lowest-order expansion (in $|A_c|^2$) of the effective ponderomotive Hamiltonian $H_{\text{eff}} = mc^2[1 + \boldsymbol{\Pi}^2/(mc)^2 - q^2|\mathcal{A}|^2/(2m^2c^4)]^{1/2} + qV_{\text{bg}}$ that a relativistic spinless particle experiences in an oscillating EM pulse (Akhiezer and Polovin, 1956; Kibble, 1966; Dodin *et al.*, 2003). In the case of nonzero $K\cdot K$, the last term of Eq. (7.69) persists and accounts for Compton scattering, much like the Bragg scattering discussed in Sec. 7.5.1.

Also, let us consider a particle that interacts with an oscillating electrostatic field so that $\mathcal{A}^\mu = (\mathcal{V}, 0)$. In this case, Eq. (7.67) gives (Appendix C.2)

$$H_{\text{eff}}(t, \mathbf{X}, \mathbf{P}) = \gamma m c^2 + q V_{\text{bg}} + \left( \frac{q^2 |\mathcal{V}|^2}{4\gamma m} \right) \frac{\mathbf{K}^2 - (\mathbf{v}_* \cdot \mathbf{K}/c)^2 - [\hbar \mathbf{K}/(2\gamma m c)]^2 (K \cdot K)}{(\Omega - \mathbf{v}_* \cdot \mathbf{K})^2 - [\hbar K \cdot K/(2\gamma m)]^2}, \tag{7.70}$$

where $\mathbf{v}_* \doteq \mathbf{\Pi}/(\gamma m)$ is the unperturbed particle velocity. The last term in Eq. (7.70) is the relativistic ponderomotive energy. [In the nonrelativistic limit, when $\gamma \simeq 1$ and $\hbar |\mathbf{K}| \ll mc$, Eq. (7.70) reduces to Eq. (7.60), as expected.] When quantum corrections are negligible, one obtains

$$H_{\text{eff}}(t, \mathbf{X}, \mathbf{P}) = \gamma m c^2 + q V_{\text{bg}} + \frac{q^2 |\mathbf{K} \mathcal{V}|^2}{4M(\Omega - \mathbf{v}_* \cdot \mathbf{K})^2}, \tag{7.71}$$

where $M \doteq m\gamma |\mathbf{K}|^2 \,/\, [\, |\mathbf{K}|^2 - (\mathbf{v}_* \cdot \mathbf{K}/c)^2]$. When $\mathbf{v}_*$ is parallel to $\mathbf{K}$, then $M = m\gamma^3$, which is interpreted as the longitudinal mass. Likewise, when $\mathbf{v}_*$ is transverse to $\mathbf{K}$, then $M = m\gamma$, which is understood as the transverse mass (Okun, 1989).

### 7.5.3 Example 3: Electrostatic wave in a density modulated plasma

As another example, let us consider an EM wave $\Psi(x)$ propagating in a density-modulated plasma. The PW dynamics is described by

$$\partial_t^2 \Psi = \boldsymbol{\nabla}^2 \Psi - \omega_p^2 \Psi, \tag{7.72}$$

where $\omega_p^2(x) \doteq 4\pi q^2 n(x)/m$ is the squared of the plasma frequency (Stix, 1992). The plasma density is modulated such that $n(x) = n_{\text{bg}}(x) + n_{\text{osc}}(x)$, where $n_{\text{bg}}(x)$ is the slowly varying background plasma density and $n_{\text{osc}}(x) \doteq \text{Re}\big[n_c(x) e^{i\Theta(x)}\big]$ is a fast modulation of small amplitude. The dispersion operator is

$$\widehat{\mathfrak{D}} = \widehat{p} \cdot \widehat{p} - \omega_p^2(\widehat{x}), \tag{7.73}$$

so the corresponding Weyl symbols are

$$D_0(x, p) = p \cdot p - \omega_{p,\text{bg}}^2(x), \tag{7.74a}$$

$$D_{\text{osc}}(x) = -\,\text{Re}\big[\omega_{p,c}^2(x) e^{i\Theta(x)}\big], \tag{7.74b}$$

where $\omega_{p,\text{bg}}^2(x) \doteq 4\pi q^2 n_{\text{bg}}(x)/m$ and $\omega_{p,c}^2(x) \doteq 4\pi q^2 n_c(x)/m$.

Upon substituting Eqs. (7.74) into Eq. (7.27), one obtains

$$D_{\text{eff}}(x,p) = p\cdot p - \omega_{p,\text{bg}}^2 + \frac{|\omega_{p,c}^2|^2 (K\cdot K)/8}{(p\cdot K)^2 - (K\cdot K)^2/4}. \tag{7.75}$$

Then, the effective Hamiltonian is given by

$$H_{\text{eff}}(t,\mathbf{x},\mathbf{p}) = \omega_0(t,\mathbf{x},\mathbf{p}) - \frac{|\omega_{p,\text{c}}^2(t,\mathbf{x})|^2}{16\omega_0^3(t,\mathbf{x},\mathbf{p})}\,\Xi(t,\mathbf{x},\mathbf{p}), \tag{7.76}$$

where $\omega_0(t,\mathbf{x},\mathbf{p}) \doteq (c^2\mathbf{p}^2 + \omega_{p,\text{bg}}^2)^{1/2}$ is the unperturbed EM wave frequency, $\Xi(t,\mathbf{x},\mathbf{p})$ is a dimensionless factor given by

$$\Xi(t,\mathbf{x},\mathbf{p}) \doteq \frac{\Omega^2 - c^2\mathbf{K}^2}{(\Omega - \mathbf{v}_*\cdot\mathbf{K})^2 - (\Omega^2 - c^2\mathbf{K}^2)^2/4\omega_0^2}, \tag{7.77}$$

and $\mathbf{v}_*(t,\mathbf{x},\mathbf{p}) = c^2\mathbf{p}/\omega_0$ is the unperturbed EM wave group velocity. (I reintroduced the missing $c$ factors for clarity.) The second term in Eq. (7.76) represents the ponderomotive frequency shift that a classical EM wave experiences in a modulated plasma.

Similarly to the previous examples, the denominator in Eq. (7.76) also contains photon recoil effects. The Bragg resonance condition is also included; i.e., for EM waves propagating in static modulated media ($\Omega = 0$), the Bragg resonance occurs at $2\mathbf{p}\cdot\mathbf{K} = \pm\mathbf{K}^2$. In the opposite limit where $\Omega \gg \mathbf{v}_*\cdot\mathbf{K}$, Eq. (7.76) becomes

$$H_{\text{eff}}(t,\mathbf{x},\mathbf{p}) = \omega_0 + \left(\frac{|\omega_{p,c}^2|^2}{16\omega_0^3}\right)\frac{N^2-1}{1-\mu^2(N^2-1)^2}, \tag{7.78}$$

where $\mu \doteq \Omega/2\omega_0$ and $N \doteq c|\mathbf{K}|/\Omega$ is the MW refraction index. Note that the GO result reported in Dodin and Fisch (2014) is recovered in the limit $\mu \ll 1$.

## 7.6 Modulational dynamics and polarizability of wave quanta

### 7.6.1 Basic equations

Knowing the effective Hamiltonian $H_{\text{eff}}$ of PWs, one can derive, without even considering the ray equations, the self-consistent dynamics of a MW when it interacts with an ensemble of PWs. (As a special case, when PWs are free charged particles, such ensemble is a plasma.) To do this, let us consider the action of the whole system in the form $\mathcal{S}_\Sigma = \mathcal{S}_{\text{mw}} + \mathcal{S}_{\text{pw}}$, where $\mathcal{S}_{\text{mw}}$ is the MW action and $\mathcal{S}_{\text{pw}}$ is the cumulative action of all PWs. Let us attribute the interaction action to $\mathcal{S}_{\text{pw}}$, so, by definition, $\mathcal{S}_{\text{mw}}$ is the system action absent PWs. Then, $\mathcal{S}_{\text{mw}}$ equals the action of the MW EM field in vacuum, $\mathcal{S}_{\text{mw}} \doteq \int \mathrm{d}^4x\,(\mathbf{E}_{\text{mw}}^2 - \mathbf{B}_{\text{mw}}^2)/(8\pi)$ (Goldstein *et al.*, 2002). Since the MW is assumed to satisfy the GO approximation, its electric and magnetic fields can be expressed as

$$\mathbf{E}_{\mathrm{mw}}(x) = \mathrm{Re}\big[\mathbf{E}_c(x)e^{i\Theta(x)}\big], \quad \mathbf{B}_{\mathrm{mw}}(x) = \mathrm{Re}\big[\mathbf{B}_c(x)e^{i\Theta(x)}\big], \tag{7.79}$$

where the envelopes $\mathbf{E}_c(x)$ and $\mathbf{B}_c(x)$ are slow compared to $\Theta(x)$. One can approximate $\mathcal{S}_{\mathrm{mw}}$ as

$$\mathcal{S}_{\mathrm{mw}} = \int \mathrm{d}^4x \left(\frac{|\mathbf{E}_c|^2}{16\pi} - \frac{c^2|\mathbf{K}\times\mathbf{E}_c|^2}{16\pi\Omega^2}\right) + \mathcal{O}(\epsilon), \tag{7.80}$$

where I substituted Faraday's law $\mathbf{B}_c \approx (c\mathbf{K}/\Omega)\times\mathbf{E}_c$.

To calculate $\mathcal{S}_{\mathrm{pw}}$, I assume that PWs are mutually incoherent and do not interact other than via the MW. (When PWs are charged particles, this is known as the collisionless-plasma approximation. In a broader context, this can be recognized as the quasilinear approximation.) Then, $\mathcal{S}_{\mathrm{pw}} = \sum_i \mathcal{S}_i$, where $\mathcal{S}_i$ are the actions of the individual PWs. After adopting $\mathcal{S}_i$ in the form of Eq. (7.45), one obtains

$$\mathcal{S}_{\mathrm{pw}} = \mathcal{S}_{\mathrm{pw},0} - \sum_i \int \mathrm{d}^4x\, \mathcal{I}_i(t,\mathbf{x})\, \Phi_i(t,\mathbf{x},\boldsymbol{\nabla}\theta_i), \tag{7.81}$$

where $\mathcal{S}_{\mathrm{pw},0} = -\sum_i \int \mathrm{d}^4x\, \mathcal{I}_i\, [\partial_t\theta_i + H_{0,i}(t,\mathbf{x},\boldsymbol{\nabla}\theta_i)]$ is independent of the MW variables, so it can be dropped. (In this section, I am only interested in ELEs for the MW, and $\mathcal{S}_{\mathrm{pw},0}$ does not contribute to those.) Let us consider PWs in groups $s$ such that, within each group, PWs have the same ponderomotive frequency shift $\Phi_s$. Then, one can rewrite $\mathcal{S}_{\mathrm{pw}}$ as

$$\mathcal{S}_{\mathrm{pw}} = -\sum_s \int \mathrm{d}^4x\, \mathrm{d}^3\mathbf{p}\, f_s(t,\mathbf{x},\mathbf{p})\, \Phi_s(t,\mathbf{x},\mathbf{p}), \tag{7.82}$$

where $f_s \doteq \sum_{i\in s} \mathcal{I}_i(t,\mathbf{x})\, \delta[\mathbf{p} - \boldsymbol{\nabla}\theta_i(t,\mathbf{x})]$. This gives $\mathcal{S}_\Sigma = \int \mathrm{d}^4x\, \mathfrak{L}$, where the Lagrangian density $\mathfrak{L}$ is

$$\mathfrak{L} = \frac{|\mathbf{E}_c|^2}{16\pi} - \frac{c^2|\mathbf{K}\times\mathbf{E}_c|^2}{16\pi\Omega^2} - \sum_s \int \mathrm{d}^3\mathbf{p}\, f_s(t,\mathbf{x},\mathbf{p})\, \Phi_s(t,\mathbf{x},\mathbf{p}). \tag{7.83}$$

The meaning of $f_s(t,\mathbf{x},\mathbf{p})$ is understood as follows. A single wave with a well-defined local momentum $\boldsymbol{\nabla}\theta_i(t,\mathbf{x})$ has a phase-space distribution that is delta-shaped along the local wavevector (momentum) coordinate, $\propto \delta[\mathbf{p} - \boldsymbol{\nabla}\theta_i(t,\mathbf{x})]$. The coefficient in front of the delta function must be the spatial probability density for the proper normalization. Here the spatial probability density is $|\psi_i(t,\mathbf{x})|^2$, which is the same as $\mathcal{I}_i(t,\mathbf{x})$. Thus, $\mathcal{I}_i(t,\mathbf{x})\, \delta[\mathbf{p} - \boldsymbol{\nabla}\theta_i(t,\mathbf{x})]$ is the phase-space density of the $i$th wave. This makes $f_s(t,\mathbf{x},\mathbf{p})$ the total phase-space density of species $s$. This formulation can also be applied, for example, to degenerate plasmas to the extent that the Hartree approximation is applicable (Lifshitz and Pitaevskii, 1981; Haas, 2011).

Specifically, if the spin–orbital interaction is negligible and particles interact with each other only through the mean EM field, their Lagrangian densities sum up (by definition of the mean-field approximation), so one recovers the same $\mathfrak{L}$ as in nondegenerate plasma. The only subtlety in this case is that $f_s(t, \mathbf{x}, \mathbf{p})$ is now restricted by the Pauli exclusion principle (or, in equilibrium, to Fermi–Dirac statistics).

Since $\Phi_s$ is bilinear in the MW field and independent on the MW phase, it can be expressed as

$$\Phi_s = -\frac{1}{4}\,\mathbf{E}_c^* \cdot \boldsymbol{\alpha}_s \cdot \mathbf{E}_c, \tag{7.84}$$

where $\boldsymbol{\alpha}_s$ is some complex tensor that can depend on $\Omega$ and $\mathbf{K}$ but not on $\mathbf{E}_c$ or $\mathbf{E}_c^*$. Explicitly, it is defined as

$$\boldsymbol{\alpha}_s(t, \mathbf{x}, \mathbf{p}, \Omega, \mathbf{K}) \doteq -4\,\frac{\partial^2}{\partial \mathbf{E}_c\, \partial \mathbf{E}_c^*}\,\Phi_s(\mathbf{E}_c, \mathbf{E}_c^*, \Omega, \mathbf{K};\, t, \mathbf{x}, \mathbf{p}), \tag{7.85}$$

or, equivalently, $\boldsymbol{\alpha}_s \doteq -4\partial^2 H_{\text{eff},s}/(\partial \mathbf{E}_c\, \partial \mathbf{E}_c^*)$. Then,

$$\mathfrak{L} = \frac{1}{16\pi}\,\mathbf{E}_c^* \cdot \boldsymbol{\varepsilon}(t, \mathbf{x}, \Omega, \mathbf{K}) \cdot \mathbf{E}_c - \frac{c^2|\mathbf{K} \times \mathbf{E}_c|^2}{16\pi\Omega^2}, \tag{7.86}$$

where I introduced $\boldsymbol{\varepsilon} \doteq 1 + \boldsymbol{\chi}$ and

$$\boldsymbol{\chi} \doteq 4\pi \sum_s \int \mathrm{d}^3\mathbf{p}\, f_s(t, \mathbf{x}, \mathbf{p})\, \boldsymbol{\alpha}_s(t, \mathbf{x}, \mathbf{p}, \Omega, \mathbf{K}). \tag{7.87}$$

By treating $(\Theta, \mathbf{E}_c, \mathbf{E}_c^*)$ as independent variables, one then obtains the following ELEs:

$$\delta\Theta: \quad \partial_t(\partial_\Omega \mathfrak{L}) - \boldsymbol{\nabla} \cdot (\partial_{\mathbf{K}} \mathfrak{L}) = 0, \tag{7.88}$$

$$\delta\mathbf{E}_c^*: \quad (\Omega/c)^2\, \boldsymbol{\varepsilon} \cdot \mathbf{E}_c + \mathbf{K} \times (\mathbf{K} \times \mathbf{E}_c) = 0, \tag{7.89}$$

plus a conjugate equation for $\mathbf{E}_c^*$. [Remember that $\Omega$ and $\mathbf{K}$ are related to $\Theta$ via Eq. (7.20).] One then recognizes these ELEs as the GO equations describing EM waves in a dispersive medium with dielectric tensor $\boldsymbol{\varepsilon}$ (Stix, 1992). Thus, $\boldsymbol{\chi}$ is the susceptibility of the medium, and $\boldsymbol{\alpha}_s$ serves as the *linear polarizability* of PWs of type $s$.[12]

Equation (7.84) can be interpreted as a fundamental relation between the ponderomotive energy and the linear polarizability. This relation is a generalization of the well-known "$K$–$\chi$ theorem" (Cary and Kaufman, 1977; Kaufman, 1987; Dodin and Fisch, 2010), which establishes this equality for classical particles, to general waves. Those include quantum particles as a special case and also photons, plasmons, phonons, etc. According to the theory presented here, any such object can be assigned a ponderomotive energy and thus has a polarizability (7.85). Some examples are discussed below.

[12] One can redo the calculation in Sec. 7.6 with $\mathcal{S}_{\text{pw}}$ in the form (7.29). Then, $\boldsymbol{\chi}$ is conveniently expressed as $\boldsymbol{\chi} = 4\pi \sum_s \int \mathrm{d}^4p\, F_s(x, p)\, \boldsymbol{\mathcal{A}}_s(x, p, K)$, where $F_s \doteq \sum_{i \in s} W_i$ the sum of the Wigner functions of species $s$, $\boldsymbol{\mathcal{A}}$ is the generalized polarizability defined via $D_{\text{NL}} = (1/4)\, \mathbf{E}_c^* \cdot \boldsymbol{\mathcal{A}} \cdot \mathbf{E}_c$, and $D_{\text{NL}}$ is the MW-dependent part of $D_{\text{eff}}$.

### 7.6.2 Example 1: Polarizability of an electron and susceptibility of an electron gas

As a first example, let us consider a nonrelativistic quantum electron with charge $q$, mass $m$, and OC momentum $\mathbf{P} \doteq m\mathbf{v}$. Suppose the electron interacts with an electrostatic MW (so $\mathbf{B}_c = 0$). Then, $H_{\text{eff}}$ can be taken from Eq. (7.60), and Eq. (7.85) readily yields that the electron polarizability is a diagonal matrix given by

$$\boldsymbol{\alpha}_e = -\mathbb{I}_3 \,\frac{q^2}{m}\left\{(\Omega - \mathbf{K}\cdot\mathbf{v})^2 - [\hbar \mathbf{K}^2/(2m)]^2\right\}^{-1}. \tag{7.90}$$

(Here, $\mathbb{I}_3$ is a $3\times 3$ unit matrix.) From Eq. (7.87), the susceptibility of the electron plasma is[13]

$$\boldsymbol{\chi}_e = -\mathbb{I}_3 \,\frac{4\pi q^2}{m}\int \mathrm{d}^3\mathbf{v}\,\frac{f(t,\mathbf{x},\mathbf{v})}{(\Omega - \mathbf{K}\cdot\mathbf{v})^2 - [\hbar \mathbf{K}^2/(2m)]^2}, \tag{7.91}$$

which is precisely the textbook result (Lifshitz and Pitaevskii, 1981; Haas, 2011). This shows that the commonly known expression for the dielectric tensor of quantum plasmas is actually a reflection of the less-known quantum ponderomotive energy.

### 7.6.3 Example 2: Polarizability of a photon and susceptibility of a photon gas

As a second example, consider an EM wave in a nonmagnetized density-modulated cold electron plasma. Upon using Gauss's law, one readily finds that $\omega_{p,c}^2 = (iq/m)\mathbf{K}\cdot\mathbf{E}_c$, where I use the same notation as in Sec. 7.5.3. Then, after using Eq. (7.76), one gets

$$\Phi = -\frac{\hbar q^2 \Xi}{16 m^2 \omega_0^3}\,(\mathbf{E}_c^* \cdot \mathbf{K}\mathbf{K}\cdot \mathbf{E}_c), \tag{7.92}$$

where $\mathbf{K}\mathbf{K}$ is a dyadic tensor and $\Xi$ is given by Eq. (7.77). (The factor $\hbar$ is introduced in order to treat $\Phi$ as a per-photon energy rather than as a classical frequency.) Hence, Eq. (7.85) gives that the photon polarizability is

$$\boldsymbol{\alpha}_{\text{ph}} = \frac{\hbar q^2 \Xi}{4 m^2 \omega_0^3}\,\mathbf{K}\mathbf{K}. \tag{7.93}$$

[13] Strictly speaking, one must only take the principal value of the integral of the susceptibility since it was assumed that the dielectric tensor $\boldsymbol{\varepsilon}$ is Hermitian. For a rigorous treatment of the resonant denominator, which gives rise to Landau damping, see the discussion in Dodin *et al.* (2017).

In principle, one must account for this polarizability when calculating $\boldsymbol{\varepsilon}$; i.e., photons contribute to the linear dielectric tensor just like electrons and ions (Bingham *et al.*, 1997). For example, Dodin and Ruiz (2017) report the correct dielectric tensor in the GO regime for Langmuir waves propagating in nonmagnetized plasma:

$$\boldsymbol{\varepsilon}_{\mathrm{LW}}(\Omega, \mathbf{K}) = \boldsymbol{\varepsilon}_0(\Omega, \mathbf{K}) + \frac{\hbar \omega_p^2 \mathbf{K}\mathbf{K}}{4 m_e} \left( \frac{n_{\mathrm{ph}}}{n_{\mathrm{bg}}} \right) \int \mathrm{d}^3\mathbf{p} \left( \frac{\mathbf{K} \cdot \boldsymbol{\nabla}_{\mathbf{p}} f_{\mathrm{ph}}(\mathbf{p})}{(\Omega - \mathbf{K} \cdot \mathbf{v}_*)\omega_0^2(\mathbf{p})} + \frac{f_{\mathrm{ph}}(\mathbf{p})}{\omega_0^3(\mathbf{p})} \right), \tag{7.94}$$

where $\boldsymbol{\varepsilon}_0(\Omega, \mathbf{K})$ is the contribution to the dielectric tensor by particles within the plasma, $n_{\mathrm{ph}}(x)$ is the density of photons, $f_{\mathrm{ph}}(\mathbf{p})$ is the normalized unperturbed photon distribution function, and the rest of the variables are defined as in Sec. 7.5.3.

As shown in Eq. (7.94), a wave bath of HF photons within the plasma can nonlinearly influence the dispersion of linear Langmuir waves. This effect is reminiscent to earlier results reported in the field of wave turbulence, where the nonlinear dielectric function was calculated for waves propagating in a homogeneous turbulent wave bath.[14] Also, note that the contribution of the HF photons can lead to certain exotic phenomena, such as photon Landau damping (Bingham *et al.*, 1997). That said, the effects from photons are relatively small, and ignoring the photon contribution to the plasma dielectric tensor is justified except at large enough photon densities (Dodin and Ruiz, 2017).

Similar calculations are also possible for dissipative dynamics and vector waves. The present formulation could also help understand the modulational dynamics of wave ensembles in a general context. However, elaborating on these topics is outside of the scope of this thesis; they will be investigated in the future.

## 7.7 Conclusions

In this Chapter, I showed that scalar waves, both classical and quantum, can experience time-averaged refraction when propagating in modulated media. This phenomenon is analogous to the ponderomotive effect encountered by charged particles in high-frequency EM fields. Here I proposed a variational theory of this *ponderomotive effect on waves* for a general nondissipative linear medium. The formulation is able to describe waves with temporal and spatial periods comparable to that of the modulation (provided that parametric resonances are avoided). This theory can be understood as a generalization of the oscillation-center theory, which is known from classical plasma physics, to any linear waves or quantum particles in particular. I also showed that any wave is, in fact, a polarizable object that contributes to the linear dielectric tensor of the ambient medium. Three examples of applications of the theory were given: a Schrödinger particle propagating in an oscillating electrostatic field, a Klein–Gordon particle interacting with modulated EM fields, and an EM wave propagating in a density-modulated plasma.

[14] For more information on systematic statistical theories of plasma turbulence and the calculation of nonlinear dielectric functions in the context of wave turbulence, see, for example the review article by Krommes (2002).

This work can be expanded in several directions. From the theoretical standpoint, one can extend the theory to dissipative waves (Dodin *et al.*, 2017) and vector waves with polarization effects (see Chapter 4), which could be important at Bragg resonances. As a second avenue of research, one could investigate the role of higher-order terms in the truncated asymptotic expansion of this theory. For example, one could consider a single quantum particle interacting with three EM waves. Upon including third-order terms in the asymptotic expansion, one would obtain an effective theory that describes three-wave interactions. This effective theory would include particle kinetic effects in the three-wave coupling coefficients. It would be interesting to investigate if these additional kinetic terms can bring new physics to the fluid-based models for Raman amplification (Malkin *et al.*, 1999). Likewise, by including fourth-order terms in the asymptotic expansion, one would also be able to describe four-wave interactions.

Finally, the systematic procedure presented here could also potentially serve as a stepping stone to study more complex nonlinear wave–wave interactions, such as modulational instabilities in general wave ensembles or wave turbulence. Concerning the latter, a very interesting yet challenging direction of future research could be the following. One of the most successful methods for studying classical turbulent systems is the theory developed by Martin *et al.* (1973). This theory is the classical version of Schwinger's functional formalism for quantum field theory. Given a nonlinear PDE, the formalism provides a self-consistent statistical description of the nonlinear interactions. However, this formalism is not very intuitive or easy to use since the dynamics of nonlinear interactions are described using physical-space representation. In contrast, there are other phase-space formulations (Dewar, 1973, 1976) of classical turbulent systems that are physically more intuitive but not as general as the formalism by Martin *et al.* (1973). As suggested by Krommes (2012), a major breakthrough could be achieved if a more geometrical and intuitive phase-space interpretation can be given to the general asymptotic theory by Martin *et al.* (1973). The systematic theory presented in this Chapter could be the starting point for developing such formalism.

# Chapter 8

# Relativistic Hamiltonian of a spin-1/2 electron in a laser field

In this Chapter, I present a ponderomotive model of a relativistic spin-1/2 electron interacting with a high-frequency (HF) electromagnetic (EM) field. Starting from the action for the Dirac electron, I derive a reduced phase-space action that describes the relativistic time-averaged dynamics of such a particle in a quasiperiodic laser pulse propagating in vacuum. The pulse is allowed to have an arbitrarily large amplitude provided that radiation damping and pair production are negligible. The model captures the Bargmann–Michel–Telegdi (BMT) spin dynamics, the Stern–Gerlach (SG) spin–orbital coupling, the conventional ponderomotive forces, and the interaction with large-scale background fields (if any). The results of this Chapter were published by Ruiz *et al.* (2015).

## 8.1 Introduction

### 8.1.1 Motivation

In recent years, many works have been focused on incorporating quantum effects into classical plasma dynamics (Melrose, 2008; Brodin *et al.*, 2008a). In particular, various models have been proposed to couple spin equations with classical equations of plasma physics. This includes the early works by Takabayasi (1955, 1957) as well as other more recent works.[1] Of particular interest in this regard is the regime wherein particles interact with HF EM radiation. In this regime, it is possible to introduce a simpler time-averaged description, in which particles experience effective time-averaged, or ponderomotive, forces (Boot and Harvie, 1957; Gaponov and Miller, 1958; Cary and Kaufman, 1977). The inclusion of spin effects can yield intriguing corrections to this time-averaged dynamics (Brodin *et al.*, 2010; Stefan *et al.*, 2011). However, current *spin-ponderomotive theories* remain limited to regimes where the radiation amplitude is small enough so that it can be treated as a perturbation. These conditions are far more restrictive than those of spinless particle theories, where non-perturbative, relativistic ponderomotive effects can be accommodated within the *effective particle mass* (Akhiezer and Polovin, 1956; Kibble, 1966; Raicher *et al.*, 2014).[2] One may wonder, then, is it possible to derive a fully relativistic, and yet transparent, theory that also accounts for the spin dynamics and the SG spin–orbital coupling?

[1]See, for example, Marklund and Brodin (2007), Brodin *et al.* (2008b), Brodin *et al.* (2011), Stefan and Brodin (2013), Dixit *et al.* (2013), Morandi *et al.* (2014), Andreev (2015), and Andreev and Kuz'menkov (2015).

Excitingly, the answer is yes, and the purpose of this Chapter is to present such a description. More specifically, I present a point-particle ponderomotive model of a Dirac electron. Starting from the action of the Dirac electron, I derive a phase-space Lagrangian (8.46) in canonical coordinates with a Hamiltonian (8.47) that describes the relativistic time-averaged dynamics of such particle in a quasiperiodic laser pulse propagating in vacuum. The pulse is allowed to have an arbitrarily large amplitude (as long as radiation damping and pair production are negligible) and, in the case of nonrelativistic interactions, a wavelength comparable to the electron de Broglie wavelength. The model captures the spin dynamics, the spin–orbital coupling, the conventional ponderomotive forces, and the interaction with large-scale background fields (if any). The ponderomotive model is then compared using numerical simulations with the non-averaged point-particle model obtained in Chapter 5. The aforementioned "effective-mass" theory for spinless particles is reproduced as a special case when the spin–orbital coupling is negligible. Also, the point-particle Lagrangian that I derive has a canonical structure, which could be helpful in simulating the corresponding dynamics using symplectic methods (Hairer *et al.*, 2006; McLachlan and Quispel, 2006; Qin *et al.*, 2015).

### 8.1.2 Overview

The rest of this Chapter is organized as follows. In Sec. 8.2, the main assumptions used throughout the work are presented. To arrive at the point-particle ponderomotive model, Secs. 8.3–8.5 apply successive approximations and reparameterizations to the Dirac action. Specifically, in Sec. 8.3 I cast the Dirac action into the phase-space representation using the Weyl symbol calculus and adopt a transformation that eliminates the HF oscillating dynamics. In Sec. 8.4, I apply the extended geometrical-optics (XGO) theory to obtain a ponderomotive model for the Dirac particle states. In Sec. 8.5, I discuss the point-particle limit of the XGO action and obtain the corresponding ELEs. In Sec. 8.6, the derived ponderomotive model is numerically compared to the non-averaged model presented in Chapter 5. In Sec. 8.7, I summarize the main results. Additional details on some of the calculations are given in Appendix C.3.

[2]Under certain conditions, the effective-mass theory is also extendable to interactions in the presence of arbitrarily strong large-scale magnetic fields (Dodin and Fisch, 2009) and plasmas with small background density (Geyko *et al.*, 2009).

## 8.2 Basic formalism

As in Chapter 5, the starting point is the action for the Dirac particle (Peskin and Schroeder, 1995)

$$\mathcal{S} = \langle\, \bar{\Psi} \mid (\hat{\not{p}} - q\widehat{\not{A}} - m\mathbb{I}_4) \mid \Psi \,\rangle\,, \tag{8.1}$$

where $\Psi(x) \doteq \langle\, x \mid \Psi \,\rangle$ is a complex-valued four-component particle wave function, $\langle\, \bar{\Psi} \mid x \,\rangle = \bar{\Psi}(x) \doteq \Psi^\dagger(x)\gamma^0$ is the Dirac conjugate of $\Psi(x)$, and the particle mass and charge are denoted by $m$ and $q$, respectively.[3] As before, the Feynman slash notation is used so that $\not{a} \doteq a_\mu \gamma^\mu$. Definitions and properties of the Dirac matrices $\gamma^\mu = (\gamma^0, \boldsymbol{\gamma})$ are given in Eqs. (5.2)–(5.5). As in Sec. 2.2.2, $\widehat{p}_\mu$ is the four-momentum operator, and $\widehat{A}^\mu = A^\mu(\widehat{x})$ is the four-potential operator describing the interaction of the particle with the surrounding EM field.

Let us consider the interaction of an electron with an EM background field and an oscillating HF laser field. The four-vector potential $A^\mu(x)$ is decomposed into two parts:

$$A^\mu(x) = A^\mu_{\text{bg}}(x) + A^\mu_{\text{osc}}(x, \Theta), \tag{8.2}$$

where $A^\mu_{\text{bg}}(x)$ describes a background field (if any) that is slowly varying[4] and $A_{\text{osc}}(x, \Theta)$ is a rapidly oscillating EM laser pulse. Let us assume that $A_{\text{osc}}(x, \Theta)$ is quasiperiodic, so[5]

$$A^\mu_{\text{osc}}(x, \Theta) = \text{Re}\big[\mathcal{A}^\mu(x) e^{i\Theta(x)}\big], \tag{8.3}$$

where $\mathcal{A}^\mu(x)$ is a slowly varying complex four-vector describing the laser envelope and $\Theta(x)$ is a rapid phase. The EM wave frequency is defined by $\Omega(x) \doteq -\partial_t \Theta$, and the wave vector is $\mathbf{K}(x) \doteq \boldsymbol{\nabla}\Theta$. Accordingly, the corresponding four-wavevector $K_\mu \doteq -\partial_\mu \Theta = (\Omega, -\mathbf{K})$. Also, let $A^\mu_{\text{osc}}(x, \Theta)$ be described within the geometrical-optics (GO) approximation and assume that the interaction takes place in vacuum. Then, the four-wavevector $K_\mu(x)$ satisfies the vacuum dispersion relation

$$K^2 = \Omega^2 - \mathbf{K}^2 = 0. \tag{8.4}$$

[3] In this Chapter, the small contribution of the anomalous magnetic moment term is neglected but could be included too, at least as a perturbation.

[4] By a slowly varying function $f(x)$ in spacetime, I mean an infinitely differentiable function that can be written as $f(\epsilon x)$ so that its $n$th derivatives are $\mathcal{O}(\epsilon)$. The precise definition of $\epsilon$ is given in Eq. (8.7).

[5] By a quasiperiodic function $f(x, \Theta)$, I mean one where $f(a, \Theta)$ is $2\pi$-periodic for any constant $a$ and where the phase $\Theta$ is a function such that its derivatives are nonzero and slowly varying.

This relation will frequently appear in the form of $\not{K}\not{K} = K^2\mathbb{I}_4 = 0$, where I used Eq. (5.3). The Lorentz-gauge condition is chosen for the oscillatory EM field so that

$$\partial_\mu A^\mu_{\text{osc}} = 0. \tag{8.5}$$

In this Chapter, I shall neglect radiation damping and assume

$$\frac{\hbar\Omega'}{mc^2} \ll 1, \tag{8.6}$$

where $\Omega'$ is the laser frequency in the electron rest frame. Thus, pair production (and annihilation) can be neglected. I also assume

$$\epsilon \doteq \max\left\{\frac{1}{\Omega T}, \frac{1}{|\mathbf{K}|\ell}\right\} \ll 1, \tag{8.7}$$

where $T$ and $\ell$ are the characteristic temporal and spatial scales of $K_\mu(x)$, $A^\mu_{\text{bg}}(x)$, and $\mathcal{A}^\mu(x)$.

Using these orderings, I aim to derive a point-particle reduced model that describes the ponderomotive ($\Theta$-averaged) dynamics of a spin-1/2 electron interacting with a slowly varying EM background field and a HF laser field. This will be done in two steps. The first step consists in adopting a transformation that eliminates the HF oscillating terms from the action. The second step consists in applying the XGO theory (similar as in Chapter 5) in order to obtain a ponderomotive point-particle model for the spin-1/2 electron.

## 8.3 Ponderomotive model

### 8.3.1 Volkov transformation

In this Section, I shall obtain a ponderomotive action for the Dirac electron. As in Chapter 7, I shall eliminate the HF oscillating terms in the dispersion operator by introducing an appropriate transformation on the Dirac wave function. Specifically, let $|\,\Psi\,\rangle = \widehat{\mathbb{V}}\,|\,\psi\,\rangle$, where $\widehat{\mathbb{V}}$ is a $4\times 4$ matrix operator, which I call the *Volkov transformation* and will be defined later. In the new variables, the action (8.1) is written as follows:

$$\mathcal{S} = \langle\,\bar{\psi}\,|\,\gamma^0\widehat{\mathbb{V}}^\dagger\gamma^0(\widehat{\not{p}} - q\widehat{\not{A}}_{\text{bg}} - q\widehat{\not{A}}_{\text{osc}} - m\mathbb{I}_4)\widehat{\mathbb{V}}\,|\,\psi\,\rangle\,, \tag{8.8}$$

where I used $\gamma^0\gamma^0 = \mathbb{I}_4$. In the phase-space representation, the action is written as [Eq. (2.42)]

$$\mathcal{S} = \text{Tr}\int \mathrm{d}^4x\,\mathrm{d}^4p\;\gamma^0\overline{D}_{\text{V}}(x,p)\,W_\psi(x,p), \tag{8.9}$$

where $W_\psi(x,p)$ is the Wigner tensor corresponding to the state $|\,\psi\,\rangle$; namely,

$$W_\psi(x,p) \doteq \frac{1}{(2\pi)^4} \int \mathrm{d}^4x\, \mathrm{d}^4p\, e^{ip\cdot s}\, \langle\, x+s/2 \,|\, \psi \,\rangle \langle\, \psi \,|\, x - s/2 \,\rangle\,. \tag{8.10}$$

Also, $\overline{D}_{\mathrm{V}}(x,p)$ is the Weyl symbol of the transformed dispersion operator. Upon using the Moyal product [Eq. (A.5)], one obtains

$$\overline{D}_{\mathrm{V}}(x,p) = \gamma^0 \mathcal{V}^\dagger(x,p)\gamma^0 \star [\not{\pi}(x,p) - \not{A}_{\mathrm{osc}}(x) - m\mathbb{I}_4] \star \mathcal{V}(x,p), \tag{8.11}$$

where $\pi_\mu(x,p) \doteq p_\mu - qA_{\mathrm{bg},\mu}(x)$ is the slowly varying particle kinetic four-momentum and $\mathcal{V}(x,p)$ is the Weyl symbol corresponding to the Volkov transformation $\widehat{\mathcal{V}}$. In the following, I shall introduce the Weyl symbol for the Volkov transformation.

For the case of a Dirac particle interacting with a homogeneous laser field in vacuum, the Dirac equation has exact solutions which are known as the *Volkov states* (Volkov, 1935). (For more information on the Volkov states, see Appendix C.3.) Based on the Volkov states, I propose to write the Weyl symbol of the Volkov transformation as follows:

$$\mathcal{V}(x,p) \doteq \Xi(x,p) \left(1 - \frac{qA_{\mathrm{osc}}\cdot K}{2(\pi\cdot K)}\right) e^{i\tilde{\theta}(x,p)}, \tag{8.12}$$

where $\Xi(x,p)$ is a $4\times 4$ matrix defined as

$$\Xi(x,p) \doteq \mathbb{I}_4 + \frac{q}{2(\pi\cdot K)} \not{K}\not{A}_{\mathrm{osc}} \tag{8.13}$$

and the real phase $\tilde{\theta}(x,p)$ is given by

$$\tilde{\theta}(x,p) \doteq \frac{q}{\pi\cdot K}\int^{\Theta} \mathrm{d}\Theta'\ [\pi\cdot A_{\mathrm{osc}}(x,\Theta')] - \frac{q^2}{2(\pi\cdot K)}\int^{\Theta} \mathrm{d}\Theta'\ \left[A^2_{\mathrm{osc}}(x,\Theta') - \langle\!\langle A^2_{\mathrm{osc}}(x,\Theta')\rangle\!\rangle\right] \tag{8.14}$$

and has the property $\langle\!\langle \tilde{\theta}(x,p)\rangle\!\rangle = 0$. Here "$\langle\!\langle \ldots \rangle\!\rangle$" is a time average over a period of the laser field. Note that $\gamma^0\Xi^\dagger\gamma^0$ is calculated as follows:

$$\gamma^0\Xi^\dagger(x,p)\gamma^0 = \gamma^0\left(\mathbb{I}_4 + \frac{q}{2(\pi\cdot K)}\not{A}^\dagger_{\mathrm{osc}}\gamma^0\gamma^0\not{K}^\dagger\right)\gamma^0 = \mathbb{I}_4 + \frac{q}{2(\pi\cdot K)}\not{A}_{\mathrm{osc}}\not{K}, \tag{8.15}$$

where I used Eqs. (5.2) and (5.5). Also note that the term $A_{\mathrm{osc}}\cdot K$ in Eq. (8.12) is $\mathcal{O}(\epsilon)$. This can be shown by using the Lorentz-gauge condition (8.5). For convenience, let us write the derivative operator $\partial_\mu$

as follows:

$$\partial_\mu f(x,\Theta) = \epsilon d_\mu f(x,\Theta) - K_\mu \partial_\Theta f(x,\Theta), \tag{8.16}$$

where $f(x,\Theta)$ is an arbitrary quasiperiodic function and $d_\mu$ indicates a derivative with respect to the slow argument of $f(x,\Theta)$. Here I added the parameter $\epsilon$ to denote that the first term on the right-hand side is small. With this notation, the Lorentz-gauge condition (8.5) is written as $K_\mu \partial_\Theta A^\mu_{\text{osc}} = \epsilon d_\mu A^\mu_{\text{osc}}$. After integrating along the eikonal phase, one obtains

$$\epsilon\chi(x,\Theta) \doteq A_{\text{osc}} \cdot K = \epsilon \int^\Theta \mathrm{d}\Theta'\, d_\mu A^\mu_{\text{osc}}(x,\Theta'). \tag{8.17}$$

Before continuing further, let us verify that the Lorentz invariant quantity $\langle\, \bar{\Psi} \mid \Psi \,\rangle = \langle\, \bar{\psi} \mid \psi \,\rangle$ is conserved under the Volkov transformation. If this quantity is to remain invariant, the Volkov operator must satisfy $\gamma^0 \widehat{\mathcal{V}}^\dagger \gamma^0 \widehat{\mathcal{V}} = \mathbb{I}_4$. In terms of Weyl symbols, this condition is written as $\gamma^0 \mathcal{V}^\dagger(x,p)\gamma^0 \star \mathcal{V}(x,p) = \mathbb{I}_4$. To the lowest order in $\epsilon$, the Moyal product becomes an ordinary product of matrices, so

$$\begin{aligned}
\gamma^0 \mathcal{V}^\dagger \gamma^0 \mathcal{V} &= \left(1 - \epsilon \frac{q\chi}{2(\pi \cdot K)}\right) \gamma^0 \Xi^\dagger \gamma^0 \Xi \left(1 - \epsilon \frac{q\chi}{2(\pi \cdot K)}\right) \\
&= \left(1 - \epsilon \frac{q\chi}{\pi \cdot K}\right) \left(\mathbb{I}_4 + \frac{q}{2(\pi \cdot K)} \not{A}_{\text{osc}} \not{K}\right) \left(\mathbb{I}_4 + \frac{q}{2(\pi \cdot K)} \not{K} \not{A}_{\text{osc}}\right) + \mathcal{O}(\epsilon^2) \\
&= \left(1 - \epsilon \frac{q\chi}{\pi \cdot K}\right) \left(\mathbb{I}_4 + \frac{q}{2(\pi \cdot K)} \left(\not{A}_{\text{osc}} \not{K} + \not{K} \not{A}_{\text{osc}}\right)\right) + \mathcal{O}(\epsilon^2) \\
&= \left(1 - \epsilon \frac{q\chi}{\pi \cdot K}\right) \left(\mathbb{I}_4 + \frac{q A_{\text{osc}} \cdot K}{\pi \cdot K}\right) + \mathcal{O}(\epsilon^2) \\
&= \left(1 - \epsilon \frac{q\chi}{\pi \cdot K}\right) \left(\mathbb{I}_4 + \epsilon \frac{q\chi}{\pi \cdot K}\right) + \mathcal{O}(\epsilon^2) \\
&= \mathbb{I}_4 + \mathcal{O}(\epsilon^2),
\end{aligned} \tag{8.18}$$

where I used Eqs. (5.3) and (8.17) so that $\not{A}_{\text{osc}} \not{K} + \not{K} \not{A}_{\text{osc}} = 2 A_{\text{osc}} \cdot K = 2\epsilon\chi$. To the next order in $\epsilon$, the Moyal product becomes the eight-dimensional phase-space Poisson bracket (Appendix A). After a series of calculations, one obtains

$$
\begin{aligned}
\frac{i}{2}\left\{\gamma^0\mathcal{V}^\dagger\gamma^0,\mathcal{V}\right\} &= \frac{i}{2}\left\{\left(1-\epsilon\frac{q\chi}{2(\pi\cdot K)}\right)\gamma^0\Xi^\dagger\gamma^0 e^{-i\tilde{\theta}},\Xi\left(1-\epsilon\frac{q\chi}{2(\pi\cdot K)}\right)e^{i\tilde{\theta}}\right\}\\
&= \left\{\tilde{\theta},\left(1-\epsilon\frac{q\chi}{2(\pi\cdot K)}\right)\gamma^0\Xi^\dagger\gamma^0\Xi\left(1-\epsilon\frac{q\chi}{2(\pi\cdot K)}\right)\right\}\\
&\quad+\frac{i}{2}\left\{\left(1-\epsilon\frac{q\chi}{2(\pi\cdot K)}\right)\gamma^0\Xi^\dagger\gamma^0,\Xi\left(1-\epsilon\frac{q\chi}{2(\pi\cdot K)}\right)\right\}\\
&= \left\{\tilde{\theta},\mathbb{I}_4\right\}+\frac{i}{2}\left\{\left(1-\epsilon\frac{q\chi}{2(\pi\cdot K)}\right)\gamma^0\Xi^\dagger\gamma^0,\Xi\left(1-\epsilon\frac{q\chi}{2(\pi\cdot K)}\right)\right\}+\mathcal{O}(\epsilon^2)\\
&= \left(\frac{i}{2}\right)\frac{q^2}{4(\pi\cdot K)^3}\not{A}_{\text{osc}}\left[(K\cdot\partial\not{K})\not{K}-\not{K}(K^2)\right]\not{A}_{\text{osc}}+\mathcal{O}(\epsilon^2).
\end{aligned}
\tag{8.19}
$$

However, note that the term $K\cdot\partial\not{K}$ is zero since

$$
K\cdot\partial\not{K}=K^\mu\partial_\mu(-\gamma^\nu\,\partial_\nu\Theta)=-K^\mu(\partial_\nu\gamma^\nu)\,\partial_\mu\Theta=K^\mu\not{\partial}K_\mu=(1/2)\not{\partial}(K\cdot K)=0,
\tag{8.20}
$$

where I used the dispersion relation (8.4) at the end. Hence, from Eqs. (8.18) and (8.19), the Volkov transformation satisfies

$$
\gamma^0\mathcal{V}^\dagger(x,p)\gamma^0\star\mathcal{V}(x,p)=\mathbb{I}_4+\mathcal{O}(\epsilon^2),
\tag{8.21}
$$

which leads to the conservation of the Lorentz invariant quantity $\langle\,\bar{\Psi}\mid\Psi\,\rangle=\langle\,\bar{\psi}\mid\psi\,\rangle$ up to $\mathcal{O}(\epsilon^2)$. This order of accuracy is sufficient for the theory that I propose in this Chapter.

### 8.3.2 Weyl symbol of the transformed dispersion operator

The next step consists in calculating explicitly the Weyl symbol (8.11) of the transformed dispersion operator. After expanding the Moyal products up to $\mathcal{O}(\epsilon)$, one obtains

$$
\begin{aligned}
\overline{D}_{\text{V}}(x,p) &= \gamma^0\mathcal{V}^\dagger\gamma^0(\not{\pi}-q\not{A}_{\text{osc}}-m\mathbb{I}_4)\mathcal{V}+\frac{i}{2}\left\{\gamma^0\mathcal{V}^\dagger(x,p)\gamma^0,\not{\pi}-q\not{A}_{\text{osc}}-m\mathbb{I}_4\right\}\mathcal{V}\\
&\quad+\frac{i}{2}\left\{\gamma^0\mathcal{V}^\dagger(x,p)\gamma^0(\not{\pi}-q\not{A}_{\text{osc}}-m\mathbb{I}_4),\mathcal{V}\right\}.
\end{aligned}
\tag{8.22}
$$

Now, let us substitute Eq. (8.12) into Eq. (8.22) and calculate the corresponding terms. For example, the first term in Eq. (8.22) is given by

$$
\begin{aligned}
\gamma^0 \mathcal{V}^\dagger \gamma^0 \not{\pi} \mathcal{V} &= \left(1 - \epsilon \frac{q\chi}{2(\pi \cdot K)}\right) \left(\mathbb{I}_4 + \frac{q}{2(\pi \cdot K)} \not{A}_{\mathrm{osc}} \not{K}\right) \not{\pi} \left(\mathbb{I}_4 + \frac{q}{2(\pi \cdot K)} \not{K} \not{A}_{\mathrm{osc}}\right) \left(1 - \epsilon \frac{q\chi}{2(\pi \cdot K)}\right) \\
&= \left(1 - \epsilon \frac{q\chi}{\pi \cdot K}\right) \left(\not{\pi} + \frac{q}{2(\pi \cdot K)} \left(\not{A}_{\mathrm{osc}} \not{K} \not{\pi} + \not{\pi} \not{K} \not{A}_{\mathrm{osc}}\right) + \frac{q^2}{4(\pi \cdot K)^2} \not{A}_{\mathrm{osc}} \not{K} \not{\pi} \not{K} \not{A}_{\mathrm{osc}}\right) + \mathcal{O}(\epsilon^2) \\
&= \left(1 - \epsilon \frac{q\chi}{\pi \cdot K}\right) \left(\not{\pi} + q\not{A}_{\mathrm{osc}} + \frac{q}{2(\pi \cdot K)} \left(-\not{A}_{\mathrm{osc}} \not{\pi} \not{K} + \not{\pi} \not{K} \not{A}_{\mathrm{osc}}\right)\right. \\
&\qquad \left. + \frac{q^2}{2(\pi \cdot K)} \not{A}_{\mathrm{osc}} \not{K} \not{A}_{\mathrm{osc}} - \frac{q^2}{4(\pi \cdot K)^2} \not{A}_{\mathrm{osc}} \not{K} \not{K} \not{\pi} \not{A}_{\mathrm{osc}}\right) + \mathcal{O}(\epsilon^2) \\
&= \left(1 - \epsilon \frac{q\chi}{\pi \cdot K}\right) \left(\not{\pi} + q\not{A}_{\mathrm{osc}} - \frac{\pi \cdot qA_{\mathrm{osc}}}{\pi \cdot K} \not{K} + \frac{q}{2(\pi \cdot K)} \left(\not{\pi} \not{A}_{\mathrm{osc}} \not{K} + \not{\pi} \not{K} \not{A}_{\mathrm{osc}}\right)\right. \\
&\qquad \left. + \frac{q^2 A_{\mathrm{osc}} \cdot K}{\pi \cdot K} \not{A}_{\mathrm{osc}} - \frac{q^2}{2(\pi \cdot K)} \not{K} \not{A}_{\mathrm{osc}} \not{A}_{\mathrm{osc}}\right) + \mathcal{O}(\epsilon^2) \\
&= \left(1 - \epsilon \frac{q\chi}{\pi \cdot K}\right) \left(\not{\pi} + q\not{A}_{\mathrm{osc}} - \frac{\pi \cdot qA_{\mathrm{osc}}}{\pi \cdot K} + \epsilon \frac{q\chi}{\pi \cdot K} \not{\pi} + \epsilon \frac{q^2 \chi}{\pi \cdot K} \not{A}_{\mathrm{osc}} - \frac{q^2 A_{\mathrm{osc}}^2}{\pi \cdot K} \not{K}\right) + \mathcal{O}(\epsilon^2) \\
&= \not{\pi} + q\not{A}_{\mathrm{osc}} - \left(1 - \epsilon \frac{q\chi}{\pi \cdot K}\right) \left(\frac{\pi \cdot qA_{\mathrm{osc}}}{\pi \cdot K} + \frac{q^2 A_{\mathrm{osc}}^2}{\pi \cdot K} \not{K}\right) + \mathcal{O}(\epsilon^2),
\end{aligned}
\tag{8.23}
$$

where I used $\not{a}\not{b} + \not{b}\not{a} = 2a \cdot b$ [Eq. (5.3)] and $\not{K}\not{K} = K^2 \mathbb{I}_4 = 0$ to commute and eliminate some of the slashed matrices.

In a similar manner, one can calculate the rest of the terms in Eq. (8.22). However, for the sake of conciseness, I shall not present the intermediate calculations in this thesis.[6] The final result is

$$
\overline{D}_V(x,p) = \not{\pi} + \frac{q^2 \langle\langle A_{\mathrm{osc}}^2 \rangle\rangle}{2(\pi \cdot K)} \not{K} - m\mathbb{I}_4 + \epsilon\, \mathcal{C}(x,p,\Theta) + \mathcal{O}(\epsilon^2),
\tag{8.24}
$$

where $\mathcal{C}(x,p,\Theta)$ is a quasiperiodic $4 \times 4$ Hermitian matrix whose explicit expression is not needed for the present purposes.

At this point, it is important to notice what was accomplished with the Volkov transformation. The original Dirac action (8.1) included a HF oscillating term $q\not{A}_{\mathrm{osc}}(x,\Theta)$, representing the interactions of the particle with the HF laser field. Using the Volkov transformation, I have replaced this HF interaction term with a slowly varying term proportional to $\langle\langle A_{\mathrm{osc}}^2 \rangle\rangle$ and with HF quasiperiodic terms of $\mathcal{O}(\epsilon)$. Unlike the method presented in Chapter 7, this was done without using any perturbative scheme based on the amplitude of $\not{A}_{\mathrm{osc}}$. In other words, the Volkov transformation is non-perturbative on the laser amplitude.

[6]See Ruiz *et al.* (2015) for more details on some of the corresponding calculations.

### 8.3.3 Ponderomotive dispersion symbol

The Weyl symbol (8.24) does not yet describe the ponderomotive dynamics, as it still depends on the fast eikonal phase $\Theta$ through the HF oscillating terms of $\mathcal{O}(\epsilon)$. One could implement a transformation similar to the one in Chapter 7 that eliminates these small HF terms. Such a transformation would be asymptotic on the small parameter $\epsilon$. In analogy to the results presented in Chapter 7, the Weyl symbol $G(x,p)$ of the transformation operator $\widehat{\mathcal{G}}$ would be proportional to $\epsilon\, \mathcal{C}(x,p,\Theta)$ so that $G(x,p,\Theta) \sim \mathcal{O}(\epsilon)$. Then, the subsequent poderomotive effect corresponding to the oscillations in $\mathcal{C}(x,p,\Theta)$ would be $\mathcal{O}(\epsilon^2)$ since the ponderomotive term would scale as $\langle\langle \epsilon\, \mathcal{C}(x,p,\Theta) G(x,p,\Theta)\rangle\rangle \sim \mathcal{O}(\epsilon^2)$. Since the goal is to develop a ponderomotive theory accurate to $\mathcal{O}(\epsilon)$, the above ponderomotive effect goes beyond the order of accuracy of the present theory.

In light of these arguments, the ponderomotive dispersion symbol $\overline{D}_{\text{pond}}(x,p)$ is simply given by the average of Eq. (8.24) so that $\overline{D}_{\text{pond}}(x,p) \doteq \langle\langle \overline{D}_V(x,p)\rangle\rangle$; or more explicitly,

$$\overline{D}_{\text{pond}}(x,p) = \not{\pi} + \frac{q^2 \langle\langle A_{\text{osc}}^2\rangle\rangle}{2(\pi\cdot K)} \not{K} - m\mathbb{I}_4 + \mathcal{O}(\epsilon^2). \tag{8.25}$$

The ponderomotive dispersion symbol $\overline{D}_{\text{pond}}(x,p)$ is slowly varying. Equation (8.25) can be cast in a more familiar form. Upon introducing the quasi four-momentum (Yakovlev, 1966)

$$\zeta_\mu(x,p) \doteq \pi_\mu(x,p) + \eta(x,p) K_\mu(x), \tag{8.26}$$

where

$$\eta(x,p) \doteq \frac{q^2 \langle\langle A_{\text{osc}}^2\rangle\rangle}{2(\pi\cdot K)}, \tag{8.27}$$

one can write $\overline{D}_{\text{pond}}(x,p)$ in a similar form as the dispersion symbol of a free-streaming Dirac particle:

$$\overline{D}_{\text{pond}}(x,p) = \not{\zeta} - m\mathbb{I}_4. \tag{8.28}$$

The quasi four-momentum $\zeta_\mu$ represents the averaged kinetic four-momentum of the particle when interacting with the HF EM field. It is composed of two parts. The first term in Eq. (8.26) is related to the particle's slowly varying oscillation-center motion. The second term in Eq. (8.26) is associated to the quiver motion of the particle. This will be further discussed in Sec. 8.5.

Upon replacing $\overline{D}_{\rm V}(x,p)$ with the ponderomotive dispersion symbol $\overline{D}_{\rm pond}(x,p)$ in Eq. (8.9), one obtains the ponderomotive action in the phase-space representation:

$$\mathcal{S}_{\rm pond} \doteq \operatorname{Tr} \int {\rm d}^4x\, {\rm d}^4p\, D_{\rm pond}(x,p)\, W_\psi(x,p), \tag{8.29}$$

where the ponderomotive dispersion symbol (without the overline) is given by

$$D_{\rm pond}(x,p) \doteq \gamma^0 \overline{D}_{\rm pond}(x,p) = \zeta_0(x,p) - H\big(\boldsymbol{\zeta}(x,p)\big) \tag{8.30}$$

and $H\big(\boldsymbol{\zeta}(x,p)\big)$ is the Hamiltonian[7]

$$H\big(\boldsymbol{\zeta}(x,p)\big) \doteq \boldsymbol{\alpha} \cdot \boldsymbol{\zeta}(x,p) + m\beta. \tag{8.31}$$

Here $\boldsymbol{\alpha} \doteq \gamma^0 \boldsymbol{\gamma}$ are $4 \times 4$ Hermitian matrices, and $\beta \doteq \gamma^0$. As a reminder, the Dirac matrices $\gamma^\mu$ are defined in Eq. (5.4).

### 8.3.4 Extended action principle

The preceding calculation concludes the first stage in obtaining a point-particle ponderomotive model for the relativistic spin-1/2 electron. With the ponderomotive dispersion symbol $D_{\rm pond}(x,p)$, one can construct a principle of stationary action from which the XGO equations can be obtained. As in Sec. 4.2, the action in the extended space is given by

$$\mathcal{S}_{\rm X} \doteq \int {\rm d}\tau\ L, \tag{8.32}$$

where $L \doteq L_\tau + L_{D_{\rm pond}}$ serves as the Lagrangian and

$$L_\tau \doteq -(i/2)\left[\langle\, \psi(\tau) \mid \partial_\tau \psi(\tau)\,\rangle - {\rm c.\,c.}\right], \tag{8.33a}$$

$$L_{D_{\rm pond}} \doteq \operatorname{Tr} \int {\rm d}^4x\, {\rm d}^4p\, D_{\rm pond}(x,p) W_\psi(x,p), \tag{8.33b}$$

As a reminder, here $\psi$ is a four-component complex-valued function that depends not only on spacetime but also on some parameter $\tau$ so that $\psi(\tau,x) = \langle\, x \mid \psi(\tau)\,\rangle$ and $\partial_\tau \psi(\tau,x) = \langle\, x \mid \partial_\tau \psi(\tau)\,\rangle$. Also, the abstract vector state $|\,\psi(\tau)\,\rangle$ belongs to the same Hilbert space with inner product defined in Eq. (4.6).

[7] Strictly speaking, Eq. (8.30) is not written in the symplectic form, so it is incorrect to refer to the function $H\big(\boldsymbol{\zeta}(x,p)\big)$ as the Hamiltonian of the system.

## 8.4 Block-diagonalizing the dispersion operator

As in Sec. 4.3.3, let us now block-diagonalize the ponderomotive dispersion symbol (8.30). Note that Eqs. (8.30) and (8.31) have a structure similar to Eqs. (5.12) and (5.13). Hence, one can readily obtain the eigenvalues and eigenmodes of the ponderomotive dispersion symbol $D_{\text{pond}}(x,p)$. The eigenvalues are

$$\lambda^{(1,2)}(x,p) = \zeta_0(x,p) - \sqrt{\boldsymbol{\zeta}^2(x,p) + m^2}, \tag{8.34a}$$

$$\lambda^{(3,4)}(x,p) = \zeta_0(x,p) + \sqrt{\boldsymbol{\zeta}^2(x,p) + m^2}. \tag{8.34b}$$

As in Sec. 5.3, the Weyl symbol $D(x,p)$ has two doubly degenerate eigenvalues. The eigenvalues $\lambda^{(1,2)}(x,p)$ correspond to the particles states, and the eigenvalues $\lambda^{(3,4)}(x,p)$ correspond to the antiparticles states. After setting the eigenvalues to zero, one obtains the lowest-order GO frequency:

$$\begin{aligned} 0 &= \zeta_0^2 - \boldsymbol{\zeta}^2 - m^2 \\ &= (\pi_\mu + \eta K_\mu)(\pi^\mu + \eta K^\mu) - m^2 \\ &= \pi^2 + 2\eta\pi\cdot K + \eta^2 K^2 - m^2 \\ &= \pi^2 + q^2 \langle\!\langle A_{\text{osc}}^2 \rangle\!\rangle - m^2, \end{aligned} \tag{8.35}$$

where I substituted Eqs. (8.4) and (8.27). Finally, upon substituting $\pi^\mu = p^\mu - qA_{\text{bg}}^\mu$ and solving for the temporal momentum $p_0$, one obtains

$$p_0 = qV_{\text{bg}} \pm \sqrt{(\mathbf{p} - q\mathbf{A}_{\text{bg}})^2 + m_{\text{eff}}^2}, \tag{8.36}$$

where $m_{\text{eff}}(x)$ is the *effective particle mass* (Akhiezer and Polovin, 1956; Kibble, 1966) given by

$$m_{\text{eff}}^2(x) \doteq m^2 - q^2 \langle\!\langle A_{\text{osc}}^2 \rangle\!\rangle(x). \tag{8.37}$$

(The minus sign is due to the chosen metric signature.) Equation (8.36) is the well-known Hamiltonian that governs the ponderomotive dynamics of a relativistic spinless particle interacting with an oscillating EM vacuum field and a slowly varying background EM field (Malka *et al.*, 1997; Quesnel and Mora, 1998; Mora and Antonsen Jr., 1997; Dodin *et al.*, 2003). The two roots in Eq. (8.36) represent solutions for the particle and the antiparticle states.

Let us consider the dynamics of the particle states. Similar to the calculations in Sec. 5.3, upon substituting $\pi_\mu \to \zeta_\mu$, one obtains $\Lambda(x,p) = (\zeta_0 - \varepsilon)\mathbb{I}_2$, where $\varepsilon(x,p) \doteq [\boldsymbol{\zeta}^2(x,p) + m^2]^{1/2}$. Similarly to Eq. (5.17), the corresponding matrix $\Xi(x,p)$ for the two particle states is

$$\Xi(x,p) = \sqrt{\frac{m+\varepsilon}{2\varepsilon}} \begin{pmatrix} \mathbb{I}_2 \\ \frac{\boldsymbol{\zeta}\cdot\boldsymbol{\sigma}}{m+\varepsilon} \end{pmatrix}. \tag{8.38}$$

It is easy to verify that $\Xi(x,p)$ diagonalizes the ponderomotive dispersion symbol (8.30). [The corresponding calculation is identical to that in Eq. (5.18).] It can also be shown that $\Xi(x,p)$ is normalized so that $\Xi^\dagger(x,p)\Xi(x,p) = \mathbb{I}_2$.

Having obtained $\Xi(x,p)$, let us now calculate the spin-coupling term $\mathcal{U}(x,p)$. Since the eigenvalues of the particle states are degenerate, one can use the expression for the spin-couplingterm given by Eq. (4.36):

$$\mathcal{U}(x,p) = \left(-\frac{\partial\lambda}{\partial p_\mu}\right)\left(\Xi^\dagger\frac{\partial\Xi}{\partial x^\mu}\right)_A + \left(\frac{\partial\lambda}{\partial x^\mu}\right)\left(\Xi^\dagger\frac{\partial\Xi}{\partial p_\mu}\right)_A + \left(\frac{\partial\Xi^\dagger}{\partial p_\mu}(D-\lambda\mathbb{I}_4)\frac{\partial\Xi}{\partial x^\mu}\right)_A, \tag{8.39}$$

where $\lambda(x,p) = \zeta_0(x,p) - \varepsilon(x,p)$. A straightforward calculation of the ponderomotive spin-coupling matrix $\mathcal{U}(x,p)$ leads to (Appendix C.3)

$$\mathcal{U}(x,p) = \frac{1}{2}\boldsymbol{\sigma}\cdot\boldsymbol{\Omega}_{\text{pond}}(x,p), \tag{8.40}$$

where $\boldsymbol{\Omega}_{\text{pond}}(x,p)$ is the ponderomotive precession frequency given by

$$\begin{aligned}\boldsymbol{\Omega}_{\text{pond}}(x,p) \doteq{}& \frac{q}{\varepsilon}\left(\mathbf{B}_{\text{bg}} - \frac{\boldsymbol{\zeta}\times\mathbf{E}_{\text{bg}}}{m+\zeta^0}\right) \\ &+ \frac{q^2}{2\varepsilon(\pi\cdot K)}\left(\mathbf{K}\times\boldsymbol{\nabla}\langle\langle A_{\text{osc}}^2\rangle\rangle - \frac{(\boldsymbol{\zeta}\times\mathbf{K})\partial_t\langle\langle A_{\text{osc}}^2\rangle\rangle}{m+\zeta^0} - \frac{K^0\boldsymbol{\zeta}\times\boldsymbol{\nabla}\langle\langle A_{\text{osc}}^2\rangle\rangle}{m+\zeta^0}\right) \\ &+ \frac{q^2\langle\langle A_{\text{osc}}^2\rangle\rangle}{2\varepsilon(\pi\cdot K)^2}\left[\left(\frac{K^0\boldsymbol{\zeta}}{m+\zeta^0} - \mathbf{K}\right)\times\left[K^0 q\mathbf{E}_{\text{bg}} + \mathbf{K}\times q\mathbf{B}_{\text{bg}} - (\pi^\mu\partial_\mu)\mathbf{K}\right]\right. \\ &\left. -\frac{\boldsymbol{\zeta}\times\mathbf{K}}{m+\zeta^0}\left[\mathbf{K}\cdot q\mathbf{E}_{\text{bg}} - (\pi^\mu\partial_\mu)K^0\right]\right]. \end{aligned} \tag{8.41}$$

An interpretation of the terms in $\boldsymbol{\Omega}_{\text{pond}}(x,p)$ will be given later.

In summary, the $2\times 2$ effective ponderomotive dispersion symbol (4.32) that governs the ponderomotive dynamics of the relativistic spin-1/2 electron is given by

$$[[D_{\text{eff}}]](x,p) = [\zeta_0(x,p) - \varepsilon(\boldsymbol{\zeta}(x,p))]\,\mathbb{I}_2 + \frac{1}{2}\boldsymbol{\sigma}\cdot\boldsymbol{\Omega}_{\text{pond}}(x,p). \tag{8.42}$$

The first term describes the ponderomotive dynamics of the relativistic spinless particle. The spin-coupling term, which is of order $\epsilon$, introduces spin-orbit coupling effects.

## 8.5 Ponderomotive point-particle model

Let us discuss the point-particle dynamics of our ponderomotive model. As in Sec. 5.5, one substitutes $\lambda(x,p)$ and $\mathcal{U}(x,p)$ into the general point-particle XGO action (4.68) in order to obtain

$$\mathcal{S} = \int \mathrm{d}\tau \left[ P \cdot \dot{X} - \frac{i}{2}\left(Z^\dagger \dot{Z} - \dot{Z}^\dagger Z\right) + \lambda(X,P) + Z^\dagger \mathcal{U}(X,P) Z \right], \tag{8.43}$$

where $X^\mu(\tau) = (X^0, \mathbf{X})$ is the particle four-position, $P^\mu(\tau) = (P_0, \mathbf{P})$ is the particle canonical four-momentum, and $Z(\tau)$ is a complex-valued two-component function that describes the particle spin state. It is normalized such that $Z^\dagger(\tau)Z(\tau) = 1$. For completeness, $\lambda(x,p) = \zeta_0(x,p) - \varepsilon(\boldsymbol{\zeta}(x,p))$, and $\mathcal{U}(x,p)$ is the total spin-coupling matrix given by Eq. (8.40). The action (5.30) is complemented by the corrected dispersion relation

$$\lambda(X,P) + Z^\dagger \mathcal{U}(X,P) Z = 0. \tag{8.44}$$

As in Sec. 4.6.3, let us define the four-momentum $p_*^\mu(t,\mathbf{x},\mathbf{p}) = (H_0(t,\mathbf{x},\mathbf{p}), \mathbf{p})$ such that $\lambda(x,p_*) = 0$. Here $H_0(t,\mathbf{x},\mathbf{p}) = (\boldsymbol{\pi}^2 + m_{\text{eff}}^2)^{1/2} + qV_{\text{bg}}$ is the lowest-order ponderomotive Hamiltonian. The XGO Hamiltonian is then given by Eq. (4.74), where

$$\begin{aligned}
\left(\frac{\partial \lambda}{\partial p_0}\right)_{p=p_*} &= \left(\frac{\partial \zeta_0}{\partial p_0} - \frac{\partial \varepsilon}{\partial p_0}\right)_{p=p_*} \\
&= 1 + K^0 \left(\frac{\partial \eta}{\partial p_0}\right)_{p=p_*} - \left(\frac{\boldsymbol{\zeta}\cdot\mathbf{K}}{\varepsilon}\frac{\partial \eta}{\partial p_0}\right)_{p=p_*} \\
&= 1 + K^0 \left(\frac{\partial \eta}{\partial p_0}\right)_{p=p_*} - \frac{\boldsymbol{\pi}\cdot\mathbf{K} + \eta(x,p_*)\mathbf{K}^2}{\pi^0(x,p_*) + \eta(x,p_*)K^0}\left(\frac{\partial \eta}{\partial p_0}\right)_{p=p_*} \\
&= \frac{1}{\pi(x,p_*) + \eta(x,p_*)K^0}\left[\pi_0(x,p_*) + \eta(x,p_*)K^0 + (\pi\cdot K)\left(\frac{\partial \eta}{\partial p_0}\right)_{p=p_*}\right] \\
&= \frac{\pi_0(x,p_*)}{\varepsilon(x,p_*)} \\
&= \frac{(\boldsymbol{\pi}^2 + m_{\text{eff}}^2)^{1/2}}{\varepsilon(x,p_*)}.
\end{aligned} \tag{8.45}$$

Here I used $K^2 = 0$ and $\partial\eta/\partial p_0 = -K^0\eta/(\pi\cdot K)$. Substituting this relation into Eq. (4.74) leads to the ponderomotive noncovariant point-particle action:[8]

$$\mathcal{S}_{\text{eff}} = \int \mathrm{d}t \left[\mathbf{P}\cdot\dot{\mathbf{X}} + \frac{i\hbar}{2}\left(Z^\dagger\dot{Z} - \dot{Z}^\dagger Z\right) - H_{\text{eff}}(t,\mathbf{X},\mathbf{P},Z,Z^\dagger)\right], \tag{8.46}$$

[8]For simplicity, from henceforth I restore the constants $c$ and $\hbar$.

where the effective Hamiltonian is given by

$$H_{\text{eff}}(t,\mathbf{X},\mathbf{P},Z,Z^\dagger) \doteq \gamma_{\text{eff}} mc^2 + q\mathbf{V}_{\text{bg}} - \frac{\hbar}{2} Z^\dagger \boldsymbol{\sigma}\cdot\boldsymbol{\Omega}_{\text{eff}}(t,\mathbf{X},\mathbf{P})Z. \tag{8.47}$$

The effective Lorentz factor associated with the particle oscillation-center motion is

$$\gamma_{\text{eff}}(t,\mathbf{X},\mathbf{P}) \doteq \sqrt{1 + a_0^2 + \left(\frac{\mathbf{P}}{mc} - \frac{q\mathbf{A}_{\text{bg}}}{mc^2}\right)^2}, \tag{8.48}$$

where

$$a_0^2(t,\mathbf{X}) \doteq -\frac{q^2 \langle\!\langle A_{\text{osc}}^2 \rangle\!\rangle}{m^2 c^4} \tag{8.49}$$

is positive under the assumed metric. For example, suppose a standard representation of the laser vector potential is $\mathbf{A}_{\text{osc}}(t,\mathbf{x}) = \text{Re}\left[\mathbf{A}_\perp(t,\mathbf{x})\, e^{i\Theta(t,\mathbf{x})}\right]$, where $\mathbf{A}_\perp \cdot \mathbf{K} = 0$ (Mora and Antonsen Jr., 1996; Malka *et al.*, 1997; Mora and Quesnel, 1998). Then, the Lorentz condition (8.5) determines the scalar potential envelope $V_{\text{osc,c}} = i(\boldsymbol{\nabla}\cdot\mathbf{A}_\perp)c^2/\Omega = \mathcal{O}(\epsilon)$. Hence, Eq. (8.49) yields

$$a_0^2 \approx \frac{q^2 |\mathbf{A}_\perp|^2}{2m^2 c^4}, \tag{8.50}$$

where I neglected a term of $\mathcal{O}(\epsilon^2)$. Note also that, loosely speaking, $a_0^2$ is the measure of the particle quiver energy in units $mc^2$. Accordingly, nonrelativistic interactions correspond to $a_0 \ll 1$. If the spin–orbital interaction is neglected, the present model yields the spinless ponderomotive model that was developed in Dodin *et al.* (2003) for a particle interacting with a laser pulse and a slow background fields simultaneously.

The effective precession frequency $\boldsymbol{\Omega}_{\text{eff}}(t,\mathbf{X},\mathbf{P})$ is given by

$$\boldsymbol{\Omega}_{\text{eff}}(t,\mathbf{X},\mathbf{P}) = \boldsymbol{\Omega}_1 + \boldsymbol{\Omega}_2 + \boldsymbol{\Omega}_3 + \mathcal{O}(\epsilon^2), \tag{8.51}$$

where

$$\boldsymbol{\Omega}_1(t,\mathbf{X},\mathbf{P}) \doteq \frac{q}{\gamma_{\text{eff}} mc}\left(\mathbf{B}_{\text{bg}} - \frac{\boldsymbol{\Lambda}\times\mathbf{E}_{\text{bg}}}{mc + \Lambda^0}\right), \tag{8.52}$$

$$\boldsymbol{\Omega}_2(t,\mathbf{X},\mathbf{P}) \doteq -\frac{mc^2}{2\gamma_{\text{eff}}(\Pi\cdot K)}\left(\mathbf{K}\times\boldsymbol{\nabla} a_0^2 - \frac{(\boldsymbol{\Lambda}\times\mathbf{K})\partial_t a_0^2}{mc^2 + \Lambda^0 c} - \frac{\Omega\boldsymbol{\Lambda}\times\boldsymbol{\nabla} a_0^2}{mc^2 + \Lambda^0 c}\right), \tag{8.53}$$

$$\begin{aligned}\boldsymbol{\Omega}_3(t,\mathbf{X},\mathbf{P}) \doteq &-\frac{mc^2 a_0^2}{2\gamma_{\text{eff}}(\Pi\cdot K)^2}\left(\frac{\Omega\boldsymbol{\Lambda}}{mc^2+\Lambda^0 c} - \mathbf{K}\right)\times\left(\frac{\Omega q\mathbf{E}_{\text{bg}}}{c^2} + \frac{\mathbf{K}\times q\mathbf{B}_{\text{bg}}}{c} - (\Pi^\mu\partial_\mu)\mathbf{K}\right)\\ &+ \frac{mc\, a_0^2}{2\gamma_{\text{eff}}(\Pi\cdot K)^2}\frac{\boldsymbol{\Pi}\times\mathbf{K}}{mc+\Lambda^0}\left[\mathbf{K}\cdot q\mathbf{E}_{\text{bg}} - (\Pi^\mu\partial_\mu)\Omega\right].\end{aligned} \tag{8.54}$$

Here $\Pi^\mu \doteq (mc\gamma_{\text{eff}}, \mathbf{P} - q\mathbf{A}_{\text{bg}}/c)$ is the particle kinetic four-momentum, $K^\mu = (\Omega/c, \mathbf{K})$ is the EM field four-wavevector, $\partial_\mu = (c^{-1}\partial_t, \boldsymbol{\nabla})$, and

$$\Lambda^\mu(t, \mathbf{X}, \mathbf{P}) = \Pi^\mu - K^\mu \frac{m^2c^2a_0^2}{2(\Pi \cdot K)} \tag{8.55}$$

is the particle quasi four-momentum. Notably, $\Lambda^\mu \to \Pi^\mu$ at $a_0 \to 0$ and $\Lambda^\mu \to \Pi^\mu - K^\mu mc^2 a_0/(2\Omega)$ at $a_0 \to +\infty$.

Let us discuss the physical interpretation of the terms appearing in $\boldsymbol{\Omega}_{\text{eff}}(t, \mathbf{X}, \mathbf{P})$. It is clear that $\boldsymbol{\Omega}_1(t, \mathbf{X}, \mathbf{P})$ describes the spin precession due to the background EM fields only. The second term $\boldsymbol{\Omega}_2(t, \mathbf{X}, \mathbf{P})$ describes the spin precession caused by the gradients of the laser field envelope. Actually, this term is quite similar to the first term in $\boldsymbol{\Omega}_{\text{eff}}$. One simply needs to introduce an effective magnetic and electric fields due to the gradients of the oscillating EM field envelope: $\mathbf{B}_{\text{osc,eff}} = \mathbf{K} \times \boldsymbol{\nabla} a_0^2$ and $\mathbf{E}_{\text{osc,eff}} = (\mathbf{K}/\Omega) \times \mathbf{B}_{\text{osc,eff}}$, respectively. With this identification, the similarities between the first and second terms of $\boldsymbol{\Omega}_{\text{eff}}$ become clear. Finally, the last term $\boldsymbol{\Omega}_3(t, \mathbf{X}, \mathbf{P})$ is harder to interpret, but it apparently causes a spin precession that is due to both the background EM fields and the HF oscillating field.

Treating $\mathbf{X}(t)$, $\mathbf{P}(t)$, $Z(t)$, and $Z^\dagger(t)$ as independent variables leads to the following ELEs:

$$\delta\mathbf{P}: \quad \frac{\mathrm{d}\mathbf{X}}{\mathrm{d}t} = \frac{\partial}{\partial \mathbf{P}}(\gamma_{\text{eff}} mc^2) - \frac{\partial}{\partial \mathbf{P}}(\mathbf{S} \cdot \boldsymbol{\Omega}_{\text{eff}}), \tag{8.56a}$$

$$\delta\mathbf{X}: \quad \frac{\mathrm{d}\mathbf{P}}{\mathrm{d}t} = -\frac{\partial}{\partial \mathbf{X}}(\gamma_{\text{eff}} mc^2 + qV_{\text{bg}}) + \frac{\partial}{\partial \mathbf{X}}(\mathbf{S} \cdot \boldsymbol{\Omega}_{\text{eff}}), \tag{8.56b}$$

$$\delta Z^\dagger: \quad \frac{\mathrm{d}Z}{\mathrm{d}t} = \frac{i}{2}\boldsymbol{\Omega}_{\text{eff}} \cdot \boldsymbol{\sigma} Z, \tag{8.56c}$$

$$\delta Z: \quad \frac{\mathrm{d}Z^\dagger}{\mathrm{d}t} = -\frac{i}{2} Z^\dagger \boldsymbol{\Omega}_{\text{eff}} \cdot \boldsymbol{\sigma}, \tag{8.56d}$$

where $\mathbf{S}(t)$ is the particle spin vector

$$\mathbf{S}(t) \doteq \frac{\hbar}{2} Z^\dagger(t) \boldsymbol{\sigma} Z(t), \tag{8.57}$$

and $|\mathbf{S}| = \hbar/2$. Equations (8.48)–(8.57) form a complete set of equations. The first terms on the right-hand side of Eqs. (8.56a) and (8.56b) describe the dynamics of a relativistic spinless particle, in agreement with earlier theories (Malka *et al.*, 1997; Quesnel and Mora, 1998; Mora and Antonsen Jr., 1997; Dodin *et al.*, 2003). The second terms describe the ponderomotive spin-orbit coupling. Equations (8.56c) and (8.56d) also yield the following ponderomotive equation for spin precession,

$$\dot{\mathbf{S}} = \mathbf{S} \times \boldsymbol{\Omega}_{\text{eff}}, \tag{8.58}$$

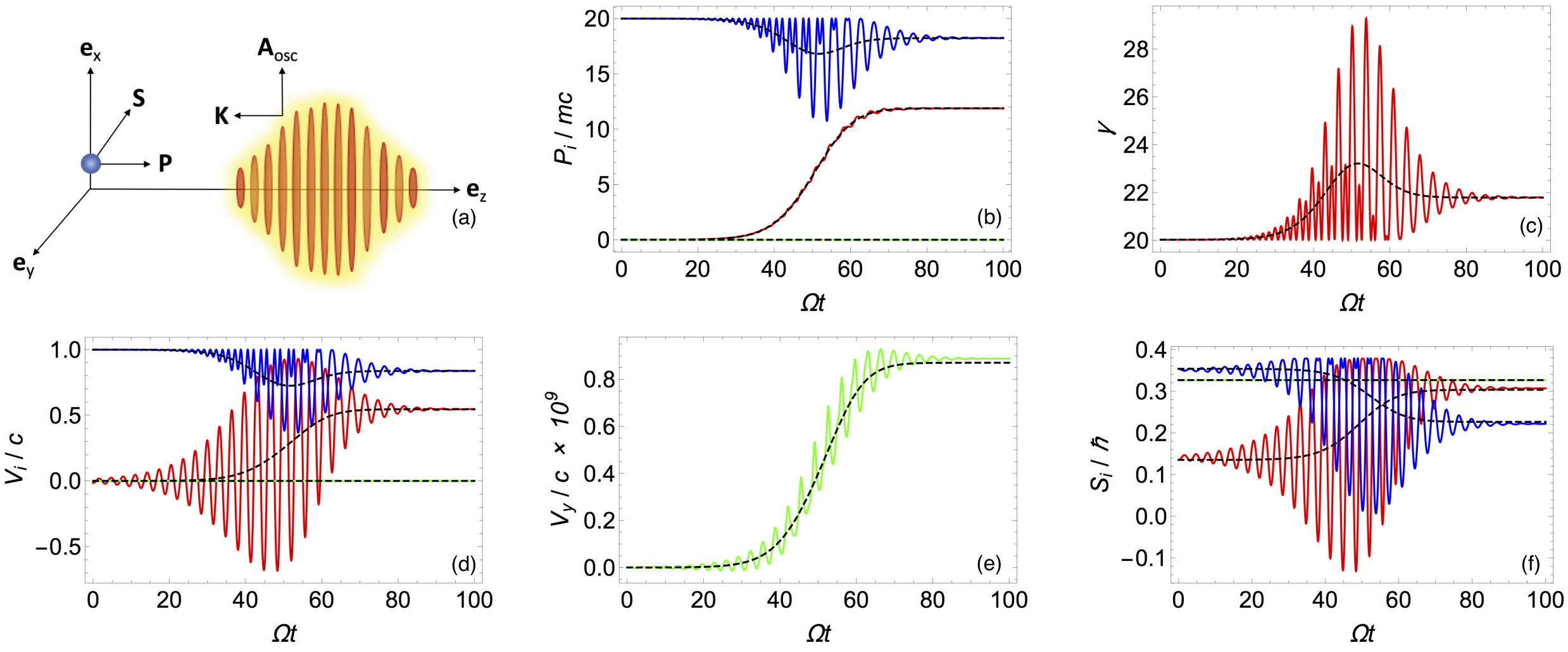

Figure 8.1: Dynamics of a relativistic spin-1/2 electron interacting with a relativistically intense laser pulse (numerical simulation): the black dashed curves correspond to the ponderomotive model described by the Lagrangian (8.46), and the colored curves correspond to the non-averaged model described by the action (5.32). (a) Schematic of the interaction; yellow and red is the laser field, blue is the particle; arrows denote the direction of the laser wave vector $\mathbf{K}$, the oscillating vector potential $\mathbf{A}_{\text{osc}}$, the particle canonical momentum $\mathbf{P}$, and the particle spin $\mathbf{S}$. The unit vectors along the reference axes are denoted by $\mathbf{e}_i$. Subfigures (b)–(f) show the components of the particle canonical momentum $\mathbf{P}$, Lorentz factor $\gamma$, velocity $\mathbf{V}$, and spin $\mathbf{S}$. The red, green, and blue lines correspond to projections on the $x$, $y$, and $z$ axes, respectively. In the figure, an electron is initially traveling along the $z$ axis and then collides with a counter-propagating laser pulse. The initial position of the particle is $\mathbf{X}_0 = (\ell/2)\mathbf{e}_x$, the initial momentum is $\mathbf{P}_0/(mc) = 20\mathbf{e}_z$, and the normalized initial spin vector is $\mathbf{S}_0/\hbar = 0.14\mathbf{e}_x + 0.33\mathbf{e}_y + 0.35\mathbf{e}_z$. The envelope of the vector potential of the laser pulse is $q\mathbf{A}_{\text{osc}}/(mc^2) = 30\ \text{sech}\left[(z - 5\ell + ct)/\ell\right] \exp\left[-(x^2 + y^2)/\ell^2\right] \mathbf{e}_x$, where $\ell = 20|\mathbf{k}|^{-1}$. These parameters correspond to a maximum intensity $I_{\max} \simeq 1.23 \times 10^{21}$ W/cm$^2$ for a 1 $\mu$m laser.

which can be checked by direct substitution. These equations are the main result of this Chapter.

## 8.6 Numerical simulations

To test the proposed ponderomotive model, Eqs. (8.56) were solved using MATHEMATICA's numerical solver `NDSolve` and compared the results with the non-averaged point-particle model presented in Chapter 5. In the first test case, I consider the dynamics of a spin-1/2 electron colliding with a counter-propagating relativistically strong ($a_0 \gg 1$) laser pulse (see Fig. 8.1). The simulation parameters are given in the caption of Fig. 8.1, and a schematic of the interaction is presented in Fig. 8.1(a). Figures 8.1(b)–8.1(e) show that the ponderomotive model accurately describes the mean evolution of the particle momentum, kinetic energy, and velocity. The main contribution to the variations in $V_x$ and $V_z$ is the ponderomotive force caused by spatial gradient of the effective mass. However, the acceleration on the $xz$ plane is caused by the SG force, as shown in Fig. 8.1(e). Also notice that the ponderomotive model is extremely accurate in describing the particle spin precession, as shown in Fig. 8.1(f).

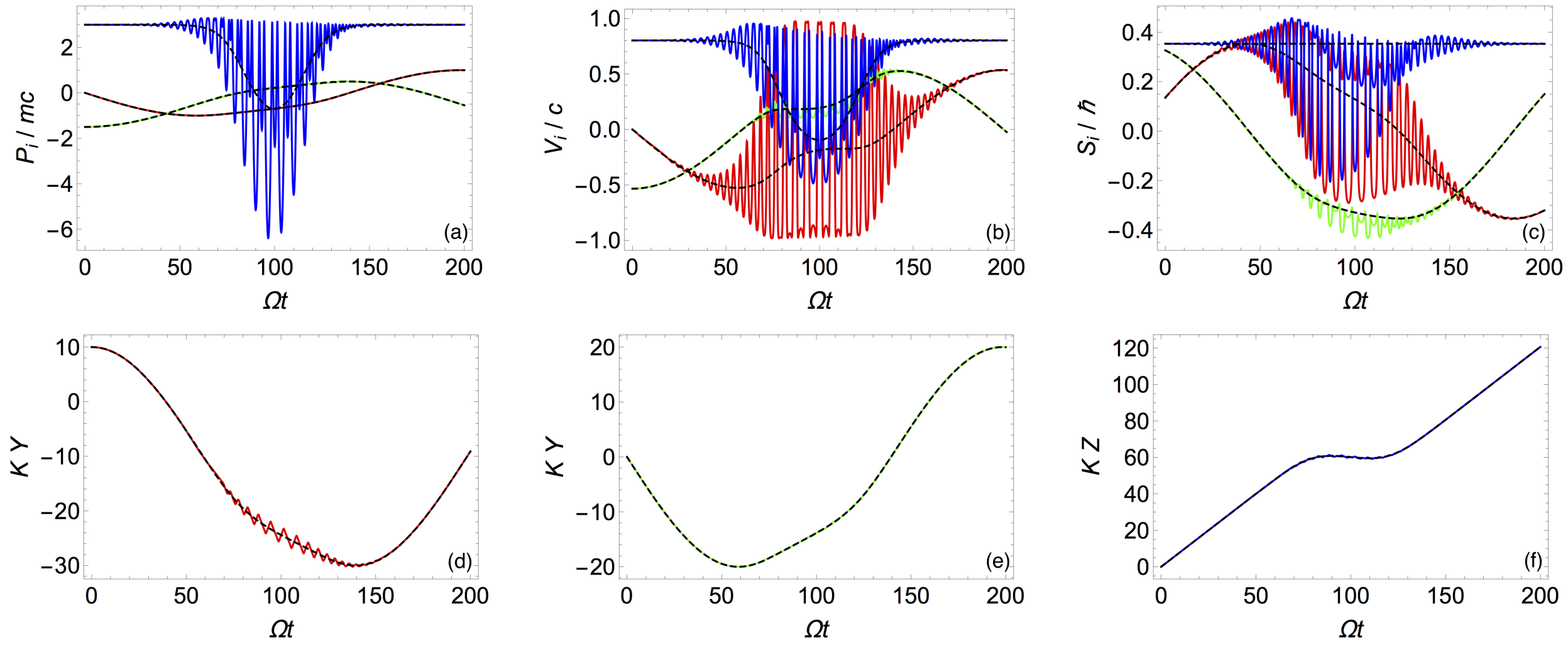

Figure 8.2: Dynamics of a relativistic spin-1/2 electron interacting with an external magnetic field and a relativistically intense laser pulse (numerical simulation): the black dashed curves correspond to the ponderomotive model described by the Lagrangian (8.46), and the colored curves correspond to the non-averaged model described by the action (5.32). Subfigures (a)–(c) show the components of the particle canonical momentum $\mathbf{P}$, velocity $\mathbf{V}$, and spin $\mathbf{S}$. The red, green, and blue lines correspond to projections on the $x$, $y$, and $z$ axes, respectively. Subfigures (d)–(f) show the components of the particle position $\mathbf{X}$. The initial position of the particle is $\mathbf{X}_0 = \mathbf{0}$, the initial momentum is $\mathbf{P}_0/(mc) = (-2\mathbf{e}_y + 3\mathbf{e}_z)$, and the normalized initial spin vector is $\mathbf{S}_0/\hbar = 0.14\mathbf{e}_x + 0.33\mathbf{e}_y + 0.35\mathbf{e}_z$. A background magnetic field is added such that $q\mathbf{A}_{\text{bg}}/(mc^2) = 0.1(-y\mathbf{e}_x + x\mathbf{e}_y)/2$, which corresponds to a static homogeneous magnetic field $B_{\text{bg}} \simeq 10.7$ MG aligned towards the $z$ axis. The envelope of the vector potential of the laser pulse is assumed to have the form $q\mathbf{A}_{\text{osc}}/(mc^2) = 10\,\text{sech}\,[(z - 8\ell + ct)/\ell]\,\mathbf{e}_x$, where $\ell = 20|\mathbf{k}|^{-1}$. These parameters correspond to a maximum laser intensity $I_{\max} \simeq 1.37 \times 10^{20}$ W/cm$^2$ for a 1 $\mu$m laser.

In the second test case, I consider a Dirac electron immersed in a background magnetic field along the $z$ axis and interacting with a laser plane wave traveling along the $z$ axis. The simulation parameters are given in Fig. 8.2. As shown in Figs. 8.2(b)–8.2(f), the ponderomotive model accurately describes the particle position, momentum, velocity, and spin. These simulations also confirm the predictions of the spinless model developed by Dodin *et al.* (2003) for a particle interacting with a relativistic laser field and a large-scale background field simultaneously.

## 8.7 Conclusions

In this Chapter, I derived a point-particle ponderomotive model of a Dirac electron interacting with a HF field. Starting from the first-principle action for the Dirac particle, I obtained a reduced phase-space action that describes the relativistic time-averaged dynamics of such particle in a quasiperiodic laser pulse in vacuum. The pulse is allowed to have an arbitrarily large amplitude (as long as radiation damping and pair production are negligible). The model captures the BMT spin dynamics, the SG spin–orbital coupling, the conventional ponderomotive forces, and the interaction with large-scale background fields (if any). Agreement with the non-averaged point-particle model presented in Chapter 5 was shown numerically. Also, the well-known theory in which ponderomotive effects are incorporated in the particle effective mass is reproduced as a special case when the spin–orbital coupling is negligible.

# Chapter 9

# Applications to wave turbulence and zonal-flow formation

In this Chapter, I present an application of the Weyl symbol calculus to study nonlinear wave phenomena such as wave turbulence. Specifically, I consider drift-wave (DW) turbulence and the formation of zonal flows (ZFs). The DW–ZF interaction has been attracting the attention of researchers for many years, particularly in connection with its effect on the transport properties of magnetically confined fusion plasmas. Among the various reduced models, the wave kinetic equation (WKE) is widely used to study DW–ZF interactions. However, this popular formulation neglects the exchange of enstrophy between DWs and ZFs and also ignores effects beyond the geometrical-optics (GO) limit. Here I present a new theory that captures both of these effects, while still treating DW quanta ("driftons") as particles in phase space. This formulation can be considered as a phase-space representation of the second-order cumulant expansion (CE2). In the GO limit, this formulation features additional terms missing in the traditional WKE that ensure exact conservation of the total enstrophy of the system, in addition to the total energy. The results presented in this Chapter were published by Ruiz *et al.* (2016).

## 9.1 Introduction

### 9.1.1 Motivation

The formation of ZFs is a problem of fundamental interest in many contexts, including physics of planetary atmospheres, astrophysics, and fusion science.[1] In particular, the interaction of ZFs and DW turbulence in laboratory plasmas significantly affects the transport of energy, momentum, and particles, so understanding it is critical to improving plasma confinement. But modeling the underlying physics remains a difficult problem. The workhorse approach to describing the DW–ZF coupling is the WKE (Diamond *et al.*, 2005; Trines *et al.*, 2005), but it is limited to the ray approximation (see Chapter 3) and, in fact, is oversimplified even as a GO model. That leads to missing essential physics, as was recently pointed out by Parker (2016) and will be elaborated below. These issues can be fixed by using the more accurate approach known as the second-order cumulant expansion, or CE2,[2] whose applications to DW–ZF physics were pursued by Farrell and Ioannou (2009), Parker and Krommes (2013), Parker and Krommes (2014), and Parker (2014). However, the CE2 is less intuitive than the WKE, and its robustness with respect to further approximations remains obscure. Having an approach as accurate as the CE2 and as intuitive as the WKE would be advantageous.

[1]See, for example, Gürcan and Diamond (2015), Vasavada and Showman (2005), Johansen *et al.* (2009), Kunz and Lesur (2013), Diamond *et al.* (2005), Fujisawa (2009), and Hillesheim *et al.* (2016).

In this Chapter, such an approach is presented for a DW turbulence model based on the generalized Hasegawa–Mima equation (Smolyakov and Diamond, 1999; Krommes and Kim, 2000). The idea is as follows. One starts by splitting the gHME into two coupled equations that describe ZFs and fluctuations, respectively, and then linearizing the equation for fluctuations, as in the CE2 approach. One then uses the techniques presented in Chapter 2 to formulate an *exact* (modulo the quasilinear approximation) Wigner–Moyal equation (Moyal, 1949; Groenewold, 1946) that describes incoherent DW fluctuations.[3]

Compared to the CE2, the Wigner–Moyal formulation is arguably more intuitive for two reasons: (i) like the traditional WKE (hereafter denoted by tWKE), it permits viewing DW quanta ("driftons") as particles,[4] except now driftons are *quantum-like* particles in phase space, i.e., have nonzero wavelengths; and (ii) the separation between Hamiltonian effects and dissipation remains transparent and unambiguous even beyond the GO approximation. Compared to the tWKE, the new approach is also more accurate because (i) it captures effects beyond the GO limit, and (ii) *even in the GO limit*, it predicts corrections to the tWKE that emerge from the newly found corrections to the drifton dispersion. [In this aspect, this approach can be understood as an expansion of the GO approximation introduced by Parker (2016).] These corrections are essential as they allow DW–ZF enstrophy exchange, which is not included in the tWKE. By deriving the GO limit from the generic principles introduced in Chapter 3, one eliminates this discrepancy and obtains a formulation that exactly conserves the total enstrophy (as opposed to the DW enstrophy conservation predicted by the tWKE) and the total energy, in precise agreement with the underlying gHME. I also illustrate the substantial difference between the GO limit of our formulation and the tWKE by using numerical simulations.

[2]For more information about the CE2 approach, see, for example, Farrell and Ioannou (2003), Farrell and Ioannou (2007), Marston *et al.* (2008), Srinivasan and Young (2012), and Ait-Chaalal *et al.* (2016).

[3]Related calculations were also proposed by Mendonça and Hizanidis (2011) and Mendonça and Benkadda (2012). However, the complete Wigner–Moyal equation for DW was not introduced explicitly in those papers, and the validity of its GO limit was not explored in detail. Here, I argue that such details are, in fact, crucial.

[4]Driftons should not be considered as real particles. They are only introduced in order to have a physical interpretation of the characteristic ray solutions of the resulting Wigner–Moyal and the WKE formulations.

### 9.1.2 Overview

The rest of this Chapter is organized as follows. In Sec. 9.2, I introduce the gHME and its quasilinear approximation. In Sec. 9.3, I derive the Wigner–Moyal formulation. In Sec. 9.4, I rederive the dispersion relation for the linear growth rate of ZFs. In Sec. 9.5, I derive a corrected WKE that, in contrast to the tWKE, conserves both the total enstrophy and energy. Numerical simulations are presented to compare the new WKE with the tWKE. In Sec. 9.6, I summarize the main results. Auxiliary calculations are included in Appendix C.4.

## 9.2 Basic model

The formulation is based on the generalized Hasegawa–Mima equation (Smolyakov and Diamond, 1999; Krommes and Kim, 2000),

$$\partial_t w + \mathbf{v} \cdot \boldsymbol{\nabla} w + \beta\, \partial_x \psi = Q, \tag{9.1}$$

which is widely used to describe electrostatic two-dimensional turbulent flows in both a magnetized plasma with a density gradient and in an atmospheric fluid on a rotating planet, where the role of DWs is played by Rossby waves (Gürcan and Diamond, 2015; Parker, 2014). Both contexts will be described on the same footing, so the results are applicable to DWs and Rossby waves equally. I assume the usual geophysical coordinate system, where $\mathbf{x} = (x, y)$ is a two-dimensional coordinate,[5] the $x$ axis is the ZF direction, and the $y$ axis is the direction of the local gradient of the plasma density or of the Coriolis parameter. (In the context of fusion plasmas, a different choice of coordinates is usually preferred in the literature, where $x$ and $y$ are swapped.) The constant $\beta$ is a measure of this gradient. The function $\psi(t, \mathbf{x})$ is the electric potential or the stream function, $\mathbf{v} = \mathbf{e}_z \times \boldsymbol{\nabla}\psi$ is the fluid velocity on the $\mathbf{x}$ plane, and $\mathbf{e}_z$ is a unit vector normal to this plane. The function $w(t, \mathbf{x})$ is the generalized vorticity given by $w \doteq (\nabla^2 - L_{\mathrm{D}}^{-2}\widehat{\alpha})\psi$, where $\nabla^2 \doteq \partial_x^2 + \partial_y^2$ is the perpendicular Laplacian, $\widehat{\alpha}$ is an operator such that $\widehat{\alpha} = 1$ in the parts of the spectrum corresponding to DWs and $\widehat{\alpha} = 0$ in those corresponding to ZFs. Also, $L_{\mathrm{D}}$ is the plasma sound radius or the deformation radius. [For plasmas, one can take $L_{\mathrm{D}} = 1$ in normalized units (Krommes and Kim, 2000). Also, the barotropic model used in geophysics is recovered in the limit $L_{\mathrm{D}} \to \infty$ (Farrell and Ioannou, 2003, 2007; Marston *et al.*, 2008; Srinivasan and Young, 2012).] The term $Q(t, \mathbf{x})$ describes external forces and dissipation. Systems with $Q = 0$ will be called isolated.

[5] Unlike the preceding Chapters of this thesis, here the spacetime four-vector notation is not used. Hence, $x$ is not the four-position but rather the $x$-axis coordinate.

Let us decompose the fields into their zonal-averaged and fluctuating components, denoted with bars and tildes, respectively. [For any function $g(t, \mathbf{x})$, its zonal average is $\bar{g}(y) \doteq \int \mathrm{d}x\, g/L_x$, where $L_x$, henceforth assumed equal to one, is the system length along the $x$ axis.] In particular, $w(t, \mathbf{x}) = \bar{w}(t, y) + \widetilde{w}(t, \mathbf{x})$, where the two components of the generalized vorticity are related to $\psi(t, \mathbf{x})$ as follows (Parker and Krommes, 2013):

$$\bar{w}(t, y) = \nabla^2 \bar{\psi} = \partial_y^2 \bar{\psi}, \qquad \widetilde{w}(t, \mathbf{x}) = \nabla_{\mathrm{D}}^2 \widetilde{\psi}, \tag{9.2}$$

and $\nabla_{\mathrm{D}}^2 \doteq \nabla^2 - L_{\mathrm{D}}^{-2}$. Equations for $\widetilde{w}(t, \mathbf{x})$ and $\bar{w}(t, y)$ are obtained by taking the zonal average and fluctuating parts of Eq. (9.1). This gives

$$\partial_t \widetilde{w} + \widetilde{\mathbf{v}} \cdot \boldsymbol{\nabla} \bar{w} + \bar{\mathbf{v}} \cdot \boldsymbol{\nabla} \widetilde{w} + \beta\, \partial_x \widetilde{\psi} + f_{\mathrm{NL}} = \widetilde{Q}, \tag{9.3a}$$

$$\partial_t \bar{w} + \overline{\widetilde{\mathbf{v}} \cdot \boldsymbol{\nabla} \widetilde{w}} = \bar{Q}, \tag{9.3b}$$

where $f_{\mathrm{NL}}(t, \mathbf{x}) \doteq \widetilde{\mathbf{v}} \cdot \boldsymbol{\nabla} \widetilde{w} - \overline{\widetilde{\mathbf{v}} \cdot \boldsymbol{\nabla} \widetilde{w}}$ is a term nonlinear with respect to fluctuations. As discussed in Srinivasan and Young (2012), this term represents eddy–eddy interactions and is responsible for the Batchelor–Kraichnan inverse energy cascade; however, it is inessential for the formation of ZFs. Since the main scope of this paper is to specifically study the interaction between eddies and ZFs, I ignore eddy–eddy interactions so $f_{\mathrm{NL}}$ will be neglected. Hence,

$$\partial_t \widetilde{w} + \widetilde{\mathbf{v}} \cdot \boldsymbol{\nabla} \bar{w} + \bar{\mathbf{v}} \cdot \boldsymbol{\nabla} \widetilde{w} + \beta\, \partial_x \widetilde{\psi} = \widetilde{Q}, \tag{9.4a}$$

$$\partial_t \bar{w} + \overline{\widetilde{\mathbf{v}} \cdot \boldsymbol{\nabla} \widetilde{w}} = \bar{Q}. \tag{9.4b}$$

Equations (9.4) compose the well-known quasilinear model (Farrell and Ioannou, 2003). In isolated systems, both sets of equations conserve the generalized enstrophy $\mathcal{Z}(t)$ and the energy$\mathcal{E}(t)$ (strictly speaking, *free* energy):

$$\mathcal{Z}(t) \doteq \frac{1}{2} \int \mathrm{d}^2 x\, w^2, \qquad \mathcal{E}(t) \doteq -\frac{1}{2} \int \mathrm{d}^2 x\, w\psi. \tag{9.5}$$

It is convenient to rewrite Eqs. (9.4) in terms of the ZF velocity $\bar{\mathbf{v}}(t, y) = \mathbf{e}_x U(t, y)$, whose only component is $U(t, y) = -\partial_y \bar{\psi}(t, y)$. Specifically, one has $\widetilde{\mathbf{v}} \cdot \boldsymbol{\nabla} \bar{w} = -(\partial_x \widetilde{\psi})(\partial_y^2 U)$, $\bar{\mathbf{v}} \cdot \boldsymbol{\nabla} \widetilde{w} = U \partial_x \widetilde{w}$, and $\overline{\widetilde{\mathbf{v}} \cdot \boldsymbol{\nabla} \widetilde{w}} = -\partial_y^2\, \overline{\widetilde{v}_x \widetilde{v}_y}$. I shall also assume $\widetilde{Q}(t, \mathbf{x}) = \widetilde{\xi}(t, \mathbf{x}) - \mu_{\mathrm{dw}} \widetilde{w}(t, \mathbf{x})$ and $\bar{Q}(t, y) = -\mu_{\mathrm{zf}} \bar{w}(t, y)$. Here, $\widetilde{\xi}(t, \mathbf{x})$ is some external force with zero zonal average (eventually, I shall assume it to be a white noise), and the constant coefficients $\mu_{\mathrm{dw}}$ and $\mu_{\rm zf}$ are intended to emulate the dissipation of DWs and ZFs caused by the external environment.[6] Then, Eqs. (9.4) become

$$\partial_t \widetilde{w} + U\partial_x \widetilde{w} + [\beta - (\partial_y^2 U)]\partial_x \widetilde{\psi} = \widetilde{\xi} - \mu_{\rm dw}\widetilde{w}, \tag{9.6a}$$

$$\partial_t U + \mu_{\rm zf} U + \partial_y \overline{\widetilde{v}_x \widetilde{v}_y} = 0. \tag{9.6b}$$

Equations (9.6) are the same model as the one that underlies the CE2. Although not exact, this model is useful because it captures key aspects of ZF dynamics, such as the formation and merging of zonal jets (Srinivasan and Young, 2012; Parker, 2014; Constantinou *et al.*, 2014). In the following, I derive a phase-space formulation to describe DW–ZF interactions.

## 9.3 Wigner–Moyal formulation

### 9.3.1 Abstract representation

As in previous Chapters of this thesis, it is convenient to relate the fluctuating vorticity field $\widetilde{w}(\mathbf{x},t)$ to an abstract state vector $|\,\widetilde{w}(t)\,\rangle$. Since the field $\widetilde{w}(\mathbf{x},t)$ only depends on two spatial coordinates, I choose the state vector $|\,\widetilde{w}(t)\,\rangle$ to belong to a Hilbert space with inner product[7]

$$\langle\, \Upsilon(t) \,|\, \Psi(t)\,\rangle = \int \mathrm{d}^2x\, \Upsilon^\dagger(t,\mathbf{x})\Psi^\dagger(t,\mathbf{x}). \tag{9.7}$$

Analogous definitions will be assumed also for $|\,\widetilde{\psi}(t)\,\rangle$ and $|\,\widetilde{\xi}(t)\,\rangle$. The original function $\widetilde{w}(\mathbf{x},t)$ is then understood as a projection of $|\,\widetilde{w}(t)\,\rangle$, namely, as its "coordinate representation" given by $\widetilde{w}(t,\mathbf{x}) = \langle\, \mathbf{x} \,|\, \widetilde{w}(t)\,\rangle$. Here, $|\,\mathbf{x}\,\rangle$ are the eigenstates of the position operator $\widehat{\mathbf{x}}$ normalized such that $\langle\, \mathbf{x}' \,|\, \widehat{\mathbf{x}} \,|\, \mathbf{x}\,\rangle = \mathbf{x}\,\langle\, \mathbf{x}' \,|\, \mathbf{x}\,\rangle = \mathbf{x}\,\delta^2(\mathbf{x}'-\mathbf{x})$. This procedure of representing the fluctuating fields to abstract vectors of a Hilbert space is equivalent to what has done in previous Chapters of this thesis. The only peculiarity of this problem in this respect is that $\tilde{w}(\mathbf{x},t)$ is real rather than complex. However, that is just a matter of initial conditions.

[6] There are various physical interpretations to the external forcing $\widetilde{\xi}(t,\mathbf{x})$ in the literature. In stochastic structural stability theories (SSST or S3T), $\widetilde{\xi}(t,\mathbf{x})$ is taken to represent the missing eddy–eddy interactions that were dropped in the quasilinear model (Farrell and Ioannou, 2003, 2009). In theories based on the CE2, the forcing represents the missing effects of extrinsic fluctuations that are not described by the gHME, such as forcing from other waves modes in the medium. More recently, St-Onge and Krommes (2017) attribute the external forcing to discrete particle noise of a weakly coupled many-body plasma. This last interpretation has led to a physically consistent model that describes the spontaneous excitation of ZFs in ion-temperature-gradient (ITG) turbulence in tokamaks.

[7] Note that this inner product differs from that introduced in Sec. 2.2.2; here the wave fields are only integrated over the $x$ and $y$ coordinates instead of the whole spacetime.

In addition to the coordinate operator $\widehat{\mathbf{x}}$, I introduce a momentum (wave-vector) operator $\widehat{\mathbf{p}}$ such that, in the coordinate representation,[8]

$$\langle\, \mathbf{x} \mid \widehat{\mathbf{p}} \mid \mathbf{x}' \,\rangle = -i \frac{\partial}{\partial \mathbf{x}} \delta^2(\mathbf{x} - \mathbf{x}'). \tag{9.8}$$

Accordingly, the relation between the fluctuating vorticity state vector and the fluctuating electric potential is $|\, \widetilde{w}(t) \,\rangle = -\widehat{p}_{\mathrm{D}}^2 \,|\, \widetilde{\psi}(t) \,\rangle$, where

$$\widehat{p}_{\mathrm{D}}^2 \doteq \widehat{p}^2 + L_{\mathrm{D}}^{-2}, \qquad \widehat{p}^2 \doteq \widehat{\mathbf{p}} \cdot \widehat{\mathbf{p}}. \tag{9.9}$$

Hence, Eq. (9.6a) can be represented as

$$i\partial_t \,|\, \widetilde{w} \,\rangle = \widehat{\mathcal{H}} \,|\, \widetilde{w} \,\rangle + i \,|\, \widetilde{\xi} \,\rangle\,, \tag{9.10}$$

where the non-Hermitian operator $\widehat{\mathcal{H}}$ is given by

$$\widehat{\mathcal{H}} \doteq -\beta \widehat{p}_x \widehat{p}_{\mathrm{D}}^{-2} + \widehat{U} \widehat{p}_x + \widehat{U}'' \widehat{p}_x \widehat{p}_{\mathrm{D}}^{-2} - i\mu_{\mathrm{dw}}. \tag{9.11}$$

Also, $\widehat{U} \doteq U(\widehat{y}, t)$, and the prime above $U$ henceforth denotes $\partial_y$; e.g., $\widehat{U}'' \doteq \partial_y^2 U(\widehat{y}, t)$. All of the state vectors are evaluated at time $t$, so I shall avoid writing the time argument inside the kets.

### 9.3.2 Generalized von Neumann equation

Let us express the operator $\widehat{\mathcal{H}}$ in Eq. (9.11) as $\widehat{\mathcal{H}} = \widehat{\mathcal{H}}_H + i\widehat{\mathcal{H}}_A$, where $\widehat{\mathcal{H}}_H \doteq (\widehat{\mathcal{H}} + \widehat{\mathcal{H}}^\dagger)/2$ and $\widehat{\mathcal{H}}_A \doteq (\widehat{\mathcal{H}} - \widehat{\mathcal{H}}^\dagger)/(2i)$ are the Hermitian and anti-Hermitian parts of $\widehat{\mathcal{H}}$, correspondingly. Explicitly, these operators are written as

$$\widehat{\mathcal{H}}_H = -\beta \widehat{p}_x \widehat{p}_{\mathrm{D}}^{-2} + \widehat{U} \widehat{p}_x + [\widehat{U}'', \widehat{p}_x \widehat{p}_{\mathrm{D}}^{-2}]_+/2, \tag{9.12a}$$

$$\widehat{\mathcal{H}}_A = [\widehat{U}'', \widehat{p}_x \widehat{p}_{\mathrm{D}}^{-2}]_-/(2i) - \mu_{\mathrm{dw}}, \tag{9.12b}$$

where $[\,\cdot\,,\,\cdot\,]_-$ denotes the commutator given by $[\widehat{A}, \widehat{B}]_- = \widehat{A}\widehat{B} - \widehat{B}\widehat{A}$ and $[\,\cdot\,,\,\cdot\,]_+$ denotes the anti-commutator given by $[\widehat{A}, \widehat{B}]_+ = \widehat{A}\widehat{B} + \widehat{B}\widehat{A}$. Let us also introduce a Hermitian operator $\widehat{W}(t) \doteq |\, \widetilde{w}(t) \,\rangle \langle\, \widetilde{w}(t) \,|$, which by analogy with quantum mechanics, is interpreted as the "fluctuating-vorticity density" operator. Using

[8] To avoid confusion with the definition given in Eq. (2.16), here I refer to the contravariant spatial component $\widehat{p}^i$ of the four-momentum operator $\widehat{p}_\mu$.

Eq. (9.10), one can easily show that $\widehat{W}(t)$ satisfies

$$i\partial_t \widehat{W} = [\widehat{\mathcal{H}}_H, \widehat{W}]_- + i[\widehat{\mathcal{H}}_A, \widehat{W}]_+ + i\widehat{\mathcal{F}}, \tag{9.13}$$

where $\widehat{\mathcal{F}} \doteq |\,\widetilde{\xi}(t)\,\rangle\,\langle\,\widetilde{w}(t)\,| + |\,\widetilde{w}(t)\,\rangle\,\langle\,\widetilde{\xi}(t)\,|$. In particular, taking the trace of this equation also gives an equation for the *total number of DW quanta*, $N \doteq \mathrm{Tr}_{\mathbf{x}}\widehat{W} = \int \mathrm{d}^2x\,\langle\,\mathbf{x}\,|\,\widehat{W}\,|\,\mathbf{x}\,\rangle = \int \mathrm{d}^2x\,\widetilde{w}^2 = \langle\,\widetilde{w}\,|\,\widetilde{w}\,\rangle$:

$$\dot{N}(t) = 2\mathrm{Tr}_{\mathbf{x}}(\widehat{\mathcal{H}}_A\widehat{W}) + \mathrm{Tr}_{\mathbf{x}}\widehat{\mathcal{F}}. \tag{9.14}$$

This indicates that $\widehat{\mathcal{H}}_A$ determines the loss of quanta, or dissipation of DWs. [In particular, the term $\mu_{\mathrm{dw}}$ in Eq. (9.12b) is responsible for DW dissipation to the external environment, whereas the term $[\widehat{U}'', \widehat{p}_x\widehat{p}_{\mathrm{D}}^{-2}]_-/(2i)$ destroys DW quanta while conserving the generalized enstrophy, as will be discussed in Sec. 9.3.5.] Also, $\widehat{\mathcal{H}}_H$ determines conservative dynamics of DWs and thus can be understood as the *drifton Hamiltonian.* (The non-Hermitian operator $\widehat{\mathcal{H}}$ will be attributed as the generalized Hamiltonian.) Notice that the distinction between dissipation and Hamiltonian effects remains unambiguous even beyond the GO approximation.

Equation (9.13) can be understood as a generalized von Neumann equation akin to the one that commonly emerges in quantum mechanics. As in the previous Chapters, I shall project this equation to the phase space using the Weyl transform. Since the inner product (9.7) is slightly different that the inner product in Sec. 2.2.2, here the Weyl transform of an arbitrary operator $\widehat{\mathcal{A}}$ is defined as

$$A(t, \mathbf{x}, \mathbf{p}) \doteq \int \mathrm{d}^2s\, e^{-i\mathbf{p}\cdot\mathbf{s}}\,\langle\,\mathbf{x} + \mathbf{s}/2\,|\,\widehat{\mathcal{A}}\,|\,\mathbf{x} - \mathbf{s}/2\,\rangle\,. \tag{9.15}$$

With this version of the Weyl transform, some of its properties in Appendix A are slightly modified, but they remain qualitatively the same.[9]

### 9.3.3 Wigner–Moyal equation

Let us introduce the Wigner function $W(t, \mathbf{x}, \mathbf{p})$ corresponding to $\widehat{W}$, i.e.,

$$W(t, \mathbf{x}, \mathbf{p}) \doteq \frac{1}{(2\pi)^2}\int \mathrm{d}^2s\, e^{-i\mathbf{p}\cdot\mathbf{s}}\,\langle\,\mathbf{x} + \mathbf{s}/2\,|\,\widehat{W}(t)\,|\,\mathbf{x} - \mathbf{s}/2\,\rangle\,, \tag{9.16}$$

which is real because $\widehat{W}$ is Hermitian. As I discussed in a broader context (see Chapters 2 and 3), in the regime when the ray approximation applies and dissipation is negligible, $W(t, \mathbf{x}, \mathbf{p})$ represents the phase-space probability density of driftons, while beyond the GO limit it can be considered as a *generalization* of this probability density. Since $\widetilde{w}(t,\mathbf{x})$ is real, one can also express $W(t,\mathbf{x},\mathbf{p})$ as

$$W(t,\mathbf{x},\mathbf{p}) \doteq \frac{1}{(2\pi)^2} \int \mathrm{d}^2 s\, e^{-i\mathbf{p}\cdot\mathbf{s}}\, \widetilde{w}(t,\mathbf{x}+\mathbf{s}/2)\widetilde{w}(t,\mathbf{x}-\mathbf{s}/2), \tag{9.17}$$

which also implies

$$W(t,\mathbf{x},\mathbf{p}) = W(t,\mathbf{x},-\mathbf{p}). \tag{9.18}$$

[9]The calculations presented in this Chapter can also be obtained by using the eight-dimensional phase-space formalism of the preceding Chapters. However, the calculations are more complicated and not much insight is gained.

One can also interpret $W(t,\mathbf{x},\mathbf{p})$ as the local spatial spectrum of the correlation function of $\widetilde{w}(t,\mathbf{x})$. Hence, $W(t,\mathbf{x},\mathbf{p})$ will also be called the *DW spectral function*.

Applying the Weyl transform (9.15) to Eq. (9.13) leads to the following equation:

$$\frac{\partial}{\partial t} W = \{\{H_H, W\}\} + [[H_A, W]] + \frac{1}{(2\pi)^2} F, \tag{9.19}$$

where the functions $H_H(t,y,\mathbf{p})$, $H_A(t,y,\mathbf{p})$, and $F(t,\mathbf{x},\mathbf{p})$ are the Weyl symbols of $\widehat{\mathcal{H}}_H$, $\widehat{\mathcal{H}}_A$, and $\widehat{\mathcal{F}}$, respectively. Also, $\{\{\cdot\,,\cdot\}\}$ and $[[\cdot\,,\cdot]]$ are the Moyal "sine" and "cosine" brackets as defined in Appendix A.[10] Upon using the Moyal product (A.5) and the fact that $U(t,y)$ is independent of $x$, one obtains the following expressions for $H_H(t,y,\mathbf{p})$ and $H_A(t,y,\mathbf{p})$:

$$H_H(t,y,\mathbf{p}) = -\beta p_x/p_{\mathrm{D}}^2 + p_x U + [[U'', p_x/p_{\mathrm{D}}^2]]/2, \tag{9.20}$$

$$H_A(t,y,\mathbf{p}) = \{\{U'', p_x/p_{\mathrm{D}}^2\}\}/2 - \mu_{\mathrm{dw}}, \tag{9.21}$$

where $p_{\mathrm{D}}^2 \doteq p^2 + L_{\mathrm{D}}^{-2}$. By analogy with quantum mechanics, Eq. (9.19) can be interpreted as a Wigner–Moyal equation for the DW spectral function.

Next, let us consider the zonal average of this equation,

$$\frac{\partial}{\partial t} \overline{W} = \{\{H_H, \overline{W}\}\} + [[H_A, \overline{W}]] + \frac{1}{(2\pi)^2} \overline{F}, \tag{9.22}$$

where $\overline{W} = \overline{W}(t,y,\mathbf{p})$. I adopt the ergodic assumption, namely, that the zonal average is equivalent to the ensemble average [denoted $\langle\langle \ldots \rangle\rangle$] over realizations of the random force $\widetilde{\xi}$ [e.g., as done by Parker and Krommes (2013)]. To calculate $\overline{F} = \langle\langle F \rangle\rangle$, let us consider integrating Eq. (9.10) on a time interval $(t_0, t)$. The result can be written as $|\,\widetilde{w}(t)\,\rangle = |\,\widetilde{w}(t_0)\,\rangle + |\,\delta\widetilde{w}(t)\,\rangle + \int_{t_0}^{t} \mathrm{d}t'\,|\,\widetilde{\xi}(t')\,\rangle$, where $|\,\delta\widetilde{w}(t)\,\rangle \doteq -i\int_{t_0}^{t} \mathrm{d}t'\widehat{\mathcal{H}}\,|\,\widetilde{w}(t')\,\rangle$. Let us assume the following correlation for the random force $\widetilde{\xi}(t,\mathbf{x})$:

$$\langle\!\langle \widetilde{\xi}(t,\mathbf{x})\widetilde{\xi}(t',\mathbf{x}')\rangle\!\rangle = \delta(t-t')\;\Xi\big((y+y')/2, \mathbf{x}-\mathbf{x}'\big), \tag{9.23}$$

where $\Xi$ is a correlation function that is homogeneous in $x$ but not necessarily in $y$ (Srinivasan and Young, 2012; Parker, 2014). Since $|\,\delta\widetilde{w}(t)\,\rangle$ can be affected by $|\,\widetilde{\xi}(t')\,\rangle$ only if $t > t'$, one has $\langle\!\langle |\,\widetilde{\xi}(t)\,\rangle\langle\,\delta\widetilde{w}(t)\,|\rangle\!\rangle = 0$. Hence,

$$\begin{aligned}
\overline{F}(y,\mathbf{p}) &= \int \mathrm{d}^2 s\, e^{-i\mathbf{p}\cdot\mathbf{s}} \langle\!\langle \langle\,\mathbf{x}+\mathbf{s}/2\,|\,(|\,\tilde{\xi}(t)\,\rangle\langle\,\tilde{w}(t)\,| + \text{h.c.})\,|\,\mathbf{x}-\mathbf{s}/2\,\rangle\rangle\!\rangle \\
&= \int \mathrm{d}^2 s\, e^{-i\mathbf{p}\cdot\mathbf{s}} \int_{t_0}^{t} \mathrm{d}t'\,\delta(t-t')\,[\Xi(y,\mathbf{s}) + \Xi(y,-\mathbf{s})] \\
&= \frac{1}{2}\int \mathrm{d}^2 s\, e^{-i\mathbf{p}\cdot\mathbf{s}}\,[\Xi(y,\mathbf{s}) + \Xi(y,-\mathbf{s})] \\
&= \int \mathrm{d}^2 s\;\Xi(y,\mathbf{s})\cos(\mathbf{p}\cdot\mathbf{s}),
\end{aligned} \tag{9.24}$$

where "h.c." denotes "Hermitian conjugate." In other words, once the correlation function $\Xi$ of $\widetilde{\xi}$ is specified, $\overline{F}$ can be readily calculated as the Fourier image of $\Xi$.

[10] Note that the derivatives involving the temporal components in the Janus operator (A.8) cancel out since the Weyl symbols in this Chapter do not depend on the temporal momentum coordinate $p_0$. Similarly, the terms involving derivatives of $p_z$ are also zero.

This concludes the calculation of the functions that determine the evolution of $\overline{W}(t,y,\mathbf{p})$ through Eq. (9.22). However, these functions generally depend on $U(t,y)$, so an additional equation for $U(t,y)$ is needed to make the theory self-consistent. This equation is derived as follows.

### 9.3.4 Equation for the zonal-flow velocity

Returning to Eq. (9.6b), one can rewrite the nonlinear term as

$$\begin{aligned}
\widetilde{v}_x(t,\mathbf{x})\widetilde{v}_y(t,\mathbf{x}) &= -(\partial_y\widetilde{\psi})(\partial_x\widetilde{\psi}) \\
&= -\,\langle\,\mathbf{x}\,|\,\widehat{p}_y\,|\,\widetilde{\psi}(t)\,\rangle\langle\,\widetilde{\psi}(t)\,|\,\widehat{p}_x\,|\,\mathbf{x}\,\rangle \\
&= -\,\langle\,\mathbf{x}\,|\,\widehat{p}_y\widehat{p}_{\mathrm{D}}^{-2}\widehat{W}(t)\widehat{p}_{\mathrm{D}}^{-2}\widehat{p}_x\,|\,\mathbf{x}\,\rangle \\
&= -\int \mathrm{d}^2 p\,\frac{p_y}{p_{\mathrm{D}}^2}\star W\star\frac{p_x}{p_{\mathrm{D}}^2}.
\end{aligned} \tag{9.25}$$

Here I used a relation similar to property (A.14) so that in this case, for any operator $\widehat{\mathcal{A}}$, one has

$$\frac{1}{(2\pi)^2}\int \mathrm{d}^2 p\, A(t,\mathbf{x},\mathbf{p}) = \frac{1}{(2\pi)^2}\int \mathrm{d}^2 p\,\mathrm{d}^2 s\, e^{-i\mathbf{p}\cdot\mathbf{s}}\,\langle\,\mathbf{x}+\mathbf{s}/2\,|\,\widehat{\mathcal{A}}\,|\,\mathbf{x}-\mathbf{s}/2\,\rangle = \langle\,\mathbf{x}\,|\,\widehat{\mathcal{A}}\,|\,\mathbf{x}\,\rangle\,. \tag{9.26}$$

Also note that the $(2\pi)^{-2}$ factor is absorbed into the definition of the Wigner function. After one introduces the averaged vorticity density $\overline{W}$, Eq. (9.6b) becomes

$$\frac{\partial}{\partial t}U + \mu_{\text{zf}}U = \frac{\partial}{\partial y}\int \mathrm{d}^2p\ \frac{p_y}{p_{\text{D}}^2} \star \overline{W} \star \frac{p_x}{p_{\text{D}}^2}. \tag{9.27}$$

Since $\overline{W}(t, y, \mathbf{p})$ is independent of $x$ and satisfies the condition (9.18), Eq. (9.27) can also be written as

$$\frac{\partial}{\partial t}U + \mu_{\text{zf}}U = \frac{\partial}{\partial y}\int \mathrm{d}^2p\ \frac{1}{p_{\text{D}}^2} \star p_x p_y \overline{W} \star \frac{1}{p_{\text{D}}^2}. \tag{9.28}$$

The combination of Eqs. (9.22) and (9.28) forms a closed set of equations that can be used to calculate the dynamics of $\overline{W}(t, y, \mathbf{p})$ and $U(t, y)$ self-consistently.

### 9.3.5 Main equations and conservation laws

Let us slightly change the notation and summarize the above equations in the form

$$\frac{\partial}{\partial t}\overline{W} = \{\{\mathcal{H}, \overline{W}\}\} + [[\Gamma, \overline{W}]] + \frac{1}{(2\pi)^2}\overline{F} - 2\mu_{\text{dw}}\overline{W}, \tag{9.29a}$$

$$\frac{\partial}{\partial t}U + \mu_{\text{zf}}U = \frac{\partial}{\partial y}\int \mathrm{d}^2p\ \frac{1}{p_{\text{D}}^2} \star p_x p_y \overline{W} \star \frac{1}{p_{\text{D}}^2}. \tag{9.29b}$$

As a reminder, $\overline{W}(t, y, \mathbf{p})$ is the zonal-averaged spectral (or Wigner) function that describes DW turbulence, and $U(y, t)$ is the ZF velocity. Also, $\overline{F}(y, \mathbf{p})$ is determined by the correlation function of the external noise $\widetilde{\xi}$ (Sec. 9.3.3). I also introduced $\mathcal{H}(t, y, \mathbf{p}) \doteq H_H$ and $\Gamma(t, y, \mathbf{p}) \doteq H_A + \mu_{\text{dw}}$; or, explicitly,

$$\mathcal{H}(t, y, \mathbf{p}) = -\beta p_x/p_{\text{D}}^2 + p_x U + [[U'', p_x/p_{\text{D}}^2]]/2, \tag{9.30a}$$

$$\Gamma(t, y, \mathbf{p}) = \{\{U'', p_x/p_{\text{D}}^2\}\}/2. \tag{9.30b}$$

Ruiz *et al.* (2016) give a spectral representation of these equations that can be used for a numerical implementation of the Wigner–Moyal formulation.

The function $\mathcal{H}(t, y, \mathbf{p})$ can be understood as the Weyl symbol of the drifton Hamiltonian, whereas $\Gamma(t, y, \mathbf{p})$ determines the dissipation of DW quanta that is caused specifically by the DW interaction with ZFs. This is explained as follows. Since Eqs. (9.29) are *exact* within the quasilinear approximation (modulo the ergodic assumption), they inherit the conservation laws of the original quasilinear model given by Eqs. (9.6). Specifically, for isolated systems ($\overline{F} = 0$ and $\mu_{\text{dw,zf}} = 0$), Eqs. (9.29) and (9.30) exactly conserve the *total* generalized enstrophy and energy [Eqs. (9.5)]

$$\mathcal{Z} = \mathcal{Z}_{\rm dw} + \mathcal{Z}_{\rm zf}, \qquad \mathcal{E} = \mathcal{E}_{\rm dw} + \mathcal{E}_{\rm zf} \tag{9.31}$$

rather than their DW and ZF components. (A direct proof is given in Appendix C.4.) For completeness, the expressions for these components are given by

$$\mathcal{Z}_{\rm dw} \doteq \frac{1}{2}\int \mathrm{d}^2x\, \widetilde{w}^2 = \frac{1}{2}\int \mathrm{d}y\, \mathrm{d}^2p\, \overline{W}, \tag{9.32a}$$

$$\mathcal{Z}_{\rm zf} \doteq \frac{1}{2}\int \mathrm{d}y\, \bar{w}^2 = \frac{1}{2}\int \mathrm{d}y\, (U')^2, \tag{9.32b}$$

$$\mathcal{E}_{\rm dw} \doteq -\frac{1}{2}\int \mathrm{d}^2x\, \widetilde{w}\widetilde{\psi} = \frac{1}{2}\int \mathrm{d}y\, \mathrm{d}^2p\, \frac{\overline{W}}{p_{\rm D}^2}, \tag{9.32c}$$

$$\mathcal{E}_{\rm zf} \doteq -\frac{1}{2}\int \mathrm{d}y\, \bar{w}\bar{\psi} = \frac{1}{2}\int \mathrm{d}y\, U^2, \tag{9.32d}$$

where I integrated Eq. (9.26) over the spatial coordinates in order to derive the second set of equalities. According to Eqs. (9.26) and (9.32a), note that $\mathcal{Z}_{\rm dw}$ and the total number of DW quanta $N(t) \doteq \mathrm{Tr}_{\mathbf{x}}\, \widehat{W}$ are the same up to a constant factor.

The conservative equations (9.29) and (9.30), which I call the *Wigner–Moyal formulation* of DW–ZF interactions, constitute the main result of this Chapter. This formulation can be understood as an alternative phase-space representation of the CE2 since it is derived from the same quasilinear model. However, the Wigner–Moyal formulation is arguably more intuitive than the CE2 for two reasons: (i) Like in the tWKE, driftons are treated as particles, except now they are *quantum-like* particles, i.e., have nonzero wavelengths; hence, one is not constrained to the GO limit. (ii) Also, the separation between Hamiltonian effects and dissipation remains transparent and unambiguous even beyond the GO approximation. The Wigner–Moyal formulation elucidates the link between the WKE formalism and the CE2 and also helps make approximations rigorous by making them systematic. Below, these and other applications are discussed in further detail.

## 9.4 Growth rate of zonal flows

To demonstrate the convenience of the Wigner–Moyal formulation, let us apply it to rederive the rate of the linear zonostropic instability, i.e., the growth rate of weak ZFs. Postulate a homogeneous equilibrium with zero ZF velocity and some DW spectral function $\mathcal{W}(\mathbf{p})$. [As pointed out in Sec. 9.3.3, the corresponding $\mathcal{W}$ represents the phase-space probability distribution of driftons.] Consider small perturbations to this equilibrium; namely,

$$U(t,y) = \delta U(t,y), \qquad \delta U(t,y) = \mathrm{Re}\,\left(U_q e^{iqy+\gamma t}\right),$$
$$\overline{W}(t,y,\mathbf{p}) = \mathcal{W}(\mathbf{p}) + \delta\overline{W}(t,y,\mathbf{p}), \qquad \delta\overline{W}(t,y,\mathbf{p}) = \mathrm{Re}\big[\overline{W}_q(\mathbf{p}) e^{iqy+\gamma t}\big].$$

Here, the constant $q$ serves as the modulation wave number, and the constant $\gamma$ is the instability rate to be found. The linearization of Eq. (9.29a) leads to

$$(\partial_t + 2\mu_{\mathrm{dw}})\delta\overline{W} + \{\!\{\beta p_x/p_{\mathrm{D}}^2, \delta\overline{W}\,\}\!\}$$
$$= \{\!\{p_x \delta U, \mathcal{W}\}\!\} + \{\!\{[[\delta U'', p_x/p_{\mathrm{D}}^2]]/2, \mathcal{W}\}\!\} + [[\{\!\{\delta U'', p_x/p_{\mathrm{D}}^2\}\!\}/2, \mathcal{W}]], \tag{9.33}$$

where I substituted Eqs. (9.30). The brackets can be calculated using Eqs. (A.17) so that

$$\{\!\{A(\mathbf{p}), e^{i\mathbf{q}\cdot\mathbf{x}}\}\!\} = \frac{1}{i}\left[A\left(\mathbf{p}+\frac{\mathbf{q}}{2}\right) - A\left(\mathbf{p}-\frac{\mathbf{q}}{2}\right)\right] e^{i\mathbf{q}\cdot\mathbf{x}}, \tag{9.34a}$$
$$[[A(\mathbf{p}), e^{i\mathbf{q}\cdot\mathbf{x}}]] = \left[A\left(\mathbf{p}+\frac{\mathbf{q}}{2}\right) + A\left(\mathbf{p}-\frac{\mathbf{q}}{2}\right)\right] e^{i\mathbf{q}\cdot\mathbf{x}}. \tag{9.34b}$$

Hence, one obtains

$$\left[i(\gamma + 2\mu_{\mathrm{dw}}) + \beta p_x\left(\frac{1}{p_{\mathrm{D},+q}^2} - \frac{1}{p_{\mathrm{D},-q}^2}\right)\right]\overline{W}_q = (\mathcal{W}_{+q} - \mathcal{W}_{-q})$$
$$\times\left[\frac{p_x q^2}{2}\left(\frac{1}{p_{\mathrm{D},+q}^2} + \frac{1}{p_{\mathrm{D},-q}^2}\right) - p_x\right] U_q + \frac{p_x q^2}{2}(\mathcal{W}_{+q} + \mathcal{W}_{-q})\left(\frac{1}{p_{\mathrm{D},+q}^2} - \frac{1}{p_{\mathrm{D},-q}^2}\right) U_q, \tag{9.35}$$

where I used the notation $A_{\pm q}(\mathbf{p}) \doteq A(\mathbf{p} \pm \mathbf{e}_y q/2)$ for any function $A(\mathbf{p})$. Solving for $\overline{W}_q(\mathbf{p})$ in terms of $U_q$ leads to

$$\overline{W}_q(\mathbf{p}) = \frac{i p_x p_{\mathrm{D},+q}^2 p_{\mathrm{D},-q}^2}{(\gamma + 2\mu_{\mathrm{dw}}) p_{\mathrm{D},+q}^2 p_{\mathrm{D},-q}^2 + 2i\beta q p_x p_y}\left[\mathcal{W}_{+q}\left(1 - \frac{q^2}{p_{\mathrm{D},+q}^2}\right) - \mathcal{W}_{-q}\left(1 - \frac{q^2}{p_{\mathrm{D},-q}^2}\right)\right] U_q.$$

Then, Eq. (9.29b) yields

$$(\gamma + \mu_{\mathrm{zf}}) e^{iqy} U_q = \frac{\partial}{\partial y}\int \mathrm{d}^2 p\, \frac{1}{p_{\mathrm{D}}^2} \star p_x p_y e^{iqy} \overline{W}_q(\mathbf{p}) \star \frac{1}{p_{\mathrm{D}}^2}.$$

Upon using Eq. (A.16) so that $A(\mathbf{p}) \star e^{i\mathbf{q}\cdot\mathbf{x}} = A(\mathbf{p} + \mathbf{q}/2) e^{i\mathbf{q}\cdot\mathbf{x}}$, one can simplify the last equation to

$$(\gamma + \mu_{\mathrm{zf}}) U_q = iq \int \mathrm{d}^2 p\, \frac{p_x p_y}{p_{\mathrm{D},+q}^2 p_{\mathrm{D},-q}^2}\, \overline{W}_q(\mathbf{p}). \tag{9.36}$$

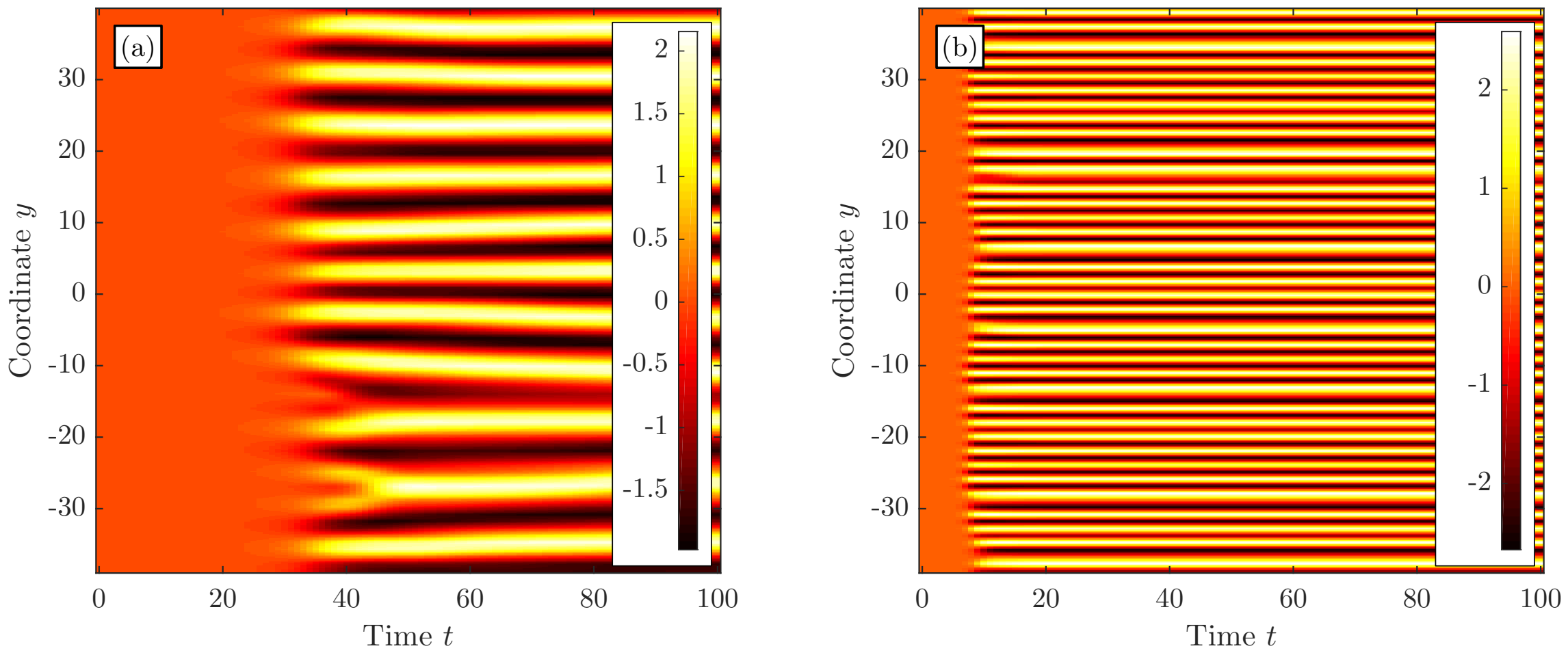

Figure 9.1: The ZF velocity $U(y,t)$ obtained by numerically integrating the WKE (9.43) for $\mathcal{H}$ and $\Gamma$ of two types: (a) the present model [Eqs. (9.44)]; (b) the tWKE model [Eqs. (9.45)]. Both simulations used the same parameters and initial conditions. Small initial values for $\overline{W}$ and $U$ were randomly assigned such that Eq. (9.18) was satisfied. The parameters used are: $\beta = 1$, $L_{\mathrm{D}} = 1$, $\mu_{\mathrm{dw,zf}} = 0.1$, and $\overline{F} = 4\pi\delta(|\mathbf{p}| - 1)$. Equation (9.43a) was discretized in a $[-39, 39] \times [-2, 2] \times [-4, 4]$ phase space using a discontinuous-Galerkin (DG) method (Liu and Shu, 2000) on a uniformly-spaced Cartesian grid with $80 \times 24 \times 48$ cells while Eq. (9.43b) was discretized on a subset of this grid. Time advancement was done using an explicit third-order strong-stability-preserving Runge–Kutta algorithm (Gottlieb *et al.*, 2001). The solution was expanded locally in each cell as a sum of piecewise polynomials of degree one. At cell interfaces, an upwind numerical flux was used in Eq. (9.43a) and a centered numerical flux was used in Eq. (9.43b). Higher-order spatial derivatives such as $U''$ and $U'''$ were computed using the Recovery-based DG method (van Leer *et al.*, 2007). For numerical stability, small hyperviscosity was added into the simulations, e.g., as done by Parker and Krommes (2014). Specifically, the terms $-2\nu(p_x^2 + p_y^2)\overline{W} + (\nu/2)\partial_y^2 \overline{W}$ and $-\nu\partial_y^4 U$ with $\nu = 0.001$ were added to the right-hand side of Eqs. (9.43a) and (9.43b), respectively.

After substituting the expression for $\overline{W}_q$, one gets

$$\gamma + \mu_{\mathrm{zf}} = \int \mathrm{d}^2 p \, \frac{q p_x^2 p_y}{(\gamma + 2\mu_{\mathrm{dw}}) p_{\mathrm{D},+q}^2 p_{\mathrm{D},-q}^2 + 2i\beta q p_x p_y} \left[ \mathcal{W}_{-q} \left( 1 - \frac{q^2}{p_{\mathrm{D},-q}^2} \right) - \mathcal{W}_{+q} \left( 1 - \frac{q^2}{p_{\mathrm{D},+q}^2} \right) \right]. \quad (9.37)$$

As expected, this dispersion relation coincides with that obtained using the CE2 formalism (Srinivasan and Young, 2012). Notably, the nonlocal dependence of the integrand on $\mathcal{W}_{\pm q} = \mathcal{W}(\mathbf{p} \pm \mathbf{e}_y q/2)$ makes the expression similar to dispersion relations that emerge in quantum mechanics; for instance, cf. Lifshitz and Pitaevskii (1981, Sec. 40).

## 9.5 Geometrical-optics limit and the wave kinetic equation

In this Section, I shall follow the procedure discussed in Chapter 3 in order to obtain the GO limit of the Wigner–Moyal formulation [Eqs. (9.29) and (9.30)]. I shall show that the governing equation for the drifton dynamics in the GO limit is the wave kinetic equation (WKE). As in Sec. 3.4, let the characteristic wavelengths for ZFs and DWs be given by $\lambda_{\text{zf}}$ and $\lambda_{\text{dw}}$, respectively, and

$$\epsilon \doteq \max\left\{\frac{\lambda_{\text{dw}}}{\lambda_{\text{zf}}}, \frac{L_{\text{D}}}{\lambda_{\text{zf}}}\right\} \ll 1. \tag{9.38}$$

The following estimates will be adopted:

$$\begin{aligned} \partial_y \overline{W} &\backsim \lambda_{\text{zf}}^{-1}\overline{W}, \qquad & \partial_{\mathbf{p}} \overline{W} &\backsim \lambda_{\text{dw}}\overline{W}, \\ \partial_y H &\backsim \lambda_{\text{zf}}^{-1} H, \qquad & \partial_{\mathbf{p}} H &\backsim L_{\text{D}} H, \end{aligned} \tag{9.39}$$

where $H$ denotes both $\mathcal{H}$ and $\Gamma$. The latter estimate is given for the *maximum* of $\partial_{\mathbf{p}} H$, which is realized at $p \sim L_D^{-1}$.[11] This gives

$$\frac{\partial^n H}{\partial y^n}\frac{\partial^n \overline{W}}{\partial p_y^n} \backsim \left(\frac{\lambda_{\text{dw}}}{\lambda_{\text{zf}}}\right)^n H\overline{W} \lesssim \epsilon^n H\overline{W}, \tag{9.40}$$

$$\frac{\partial^n H}{\partial p_y^n}\frac{\partial^n \overline{W}}{\partial y^n} \backsim \left(\frac{L_{\text{D}}}{\lambda_{\text{zf}}}\right)^n H\overline{W} \lesssim \epsilon^n H\overline{W}. \tag{9.41}$$

The lowest-order approximations of the Moyal brackets are

$$\{\{A, B\}\} \simeq \{A, B\}, \qquad [[A, B]] \simeq 2AB, \tag{9.42}$$

where $\{\,\cdot\,,\,\cdot\,\}$ denotes the canonical Poisson bracket so that $\{A, B\} = (\partial_{\mathbf{x}} A)\cdot(\partial_{\mathbf{p}} B) - (\partial_{\mathbf{p}} A)\cdot(\partial_{\mathbf{x}} B)$ is the canonical Poisson bracket. Hence, to the lowest-order approximation, Eqs. (9.29) reduce to

$$\frac{\partial}{\partial t}\overline{W} = \{\mathcal{H}, \overline{W}\} + 2\Gamma\overline{W} + \frac{1}{(2\pi)^2}\overline{F} - 2\mu_{\text{dw}}\overline{W}, \tag{9.43a}$$

$$\frac{\partial}{\partial t}U + \mu_{\text{zf}}\, U = \frac{\partial}{\partial y}\int \mathrm{d}^2 p\, \frac{p_x p_y \overline{W}}{p_{\text{D}}^4}, \tag{9.43b}$$

where

$$\mathcal{H}(t, y, \mathbf{p}) \simeq -\beta p_x/p_{\text{D}}^2 + p_x U + p_x U''/p_{\text{D}}^2, \tag{9.44a}$$

$$\Gamma(t, y, \mathbf{p}) \simeq \{U'', p_x/p_{\text{D}}^2\}/2 = -p_x p_y U'''/p_{\text{D}}^4. \tag{9.44b}$$

One may recognize Eq. (9.43a) as a variation of the WKE, so one can consider Eqs. (9.43) and (9.44) to be the *WKE limit* of the Wigner–Moyal formulation. Clearly, $\mathcal{H}(t, y, \mathbf{p})$ acts as the drifton ray Hamiltonian while $\Gamma(t, y, \mathbf{p})$ acts as the corresponding dissipation rate. [The factors of two in Eq. (9.43a) are due to the fact that $\overline{W}$ is quadratic in the DW amplitude.] In other words, $\omega(t, y, \mathbf{p}) \doteq \mathcal{H} + i\Gamma - i\mu_{\text{dw}}$ can be viewed as the local complex frequency of DWs with a given wave vector $\mathbf{p}$.

[11] In the barotropic limit ($L_{\text{D}} \to \infty$), the ordering (9.38) can no longer be satisfied. This means that a GO approximation of Eqs. (9.29) and (9.30) is not possible in this limit. This problem originates from the fact that the Hamiltonian is divergent at $p = 0$. Only if $\overline{W} = 0$ for $p < p_0$, one can estimate $\partial_{\mathbf{p}} H \lesssim H/p_0$, and the GO approximation applies at $(\lambda_{\text{zf}} p_0)^{-1} \ll 1$.

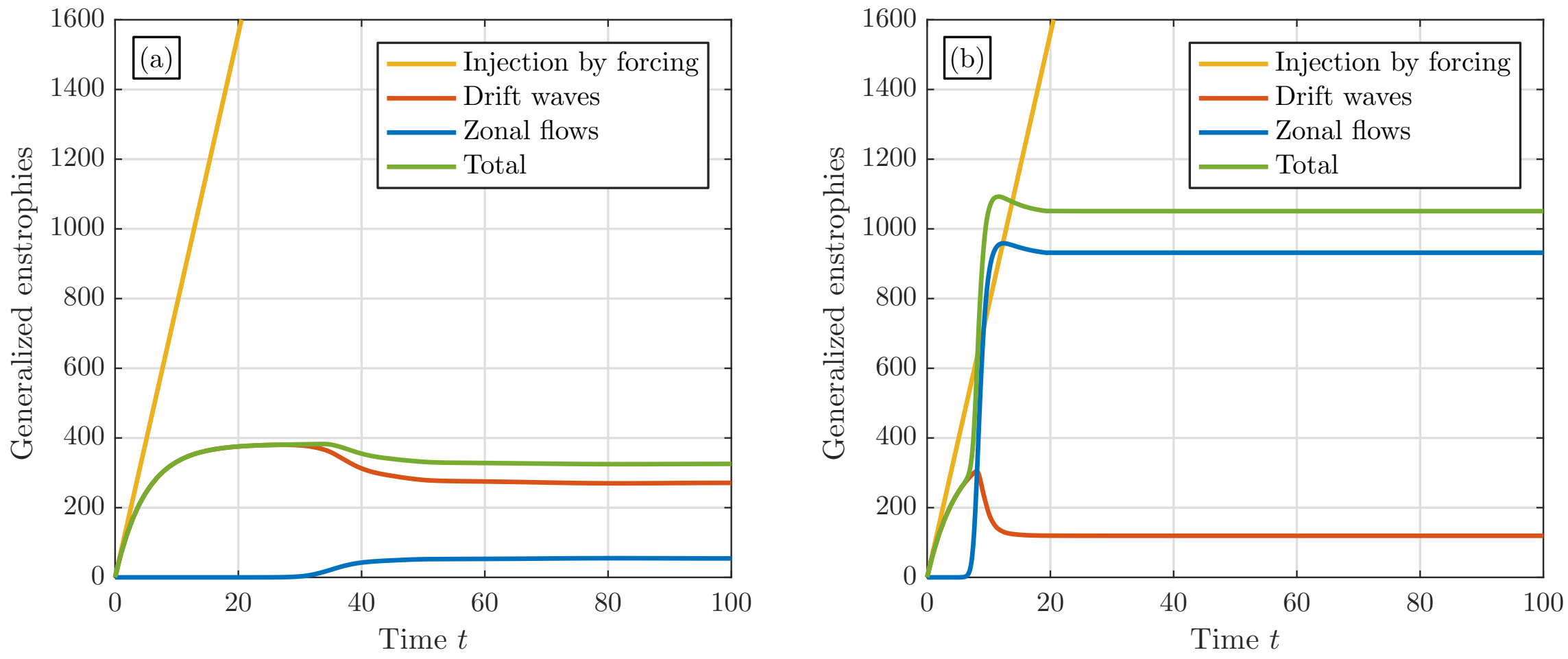

Figure 9.2: The total, DW, and ZF generalized enstrophies obtained by numerically integrating the WKE (9.43) for $\mathcal{H}$ and $\Gamma$ of two types: (a) the present model [Eqs. (9.44)]; (b) the tWKE model [Eqs. (9.45)]. The yellow lines show the total generalized enstrophy injected into the system by the external forcing $\overline{F}$. The initial conditions and simulation parameters are the same as in Fig. 9.1.

The present WKE differs from the tWKE, which assumes a simpler dispersion of DWs:

$$\mathcal{H}(t, y, \mathbf{p}) = -\beta p_x / p_{\text{D}}^2 + p_x U, \tag{9.45a}$$

$$\Gamma(t, y, \mathbf{p}) = 0. \tag{9.45b}$$

Although the difference is only in the higher-order derivatives of $U$, these terms remain important for various reasons. For example, in the Hamiltonian $\mathcal{H}$, $U''$ can be comparable to $\beta$ [as is sometimes the case in geophysics (Vasavada and Showman, 2005)]. This affects the topology of the ray phase space and the stability of ray trajectories, as will be reported in future work. Also, consider the following. In isolated systems, the tWKE is $\partial_t \overline{W} = \{\mathcal{H}, \overline{W}\}$, so it conserves DW quanta, or, in other words, the DW generalized enstrophy $\mathcal{Z}_{\text{dw}}$ [Eq. (9.32a)]. At the same time, $\mathcal{Z}_{\text{zf}}$ [Eq. (9.32b)] generally evolves, so the total generalized enstrophy $\mathcal{Z} = \mathcal{Z}_{\text{dw}} + \mathcal{Z}_{\text{zf}}$ does too. This is in contradiction with the gHME, which conserves $\mathcal{Z}$, and can lead to overestimating the ZF velocity and shear generated by DW turbulence.[12] In contrast to the tWKE, the present formulation is free from such issues because Eqs. (9.43) and (9.44) exactly conserve both $\mathcal{Z}(t)$ and $\mathcal{E}(t)$ (Appendix C.4). Note that, in order to retain this conservation property, it is necessary to keep both $U'''$ and $U''$ in Eqs. (9.44). In this sense, *Eqs. (9.43) and (9.44) represent the simplest GO model that is physically meaningful in the nonlinear regime.* This is in agreement with Parker (2016), where a similar conclusion was made based on comparing the linear zonostrophic instability rate predicted by the CE2. [As a note on terminology, Parker (2016) refers to the tWKE [Eqs. (9.43) and (9.45)] as the "Asymptotic WKE," i.e., the limit obtained when one assumes the ZFs are asymptotically large scale. Also, Parker (2016) refers to Eqs. (9.43) and (9.44) as "CE2-GO."]

[12]The fundamental reason for this discrepancy is that the tWKE assumes $\overline{W}$ to be the true distribution probability of driftons, but $\overline{W}$ is not. The true probability distribution would satisfy a strictly conservative equation. In the GO limit, it can always be found, at least numerically, as will be discussed elsewhere. The special case when $\partial_t U''$ is negligible is discussed in Parker (2016).

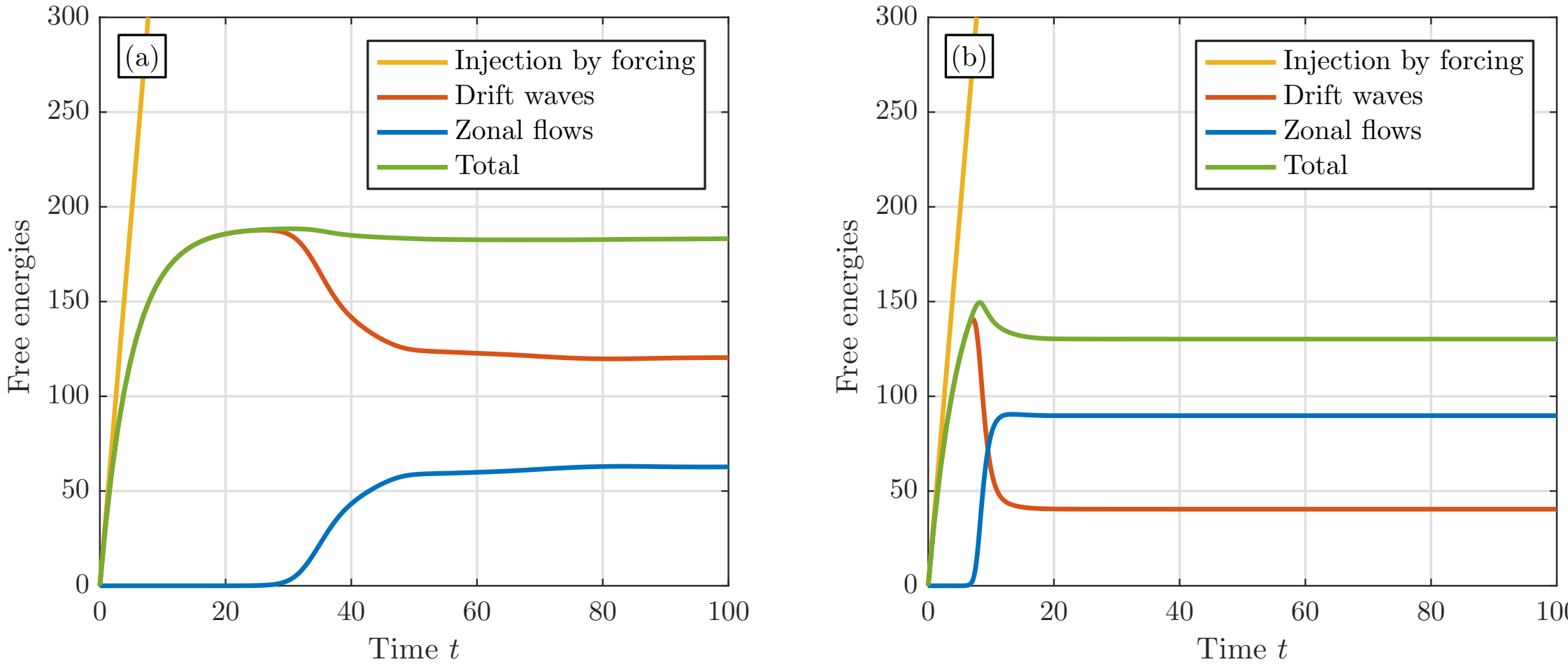

Figure 9.3: The total, DW, and ZF energies obtained by numerically integrating the WKE (9.43) for $\mathcal{H}$ and $\Gamma$ of two types: (a) the present model [Eqs. (9.44)]; (b) the tWKE model [Eqs. (9.45)]. The yellow lines show the total energy injected into the system by the external forcing $\overline{F}$ The initial conditions and simulation parameters are the same as in Fig. 9.1.

The numerical results presented in Figs. 9.1–9.3 illustrate the different dynamics predicted by the present WKE and the tWKE models [subfigures (a) and (b), respectively]. As shown in Fig. 9.1, while the present WKE model predicts ZFs with a particular $\lambda_{\rm zf}$, the scale of ZFs predicted by tWKE is determined by nothing but the grid size that is used in simulations. This is because the tWKE predicts that the rate of the zonostrophic instability $\gamma$ (Sec. 9.4) scales linearly with the ZF wave number $q$, so ZFs are produced at the largest $q$ that is allowed in the simulation (Parker, 2016).

Consider also the generalized enstrophy plots in Fig. 9.2. To aid the discussion, I added plots of the generalized enstrophy $\mathcal{Z}_{\rm ext}$ that the external forcing $\overline{F}$ injects into the DW–ZF system: $\mathcal{Z}_{\rm ext} = (t/2)(2\pi)^{-2}\int {\rm d}y\, {\rm d}^2p\, \overline{F}$. Within the present model, $\mathcal{Z}$ remains always smaller than $\mathcal{Z}_{\rm ext}$, which is natural, since the simulation is done for $\mu_{\rm dw,zf} > 0$. In contrast, the tWKE model predicts that $\mathcal{Z}$ can surpass $\mathcal{Z}_{\rm ext}$, which is unphysical. In addition, the values of the ZF and total enstrophies predicted by the tWKE are several times larger than those predicted by the present model.

For the sake of completeness, Fig. 9.3 also presents the corresponding energies and the energy $\mathcal{E}_{\rm ext}$ introduced by the external force, $\mathcal{E}_{\rm ext} = (t/2)(2\pi)^{-2}\int {\rm d}y\, {\rm d}^2p\, \overline{F}/p_{\rm D}^2$. In both cases, $\mathcal{E}(t) \leq \mathcal{E}_{\rm ext}(t)$, which is in agreement with the fact that both models conserve the total energy of an isolated system. Still, the tWKE predicts very different results quantitatively even though the tWKE model [Eqs. (9.45)] is seemingly close to the present model [Eqs. (9.44)].

## 9.6 Conclusions

In this Chapter, I presented an application of the Weyl symbol calculus to study nonlinear wave phenomena. Specifically, I proposed a new formulation of DW–ZF interactions that is more accurate than the tWKE and, simultaneously, more intuitive than the CE2. I adopted the same quasilinear model [Eqs. (9.6)] that was previously applied to derive the CE2. Then, I manipulated it by using the Weyl calculus in order to produce a phase-space formulation of DW–ZF interactions. The resulting formulation [Eqs. (9.29) and (9.30)] is akin to a quantum kinetic theory and involves a pseudodifferential Wigner–Moyal equation.

On one hand, the Wigner–Moyal formulation can be understood as an alternative representation to the CE2 since both models use the same assumptions. For example, I showed that it leads to the same linear growth rate of weak ZFs as that obtained from the CE2 (Sec. 9.4). On the other hand, the Wigner–Moyal formulation is arguably more intuitive than the CE2 for two reasons: (i) it permits treating driftons as particles (i.e., as objects traveling in phase space), except now they are *quantum-like* particles with nonzero wavelengths; and (ii) the separation between Hamiltonian effects and dissipation remains unambiguous even beyond the GO limit.

Compared to the tWKE, the new approach is also more precise because (i) it captures effects beyond the GO limit; and (ii) even in the GO limit, it predicts corrections to the tWKE that emerge from the newly found corrections to the drifton dispersion (Sec. 9.5). These corrections are essential as they allow DW–ZF enstrophy exchange, which is not included in the tWKE. By deriving the GO limit from first principles, one eliminates this discrepancy, and one arrives at a model that exactly conserves the total enstrophy (as opposed to the DW enstrophy conservation predicted by the tWKE) and the total energy, in agreement with the underlying gHME. I also illustrated the substantial difference between the GO limit of the present WKE model and the tWKE using numerical simulations.

This work can be expanded at least in two directions. First, the difference between the Wigner–Moyal formulation and the newly proposed WKE can be assessed quantitatively by using numerical simulations. Second, the analytic methods that I proposed here can be extended to other turbulence models, such as those in Anderson *et al.* (2002) and Anderson *et al.* (2006). The anticipated benefit is that more accurate equations would be derived that would respect fundamental conservation laws that existing theories may be missing otherwise.

# Part IV

# Conclusions

# Chapter 10

# Conclusions

In this final Chapter, I present a brief summary of the main results obtained in this dissertation. Conclusions, as well as recommendations for future research, are given. Since conclusions were already presented at the end of each Chapter, the discussion here is meant to be brief.

## 10.1 Thesis summary

This thesis was concentrated on developing a systematic theory of waves based on variational principles. Although the main motivation of this research program was to better understand wave dynamics in plasmas, the present thesis was not solely focused on this particular class of waves. Instead of focusing on waves satisfying specific equations, this dissertation considered waves as abstract objects of a Lagrangian theory. *Phase-space variational methods* were then used to obtain reduced models for describing general nondissipative waves. This variational phase-space description of waves helped simplify calculations, highlight the underlying wave symmetries, and lead to improved reduced modeling of wave dynamics. This research methodology allowed one to explore fundamental properties that are universal to all types of waves.

This thesis presents two important breakthroughs in the general theory of waves. First, an extension of the theory of geometrical optics (GO) was proposed. This new theory includes the effects of wave polarization. Polarization effects can be manifested as the polarization precession of waves and as a polarization-driven bending of ray trajectories. Second, it was shown that phase-space methods are potentially useful for studying a variety of problems involving nonlinear wave–wave interactions. Specifically, a general theory was developed that describes the ponderomotive refraction that a wave can experience when interacting with another wave. It was also shown that phase-space methods can be useful to study problems in the field of wave turbulence, such as the nonlinear interaction of high-frequency waves with large-scale structures.

## 10.2 Main results of the thesis

### 10.2.1 Extended geometrical optics

Even when one ignores diffraction, the GO ray equations are not entirely accurate. This occurs because GO treats wave rays as classical particles that are described by their position and momentum coordinates. However, vector waves have another degree of freedom, namely, their polarization. Polarization dynamics are manifested in two forms: (i) mode conversion (Friedland *et al.*, 1987), which is the transfer of wave action between resonant eigenmodes and can be understood as the precession of the wave polarization; and (ii) polarization-driven bending of ray trajectories, which refers to deviations of the GO ray trajectories arising from first-order corrections to the GO dispersion relation. In Chapter 4, **I presented an extension and reformulation of GO as a Lagrangian theory, whose action includes the aforementioned effects of wave polarization.** This newly developed theory is called *extended geometrical optics* (XGO).

In Chapter 5, **I applied the XGO theory to obtain the first-ever point-particle Lagrangian model for the relativistic spin-1/2 particle.** This model captures the well-known classical relativistic dynamics and the spin dynamics, such as the Stern–Gerlach force and the particle spin precession. Surprisingly, this model agrees with other previously developed classical models that were obtained using the Foldy–Wouthuysen transformation (Foldy and Wouthuysen, 1950).

In Chapter 6, **I applied the developed XGO theory to study polarization effects on radio-frequency (RF) waves propagating in magnetized plasma.** For waves in weakly magnetized plasma, it was analytically shown that polarization effects are manifested as a precession of the wave polarization and as a polarization-driven bending of the GO ray trajectories. Numerical simulations were presented for the case of waves in strongly magnetized plasma. In particular, it is speculated that polarization effects are important to consider when calculating the ray trajectories of RF waves propagating in tokamaks. These effects might be more important for RF waves used for ion-cyclotron heating or for lower-hybrid current drive.

### 10.2.2 Phase-space methods and nonlinear wave–wave interactions

As a second main contribution of this thesis, it was shown that phase-space methods are potentially useful for studying nonlinear wave–wave interactions. Specifically, I showed that scalar waves, both classical and quantum, can experience a time-averaged refraction when propagating in modulated media. This phenomenon is analogous to the ponderomotive effect experienced by charged particles interacting with high-frequency electromagnetic fields. In Chapter 7, **I proposed a general variational theory that** **can describe this ponderomotive effect on waves.** The formulation is able to describe waves with temporal and spatial periods comparable to that of the modulation. This theory is a generalization of the oscillation-center theory (Dewar, 1973), which is known from classical plasma physics, to any linear waves or quantum particles in particular.

In Chapter 7, I provided an explicit expression for the ponderomotive energy corresponding to the time-averaged refraction of waves. This ponderomotive energy was associated to the linear polarizability of the wave. The obtained relation is a generalization of the well-known "$K$–$\chi$ theorem" (Kaufman, 1987). **The main consequence of this generalization is that any wave is, in fact, a polarizable object that can contribute to the linear dielectric tensor of the ambient medium.** Hence, classical particles and wave quanta, such as photons, phonons, and plasmons, can all contribute to the dielectric tensor of the medium. The contribution of waves to the dielectric tensor can lead to exotic effects such as photon Landau damping (Bingham *et al.*, 1997).

In Chapter 8, **I obtained a point-particle Lagrangian model that describes the ponderomotive dynamics of a relativistic spin-1/2 electron interacting with a laser pulse propagating in vacuum.** The novelty of this result is that the EM pulse is allowed to have an arbitrarily large amplitude provided that radiation damping and pair production are negligible. This non-perturbative model captures the spin dynamics, the Stern–Gerlach spin–orbital coupling, the conventional ponderomotive forces, and the interaction with large-scale background fields (if any). This ponderomotive model was numerically compared to the non-averaged point-particle model given in Chapter 5. Excellent agreement was obtained between the two models.

In Chapter 9, I presented an application of the Weyl symbol calculus to study the collective behavior of incoherent high-frequency waves in the particular context of zonal-flow (ZF) formation in wave turbulence. In recent years, the appearance of ZFs in drift-wave (DW) turbulence inside tokamaks has captured much attention. Among the various reduced models, the wave kinetic equation (WKE) is widely used to study this phenomenon. However, this formulation neglects the exchange of enstrophy between DWs and ZFs and also ignores effects that go beyond the GO approximation. In this Chapter, **I proposed a new theory that captures both of these effects while still treating DW quanta ("driftons") as particles in phase space.** This formulation can be considered as a phase-space representation of the second-order cumulant expansion, or CE2. This formulation could potentially serve as a stepping stone to improving our understanding of ZF formation in more complicated turbulent systems.

## 10.3 Future work

Life, as well as research, is a never ending learning process. Inevitably, a wide variety of questions remain to be answered in this dissertation. In the following, I discuss some avenues of future research that I consider most important. The interested reader can refer to the conclusions at the end of each Chapter for a more thorough discussion.

The XGO theory presented in Chapter 4 can be extended in several directions. First, XGO considers a wave propagating in a flat spacetime Minkowski metric. It would be interesting to extend the theory in order to consider waves propagating in curved spacetime manifolds (or as a special case, curvilinear coordinates). Such generalized theory would cover a wider class of waves, for example electromagnetic waves propagating in the vicinity of black holes. This formalism would be a natural extension to the axiomatic GO theory developed by Dodin and Fisch (2012), as it would include the effects due to wave polarization.

Another opportunity of future research is the following. Currently, XGO theory only considers nondissipative vector waves. Future work could be done to develop a variational XGO theory that also describes dissipative waves. This variational theory could perhaps be obtained by using the recent technique proposed by Dodin *et al.* (2017), where dissipation is included into variational principles for linear waves.

If the previously proposed theories are developed, one could potentially describe dissipative waves or quantum particles in exotic environments. For example, one could develop a theory for spin-1/2 particles located in the vicinity of black holes, where the effects from general relativity are expected to be important. A self-consistent variational theory with radiation damping could be developed.

Another ambitious project for future research could be to further expand the XGO theory in order to include wave diffraction. This more accurate theory could be obtained by expanding the action functional to the second order in the GO parameter. Diffraction effects are expected to be contained in the second-order terms of the action functional. A potential application of this new theory would be to describe the propagation of optical vortex beams through inhomogeneous dielectric media. The effects of the wave polarization, as well as the effects of the intrinsic angular momentum of the beam, could be included in the calculation of the ray trajectories. Diffraction effects on the propagation of RF waves in tokamaks could be described with the proposed theory.

From the practical standpoint, the results presented in Chapter 6 could be extended in order to assess the importance of polarization effects on the propagation of RF waves in tokamaks. Based on estimates of the GO parameter, it is speculated that polarization effects could be important for waves with lower frequencies; for example, waves used for ion-cyclotron heating or for lower-hybrid current drive. It would be useful for future RF applications to develop a ray tracing code that captures the polarization-driven bending of ray trajectories, as well as mode conversion, simultaneously.

In Chapter 7, I presented a systematic method to calculate the effective dispersion operator of a wave propagating in a medium modulated by another wave. By truncating the asymptotic theory to the second order in the modulation amplitude, I showed that the resulting effective dispersion operator describes the wave ponderomotive dynamics. However, more terms in the asymptotic expansion could be included. For example, one could consider a single quantum particle interacting with three EM waves. By following the technique presented in Chapter 7, one could obtain an effective dispersion operator describing the particle time-averaged dynamics. Upon including third-order terms in the asymptotic expansion, one would obtain an effective theory that describes three-wave interactions. This effective theory would include particle kinetic effects. It would be interesting to investigate if such kinetic effects could bring new physics to the fluid-based models for Raman amplification (Malkin *et al.*, 1999). Likewise, by including fourth-order terms in the asymptotic expansion, one would also be able to describe four-wave interactions.

The systematic procedure presented in Chapter 7 could also potentially serve as a stepping stone to study more complex nonlinear wave–wave interactions, such as modulational instabilities in general wave ensembles or wave turbulence. Concerning the latter, a possible direction of future research could be the following. One of the most successful methods for studying classical turbulent systems is the theory developed by Martin *et al.* (1973). This theory is the classical version of Schwinger's functional formalism for quantum field theory. Given a nonlinear PDE, the formalism provides a self-consistent statistical description of the nonlinear interactions. However, this formalism is not very intuitive or easy to use since the dynamics of nonlinear interactions are described using the physical-space representation. In contrast, there are other phase-space formulations (Dewar, 1973, 1976) of classical turbulent systems that are not as general as the formalism by Martin *et al.* (1973) but are more physically intuitive. As suggested by Krommes (2012), a major breakthrough could be achieved if a more geometrical and intuitive phase-space interpretation can be given to the general asymptotic theory by Martin *et al.* (1973). The basic ideas presented in Chapter 7 could serve as a starting point for the above proposed work.

Another possible direction of future research in the field of turbulence is the following. The analytic methods proposed in Chapter 9 could be extended to study the excitation of zonal flows in other more complicated turbulence systems, such as the Hasegawa–Wakatani model (Hasegawa and Wakatani, 1983; Numata *et al.*, 2007) and other models describing ion-temperature-gradient turbulence (Anderson *et al.*, 2002, 2006). In contrast to the scalar generalized-Hasegawa–Mima equation used in Chapter 9, these models are based on coupled partial differential equations which makes the Wigner–Moyal formulation somewhat more complicated. If such phase-space models are developed, the anticipated benefit is that more accurate equations would be derived that would respect the conservation laws of the parent models.

Of course, the subjects that are mentioned above are only a few starting points for future avenues of research. In other words, much works remains to be done in the field of wave physics.

## 10.4 Final remarks

This thesis was focused on developing a systematic theory of waves based on phase-space variational principles. Two important breakthroughs were achieved. First, an extension of the theory of geometrical optics was proposed in order to include polarization effects. Second, it was shown that phase-space methods are potentially useful for studying a variety of problems involving nonlinear wave–wave interactions, such as ponderomotive wave refraction and the excitation of large-scale structures due to the interaction of high-frequency, small-scale waves.

In a broader perspective, the underlying essence of this dissertation is that both particles and fields can be described under the same footing, as waves. Upon considering the recent advances in the field of wave physics, I believe that much of plasma physics can be constructed as a unified geometric theory of waves based on phase-space variational principles. This theory would based on a single object, which is the action functional. I believe that the development of this field theoretical approach based on the phase-space representation may have far-reaching implications for the theory of plasmas. The anticipated benefits are the following: the development of systematic and self-consistent reduced models for plasma dynamics, the expansion of current models in order to include higher-order effects, and the incorporation of relativistic and quantum effects without the necessity of manually adding patches to classical plasma models. If such a formalism is developed, exciting rewards may be awaiting in the future.

# Part V

# Appendices

# Appendix A

# Main properties of the Weyl symbol calculus

This appendix summarizes the conventions for the Weyl symbol calculus. For the mathematical proofs of the relations shown below, please refer to Appendix B. Also, for more information, see the excellent reviews by Tracy *et al.* (2014), Imre *et al.* (1967), Baker Jr. (1958), Curtright *et al.* (1998), McDonald (1988), and Weyl (1931).

Let $\widehat{\mathcal{A}}$ be a matrix, whose components are composed of linear scalar operators. The Weyl symbol $A(x,p)$ is defined as the Weyl transform $\mathsf{W}[\widehat{\mathcal{A}}]$ of the linear operator $\widehat{\mathcal{A}}$; namely,

$$A(x,p) \doteq \mathsf{W}[\widehat{\mathcal{A}}](x,p) = \int \mathrm{d}^4 s \, e^{ip\cdot s} \, \langle \, x + s/2 \, | \, \widehat{\mathcal{A}} \, | \, x - s/2 \, \rangle \, , \tag{A.1}$$

where $|\, x \,\rangle$ are the eigenstates of the position operator $\widehat{x}^\mu$ such that $\langle \, x \, | \, \widehat{x}^\mu \, | \, x' \, \rangle = x^\mu \delta^4(x - x')$. Also, $p \cdot s = p_0 s_0 - \mathbf{p} \cdot \mathbf{s}$, and the integrals span $\mathbb{R}^4$. This description of the operators is known as the *phase-space representation* since the Weyl symbols are functions of the eight-dimensional phase space. Conversely, the inverse Weyl transformation is given by

$$\widehat{\mathcal{A}} = \frac{1}{(2\pi)^4} \int \mathrm{d}^4 x \, \mathrm{d}^4 p \, \mathrm{d}^4 s \, e^{ip\cdot s} A(x,p) \, | \, x - s/2 \, \rangle \, \langle \, x + s/2 \, | \, . \tag{A.2}$$

The projection of the operator $\widehat{\mathcal{A}}$ on the position eigenstates, $\mathcal{A}(x,x') = \langle \, x \, | \, \widehat{\mathcal{A}} \, | \, x' \, \rangle$, is

$$\mathcal{A}(x,x') = \frac{1}{(2\pi)^4} \int \mathrm{d}^4 p \, e^{-ip\cdot(x-x')} A\left( \frac{x+x'}{2}, p \right) . \tag{A.3}$$

In the following, I shall outline a number of useful properties of the Weyl transform.

- For any operator $\widehat{\mathcal{A}}$, the trace $\mathrm{Tr}_x[\widehat{\mathcal{A}}] \doteq \int \mathrm{d}^4x \, \langle \, x \mid \widehat{\mathcal{A}} \mid x \, \rangle$ can be expressed as

$$\mathrm{Tr}_x[\widehat{\mathcal{A}}] = \frac{1}{(2\pi)^4} \int \mathrm{d}^4x \, \mathrm{d}^4p \, A(x,p). \tag{A.4}$$

- If $A(x,p)$ is the Weyl symbol of $\widehat{\mathcal{A}}$, then $A^\dagger(x,p)$ is the Weyl symbol of $\widehat{\mathcal{A}}^\dagger$. As a corollary, if the operator $\widehat{\mathcal{A}}$ is Hermitian ($\widehat{\mathcal{A}} = \widehat{\mathcal{A}}^\dagger$), then the Weyl symbol satisfies $A(x,p) = A^\dagger(x,p)$.

- For linear operators $\widehat{\mathcal{A}}$, $\widehat{\mathcal{B}}$ and $\widehat{\mathcal{C}}$ where $\widehat{\mathcal{C}} = \widehat{\mathcal{A}}\widehat{\mathcal{B}}$, the corresponding Weyl symbols satisfy

$$C(x,p) = A(x,p) \star B(x,p). \tag{A.5}$$

Here "$\star$" refers to the *Moyal product* (Moyal, 1949), which is given by

$$A(x,p) \star B(x,p) \doteq A(x,p) e^{i\widehat{\mathcal{L}}/2} B(x,p). \tag{A.6}$$

Also, $\widehat{\mathcal{L}}$ is the *Janus operator*

$$\widehat{\mathcal{L}} \doteq \overleftarrow{\partial_p} \cdot \overrightarrow{\partial_x} - \overleftarrow{\partial_x} \cdot \overrightarrow{\partial_p} = \{ \cdot , \cdot \}. \tag{A.7}$$

The arrows indicate the direction in which the derivatives act, and $A\widehat{\mathcal{L}}B = \{A, B\}$ is the canonical Poisson bracket in the eight-dimensional phase space, namely,

$$\widehat{\mathcal{L}} = \frac{\overleftarrow{\partial}}{\partial p^0} \frac{\overrightarrow{\partial}}{\partial x^0} - \frac{\overleftarrow{\partial}}{\partial x^0} \frac{\overrightarrow{\partial}}{\partial p^0} + \frac{\overleftarrow{\partial}}{\partial \mathbf{x}} \cdot \frac{\overrightarrow{\partial}}{\partial \mathbf{p}} - \frac{\overleftarrow{\partial}}{\partial \mathbf{p}} \cdot \frac{\overrightarrow{\partial}}{\partial \mathbf{x}}. \tag{A.8}$$

- The Moyal product is associative; i.e., for arbitrary symbols $A$, $B$, and $C$, one has

$$A \star B \star C = (A \star B) \star C = A \star (B \star C). \tag{A.9}$$

- The anti-symmetrized Moyal product defines the so-called *Moyal bracket*, namely,

$$\{\{A, B\}\} \doteq -i \, (A \star B - B \star A) \, . \tag{A.10}$$

Likewise, the symmetrized Moyal product is defined as

$$[[A, B]] \doteq A \star B + B \star A. \tag{A.11}$$

- In the special case where the $A(x,p)$ and $B(x,p)$ are scalar functions, then

$$\{\{A,B\}\} = 2A\sin\left(\frac{\widehat{\mathcal{L}}}{2}\right)B, \tag{A.12a}$$

$$[[A,B]] = 2A\cos\left(\frac{\widehat{\mathcal{L}}}{2}\right)B. \tag{A.12b}$$

Because of the latter equalities, these brackets are also referred as the *sine* and *cosine* brackets, respectively. To the lowest order in the Taylor expansion, the brackets reduce to

$$\{\{A,B\}\} \simeq \{A,B\}, \tag{A.13a}$$

$$[[A,B]] \simeq 2AB. \tag{A.13b}$$

In a wave equation involving the sine and cosine brackets, neglecting higher-order phase-space derivatives in the brackets leads to the wave kinetic equation. Higher-order wave effects, such as diffraction and tunneling, are lost in this limit. For this reason, it is called the GO approximation or the ray approximation. For more information, see Chapter 3.

- For fields that vanish rapidly enough at infinity, when integrated over all phase space the Moyal product of two symbols equals the regular matrix product; i.e.,

$$\int \mathrm{d}^4x\,\mathrm{d}^4p\,A\star B = \int \mathrm{d}^4x\,\mathrm{d}^4p\,AB. \tag{A.14}$$

As a corollary,

$$\int \mathrm{d}^4x\,\mathrm{d}^4p\,\{\{A,B\}\} = -i\int \mathrm{d}^4x\,\mathrm{d}^4p\,(AB-BA), \tag{A.15a}$$

$$\int \mathrm{d}^4x\,\mathrm{d}^4p\,[[A,B]] = \int \mathrm{d}^4x\,\mathrm{d}^4p\,(AB+BA). \tag{A.15b}$$

- For any constant four-vector $q$, one has

$$A(p)\star e^{-iq\cdot x} = A(p)e^{(i/2)\overleftarrow{\partial}_p\cdot\overrightarrow{\partial}_x}e^{-iq\cdot x} = A(p)e^{(q/2)\cdot\overleftarrow{\partial}_p}e^{-iq\cdot x} = A(p+q/2)e^{-iq\cdot x}. \tag{A.16}$$

As a corollary, one has

$$\{\{A(p), e^{-iq\cdot x}\}\} = -i\left[A\left(p + q/2\right) - A\left(p - q/2\right)\right] e^{-iq\cdot x}, \tag{A.17a}$$

$$[[A(p), e^{-iq\cdot x}]] = \left[A\left(p + q/2\right) + A\left(p - q/2\right)\right] e^{-iq\cdot x}. \tag{A.17b}$$

- For any constants four-vectors $k$ and $q$, one can also show that

$$A(p)e^{-ik\cdot x} \star B(p)e^{-iq\cdot x} = A(p + q/2)B(p - k/2)e^{-i(k+q)\cdot x}. \tag{A.18}$$

- Now I tabulate some Weyl transforms of various operators. The Weyl transforms of the identity, position, and momentum operators are given by (see Sec. 2.3 for details)

$$\mathsf{W}[\,\widehat{1}\,] = 1, \qquad \mathsf{W}[\widehat{x}^{\mu}] = x^{\mu}, \qquad \mathsf{W}[\widehat{p}_{\mu}] = p_{\mu}. \tag{A.19}$$

For any two operators $f(\widehat{x})$ and $g(\widehat{p})$, one has

$$\mathsf{W}[\,f(\widehat{x})\,] = f(x), \qquad \mathsf{W}[\,g(\widehat{p})\,] = g(p). \tag{A.20}$$

Similarly, upon using the Moyal product (A.6), one has

$$\mathsf{W}[\,\widehat{p}_{\mu} f(\widehat{x})\,] = p_{\mu} f(x) + (i/2)\partial_{\mu} f(x), \tag{A.21}$$

$$\mathsf{W}[\,f(\widehat{x})\widehat{p}_{\mu}\,] = p_{\mu} f(x) - (i/2)\partial_{\mu} f(x). \tag{A.22}$$

# Appendix B

# Mathematical foundations of the Weyl symbol calculus

This Appendix presents a direct construction of the mathematical formalism behind the Weyl symbol calculus. The approach adopted here follows closely those given in Balazs and Jennings (1984), Serimaa *et al.* (1986), and Müller (1999). The proofs are given with considerable detail in order to facilitate the interested readers' introduction to the Weyl symbol calculus.

## B.1 Basic formalism

Let $\Psi(x)$ be a square-integrable $N$-component wave function depending on $\mathbb{R}^4$. In the Dirac notation, $\Psi(x)$ is denoted as a ket state $|\,\Psi\,\rangle$ of the Hilbert space with inner product

$$\langle\,\Upsilon\mid\Psi\,\rangle \doteq \int \mathrm{d}^4x\,\Upsilon^\dagger(x)\Psi(x), \tag{B.1}$$

where $\mathrm{d}^4x \doteq \mathrm{d}t\,\mathrm{d}^3x$ and the integrals span $\mathbb{R}^4$. The physical-space representation of the wave function is recovered from the inner product $\Psi(x) = \langle\,x\mid\Psi\,\rangle$. Here $|\,x\,\rangle \doteq |\,(t,\mathbf{x})\,\rangle$ is an eigenstate of the coordinate operator $\widehat{x}^\mu$ such that

$$\langle\,x\mid\widehat{x}^\mu\mid x'\,\rangle = x^\mu\,\langle\,x\mid x'\,\rangle = x^\mu\delta^4(x-x'). \tag{B.2}$$

A linear operator $\widehat{\mathcal{A}}$ is a mathematical object that maps one state $|\,\Psi\,\rangle$ into another state $|\,\Phi\,\rangle$; i.e., $|\,\Phi\,\rangle = \widehat{\mathcal{A}}\,|\,\Psi\,\rangle$. As an example, in the physical-space representation the momentum operator $\widehat{p}_\mu$ is defined as

$$\langle\, x \mid \widehat{p}_\mu \mid x'\,\rangle \doteq i\frac{\partial}{\partial x^\mu}\delta^4(x-x'). \tag{B.3}$$

Since I consider multi-component wave functions $\Psi(x)$, the linear operators $\widehat{\mathcal{A}}$ will be matrices in general, whose components are operators depending on $\widehat{x}$ and $\widehat{p}$.

**Proposition 1.** *Let $u, v \in \mathbb{R}^4$ be constants and $n \in \mathbb{N}$. Then,*

$$\langle\, x \mid (v\cdot\widehat{x})^n \mid x'\,\rangle = (v\cdot x)^n\delta^4(x-x'), \tag{B.4}$$

$$\langle\, x \mid (u\cdot\widehat{p})^n \mid x'\,\rangle = \left(iu\cdot\frac{\partial}{\partial x}\right)^n\delta^4(x-x'). \tag{B.5}$$

*Proof.* Let us prove by induction. The eigenstates $|\,x\,\rangle$ form an orthonormal basis so they satisfy $\int \mathrm{d}^4x\ |\,x\,\rangle\langle\,x\,| = \widehat{1}$, where $\widehat{1}$ is the identity operator. Upon using this relation, one obtains

$$\begin{aligned}
\langle\, x \mid (v\cdot\widehat{x})^{n+1} \mid x'\,\rangle &= \langle\, x \mid (v\cdot\widehat{x})^n(v\cdot\widehat{x}) \mid x'\,\rangle \\
&= \int \mathrm{d}^4x''\ \langle\, x \mid (v\cdot\widehat{x})^n \mid x''\,\rangle\langle\, x'' \mid (v\cdot\widehat{x}) \mid x'\,\rangle \\
&= \int \mathrm{d}^4x''\ (v\cdot x)^n\delta^4(x-x'')\,(v\cdot x'')\delta^4(x''-x') \\
&= (v\cdot x)^n(v\cdot x')\delta^4(x-x') \\
&= (v\cdot x)^{n+1}\delta^4(x-x'),
\end{aligned}$$

where the last equality is in the distribution sense. For the momentum operator, one has

$$\begin{aligned}
\langle\, x \mid (u\cdot\widehat{p})^{n+1} \mid x'\,\rangle &= \int \mathrm{d}^4x''\ \langle\, x \mid (u\cdot\widehat{p})^n \mid x''\,\rangle\langle\, x'' \mid (u\cdot\widehat{p}) \mid x'\,\rangle \\
&= \int \mathrm{d}^4x''\ \left[\left(iu\cdot\frac{\partial}{\partial x}\right)^n\delta^4(x-x'')\right]\left[iu\cdot\frac{\partial}{\partial x''}\delta^4(x''-x')\right] \\
&= -\int \mathrm{d}^4x''\ \left[\left(iu\cdot\frac{\partial}{\partial x}\right)^n\left(iu\cdot\frac{\partial}{\partial x''}\right)\delta^4(x-x'')\right]\delta^4(x''-x') \\
&= -\left(iu\cdot\frac{\partial}{\partial x}\right)^n\left(iu\cdot\frac{\partial}{\partial x'}\right)\delta^4(x-x') \\
&= \left(iu\cdot\frac{\partial}{\partial x}\right)^{n+1}\delta^4(x-x').
\end{aligned}$$

□

**Proposition 2.** *Let $u, v \in \mathbb{R}^4$ be constants. Then,*

$$e^{-iv\cdot\widehat{x}}\,|\,x\,\rangle = e^{-iv\cdot x}\,|\,x\,\rangle\,, \tag{B.6}$$

$$e^{-iu\cdot\widehat{p}}\,|\,x\,\rangle = |\,x-u\,\rangle\,. \tag{B.7}$$

*Proof.* The proof consists in Taylor expanding the exponential operator, inserting the identity operator $\int \mathrm{d}^4x\ |\,x\,\rangle\langle\,x\,| = \widehat{1}$, and using Proposition 1. One obtains

$$\begin{aligned}
e^{-iv\cdot\widehat{x}}\,|\,x\,\rangle &= \int \mathrm{d}^4x'\ |\,x'\,\rangle\langle\,x'\,|\ e^{-iv\cdot\widehat{x}}\ |\,x\,\rangle \\
&= \sum_n \frac{1}{n!}\int \mathrm{d}^4x'\ |\,x'\,\rangle\langle\,x'\,|\ (-iv\cdot\widehat{x})^n\ |\,x\,\rangle \\
&= \sum_n \frac{1}{n!}\int \mathrm{d}^4x'\ |\,x'\,\rangle\,(-iv\cdot x')^n\delta^4(x'-x) \\
&= e^{-iv\cdot x}\,|\,x\,\rangle\,.
\end{aligned}$$

Similarly, for the exponentiated momentum operator, the calculation is given by

$$\begin{aligned}
e^{-iu\cdot\widehat{p}}\,|\,x\,\rangle &= \int \mathrm{d}^4x'\ |\,x'\,\rangle\langle\,x'\,|\ e^{-iu\cdot\widehat{p}}\ |\,x\,\rangle \\
&= \sum_n \frac{1}{n!}\int \mathrm{d}^4x'\ |\,x'\,\rangle\langle\,x'\,|\ (-iu\cdot\widehat{p})^n\ |\,x\,\rangle \\
&= \sum_n \frac{1}{n!}\int \mathrm{d}^4x'\ |\,x'\,\rangle\left(u\cdot\frac{\partial}{\partial x'}\right)^n\delta^4(x'-x) \\
&= \int \mathrm{d}^4x'\ |\,x'\,\rangle\,\delta^4(x'+u-x) \\
&= |\,x-u\,\rangle\,.
\end{aligned}$$

□

## B.2 Heisenberg generating operator

**Definition 1 (Heisenberg generating operator).** *The Heisenberg generating operator $\widehat{\mathcal{T}}(u,v)$ is defined as*

$$\widehat{\mathcal{T}}(u,v) \doteq \exp\left(-iu\cdot\widehat{p} - iv\cdot\widehat{x}\right), \tag{B.8}$$

*where $u, v \in \mathbb{R}^4$, $\widehat{x}^\mu$ is the position operator in $\mathbb{R}^4$, and $\widehat{p}_\mu$ is the momentum operator in $\mathbb{R}^4$. Here the four-dimensional Minkowski metric is used; e.g., $u\cdot\widehat{p} = u^0\widehat{p}^0 - \mathbf{u}\cdot\widehat{\mathbf{p}}$.*

In the literature, the Heisenberg generating operators are typically defined using the position and momentum operators in $\mathbb{R}^3$ [see, for example, Balazs and Jennings (1984), Serimaa *et al.* (1986), Müller (1999), Tracy *et al.* (2014), and McDonald (1988)]. In other words, the generating operators are $\widehat{\mathfrak{T}}(\mathbf{u},\mathbf{v}) \doteq \exp(i\mathbf{u}\cdot\widehat{\mathbf{p}}+i\mathbf{v}\cdot\widehat{\mathbf{x}})$. Here I construct a Lorentz covariant theory so the Heisenberg operators are defined using the position and momentum operators in $\mathbb{R}^4$.

**Proposition 3.** *The projection of the Heisenberg generating operator* $\widehat{\mathfrak{T}}(u,v)$ *on the position eigenstates* $|\,x\,\rangle$ *is as follows:*

$$\langle\, x' \mid \widehat{\mathfrak{T}}(u,v) \mid x'' \,\rangle = \delta^4(x' - x'' + u) \exp\left(-iv\cdot x'' + iu\cdot v/2\right). \tag{B.9}$$

*Proof.* Let us use the Baker–Campbell–Hausdorff formula: if $\widehat{A}$ and $\widehat{B}$ are two operators that commute with their commutator, then $\log(e^{\widehat{A}}e^{\widehat{B}}) = \widehat{A} + \widehat{B} + [\widehat{A},\widehat{B}]/2$. Upon using the fact that $[\widehat{p}_\mu, \widehat{x}^\nu] = i\delta^\nu_\mu$, one has $\log(e^{-iu\cdot\widehat{p}}e^{-iv\cdot\widehat{x}}) = -iu\cdot\widehat{p}-iv\cdot\widehat{x}-[u\cdot\widehat{p}, v\cdot\widehat{x}]/2 = -iu\cdot\widehat{p}-iv\cdot\widehat{x}-iu\cdot v/2$. Hence, $e^{-iu\cdot\widehat{p}-iv\cdot\widehat{x}} = e^{-iu\cdot\widehat{p}}e^{-iv\cdot\widehat{x}}e^{iu\cdot v/2}$. Then with Proposition (2), one obtains

$$\begin{aligned}\langle\, x' \mid \widehat{\mathfrak{T}}(u,v) \mid x'' \,\rangle &= \langle\, x' \mid e^{-iu\cdot\widehat{p}-iv\cdot\widehat{x}} \mid x'' \,\rangle \\ &= \langle\, x' \mid e^{-iu\cdot\widehat{p}}e^{-iv\cdot\widehat{x}}e^{iu\cdot v/2} \mid x'' \,\rangle \\ &= \langle\, x' \mid e^{-iu\cdot\widehat{p}} \mid x'' \,\rangle\, e^{-iv\cdot x''+iu\cdot v/2} \\ &= \langle\, x' \mid x'' - u \,\rangle\, e^{-iv\cdot x''+iu\cdot v/2} \\ &= \delta^4(x' - x'' + u)\, e^{-iv\cdot x''+iu\cdot v/2}.\end{aligned}$$

□

**Definition 2.** *Let the trace* $\mathrm{Tr}_x$ *of a linear operator* $\widehat{\mathcal{A}}$ *be defined as* $\mathrm{Tr}_x[\widehat{\mathcal{A}}] \doteq \int \mathrm{d}^4x\, \langle\, x \mid \widehat{\mathcal{A}} \mid x \,\rangle$.

Here I added the index $x$ here to emphasize that the summation (integration) is done over $x$ only, while $\langle\, x \mid \widehat{\mathcal{A}} \mid x \,\rangle$ by itself can be a matrix.

**Corollary 1.** *The trace of the Heisenberg generating operator* $\widehat{\mathfrak{T}}(u,v)$ *is*

$$\mathrm{Tr}_x[\widehat{\mathfrak{T}}(u,v)] = (2\pi)^4\, \delta^4(u)\, \delta^4(v). \tag{B.10}$$

*Proof.* After substituting Proposition 3, one obtains

$$\begin{aligned}\mathrm{Tr}_x[\widehat{\mathfrak{T}}(u,v)] &= \int \mathrm{d}^4x\, \langle\, x \mid \widehat{\mathfrak{T}}(u,v) \mid x \,\rangle \\ &= \int \mathrm{d}^4x\, \delta^4(u) e^{-iv\cdot x+iu\cdot v/2} \\ &= (2\pi)^4\, \delta^4(u)\, \delta^4(v) e^{iu\cdot v/2} \\ &= (2\pi)^4\, \delta^4(u)\, \delta^4(v),\end{aligned}$$

where the last equality is in the distribution sense. $\square$

**Proposition 4.** *Let $\widehat{\mathfrak{T}}(u,v)$ be a Heisenberg generating operator. Then,*

$$\widehat{\mathfrak{T}}(u,v)\widehat{\mathfrak{T}}(u',v') = \widehat{\mathfrak{T}}(u+u',v+v')\exp[i(v\cdot u' - u\cdot v')/2]. \tag{B.11}$$

*Proof.* Upon using the Baker–Campbell–Hausdorff formula introduced in Proposition 3, one has

$$\begin{aligned}
\widehat{\mathfrak{T}}(u,v)\widehat{\mathfrak{T}}(u',v') &= \exp\left(-iu\cdot\widehat{p} - iv\cdot\widehat{x}\right)\,\exp\left(-iu'\cdot\widehat{p} - iv'\cdot\widehat{x}\right)\\
&= \exp\left\{-i(u+u')\cdot\widehat{p} - i(v+v')\cdot\widehat{x} - \left[u\cdot\widehat{p}+v\cdot\widehat{x}, u'\cdot\widehat{p}+v'\cdot\widehat{x}\right]/2\right\}\\
&= \exp\left[-i(u+u')\cdot\widehat{p} - i(v+v')\cdot\widehat{x}\right]\exp\left[i(v\cdot u' - u\cdot v')/2\right]\\
&= \widehat{\mathfrak{T}}(u+u',v+v')\exp[i(v\cdot u' - u\cdot v')/2].
\end{aligned}$$ $\square$

Note that the multiplication of two generating Heisenberg operators is another Heisenberg generating operator up to a constant phase. This is the defining property of the Weyl–Heisenberg group. For the interested reader, Tracy *et al.* (2014) present an alternate construction of the Weyl symbol calculus that is based on theory of the Weyl–Heisenberg group.

## B.3 Wigner operator

**Definition 3** (**Wigner operator**). *Let $x,p\in\mathbb{R}^4$. The Wigner operator $\widehat{\Delta}(x,p)$ is defined as the Fourier transform of the Heisenberg generating operator so that*

$$\widehat{\Delta}(x,p) \doteq (2\pi)^{-8}\int \mathrm{d}^4u\,\mathrm{d}^4v\,\widehat{\mathfrak{T}}(u,v)e^{iu\cdot p + iv\cdot x}. \tag{B.12}$$

**Proposition 5.** *The projection of the Wigner operator $\widehat{\Delta}(x,p)$ on the position eigenstates $|\,x\,\rangle$ is given by*

$$\langle\, x' \mid \widehat{\Delta}(x,p) \mid x''\,\rangle = (2\pi)^{-4}\int \mathrm{d}^4u\,\delta^4(x'-x''+u)\,\delta^4(x-x''+u/2)e^{iu\cdot p}. \tag{B.13}$$

*Proof.* Upon substituting Proposition 3 into Eq. (B.12), one obtains

$$\begin{aligned}\langle\, x' \mid \widehat{\Delta}(x,p) \mid x''\,\rangle &= (2\pi)^{-8}\int \mathrm{d}^4u\,\mathrm{d}^4v\;\langle\, x' \mid \widehat{\mathfrak{T}}(x,p) \mid x''\,\rangle\, e^{iu\cdot p+iv\cdot x}\\ &= (2\pi)^{-8}\int \mathrm{d}^4u\,\mathrm{d}^4v\,\delta^4(x'-x''+u)e^{-iv\cdot x''+iu\cdot v/2}e^{iu\cdot p+iv\cdot x}\\ &= (2\pi)^{-4}\int \mathrm{d}^4u\,\delta^4(x'-x''+u)\,\delta^4(x-x''+u/2)e^{iu\cdot p}.\end{aligned}$$

□

**Lemma 1.** *The Wigner operators $\widehat{\Delta}(x,p)$ are complete in the following sense:*

$$\int \mathrm{d}^4x\,\mathrm{d}^4p\,\widehat{\Delta}(x,p) = \widehat{1}. \tag{B.14}$$

*Proof.* After substituting the definition of the Wigner operator and integrating, one has

$$\begin{aligned}\int \mathrm{d}^4x\,\mathrm{d}^4p\,\widehat{\Delta}(x,p) &= (2\pi)^{-8}\int \mathrm{d}^4x\,\mathrm{d}^4p\,\mathrm{d}^4u\,\mathrm{d}^4v\,\widehat{\mathfrak{T}}(u,v)e^{iu\cdot p+iv\cdot x}\\ &= \int \mathrm{d}^4u\,\mathrm{d}^4v\,\widehat{\mathfrak{T}}(u,v)\,\delta^4(u)\,\delta^4(v)\\ &= \widehat{\mathfrak{T}}(0,0)\\ &= \widehat{1}.\end{aligned}$$

□

**Lemma 2.** *The Wigner operators $\widehat{\Delta}(x,p)$ are orthogonal in the following sense:*

$$\mathrm{Tr}_x[\widehat{\Delta}(x,p)\widehat{\Delta}(x',p')] = (2\pi)^{-4}\,\delta^4(x-x')\,\delta^4(p-p'). \tag{B.15}$$

*Proof.* This is proven by using Proposition 4:

$$\begin{aligned}&\mathrm{Tr}_x[\widehat{\Delta}(x,p)\widehat{\Delta}(x',p')]\\ &\quad= (2\pi)^{-16}\int \mathrm{d}^4u\,\mathrm{d}^4v\,\mathrm{d}^4u'\,\mathrm{d}^4v'\;\mathrm{Tr}_x[\widehat{\mathfrak{T}}(u,v)e^{iu\cdot p+iv\cdot x}\widehat{\mathfrak{T}}(u',v')e^{iu'\cdot p'+iv'\cdot x'}]\\ &\quad= (2\pi)^{-16}\int \mathrm{d}^4u\,\mathrm{d}^4v\,\mathrm{d}^4u'\,\mathrm{d}^4v'\;\mathrm{Tr}_x[\widehat{\mathfrak{T}}(u,v)\widehat{\mathfrak{T}}(u',v')]e^{iu\cdot p+iv\cdot x}e^{iu'\cdot p'+iv'\cdot x'}\\ &\quad= (2\pi)^{-16}\int \mathrm{d}^4u\,\mathrm{d}^4v\,\mathrm{d}^4u'\,\mathrm{d}^4v'\;\mathrm{Tr}_x[\widehat{\mathfrak{T}}(u+u',v+v')]e^{i(v\cdot u'-u\cdot v')/2}e^{iu\cdot p+iv\cdot x}e^{iu'\cdot p'+iv'\cdot x'}.\end{aligned}$$

Then, substituting the result in Corollary 1 leads to

$$
\begin{aligned}
&\mathrm{Tr}_x[\widehat{\Delta}(x,p)\widehat{\Delta}(x',p')] \\
&\quad = (2\pi)^{-12}\int \mathrm{d}^4u\,\mathrm{d}^4v\,\mathrm{d}^4u'\,\mathrm{d}^4v'\,\delta^4(u+u')\delta^4(v+v')e^{i(v\cdot u'-u\cdot v')/2}e^{iu\cdot p+iv\cdot x}e^{iu'\cdot p'+iv'\cdot x'} \\
&\quad = (2\pi)^{-12}\int \mathrm{d}^4u\,\mathrm{d}^4v\,e^{iu\cdot(p-p')}e^{iv\cdot(x-x')} \\
&\quad = (2\pi)^{-4}\,\delta^4(x-x')\,\delta^4(p-p'). \qquad \square
\end{aligned}
$$

**Proposition 6.** *The Wigner operator $\widehat{\Delta}(x,p)$ can alternatively be expressed as*

$$
\widehat{\Delta}(x,p) = (2\pi)^{-4}\int \mathrm{d}^4s\,e^{is\cdot p}\,|\,x-s/2\,\rangle\,\langle\,x+s/2\,|\,. \tag{B.16}
$$

*Proof.* Using the completeness relation and Proposition 5 leads to

$$
\begin{aligned}
\widehat{\Delta}(x,p) &= \left(\int \mathrm{d}^4x'\;|\,x'\,\rangle\,\langle\,x'\,|\right)\widehat{\Delta}(x,p)\left(\int \mathrm{d}^4x''\;|\,x''\,\rangle\,\langle\,x''\,|\right) \\
&= \int \mathrm{d}^4x'\,\mathrm{d}^4x''\;|\,x'\,\rangle\,\langle\,x'\,|\,\widehat{\Delta}(x,p)\,|\,x''\,\rangle\,\langle\,x''\,| \\
&= (2\pi)^{-4}\int \mathrm{d}^4x'\,\mathrm{d}^4x''\,\mathrm{d}^4u\,e^{iu\cdot p}\delta^4(x'-x''+u)\,\delta^4(x-x''+u/2)\,|\,x'\,\rangle\,\langle\,x''\,| \\
&= (2\pi)^{-4}\int \mathrm{d}^4x''\,\mathrm{d}^4u\,e^{iu\cdot p}\delta^4(x-x''+u/2)\,|\,x''-u\,\rangle\,\langle\,x''\,| \\
&= (2\pi)^{-4}\int \mathrm{d}^4u\,e^{iu\cdot p}\,|\,x-u/2\,\rangle\,\langle\,x+u/2\,|\,. \qquad \square
\end{aligned}
$$

Now we have the tools to define the Weyl transformation. Below, it will also be shown that a linear operator $\widehat{\mathcal{A}}$ can be written as a linear combination of Wigner operators.

## B.4 Weyl transformation

**Definition 4 (Weyl transformation).** *The Weyl transform of an arbitrary linear operator $\widehat{\mathcal{A}}$ is defined as*

$$
\mathsf{W}[\widehat{\mathcal{A}}](x,p) \doteq (2\pi)^4\mathrm{Tr}_x[\widehat{\Delta}(x,p)\widehat{\mathcal{A}}]. \tag{B.17}
$$

*In general, $A(x,p) \doteq \mathsf{W}[\widehat{\mathcal{A}}](x,p)$ is a matrix, whose components are functions of the eight-dimensional phase space. The function $A(x,p)$ is called the "Weyl symbol" of the operator $\widehat{\mathcal{A}}$.*

**Remark 1.** *The possible issues regarding the existence of the Weyl transformation or its inverse are not considered here.*

**Proposition 7.** *The Weyl symbol $A(x,p)$ corresponding to the operator $\widehat{\mathcal{A}}$ is also given by*

$$A(x,p) = \int \mathrm{d}^4 s\, e^{is\cdot p} \,\langle\, x + s/2 \mid \widehat{\mathcal{A}} \mid x - s/2 \,\rangle\,. \tag{B.18}$$

*Proof.* Upon using Proposition 6, one obtains

$$\begin{aligned}
A(x,p) &= (2\pi)^4 \mathrm{Tr}_x[\widehat{\Delta}(x,p)\widehat{\mathcal{A}}] \\
&= \mathrm{Tr}_x \left[\int \mathrm{d}^4 s e^{ip\cdot s} \,|\, x - s/2 \,\rangle \,\langle\, x + s/2 \,|\, \widehat{\mathcal{A}}\right] \\
&= \int \mathrm{d}^4 x'\, \mathrm{d}^4 s\, e^{ip\cdot s} \,\langle\, x' \mid x - s/2 \,\rangle \,\langle\, x + s/2 \mid \widehat{\mathcal{A}} \mid x' \,\rangle \\
&= \int \mathrm{d}^4 x'\, \mathrm{d}^4 s\, e^{ip\cdot s} \delta^4(x' - x + s/2) \,\langle\, x + s/2 \mid \widehat{\mathcal{A}} \mid x' \,\rangle \\
&= \int \mathrm{d}^4 s\, e^{ip\cdot s} \,\langle\, x + s/2 \mid \widehat{\mathcal{A}} \mid x - s/2 \,\rangle\,.
\end{aligned}$$

□

**Corollary 2.** *Let $\widehat{\mathcal{A}}_H$ be a Hermitian operator such that $\widehat{\mathcal{A}}_H = \widehat{\mathcal{A}}_H^\dagger$. Then, the corresponding Weyl symbol $A_H(x,p)$ is Hermitian; i.e., $[A_H(x,p)]^\dagger = A_H(x,p)$, where $[A_H(x,p)]^\dagger$ is the conjugate transpose of the Weyl symbol $A_H(x,p)$.*

*Proof.* From Proposition 7 and the definition of Hermitian operators, one obtains

$$\begin{aligned}
[A_H(x,p)]^\dagger &= \int \mathrm{d}^4 s\, e^{-is\cdot p} \left(\langle\, x + s/2 \mid \widehat{\mathcal{A}}_H \mid x - s/2 \,\rangle\right)^\dagger \\
&= \int \mathrm{d}^4 s\, e^{-is\cdot p} \,\langle\, x - s/2 \mid \widehat{\mathcal{A}}_H^\dagger \mid x + s/2 \,\rangle \\
&= \int \mathrm{d}^4 u\, e^{iu\cdot p} \,\langle\, x + u/2 \mid \widehat{\mathcal{A}}_H \mid x - u/2 \,\rangle \\
&= A_H(x,p).
\end{aligned}$$

□

**Corollary 3.** *For an operator $\widehat{\mathcal{A}}$, its trace $\mathrm{Tr}_x[\widehat{\mathcal{A}}] = \int \mathrm{d}^4 x\, \langle\, x \mid \widehat{\mathcal{A}} \mid x \,\rangle$ can be expressed in terms of the corresponding Weyl symbol $A(x,p)$ as follows:*

$$\mathrm{Tr}_x[\widehat{\mathcal{A}}] = (2\pi)^{-4} \int \mathrm{d}^4 x\, \mathrm{d}^4 p\, A(x,p). \tag{B.19}$$

*Proof.* After using Proposition 7, one integrates $A(x,p)$ over all phase space in order to obtain

$$\begin{aligned}(2\pi)^{-4}\int \mathrm{d}^4x\,\mathrm{d}^4p\,A(x,p) &= (2\pi)^{-4}\int \mathrm{d}^4x\,\mathrm{d}^4p\,\mathrm{d}^4s\,e^{is\cdot p}\,\langle\, x+s/2\mid \widehat{\mathcal{A}}\mid x-s/2\,\rangle\\ &= \int \mathrm{d}^4x\,\mathrm{d}^4s\,\delta^4(s)\,\langle\, x+s/2\mid \widehat{\mathcal{A}}\mid x-s/2\,\rangle\\ &= \mathrm{Tr}_x[\widehat{\mathcal{A}}].\end{aligned}$$

□

**Theorem 1 (Inverse Weyl transformation).** *Any linear operator $\widehat{\mathcal{A}}$ can be represented by a linear combination of Wigner operators $\widehat{\Delta}(x,p)$ so that*

$$\widehat{\mathcal{A}} = \mathsf{W}^{-1}[A(x,p)] \doteq \int \mathrm{d}^4x\,\mathrm{d}^4p\,A(x,p)\widehat{\Delta}(x,p), \tag{B.20}$$

*where $A(x,p)$ is the Weyl symbol (B.17) of the operator $\widehat{\mathcal{A}}$.*

*Proof.* Let us first show that Eq. (B.20) is self-consistent with the definition of the Weyl symbol (B.17). One first multiplies Eq. (B.20) by the Wigner operator $\widehat{\Delta}(x,p)$ from the right and calculates the trace. Using Lemma 2 leads to

$$\begin{aligned}(2\pi)^4\mathrm{Tr}_x[\widehat{\mathcal{A}}\widehat{\Delta}(x,p)] &= (2\pi)^4\mathrm{Tr}_x[\widehat{\Delta}(x,p)\int \mathrm{d}^4x'\,\mathrm{d}^4p'\,A(x',p')\widehat{\Delta}(x',p')]\\ &= (2\pi)^4\int \mathrm{d}^4x'\,\mathrm{d}^4p'\,A(x',p')\mathrm{Tr}_x[\widehat{\Delta}(x,p)\widehat{\Delta}(x',p')]\\ &= \int \mathrm{d}^4x'\,\mathrm{d}^4p'\,A(x',p')\,\delta^4(x-x')\,\delta^4(p-p')\\ &= A(x,p).\end{aligned}$$

Thus, Eq. (B.20) is consistent with the definition of the Weyl symbol (B.17). Upon inserting the Wigner function (B.12) and using Proposition 7, one obtains

$$\begin{aligned}\widehat{\mathcal{A}} &= \int \mathrm{d}^4x\,\mathrm{d}^4p\,A(x,p)\widehat{\Delta}(x,p)\\ &= \int \mathrm{d}^4x\,\mathrm{d}^4p\,\mathrm{d}^4s\,e^{is\cdot p}\,\langle\, x+s/2\mid \widehat{\mathcal{A}}\mid x-s/2\,\rangle\,\widehat{\Delta}(x,p)\\ &= (2\pi)^{-8}\int \mathrm{d}^4x\,\mathrm{d}^4p\,\mathrm{d}^4s\,\mathrm{d}^4u\,\mathrm{d}^4v\,e^{is\cdot p}\,\langle\, x+s/2\mid \widehat{\mathcal{A}}\mid x-s/2\,\rangle\,\widehat{\mathfrak{T}}(u,v)e^{iu\cdot p+v\cdot x}\\ &= (2\pi)^{-4}\int \mathrm{d}^4x\,\mathrm{d}^4s\,\mathrm{d}^4u\,\mathrm{d}^4v\,\delta^4(s+u)\,\langle\, x+s/2\mid \widehat{\mathcal{A}}\mid x-s/2\,\rangle\,\widehat{\mathfrak{T}}(u,v)e^{iv\cdot x}\\ &= (2\pi)^{-4}\int \mathrm{d}^4x\,\mathrm{d}^4u\,\mathrm{d}^4v\,\langle\, x-u/2\mid \widehat{\mathcal{A}}\mid x+u/2\,\rangle\,\widehat{\mathfrak{T}}(u,v)e^{iv\cdot x}.\end{aligned} \tag{B.21}$$

Now, let us use the following representation for the Heisenberg generating operator

$$\begin{aligned}\widehat{\mathcal{T}}(u,v) &= \int \mathrm{d}^4x'\,\mathrm{d}^4x''\,\,|\,x'\,\rangle\,\langle\,x'\,|\,\widehat{\mathcal{T}}(u,v)\,|\,x''\,\rangle\,\langle\,x''\,| \\ &= \int \mathrm{d}^4x'\,\mathrm{d}^4x''\,\delta^4(x'-x''+u)e^{-iv\cdot x''+iu\cdot v/2}\,|\,x'\,\rangle\,\langle\,x''\,|\,,\end{aligned}$$

which can be obtained by using Proposition 3. Substituting into Eq. (B.21) leads to

$$\begin{aligned}\widehat{\mathcal{A}} = (2\pi)^{-4} &\int \mathrm{d}^4x\,\mathrm{d}^4x'\,\mathrm{d}^4x''\,\mathrm{d}^4u\,\mathrm{d}^4v\,\langle\,x-u/2\,|\,\widehat{\mathcal{A}}\,|\,x+u/2\,\rangle \\ &\qquad\times\delta^4(x'-x''+u)\,|\,x'\,\rangle\,\langle\,x''\,|\,e^{iv\cdot(x-x''+u/2)} \\ &= \int \mathrm{d}^4x\,\mathrm{d}^4x'\,\mathrm{d}^4x''\,\mathrm{d}^4u\,\delta^4(x'-x''+u)\,\delta^4(x-x''+u/2)\,\langle\,x-u/2\,|\,\widehat{\mathcal{A}}\,|\,x+u/2\,\rangle\,|\,x'\,\rangle\,\langle\,x''\,| \\ &= \int \mathrm{d}^4x\,\mathrm{d}^4x'\,\mathrm{d}^4x''\,\delta^4[x-(x''+x')/2]\,\langle\,x-(x''-x')/2\,|\,\widehat{\mathcal{A}}\,|\,x+(x''-x')/2\,\rangle\,|\,x'\,\rangle\,\langle\,x''\,| \\ &= \int \mathrm{d}^4x'\,\mathrm{d}^4x''\,\,|\,x'\,\rangle\,\langle\,x'\,|\,\widehat{\mathcal{A}}\,|\,x''\,\rangle\,\langle\,x''\,| \\ &= \widehat{\mathcal{A}}.\end{aligned}$$

□

**Corollary 4.** *The inverse Weyl transform can also be written alternatively as*

$$\widehat{\mathcal{A}} = (2\pi)^{-4}\int \mathrm{d}^4x\,\mathrm{d}^4p\,\mathrm{d}^4s\,e^{is\cdot p}\,A(x,p)\,\,|\,x-s/2\,\rangle\,\langle\,x+s/2\,|\,. \tag{B.22}$$

*Proof.* Upon using Proposition 6, the Wigner operator can be alternatively written as

$$\widehat{\Delta}(x,p) = (2\pi)^{-4}\int \mathrm{d}^4s\,e^{is\cdot p}\,|\,x-s/2\,\rangle\,\langle\,x+s/2\,|$$

Inserting this representation of the Wigner operator into Eq. (B.20) leads to the desired result. □

**Corollary 5.** *In terms of the corresponding Weyl symbol $A(x,p)$, the projection of the operator $\widehat{\mathcal{A}}$ on the position eigenstates, $\mathcal{A}(x',x'') \doteq \langle\,x'\,|\,\widehat{\mathcal{A}}\,|\,x''\,\rangle$, is given by*

$$\mathcal{A}(x',x'') \doteq \langle\,x'\,|\,\widehat{\mathcal{A}}\,|\,x''\,\rangle = (2\pi)^{-4}\int \mathrm{d}^4p\,e^{-i(x'-x'')\cdot p}\,A\left(\frac{x'+x''}{2},p\right). \tag{B.23}$$

*Proof.* Upon using Proposition 5 and Theorem 1, one obtains

$$\begin{aligned}\mathcal{A}(x',x'') &= \int \mathrm{d}^4x\,\mathrm{d}^4p\,A(x,p)\,\langle\, x' \mid \widehat{\Delta}(x,p) \mid x''\,\rangle \\ &= (2\pi)^{-4}\int \mathrm{d}^4x\,\mathrm{d}^4p\,\mathrm{d}^4u\,A(x,p)\delta^4(x'-x''+u)\,\delta^4(x-x''+u/2)\,e^{iu\cdot p} \\ &= (2\pi)^{-4}\int \mathrm{d}^4x\,\mathrm{d}^4p\,A(x,p)\,\delta^4\big(x-(x'+x'')/2\big)\,e^{-i(x'-x'')\cdot p} \\ &= (2\pi)^{-4}\int \mathrm{d}^4p\,e^{-i(x'-x'')\cdot p}A\left(\frac{x'+x''}{2},p\right). \qquad \square\end{aligned}$$

So far, I have constructed the basis of the mathematical machinery of the Weyl symbol calculus. In the following, I shall present perhaps the most valuable tool yet, the Moyal product.

## B.5 Moyal product

**Lemma 3.** *The Wigner operators satisfy the following property:*

$$\begin{aligned}&\mathrm{Tr}_x[\widehat{\Delta}(x,p)\widehat{\Delta}(x',p')\widehat{\Delta}(x'',p'')] \\ &\qquad = (2\pi)^{-12}\int \mathrm{d}^4u\,\mathrm{d}^4v\,\delta^4(p'-p''+v/2)\delta^4(x'-x''-u/2)e^{iu\cdot(p-p'')+iv\cdot(x-x'')}. \qquad \text{(B.24)}\end{aligned}$$

*Proof.* Inserting the Wigner operators (B.12) and using Proposition 4 leads to

$$\begin{aligned}&\mathrm{Tr}_x[\widehat{\Delta}(x,p)\widehat{\Delta}(x',p')\widehat{\Delta}(x'',p'')] \\ &\quad = (2\pi)^{-24}\int \mathrm{d}^4u\,\mathrm{d}^4v\,\mathrm{d}^4u'\,\mathrm{d}^4v'\,\mathrm{d}^4u''\,\mathrm{d}^4v''\,e^{iu\cdot p+iv\cdot x}e^{iu'\cdot p'+iv'\cdot x'}e^{iu''\cdot p''+iv''\cdot x''} \\ &\qquad\qquad \times \mathrm{Tr}_x[\widehat{\mathfrak{T}}(u,v)\widehat{\mathfrak{T}}(u',v')\widehat{\mathfrak{T}}(u'',v'')] \\ &\quad = (2\pi)^{-24}\int \mathrm{d}^4u\,\mathrm{d}^4v\,\mathrm{d}^4u'\,\mathrm{d}^4v'\,\mathrm{d}^4u''\,\mathrm{d}^4v''\,e^{iu\cdot p+iv\cdot x}e^{iu'\cdot p'+iv'\cdot x'}e^{iu''\cdot p''+iv''\cdot x''} \\ &\qquad\qquad \times \mathrm{Tr}_x[\widehat{\mathfrak{T}}(u+u'+u'',v+v'+v'')]e^{i[v\cdot(u'+u'')-u\cdot(v'+v'')]/2}e^{i(v'\cdot u''-u'\cdot v'')/2}.\end{aligned}$$

Upon using Corollary 1 to evaluate the trace, one then obtains

$$
\begin{aligned}
&= (2\pi)^{-20}\int \mathrm{d}^4u\,\mathrm{d}^4v\,\mathrm{d}^4u'\,\mathrm{d}^4v'\,\mathrm{d}^4u''\,\mathrm{d}^4v''\,e^{iu\cdot p+iv\cdot x}e^{iu'\cdot p'+iv'\cdot x'}e^{iu''\cdot p''+iv''\cdot x''}\\
&\qquad\qquad \times \delta^4(u+u'+u'')\delta^4(v+v'+v'')e^{i[v\cdot(u'+u'')-u\cdot(v'+v'')]/2}e^{i(v'\cdot u''-u'\cdot v'')/2}\\
&= (2\pi)^{-20}\int \mathrm{d}^4u\,\mathrm{d}^4v\,\mathrm{d}^4u'\,\mathrm{d}^4v'\,e^{iu\cdot p+iv\cdot x}e^{iu'\cdot p'+iv'\cdot x'}e^{-i(u+u')\cdot p''-i(v+v')\cdot x''}e^{iu'\cdot v/2-iv'\cdot u/2}\\
&= (2\pi)^{-20}\int \mathrm{d}^4u\,\mathrm{d}^4v\,\mathrm{d}^4u'\,\mathrm{d}^4v'\,e^{iu'\cdot(p'-p''+v/2)}e^{iv'\cdot(x'-x''-u/2)}e^{iu\cdot(p-p'')+iv\cdot(x-x'')}\\
&= (2\pi)^{-12}\int \mathrm{d}^4u\,\mathrm{d}^4v\,\delta^4(p'-p''+v/2)\delta^4(x'-x''-u/2)e^{iu\cdot(p-p'')+iv\cdot(x-x'')}. \qquad \square
\end{aligned}
$$

**Theorem 2.** *Let $\widehat{\mathcal{A}}$, $\widehat{\mathcal{B}}$, and $\widehat{\mathcal{C}}$ be linear operators such that $\widehat{\mathcal{C}} = \widehat{\mathcal{A}}\widehat{\mathcal{B}}$. Then, the associated Weyl symbols $A(x,p)$, $B(x,p)$, and $C(x,p)$ satisfy*

$$C(x,p) = A(x,p) \star B(x,p). \tag{B.25}$$

*Here "$\star$" is the Moyal product, which is given by*

$$A(x,p)\star B(x,p) \doteq A(x,p)e^{i\widehat{\mathcal{L}}/2}B(x,p). \tag{B.26}$$

*Also, $\widehat{\mathcal{L}}$ is the Janus operator*

$$\widehat{\mathcal{L}} \doteq \frac{\overleftarrow{\partial}}{\partial p_\mu}\frac{\overrightarrow{\partial}}{\partial x^\mu} - \frac{\overleftarrow{\partial}}{\partial x^\mu}\frac{\overrightarrow{\partial}}{\partial p_\mu},$$

*where the arrows indicate the direction in which the derivatives act.*

*Proof.* Let us start from the definition of the Weyl symbol (B.17). One then uses Theorem 1 to express the operators $\widehat{\mathcal{B}}$ and $\widehat{\mathcal{C}}$ in terms of the Wigner operators. Then, one obtains

$$
\begin{aligned}
C(x,p) &= (2\pi)^4\mathrm{Tr}_x[\widehat{\Delta}(x,p)\widehat{\mathcal{B}}\widehat{\mathcal{C}}]\\
&= (2\pi)^4\int \mathrm{d}^4x'\,\mathrm{d}^4p'\,\mathrm{d}^4x''\,\mathrm{d}^4p''\,A(x',p')B(x'',p'')\mathrm{Tr}_x[\widehat{\Delta}(x,p)\widehat{\Delta}(x',p')\widehat{\Delta}(x'',p'')].
\end{aligned}
$$

Upon substituting the result in Lemma 3, one then obtains

$$\begin{aligned}
C(x,p) &= (2\pi)^{-8} \int \mathrm{d}^4x'\,\mathrm{d}^4p'\,\mathrm{d}^4x''\,\mathrm{d}^4p''\,\mathrm{d}^4u\,\mathrm{d}^4v\, A(x',p')B(x'',p'') \\
&\qquad \times \delta^4(p'-p''+v/2)\delta^4(x'-x''-u/2)e^{iu\cdot(p-p'')+iv\cdot(x-x'')} \\
&= (2\pi)^{-8} \int \mathrm{d}^4x'\,\mathrm{d}^4p'\,\mathrm{d}^4u\,\mathrm{d}^4v\, A(x',p')B(x'-u/2,p'+v/2)e^{iu\cdot(p-p')+iv\cdot(x-x')} \\
&= (2\pi)^{-8} \int \mathrm{d}^4x'\,\mathrm{d}^4p'\,\mathrm{d}^4u\,\mathrm{d}^4v\, A(x',p')e^{iu\cdot(p-p')+iv\cdot(x-x')} \\
&\qquad \times \sum_{n=0}^{\infty} \frac{1}{n!}\left(-\frac{u}{2}\cdot\frac{\partial}{\partial x'}+\frac{v}{2}\cdot\frac{\partial}{\partial p'}\right)^n B(x',p').
\end{aligned}$$

The next step consists in using the following Fourier integral

$$\begin{aligned}
&\int \mathrm{d}^4u\,\mathrm{d}^4v\, e^{iu\cdot(p-p')+iv\cdot(x-x')}\left(-\frac{u}{2}\cdot\frac{\partial}{\partial x'}+\frac{v}{2}\cdot\frac{\partial}{\partial p'}\right)^n B(x',p') \\
&\quad = \int \mathrm{d}^4u\,\mathrm{d}^4v \left[e^{iu\cdot(p-p')+iv\cdot(x-x')}\left(-\frac{i}{2}\frac{\overleftarrow{\partial}}{\partial p'}\cdot\frac{\overrightarrow{\partial}}{\partial x'}+\frac{i}{2}\frac{\overleftarrow{\partial}}{\partial x'}\cdot\frac{\overrightarrow{\partial}}{\partial p'}\right)^n B(x',p')\right] \\
&\quad = (2\pi)^8\left[\delta^4(x-x')\delta^4(p-p')\left(-\frac{i}{2}\frac{\overleftarrow{\partial}}{\partial p'}\cdot\frac{\overrightarrow{\partial}}{\partial x'}+\frac{i}{2}\frac{\overleftarrow{\partial}}{\partial x'}\cdot\frac{\overrightarrow{\partial}}{\partial p'}\right)^n B(x',p')\right],
\end{aligned}$$

where the arrows indicate the direction in which the derivatives act. Substituting the above gives

$$\begin{aligned}
C(x,p) &= \int \mathrm{d}^4x'\,\mathrm{d}^4p'\, A(x',p') \\
&\qquad \times \sum_{n=0}^{\infty}\frac{1}{n!}\left(\frac{i}{2}\right)^n\left[\delta^4(x-x')\delta^4(p-p')\left(-\frac{\overleftarrow{\partial}}{\partial p'}\cdot\frac{\overrightarrow{\partial}}{\partial x'}+\frac{\overleftarrow{\partial}}{\partial x'}\cdot\frac{\overrightarrow{\partial}}{\partial p'}\right)^n B(x',p')\right].
\end{aligned}$$

Integrating by parts leads to

$$\begin{aligned}
C(x,p) &= \int \mathrm{d}^4x'\,\mathrm{d}^4p'\,\delta^4(x-x')\delta^4(p-p') \\
&\qquad \times \sum_{n=0}^{\infty}\frac{1}{n!}\left(\frac{i}{2}\right)^n\left[A(x',p')\left(\frac{\overleftarrow{\partial}}{\partial p'}\cdot\frac{\overrightarrow{\partial}}{\partial x'}-\frac{\overleftarrow{\partial}}{\partial x'}\cdot\frac{\overrightarrow{\partial}}{\partial p'}\right)^n B(x',p')\right].
\end{aligned}$$

Finally, after integrating over all phase space, one obtains the Moyal product:

$$C(x,p) = A(x,p)\exp\left[\frac{i}{2}\left(\frac{\overleftarrow{\partial}}{\partial p}\cdot\frac{\overrightarrow{\partial}}{\partial x}-\frac{\overleftarrow{\partial}}{\partial x}\cdot\frac{\overrightarrow{\partial}}{\partial p}\right)\right]B(x,p). \qquad \square$$

**Corollary 6.** *The Moyal product is associative; i.e., for arbitrary Weyl symbols $A(x,p)$, $B(x,p)$, and $C(x,p)$, then*

$$A \star B \star C = (A \star B) \star C = A \star (B \star C). \tag{B.27}$$

*Proof.* One calculates the Weyl symbol of $\widehat{\mathcal{D}} \doteq \widehat{\mathcal{A}}\widehat{\mathcal{B}}\widehat{\mathcal{C}}$ and defines the operators $\widehat{\mathcal{E}} \doteq \widehat{\mathcal{A}}\widehat{\mathcal{B}}$ and $\widehat{\mathcal{F}} \doteq \widehat{\mathcal{B}}\widehat{\mathcal{C}}$. Hence, one writes $\widehat{\mathcal{D}} = \widehat{\mathcal{E}}\widehat{\mathcal{C}} = \widehat{\mathcal{A}}\widehat{\mathcal{F}}$. Upon using the Moyal product, one writes the Weyl transform of $\widehat{\mathcal{D}}$ as $D = E \star C = A \star F$. Substituting $E = A \star B$ and $F = B \star C$ leads to the claimed result. □

**Corollary 7.** *Let $A(x,p)$ and $B(x,p)$ be two Weyl symbols that vanish rapidly enough at infinity. The phase-space integral of the Moyal product of the two symbols equals the integral of the matrix product of these symbols:*

$$\int \mathrm{d}^4x\, \mathrm{d}^4p\, A(x,p) \star B(x,p) = \int \mathrm{d}^4x\, \mathrm{d}^4p\, A(x,p) B(x,p). \tag{B.28}$$

*Proof.* Starting from Theorem 2, one writes $A(x,p) \star B(x,p) = (2\pi)^4 \mathrm{Tr}_x[\widehat{\Delta}(x,p)\widehat{\mathcal{B}}\widehat{\mathcal{C}}]$, where $\widehat{\mathcal{A}}$ and $\widehat{\mathcal{B}}$ are the operators corresponding to $A(x,p)$ and $B(x,p)$, respectively. One uses Theorem 1 to express the operators $\widehat{\mathcal{A}}$ and $\widehat{\mathcal{B}}$ in terms of the Wigner operators. One obtains

$$\begin{aligned}\int \mathrm{d}^4x\, \mathrm{d}^4p\, A(x,p) \star B(x,p) = (2\pi)^4 \int & \mathrm{d}^4x\, \mathrm{d}^4p\, \mathrm{d}^4x'\, \mathrm{d}^4p'\, \mathrm{d}^4x''\, \mathrm{d}^4p'' \\ & \times A(x',p')B(x'',p'')\mathrm{Tr}_x[\widehat{\Delta}(x,p)\widehat{\Delta}(x',p')\widehat{\Delta}(x'',p'')].\end{aligned}$$

Assuming that the fields vanish rapidly enough at infinity so that the integrals over phase space converge to a finite number, one can use Fubini's theorem to change the order of integration. Hence, one can write

$$\begin{aligned}\int \mathrm{d}^4x\, \mathrm{d}^4p\, A(x,p) \star B(x,p) = (2\pi)^4 \int & \mathrm{d}^4x'\, \mathrm{d}^4p'\, \mathrm{d}^4x''\, \mathrm{d}^4p''\, A(x',p')B(x'',p'') \\ & \times \int \mathrm{d}^4x\, \mathrm{d}^4p\, \mathrm{Tr}_x[\widehat{\Delta}(x,p)\widehat{\Delta}(x',p')\widehat{\Delta}(x'',p'')].\end{aligned}$$

From Lemma 3, one has

$$\int \mathrm{d}^4x\, \mathrm{d}^4p\, \mathrm{Tr}_x[\widehat{\Delta}(x,p)\widehat{\Delta}(x',p')\widehat{\Delta}(x'',p'')] = (2\pi)^{-4}\delta^4(x'-x'')\,\delta^4(p'-p'').$$

After substituting into the above, one obtains

$$\int \mathrm{d}^4x\, \mathrm{d}^4p\, A(x,p) \star B(x,p) = \int \mathrm{d}^4x'\, \mathrm{d}^4p'\, \mathrm{d}^4x''\, \mathrm{d}^4p'' \delta^4(x'-x'')\,\delta^4(p'-p'')A(x',p')B(x'',p'').$$

Integrating over $x''$ and $p''$ leads to the desired result. □

## B.6 Moyal brackets

Now that the phase-space Moyal product has been obtained, I now introduce the Moyal brackets.

**Definition 5.** *The Moyal bracket is a suitably normalized antisymmetrization of the phase-space Moyal product "$\star$". For two arbitrary symbols $A(x,p)$ and $B(x,p)$, it is defined as*

$$\{\{A, B\}\} \doteq -i\,(A \star B - B \star A)\,. \tag{B.29}$$

**Proposition 8.** *Let $A(x,p)$ and $B(x,p)$ be two scalar Weyl symbols [i.e., $A(x,p)$ and $B(x,p)$ are not matrices]. The Moyal bracket $\{\{A, B\}\}$ can also be written as*

$$\{\{A, B\}\} = 2A(x,p)\sin\left(\frac{\widehat{\mathcal{L}}}{2}\right)B(x,p). \tag{B.30}$$

*For this reason, the Moyal bracket is often referred as the "sine bracket."*

*Proof.* Upon inserting the definition of the Moyal product (B.26) into Eq. (B.29), one obtains $\{\{A, B\}\} = -i(Ae^{i\widehat{\mathcal{L}}/2}B - Be^{i\widehat{\mathcal{L}}/2}A)$. From the definition of the Janus operator, $Be^{i\widehat{\mathcal{L}}/2}A = Ae^{-i\widehat{\mathcal{L}}/2}B$. Hence, $\{\{A, B\}\} = -iA(e^{i\widehat{\mathcal{L}}/2} - e^{-i\widehat{\mathcal{L}}/2})B$. Using the relation between the sine and exponential functions leads to $\{\{A, B\}\} = 2A\sin(\widehat{\mathcal{L}}/2)B$. □

**Definition 6.** *The "cosine bracket" is a suitably normalized symmetrization of the phase-space Moyal product "$\star$". For two arbitrary symbols $A(x,p)$ and $B(x,p)$, it is defined as*

$$[[A, B]] \doteq A \star B + B \star A. \tag{B.31}$$

**Proposition 9.** *Let $A(x,p)$ and $B(x,p)$ be two scalar Weyl symbols. The symmetrized Moyal bracket $[[A, B]]$ can also be written as*

$$[[A, B]] = 2A(x,p)\cos\left(\frac{\widehat{\mathcal{L}}}{2}\right)B(x,p). \tag{B.32}$$

*Proof.* The proof is identical to that of Proposition 8. □

## B.7 Applications to wave dynamics

In this following, I shall derive some basic results regarding wave dynamics. I shall use the machinery that I have presented so far.

**Proposition 10.** *Let $|\,\Psi\,\rangle$ represent a scalar wave that satisfies the linear wave equation*

$$\widehat{\mathfrak{D}}\,|\,\Psi\,\rangle = 0, \tag{B.33}$$

*where $\widehat{\mathfrak{D}}$ is a scalar operator, called the dispersion operator. Let $\widehat{\mathfrak{D}}_H$ and $\widehat{\mathfrak{D}}_A$ be the Hermitian and anti-Hermitian parts of the dispersion operator, respectively. Let $\widehat{\mathfrak{N}} \doteq |\,\Psi\,\rangle\langle\,\Psi\,|$ be the density operator. Then, the Wigner function $W_\Psi(x,p)$, which is defined as a properly normalized Weyl symbol of the density operator,*

$$W_\Psi(x,p) \doteq \mathrm{Tr}_x[\widehat{\Delta}(x,p)\widehat{\mathfrak{N}}] = \frac{1}{(2\pi)^4}\int \mathrm{d}^4s\, e^{ip\cdot s}\,\langle\, x+s/2\,|\,\Psi\,\rangle\langle\,\Psi\,|\,x-s/2\,\rangle \tag{B.34}$$

*satisfies*

$$\{\{D_H, W_\Psi\}\} = -[[D_A, W_\Psi]], \tag{B.35}$$

*where $D_H(x,p)$ and $D_A(x,p)$ are the Weyl symbols of $\widehat{\mathfrak{D}}_H$ and $\widehat{\mathfrak{D}}_A$, respectively.*

*Proof.* Multiplying Eq. (B.33) by the ket state $\langle\,\Psi\,|$ leads to $(\widehat{\mathfrak{D}}_H + i\widehat{\mathfrak{D}}_A)\,\widehat{\mathfrak{N}} = 0$. The adjoint equation is $\widehat{\mathfrak{N}}\,(\widehat{\mathfrak{D}}_H - i\widehat{\mathfrak{D}}_A) = 0$. After subtracting both equations, one obtains $-i(\widehat{\mathfrak{D}}_H\,\widehat{\mathfrak{N}} - \widehat{\mathfrak{N}}\,\widehat{\mathfrak{D}}_H) = -(\widehat{\mathfrak{D}}_A\,\widehat{\mathfrak{N}} + \widehat{\mathfrak{N}}\,\widehat{\mathfrak{D}}_A)$. Upon using the Moyal product (Theorem 2) to calculate the Weyl transform, one obtains $-i(D_H \star W_\Psi - W_\Psi \star D_H) = -(D_A \star W_\Psi + W_\Psi \star D_A)$. Substituting Definitions (5) and (6) leads to the desired result. □

**Theorem 3.** *Let $|\,\Psi\,\rangle$ represent a scalar wave that satisfies the linear wave equation*

$$\widehat{\mathfrak{D}}\,|\,\Psi\,\rangle = 0,$$

*where $\widehat{\mathfrak{D}}$ is the dispersion operator. Let $\widehat{\mathfrak{D}}$ be in the symplectic form so that its corresponding symbol is written as $D(x,p) = p_0 - H(t,\mathbf{x},\mathbf{p})$. The Weyl symbol $H(t,\mathbf{x},\mathbf{p})$ is called the Hamiltonian of the system and is independent of the temporal momentum coordinate $p_0$. The Wigner function in the six-dimensional phase-space,*

$$F_\Psi(t,\mathbf{x},\mathbf{p}) \doteq \int \mathrm{d}p_0\, W_\Psi(x,p) = \frac{1}{(2\pi)^3}\int \mathrm{d}^3\mathbf{s}\, e^{-i\mathbf{p}\cdot\mathbf{s}}\,\langle\,(t,\mathbf{x}+\mathbf{s}/2)\,|\,\Psi\,\rangle\langle\,\Psi\,|\,(t,\mathbf{x}-\mathbf{s}/2)\,\rangle\,, \tag{B.36}$$

*satisfies the "Quantum Liouville," or "Wigner–Moyal" equation*

$$\frac{\partial}{\partial t}F_\Psi(t,\mathbf{x},\mathbf{p}) + \{\{F_\Psi(t,\mathbf{x},\mathbf{p}), H_H(t,\mathbf{x},\mathbf{p})\}\} = [[F_\Psi(t,\mathbf{x},\mathbf{p}), H_A(t,\mathbf{x},\mathbf{p})]], \tag{B.37}$$

*where $H_H(t,\mathbf{x},\mathbf{p})$ and $H_A(t,\mathbf{x},\mathbf{p})$ are the Hermitian and anti-Hermitian components of $H(t,\mathbf{x},\mathbf{p})$.*

*Proof.* From Proposition 10, the eight-dimensional Wigner function satisfies

$$\{\{p_0, W_\Psi\}\} - \{\{H_H, W_\Psi\}\} = [[H_A, W_\Psi]], \tag{B.38}$$

One then integrates the equation over the temporal momentum coordinate $p_0$. Upon using Proposition 8 and calculating the Moyal sine bracket, ones writes the first term as

$$\int \mathrm{d}p_0 \, \{\{p_0, W_\Psi\}\} = \int \mathrm{d}p_0 \, \frac{\partial}{\partial t} W_\Psi(x,p) = \frac{\partial}{\partial t} F_\Psi(t, \mathbf{x}, \mathbf{p}).$$

For the second term on the left-hand side of Eq. (B.38), note that $H_H(t, \mathbf{x}, \mathbf{p})$ does not depend on $p_0$. Upon using the anti-symmetry of the sine bracket, one obtains

$$\begin{aligned} - \int \mathrm{d}p_0 \, \{\{H_H(t, \mathbf{x}, \mathbf{p}), W_\Psi(x,p)\}\} &= \{\{ \, [ \int \mathrm{d}p_0 \, W_\Psi(x,p) ] \, , \, H_H(t, \mathbf{x}, \mathbf{p}) \, \}\} \\ &= \{\{F_\Psi(t, \mathbf{x}, \mathbf{p}), H_H(t, \mathbf{x}, \mathbf{p})\}\}, \end{aligned}$$

where I assumed that $W_\Psi(x,p)$ and its derivatives vanish at $p_0 \to \pm\infty$. Similar arguments follow for the right-hand side of Eq. (B.38). This gives the desired result in Eq. (B.35). □

**Remark 2.** *The Wigner–Moyal equation (B.35) is an evolution equation for the Wigner function* $F_\Psi(t, \mathbf{x}, \mathbf{p})$ *in the six-dimensional phase-space, and it is the quantum analogue to the well-known Liouville equation of classical physics. In certain limits, this equation can be reduced to the wave kinetic equation.*

# Appendix C

# Auxiliary calculations

## C.1 Point-particle Lagrangian model for the Dirac electron

Here I include additional details on some of the calculations reported in Chapter 5. First, I present the calculation of the spin-coupling matrix $\mathcal{U}$. Afterwards, I show how to obtain the contributions of the anomalous magnetic moment to the action functional.

### C.1.1 Calculation of the spin-coupling matrix

As a reminder from Chapter 5, the eigenvalue corresponding to the particle states is given by $\lambda(x,p) = \pi_0(t,\mathbf{x},p_0) - \varepsilon(t,\mathbf{x},\mathbf{p})$, where $\varepsilon(t,\mathbf{x},\mathbf{p}) = (\boldsymbol{\pi}^2 + m^2)^{1/2}$ is the particle kinetic energy and $\pi_\mu(x,p) = p_\mu - qA_\mu(x)$ is the particle kinetic four-momentum. The dispersion symbol is $D(x,p) = p_0\mathbb{I}_4 - H(t,\mathbf{x},\mathbf{p})$, where the Hamiltonian is given by $H(t,\mathbf{x},\mathbf{p}) = \boldsymbol{\alpha}\cdot\boldsymbol{\pi} + \beta m + qV\mathbb{I}_4$. The $4\times 4$ Hermitian matrices $\boldsymbol{\alpha}$ and $\beta$ are introduced in Sec. 5.2.1. The eigenmode matrix $\Xi(t,\mathbf{x},\mathbf{p})$ is

$$\Xi(t,\mathbf{x},\mathbf{p}) = C(t,\mathbf{x},\mathbf{p}) \begin{pmatrix} \mathbb{I}_2 \\ \frac{\boldsymbol{\pi}\cdot\boldsymbol{\sigma}}{m+\varepsilon} \end{pmatrix}, \tag{C.1}$$

where the normalization term is $C(t,\mathbf{x},\mathbf{p}) \doteq [(m+\varepsilon)/(2\varepsilon)]^{1/2}$. Now, let us calculate the corresponding spin-coupling matrix (4.35). The first two terms can be concisely written by using the eight-dimensional phase-space Poisson bracket. One obtains

$$
\begin{aligned}
&\left(-\frac{\partial\lambda}{\partial p_\mu}\right)\left(\Xi^\dagger\frac{\partial\Xi}{\partial x^\mu}\right)_A+\left(\frac{\partial\lambda}{\partial x^\mu}\right)\left(\Xi^\dagger\frac{\partial\Xi}{\partial p_\mu}\right)_A\\
&\qquad=-\left(\Xi^\dagger\{\lambda,\Xi\}\right)_A\\
&\qquad=-\left(\Xi^\dagger\left\{\lambda,\left[C\begin{pmatrix}\mathbb{I}_2, & \frac{\boldsymbol{\pi}\cdot\boldsymbol{\sigma}}{m+\varepsilon}\end{pmatrix}^{\mathrm{T}}\right]\right\}\right)_A\\
&\qquad=-\left(C\begin{pmatrix}\mathbb{I}_2, & \frac{\boldsymbol{\pi}\cdot\boldsymbol{\sigma}}{m+\varepsilon}\end{pmatrix}\left[\{\lambda,C\}\begin{pmatrix}\mathbb{I}_2, & \frac{\boldsymbol{\pi}\cdot\boldsymbol{\sigma}}{m+\varepsilon}\end{pmatrix}^{\mathrm{T}}+C\left\{\lambda,\begin{pmatrix}\mathbb{I}_2, & \frac{\boldsymbol{\pi}\cdot\boldsymbol{\sigma}}{m+\varepsilon}\end{pmatrix}^{\mathrm{T}}\right\}\right]\right)_A\\
&\qquad=-C\left\{\lambda,C\right\}\left(\mathbb{I}_2+\frac{\boldsymbol{\pi}\cdot\boldsymbol{\sigma}}{m+\varepsilon}\frac{\boldsymbol{\pi}\cdot\boldsymbol{\sigma}}{m+\varepsilon}\right)_A-C^2\left(\frac{\boldsymbol{\pi}\cdot\boldsymbol{\sigma}}{m+\varepsilon}\left\{\lambda,\frac{\boldsymbol{\pi}\cdot\boldsymbol{\sigma}}{m+\varepsilon}\right\}\right)_A. \qquad \text{(C.2)}
\end{aligned}
$$

The Pauli matrices satisfy the following property: if $\mathbf{a}$ and $\mathbf{b}$ are two arbitrary three-component vectors; then $(\mathbf{a}\cdot\boldsymbol{\sigma})(\mathbf{b}\cdot\boldsymbol{\sigma})=(\mathbf{a}\cdot\mathbf{b})\mathbb{I}_2+i(\mathbf{a}\times\mathbf{b})\cdot\boldsymbol{\sigma}$. Upon substituting this property, one obtains

$$
\begin{aligned}
-\left(\Xi^\dagger\{\lambda,\Xi\}\right)_A=&-\frac{1}{2}\left\{\lambda,C^2\right\}\left(1+\frac{\boldsymbol{\pi}^2}{(m+\varepsilon)^2}\right)_A\mathbb{I}_2\\
&-C^2\left(\frac{\boldsymbol{\pi}}{m+\varepsilon}\cdot\left\{\lambda,\frac{\boldsymbol{\pi}}{m+\varepsilon}\right\}\mathbb{I}_2+i\frac{\boldsymbol{\pi}}{m+\varepsilon}\times\left\{\lambda,\frac{\boldsymbol{\pi}}{m+\varepsilon}\right\}\cdot\boldsymbol{\sigma}\right)_A. \qquad \text{(C.3)}
\end{aligned}
$$

Here only the last the last term inside the brackets of the previous equation can lead to an anti-Hermitian term. Thus, one obtains

$$
-\left(\Xi^\dagger\{\lambda,\Xi\}\right)_A=-\frac{\boldsymbol{\pi}\times\{\pi_0-\varepsilon,\boldsymbol{\pi}\}}{2\varepsilon(m+\varepsilon)}\cdot\boldsymbol{\sigma}, \qquad \text{(C.4)}
$$

where I substituted the expressions for $C(t,\mathbf{x},\mathbf{p})$ and $\lambda(x,p)$.

Let us now calculate the last term in Eq. (4.35). First, note that the matrix $D(x,p)-\lambda(x,p)\mathbb{I}_4$ in Eq. (4.35) is explicitly written as

$$
D(x,p)-\lambda(x,p)\mathbb{I}_4=\begin{pmatrix}(\varepsilon-m)\mathbb{I}_2 & -\boldsymbol{\pi}\cdot\boldsymbol{\sigma}\\ -\boldsymbol{\pi}\cdot\boldsymbol{\sigma} & (\varepsilon+m)\mathbb{I}_2\end{pmatrix}. \qquad \text{(C.5)}
$$

When substituting into the last term in Eq. (4.35), one obtains

$$\left(\frac{\partial\Xi^\dagger}{\partial p_\mu}(D-\lambda\mathbb{I}_4)\frac{\partial\Xi}{\partial x^\mu}\right)_A = \left\{\frac{\partial\Xi^\dagger}{\partial p_\mu}\begin{pmatrix}(\varepsilon-m)\mathbb{I}_2 & -\boldsymbol{\pi}\cdot\boldsymbol{\sigma}\\ -\boldsymbol{\pi}\cdot\boldsymbol{\sigma} & (\varepsilon+m)\mathbb{I}_2\end{pmatrix}\left[\frac{\partial C}{\partial x^\mu}\begin{pmatrix}\mathbb{I}_2\\ \frac{\boldsymbol{\pi}\cdot\boldsymbol{\sigma}}{m+\varepsilon}\end{pmatrix}+C\frac{\partial}{\partial x^\mu}\begin{pmatrix}\mathbb{I}_2\\ \frac{\boldsymbol{\pi}\cdot\boldsymbol{\sigma}}{m+\varepsilon}\end{pmatrix}\right]\right\}_A$$

$$= \frac{\partial C}{\partial x^\mu}\left[\frac{\partial\Xi^\dagger}{\partial p_\mu}\begin{pmatrix}\left(\varepsilon-m-\frac{\boldsymbol{\pi}^2}{m+\varepsilon}\right)\mathbb{I}_2\\ 0\end{pmatrix}\right]_A + C\left[\frac{\partial\Xi^\dagger}{\partial p_\mu}\begin{pmatrix}-(\boldsymbol{\pi}\cdot\boldsymbol{\sigma})\frac{\partial}{\partial x^\mu}\frac{\boldsymbol{\pi}\cdot\boldsymbol{\sigma}}{m+\varepsilon}\\ (m+\varepsilon)\frac{\partial}{\partial x^\mu}\frac{\boldsymbol{\pi}\cdot\boldsymbol{\sigma}}{m+\varepsilon}\end{pmatrix}\right]_A$$

$$= C\left[\frac{\partial\Xi^\dagger}{\partial p_\mu}\begin{pmatrix}-(\boldsymbol{\pi}\cdot\boldsymbol{\sigma})\frac{\partial}{\partial x^\mu}\frac{\boldsymbol{\pi}\cdot\boldsymbol{\sigma}}{m+\varepsilon}\\ (m+\varepsilon)\frac{\partial}{\partial x^\mu}\frac{\boldsymbol{\pi}\cdot\boldsymbol{\sigma}}{m+\varepsilon}\end{pmatrix}\right]_A$$

$$= C\left\{\left[\frac{\partial C}{\partial p_\mu}\begin{pmatrix}\mathbb{I}_2, & \frac{\boldsymbol{\pi}\cdot\boldsymbol{\sigma}}{m+\varepsilon}\end{pmatrix}+C\frac{\partial}{\partial p_\mu}\begin{pmatrix}\mathbb{I}_2, & \frac{\boldsymbol{\pi}\cdot\boldsymbol{\sigma}}{m+\varepsilon}\end{pmatrix}\right]\begin{pmatrix}-(\boldsymbol{\pi}\cdot\boldsymbol{\sigma})\frac{\partial}{\partial x^\mu}\frac{\boldsymbol{\pi}\cdot\boldsymbol{\sigma}}{m+\varepsilon}\\ (m+\varepsilon)\frac{\partial}{\partial x^\mu}\frac{\boldsymbol{\pi}\cdot\boldsymbol{\sigma}}{m+\varepsilon}\end{pmatrix}\right\}_A$$

$$= \frac{(m+\varepsilon)^2}{2\varepsilon}\left[\frac{\partial}{\partial p_\mu}\left(\frac{\boldsymbol{\pi}\cdot\boldsymbol{\sigma}}{m+\varepsilon}\right)\frac{\partial}{\partial x^\mu}\left(\frac{\boldsymbol{\pi}\cdot\boldsymbol{\sigma}}{m+\varepsilon}\right)\right]_A$$

$$= \frac{(m+\varepsilon)^2}{2\varepsilon}\left\{\left(\frac{1}{m+\varepsilon}\frac{\partial}{\partial p_\mu}(\boldsymbol{\pi}\cdot\boldsymbol{\sigma})-\frac{\boldsymbol{\pi}\cdot\boldsymbol{\sigma}}{(m+\varepsilon)^2}\frac{\partial\varepsilon}{\partial p_\mu}\right)\frac{\partial}{\partial x^\mu}\left(\frac{\boldsymbol{\pi}\cdot\boldsymbol{\sigma}}{m+\varepsilon}\right)\right\}_A$$

$$= \frac{1}{2\varepsilon}\left(\frac{\partial}{\partial p_\mu}(\boldsymbol{\pi}\cdot\boldsymbol{\sigma})\frac{\partial}{\partial x^\mu}(\boldsymbol{\pi}\cdot\boldsymbol{\sigma})\right)_A - \frac{1}{2\varepsilon(m+\varepsilon)}\left(\frac{\partial}{\partial p_\mu}(\boldsymbol{\pi}\cdot\boldsymbol{\sigma})\frac{\partial\varepsilon}{\partial x^\mu}(\boldsymbol{\pi}\cdot\boldsymbol{\sigma})\right)_A$$

$$- \frac{1}{2\varepsilon(m+\varepsilon)}\left((\boldsymbol{\pi}\cdot\boldsymbol{\sigma})\frac{\partial\varepsilon}{\partial p_\mu}\frac{\partial}{\partial x^\mu}(\boldsymbol{\pi}\cdot\boldsymbol{\sigma})\right)_A$$

$$= \frac{1}{2\varepsilon}\left(\frac{\partial\boldsymbol{\pi}}{\partial p_\mu}\times\frac{\partial\boldsymbol{\pi}}{\partial x^\mu}\right)\cdot\boldsymbol{\sigma} - \frac{1}{2\varepsilon(m+\varepsilon)}\left(\frac{\partial\varepsilon}{\partial x^\mu}\frac{\partial\boldsymbol{\pi}}{\partial p_\mu}\times\boldsymbol{\pi}\right)\cdot\boldsymbol{\sigma}$$

$$- \frac{1}{2\varepsilon(m+\varepsilon)}\left(\frac{\partial\varepsilon}{\partial p_\mu}\boldsymbol{\pi}\times\frac{\partial\boldsymbol{\pi}}{\partial x^\mu}\right)\cdot\boldsymbol{\sigma}$$

$$= \frac{1}{2\varepsilon}\left(\frac{\partial\boldsymbol{\pi}}{\partial p_\mu}\times\frac{\partial\boldsymbol{\pi}}{\partial x^\mu}\right)\cdot\boldsymbol{\sigma} - \frac{1}{2\varepsilon(m+\varepsilon)}(\boldsymbol{\pi}\times\{\varepsilon,\boldsymbol{\pi}\})\cdot\boldsymbol{\sigma}. \tag{C.6}$$

Summing Eqs. (C.4) and (C.6) leads to the spin-coupling matrix

$$\mathcal{U}(t,\mathbf{x},\mathbf{p}) = -\frac{\boldsymbol{\pi}\times\{\pi_0,\boldsymbol{\pi}\}}{2\varepsilon(m+\varepsilon)}\cdot\boldsymbol{\sigma} + \frac{1}{2\varepsilon}\left(\frac{\partial\boldsymbol{\pi}}{\partial p_\mu}\times\frac{\partial\boldsymbol{\pi}}{\partial x^\mu}\right)\cdot\boldsymbol{\sigma}. \tag{C.7}$$

Let us now calculate the remaining Poisson bracket. In terms of components, one has

$$\{\pi_0,\boldsymbol{\pi}\}^i = \frac{\partial\pi_0}{\partial\pi_\mu}\frac{\partial\pi^i}{\partial x^\mu} - \frac{\partial\pi^0}{\partial x_\mu}\frac{\partial\pi^i}{\partial p^\mu} = \frac{\partial\pi^i}{\partial x^0} + \frac{\partial\pi^0}{\partial x^j}\frac{\partial\pi^i}{\partial p^j} = \frac{\partial\pi^i}{\partial t} + \frac{\partial\pi^0}{\partial x^i} = q(-\partial_t\mathbf{A} - q\boldsymbol{\nabla}V)^i. \tag{C.8}$$

In a similar manner, one obtains

$$\left(\frac{\partial\boldsymbol{\pi}}{\partial p_\mu}\times\frac{\partial\boldsymbol{\pi}}{\partial x^\mu}\right)^i = -\varepsilon^{ijk}\frac{\partial\pi^j}{\partial p^l}\frac{\partial\pi^k}{\partial x^l} = -\varepsilon^{ijk}q\frac{\partial\pi^k}{\partial x^j} = \varepsilon^{ijk}q\frac{\partial A^k}{\partial x^j} = q(\boldsymbol{\nabla}\times\mathbf{A})^i. \tag{C.9}$$

Hence, substituting Eqs. (C.8) and (C.9) into Eq. (C.7) leads to

$$\mathcal{U}(t,\mathbf{x},\mathbf{p}) = \frac{q}{2\varepsilon}\left(\mathbf{B} - \frac{\boldsymbol{\pi}\times\mathbf{E}}{m+\varepsilon}\right)\cdot\boldsymbol{\sigma}, \tag{C.10}$$

where $\mathbf{E}(t,\mathbf{x}) \doteq -\partial_t\mathbf{A} - \boldsymbol{\nabla}V$ is the electric field and $\mathbf{B}(t,\mathbf{x}) \doteq \boldsymbol{\nabla}\times\mathbf{A}$ is the magnetic field.

### C.1.2 Anomalous magnetic moment

Below, I present the calculation of the term $\Xi^\dagger\beta\sigma_{\mu\nu}F^{\mu\nu}\Xi$ in Eq. (5.27):

$$\begin{aligned}
&\Xi^\dagger(t,\mathbf{x},\mathbf{p})\beta\sigma_{\mu\nu}F^{\mu\nu}(t,\mathbf{x})\Xi(t,\mathbf{x},\mathbf{p})\\
&\quad= \frac{m+\varepsilon}{\varepsilon}\begin{pmatrix}\mathbb{I}_2 & \frac{\boldsymbol{\pi}\cdot\boldsymbol{\sigma}}{m+\varepsilon}\end{pmatrix}\begin{pmatrix}\mathbb{I}_2 & 0\\ 0 & -\mathbb{I}_2\end{pmatrix}\begin{pmatrix}-\mathbf{B}\cdot\boldsymbol{\sigma} & i\mathbf{E}\cdot\boldsymbol{\sigma}\\ i\mathbf{E}\cdot\boldsymbol{\sigma} & -\mathbf{B}\cdot\boldsymbol{\sigma}\end{pmatrix}\begin{pmatrix}\mathbb{I}_2\\ \frac{\boldsymbol{\pi}\cdot\boldsymbol{\sigma}}{m+\varepsilon}\end{pmatrix}\\
&\quad= -\frac{m+\varepsilon}{\varepsilon}\begin{pmatrix}\mathbb{I}_2 & \frac{\boldsymbol{\pi}\cdot\boldsymbol{\sigma}}{m+\varepsilon}\end{pmatrix}\begin{pmatrix}\mathbf{B}\cdot\boldsymbol{\sigma} - i\mathbf{E}\cdot\boldsymbol{\sigma}\,\frac{\boldsymbol{\pi}\cdot\boldsymbol{\sigma}}{m+\varepsilon}\\ i\mathbf{E}\cdot\boldsymbol{\sigma} - \mathbf{B}\cdot\boldsymbol{\sigma}\,\frac{\boldsymbol{\pi}\cdot\boldsymbol{\sigma}}{m+\varepsilon}\end{pmatrix}\\
&\quad= -\frac{m+\varepsilon}{\varepsilon}\begin{pmatrix}\mathbb{I}_2 & \frac{\boldsymbol{\pi}\cdot\boldsymbol{\sigma}}{m+\varepsilon}\end{pmatrix}\begin{pmatrix}\mathbf{B}\cdot\boldsymbol{\sigma} - \frac{i\mathbf{E}\cdot\boldsymbol{\pi}}{m+\varepsilon} + \frac{(\mathbf{E}\times\boldsymbol{\pi})\cdot\boldsymbol{\sigma}}{m+\varepsilon}\\ i\mathbf{E}\cdot\boldsymbol{\sigma} - \frac{\mathbf{B}\cdot\boldsymbol{\pi}}{m+\varepsilon} - \frac{i(\mathbf{B}\times\boldsymbol{\pi})\cdot\boldsymbol{\sigma}}{m+\varepsilon}\end{pmatrix}\\
&\quad= -\frac{m+\varepsilon}{\varepsilon}\left(\mathbf{B}\cdot\boldsymbol{\sigma} - \frac{i\mathbf{E}\cdot\boldsymbol{\pi}}{m+\varepsilon} + \frac{(\mathbf{E}\times\boldsymbol{\pi})\cdot\boldsymbol{\sigma}}{m+\varepsilon} + \frac{i\boldsymbol{\pi}\cdot\boldsymbol{\sigma}}{m+\varepsilon}\mathbf{E}\cdot\boldsymbol{\sigma}\right.\\
&\qquad\qquad\left. - \frac{(\boldsymbol{\pi}\cdot\mathbf{B})(\boldsymbol{\pi}\cdot\boldsymbol{\sigma})}{(m+\varepsilon)^2} - \frac{i\boldsymbol{\pi}\cdot\boldsymbol{\sigma}}{m+\varepsilon}\frac{(\mathbf{B}\times\boldsymbol{\pi})\cdot\boldsymbol{\sigma}}{m+\varepsilon}\right)\\
&\quad= -\frac{m+\varepsilon}{\varepsilon}\left(\mathbf{B}\cdot\boldsymbol{\sigma} - 2\frac{(\boldsymbol{\pi}\times\mathbf{E})\cdot\boldsymbol{\sigma}}{m+\varepsilon} - \frac{(\boldsymbol{\pi}\cdot\mathbf{B})(\boldsymbol{\pi}\cdot\boldsymbol{\sigma})}{(m+\varepsilon)^2} + \frac{[\boldsymbol{\pi}\times(\mathbf{B}\times\boldsymbol{\pi})]\cdot\boldsymbol{\sigma}}{(m+\varepsilon)^2}\right)\\
&\quad= -\frac{m+\varepsilon}{\varepsilon}\left(\mathbf{B}\cdot\boldsymbol{\sigma} - 2\frac{(\boldsymbol{\pi}\times\mathbf{E})\cdot\boldsymbol{\sigma}}{m+\varepsilon} - \frac{(\boldsymbol{\pi}\cdot\mathbf{B})(\boldsymbol{\pi}\cdot\boldsymbol{\sigma})}{(m+\varepsilon)^2} + \frac{\boldsymbol{\pi}^2(\mathbf{B}\cdot\boldsymbol{\sigma}) - (\boldsymbol{\pi}\cdot\mathbf{B})(\boldsymbol{\pi}\cdot\boldsymbol{\sigma})}{(m+\varepsilon)^2}\right)\\
&\quad= -\frac{m+\varepsilon}{\varepsilon}\left(\mathbf{B}\cdot\boldsymbol{\sigma} - 2\frac{(\boldsymbol{\pi}\times\mathbf{E})\cdot\boldsymbol{\sigma}}{m+\varepsilon} - 2\frac{(\boldsymbol{\pi}\cdot\mathbf{B})(\boldsymbol{\pi}\cdot\boldsymbol{\sigma})}{(m+\varepsilon)^2} + \frac{\boldsymbol{\pi}^2(\mathbf{B}\cdot\boldsymbol{\sigma})}{(m+\varepsilon)^2}\right)\\
&\quad= -\frac{m+\varepsilon}{\varepsilon}\left(\frac{2\varepsilon}{m+\varepsilon}\mathbf{B}\cdot\boldsymbol{\sigma} - 2\frac{(\boldsymbol{\pi}\times\mathbf{E})\cdot\boldsymbol{\sigma}}{m+\varepsilon} - 2\frac{(\boldsymbol{\pi}\cdot\mathbf{B})(\boldsymbol{\pi}\cdot\boldsymbol{\sigma})}{(m+\varepsilon)^2}\right)\\
&\quad= -2\left(\mathbf{B} - \frac{\boldsymbol{\pi}\times\mathbf{E}}{\varepsilon} - \frac{(\boldsymbol{\pi}\cdot\mathbf{B})\boldsymbol{\pi}}{\varepsilon(m+\varepsilon)}\right)\cdot\boldsymbol{\sigma}.
\end{aligned} \tag{C.11}$$

## C.2 Ponderomotive dynamics of waves

Here I include additional details on some of the calculations reported in Chapter 7.

### C.2.1 Effective dispersion symbol

Here I present some intermediate results needed in order to obtain Eq. (7.23). Let $A(x,p)$ be a slowly varying function in spacetime. Also, let $\Theta(x)$ be a fast oscillating function and $K_\mu(x) \doteq -\partial_\mu \Theta$ be a slowly varying function. Upon substituting the expression for the Moyal product, one obtains

$$\begin{aligned}
A(x,p) \star e^{i\Theta} &= A(x,p) e^{i\widehat{\mathcal{L}}/2} e^{i\Theta} \\
&= A\left(x,p\right) \left( \sum_{n=0}^{\infty} \frac{i^n}{2^n n!} \frac{\overleftarrow{\partial}^n}{\partial p^n} \cdot \frac{\overrightarrow{\partial}^n}{\partial x^n} \right) e^{i\Theta} \\
&= A\left( x, p + \frac{i}{2} \frac{\overrightarrow{\partial}}{\partial x} \right) e^{i\Theta} \\
&= A\left( x, p - \frac{\partial \Theta}{\partial x} \right) e^{i\Theta} + \mathcal{O}(\epsilon_{\mathrm{mw}}) \\
&= A(x, p + K/2) e^{i\Theta} + \mathcal{O}(\epsilon_{\mathrm{mw}}),
\end{aligned} \tag{C.12}$$

where the symbol "·" denotes contraction. Similarly,

$$e^{i\Theta} \star A(x,p) = A(x, p - K/2) e^{i\Theta} + \mathcal{O}(\epsilon_{\mathrm{mw}}). \tag{C.13}$$

Upon substituting these equations, one obtains the result in Eq. (7.23).

For the calculation shown in Eq. (7.25), one needs the following result:

$$\begin{aligned}
& A(x,p) e^{i\Theta_1} \star B(x,p) e^{i\Theta_2} \\
&= A(x,p) e^{i\Theta_1} e^{i\widehat{\mathcal{L}}/2} B(x,p) e^{i\Theta_2} \\
&= A(x,p) e^{i\Theta_1} e^{i(\overleftarrow{\partial}_p \cdot \overrightarrow{\partial}_x - \overleftarrow{\partial}_x \cdot \overrightarrow{\partial}_p)/2} e^{i\Theta_2} B(x,p) \\
&= A(x,p) e^{i\Theta_1} e^{-(\overleftarrow{\partial}_p \cdot \partial_x \Theta_2 - \partial_x \Theta_1 \cdot \overrightarrow{\partial}_p)/2} e^{i\Theta_2} B(x,p) + \mathcal{O}(\epsilon_{\mathrm{mw}}) \\
&= A(x,p) e^{i\Theta_1} e^{\overleftarrow{\partial}_p \cdot (K_2/2)} e^{-(K_1/2) \cdot \overrightarrow{\partial}_p} e^{i\Theta_2} B(x,p) + \mathcal{O}(\epsilon_{\mathrm{mw}}) \\
&= A(x, p + K_2/2) B(x, p - K_1/2) e^{i(\Theta_1 + \Theta_2)} + \mathcal{O}(\epsilon_{\mathrm{mw}}).
\end{aligned} \tag{C.14}$$

Substituting this result into $C_2(x,p)$ leads to

$$
\begin{aligned}
C_2 =& \{\{G_1, D_1\}\}/2 = \{\{\mathrm{Re}\big[\mathcal{G}_1(x,p)e^{i\Theta}\big], \mathrm{Re}\big[\mathcal{D}_1(x,p)e^{i\Theta}\big]\}\}/2 \\
=& \mathrm{Re}\big[\ \{\{\mathcal{G}_1(x,p)e^{i\Theta}, \mathcal{D}_1^*(x,p)e^{-i\Theta}\}\}\ \big]/4 + \mathrm{Re}\big[\ \{\{\mathcal{G}_1(x,p)e^{i\Theta}, \mathcal{D}_1(x,p)e^{i\Theta}\}\}\ \big]/4 \\
=& \mathrm{Re}\big[\mathcal{G}_1(x,p-K/2)\mathcal{D}_1^*(x,p-K/2)/(4i)\big] - \mathrm{Re}\big[\mathcal{G}_1(x,p+K/2)\mathcal{D}_1^*(x,p+K/2)/(4i)\big] \\
&+ \mathrm{Re}\big[\mathcal{C}_2(x,p)e^{2i\Theta(x)}\big] + \mathcal{O}(\epsilon_{\mathrm{mw}}),
\end{aligned}
\tag{C.15}
$$

where $\mathcal{C}_2(x,p)$ is some function, whose explicit expression is not important for our purposes. After substituting Eq. (7.24) into Eq. (C.15), one obtains

$$
\begin{aligned}
C_2 =& -\frac{1}{4}\left(\frac{|\mathcal{D}_1(x,p+K/2)|^2}{D_0(x,p+K)-D_0(x,p)} + \frac{|\mathcal{D}_1(x,p-K/2)|^2}{D_0(x,p-K)-D_0(x,p)}\right) + \mathrm{Re}\big[\mathcal{C}_2(x,p)e^{2i\Theta(x)}\big] + \mathcal{O}(\epsilon_{\mathrm{mw}}) \\
=& -\frac{1}{4}\sum_{n=\pm 1}\frac{|\mathcal{D}_1(x,p+nK/2)|^2}{D_0(x,p+nK)-D_0(x,p)} + \mathrm{Re}\big[\mathcal{C}_2(x,p)e^{2i\Theta(x)}\big] + \mathcal{O}(\epsilon_{\mathrm{mw}}).
\end{aligned}
\tag{C.16}
$$

### C.2.2 Obtaining the original wave function

The calculation of Eq. (7.53) is presented below. From Eq. (7.22), $G_1(x,p)$ is composed by two parts: $G_1 = (1/2)\big[\mathcal{G}_1 e^{i\Theta} + \mathcal{G}_1^* e^{-i\Theta}\big]$. Here I present only the calculation of the first term in Eq. (7.52). Taylor expanding $\mathcal{G}_1(x,p)$ on the momentum component and integrating by parts leads to

$$
\begin{aligned}
&\frac{1}{(2\pi)^4}\int \mathrm{d}^4y\,\mathrm{d}^4p\, e^{-ip\cdot(x-y)}\,\mathcal{G}_1\left(\frac{x+y}{2},p\right) e^{i\Theta((x+y)/2)}\psi(y) \\
&=\frac{1}{(2\pi)^4}\sum_{n=0}^{\infty}\int \mathrm{d}^4y\,\mathrm{d}^4p\, e^{-ip\cdot(x-y)}\left(p\cdot\frac{\partial}{\partial p}\right)^n \mathcal{G}_1\left(\frac{x+y}{2},0\right) e^{i\Theta((x+y)/2)}\psi(y) \\
&=\frac{1}{(2\pi)^4}\sum_{n=0}^{\infty}\int \mathrm{d}^4y\,\mathrm{d}^4p\, e^{-ip\cdot(x-y)}\left(-i\frac{\overleftarrow{\partial}}{\partial y}\cdot\frac{\partial}{\partial p}\right)^n \mathcal{G}_1\left(\frac{x+y}{2},0\right) e^{i\Theta((x+y)/2)}\psi(y) \\
&=\sum_{n=0}^{\infty}\int \mathrm{d}^4y\,\delta^4(x-y)\left(-i\frac{\overleftarrow{\partial}}{\partial y}\cdot\frac{\partial}{\partial p}\right)^n \mathcal{G}_1\left(\frac{x+y}{2},0\right) e^{i\Theta((x+y)/2)}\psi(y) \\
&=\sum_{n=0}^{\infty}\int \mathrm{d}^4y\,\delta^4(x-y)\left(i\frac{\overrightarrow{\partial}}{\partial y}\cdot\frac{\partial}{\partial p}\right)^n \mathcal{G}_1\left(\frac{x+y}{2},0\right) e^{i\Theta((x+y)/2)}\psi(y),
\end{aligned}
$$

where the arrow indicates the direction in which the derivative acts. Now, one substitutes $\psi(x) = a(x)\,e^{i\theta(x)}$. After noting that $\Theta(x)$ and $\theta(x)$ are fast compared to $\mathcal{G}_1(x,p)$, $\mathcal{I}_0(x)$, $k_\mu(x) = -\partial_\mu\theta$, and $K_\mu(x) = -\partial_\mu\Theta$, one approximates the integral by

$$
\begin{aligned}
&\int \mathrm{d}^4 y\, \mathrm{d}^4 p\, e^{-ip\cdot(x-y)}\, \mathcal{G}_1\left(\frac{x+y}{2}, p\right) e^{i\Theta((x+y)/2)} \psi(y) \\
&\simeq \sum_{n=0}^{\infty} \int \mathrm{d}^4 y\, \delta^4(x-y)\, a(y)\, \mathcal{G}_1\left(\frac{x+y}{2}, 0\right) \left( i \frac{\overleftarrow{\partial}}{\partial p} \cdot \frac{\overrightarrow{\partial}}{\partial y} \right)^n e^{i\Theta((x+y)/2)} e^{i\theta(y)} \\
&\simeq \sum_{n=0}^{\infty} \int \mathrm{d}^4 y\, \delta^4(x-y)\, a(y)\, \mathcal{G}_1\left(\frac{x+y}{2}, 0\right) \left[ -\frac{\overleftarrow{\partial}}{\partial p} \cdot \left( \frac{\partial \theta}{\partial y} + \frac{1}{2}\frac{\partial \Theta}{\partial y} \right) \right]^n e^{i\Theta((x+y)/2)} e^{i\theta(y)} \\
&= \int \mathrm{d}^4 y\, \delta^4(x-y)\, a(y)\, \mathcal{G}_1\left(\frac{x+y}{2}, -\frac{\partial \theta}{\partial y} - \frac{1}{2}\frac{\partial \Theta}{\partial y}\right) e^{i\Theta((x+y)/2)} e^{i\theta(y)} \\
&= a(x)\, e^{i\theta(x)}\, \mathcal{G}_1\big(x, k(x) + K(x)/2\big)\, e^{i\Theta(x)}.
\end{aligned} \tag{C.17}
$$

Repeating a similar calculation for the $\exp(-i\Theta)$ component of $G_1(x,p)$ leads to Eq. (7.53).

### C.2.3 Ponderomotive Hamiltonian of a relativistic particle

The calculation of Eq. (7.70) is presented below. Substituting $\mathcal{A}^\mu = (\mathcal{V}, 0)$ into Eq. (7.67) leads to

$$
\begin{aligned}
&H_{\text{eff}}(t, \mathbf{X}, \mathbf{P}) - \gamma m c^2 - q V_{\text{bg}} \\
&\quad = -\frac{q^2 (\mathcal{A}\cdot\mathcal{A}^*)}{4\gamma m c^2} + \frac{1}{2\gamma m c^2} \sum_{n=\pm 1} \frac{q^2 |\mathcal{A}\cdot(\Pi_* + n\hbar K/2)|^2}{2n\Pi_*\cdot\hbar K + n^2\hbar^2 K\cdot K}, \\
&\quad = -\frac{q^2|\mathcal{V}|^2}{4\gamma m c^2} + \frac{q^2|\mathcal{V}|^2}{2\gamma m c^2} \left( \frac{[\gamma m c + \hbar\Omega/(2c)]^2}{2\Pi_*\cdot\hbar K + \hbar^2 K\cdot K} - \frac{[\gamma m c - \hbar\Omega/(2c)]^2}{2\Pi_*\cdot\hbar K - \hbar^2 K\cdot K} \right) \\
&\quad = -\frac{q^2|\mathcal{V}|^2}{4\gamma m c^2} + \left(\frac{q^2|\mathcal{V}|^2}{2\gamma m c^2}\right) \sum_{n=\pm 1} \frac{n[\gamma m c + n\hbar\Omega/(2c)]^2 (2\Pi_*\cdot\hbar K - n\hbar^2 K\cdot K)}{4(\Pi_*\cdot\hbar K)^2 - (\hbar^2 K\cdot K)^2} \\
&\quad = -\frac{q^2|\mathcal{V}|^2}{4\gamma m c^2} + \left(\frac{q^2|\mathcal{V}|^2}{8\gamma m c^2}\right) \frac{4\gamma m\Omega(\Pi_*\cdot K) - 2\gamma^2 m^2 c^2 (K\cdot K) - \hbar^2\Omega^2 (K\cdot K)/(2c^2)}{(\Pi_*\cdot K)^2 - (\hbar K\cdot K/2)^2} \\
&\quad = \left(\frac{q^2|\mathcal{V}|^2}{4\gamma m}\right) \frac{\mathbf{K}^2 - [(\mathbf{\Pi}\cdot\mathbf{K}/(\gamma m c)]^2 - [\hbar\mathbf{K}/(2\gamma m c)]^2 (K\cdot K)}{[\Omega - \mathbf{\Pi}\cdot\mathbf{K}/(\gamma m)]^2 - [\hbar K\cdot K/(2\gamma m)]^2} \\
&\quad = \left(\frac{q^2|\mathcal{V}|^2}{4\gamma m}\right) \frac{\mathbf{K}^2 - [\mathbf{\Pi}\cdot\mathbf{K}/(\gamma m c)]^2}{[\Omega - \mathbf{\Pi}\cdot\mathbf{K}/(\gamma m)]^2} + \mathcal{O}\left(\hbar^2\right).
\end{aligned} \tag{C.18}
$$

## C.3 Relativistic Hamiltonian of a spin-1/2 electron in a laser field

In this Section, I include additional details on some of the calculations reported in Chapter 8. First, I shall give a brief introduction to the semiclassical Volkov states. Afterwards, I shall present the calculation of the spin-coupling matrix $\mathcal{U}$.

### C.3.1 Volkov states

The Volkov states are eigenstates of the Dirac equation with an homogeneous EM vacuum field.[1] Here I present a derivation of these states. Consider the second-order Dirac equation

$$\left(D_\mu D^\mu + m^2 + \frac{1}{2} q \sigma_{\mu\nu} F^{\mu\nu}\right) \Psi = 0, \tag{C.19}$$

where $iD_\mu \doteq i\partial_\mu - qA_\mu$ is the covariant derivative, $\sigma_{\mu\nu} \doteq i[\gamma_\mu, \gamma_\nu]/2$ is twice the (relativistic) spin operator, and $F^{\mu\nu} = \partial^\mu A^\nu - \partial^\nu A^\mu$ is the EM tensor. Let the four-potential $A^\mu(x)$ be composed of two parts such that $A^\mu = A^\mu_{\text{bg}} + A^\mu_{\text{osc}}$, where $A_{\text{bg}}$ is a constant and $A_{\text{osc}}(\Theta)$ is strictly periodic. I assume that the four-wavevector $K_\mu \doteq -\partial_\mu \Theta$ corresponding to the oscillating field is constant and satisfies the dispersion relation for an EM field in vacuum $K^2 = 0$. I also choose the Lorentz gauge so that $\partial_\mu A^\mu_{\text{osc}} = -K_\mu A^\mu_{\text{osc}} = 0$.

Let us search for $\Psi(x)$ in the Floquet-Bloch form. Specifically, let us consider $\Psi = u(\Theta) e^{i\theta(x)}$, where $u(\Theta)$ is a periodic four-component function of $\Theta$ and $p_\mu \doteq -\partial_\mu \theta$ is constant. One writes $u(\Theta)$ as follows: $u(\Theta) = \Xi(\Theta) e^{i\tilde{\theta}(\Theta)} \varphi$, where $\Xi(\Theta)$ is a $4 \times 4$ matrix, $\tilde{\theta}(\Theta)$ is a real scalar function, and $\varphi$ is a constant four-component function. Substituting this ansatz leads to

$$\left[\pi^2 - m^2 + 2(\pi \cdot K)\partial_\Theta \tilde{\theta} - 2q(\pi \cdot A_{\text{osc}}) + q^2 A^2_{\text{osc}}\right] \Xi \varphi - 2i(\pi \cdot K)(\partial_\Theta \Xi)\varphi - \frac{1}{2} q \sigma_{\mu\nu} F^{\mu\nu} \Xi \varphi = 0, \tag{C.20}$$

where $\pi^\mu \doteq p^\mu - qA^\mu_{\text{bg}}$. Let $\tilde{\theta}(\Theta)$ and $\Xi(\Theta)$ satisfy the following equations:

$$\pi^2 - m^2 + 2(\pi \cdot K)\partial_\Theta \tilde{\theta} - 2q(\pi \cdot A_{\text{osc}}) + q^2 A^2_{\text{osc}} = 0, \tag{C.21}$$

$$-2i(\pi \cdot K)(\partial_\Theta \Xi) - \frac{1}{2} q \sigma_{\mu\nu} F^{\mu\nu} \Xi = 0. \tag{C.22}$$

The integration constants can be chosen arbitrarily since they merely redefine $\varphi$. Hence, let us require $\Xi(\Theta) \to \mathbb{I}_4$ at vanishing $A_{\text{osc}}$ and $\langle\langle \tilde{\theta} \rangle\rangle = 0$ so that $\tilde{\theta}$ represents a phase shift due to the oscillating EM field. For $\Xi(\Theta)$, one obtains

$$\begin{aligned} \Xi(\Theta) &= \mathcal{T} \exp\left[\frac{iq}{4(\pi \cdot K)} \int^\Theta \sigma_{\mu\nu} F^{\mu\nu}(\Theta')\, d\Theta'\right] \\ &= \mathbb{I}_4 + \frac{q}{2(\pi \cdot K)} \not{K} \not{A}_{\text{osc}}(\Theta), \end{aligned} \tag{C.23}$$

[1]For more information, see Volkov (1935), Bergou and Varro (1980), and Raicher and Eliezer (2013).

where I used $\sigma_{\mu\nu}F^{\mu\nu} = \sigma_{\mu\nu}(\partial^\mu A^\nu - \partial^\nu A^\mu)$. Since $A^\mu_{\rm osc}$ only depends on $\Theta$, then

$$\sigma_{\mu\nu}F^{\mu\nu} = -\sigma_{\mu\nu}(K^\mu \partial_\Theta A^\nu_{\rm osc} - K^\nu \partial_\Theta A^\mu_{\rm osc}) = -i(\not{K}\partial_\Theta \not{A}_{\rm osc} - \partial_\Theta \not{A}_{\rm osc}\not{K}) = -2i\not{K}\partial_\Theta \not{A}_{\rm osc}. \tag{C.24}$$

Here $\not{K}\not{A}_{\rm osc} + \not{A}_{\rm osc}\not{K} = 0$ since $K \cdot A_{\rm osc} = 0$ from the Lorentz gauge. Also, note that the ordered exponential [denoted by $\mathcal{T}\exp(\cdots)$] becomes an ordinary exponential due to

$$\sigma_{\mu\nu}F^{\mu\nu}(\Theta_1)\sigma_{\alpha\xi}F^{\alpha\xi}(\Theta_2) = -4\not{K}[\partial_\Theta \not{A}_{\rm osc}(\Theta_1)]\not{K}[\partial_\Theta \not{A}_{\rm osc}(\Theta_2)] = 4\not{K}\not{K}\partial_\Theta \not{A}_{\rm osc}(\Theta_1)\not{A}_{\rm osc}(\Theta_2) = 0, \tag{C.25}$$

where I used $\not{K}\not{K} = K\cdot K = 0$. To obtain $\tilde{\theta}(\Theta)$, let us average Eq. (C.21). This leads to $\pi^2 - m^2 + q^2\langle\langle A^2_{\rm osc}\rangle\rangle = 0$, which serves as the dispersion relation for $\pi_\mu$. Subtracting this relation from Eq. (C.21) and solving for $\tilde{\theta}$ leads to Eq. (8.14). Finally, if one substitutes $\psi = \Xi e^{i\theta + \tilde{\theta}}\varphi$ into the first-order Dirac equation, one finds the constant four-component spinor $\varphi$ that satisfies the equation.

### C.3.2 Calculation of the spin-coupling matrix

Here I calculate the spin-coupling matrix for the ponderomotive model of the spin-1/2 electron. As a reminder from Chapter 8, the eigenvalue is given by $\lambda(x,p) = \zeta_0(x,p) - \varepsilon(x,p)$, where $\varepsilon(x,p) = [\boldsymbol{\zeta}^2(x,p) + m^2]^{1/2}$ and $\zeta_\mu(x,p) = \pi_\mu + \eta K_\mu$. Also, $\eta(x,p) = q^2\langle\langle A^2_{\rm osc}\rangle\rangle/[2(\pi\cdot K)]$ and $\pi_\mu = p_\mu - qA_{{\rm bg},\mu}$. The dispersion symbol is $D(x,p) = \zeta_0\mathbb{I}_4 - H(x,p)$, where $H(x,p) = \boldsymbol{\alpha}\cdot\boldsymbol{\zeta} + \beta m$. The $4\times4$ Hermitian matrices $\boldsymbol{\alpha}$ and $\beta$ are given in Sec. 5.2.1. The eigenmode matrix $\Xi(x,p)$ is

$$\Xi(x,p) = C(x,p)\begin{pmatrix} \mathbb{I}_2 \\ \frac{\boldsymbol{\zeta}\cdot\boldsymbol{\sigma}}{m+\varepsilon} \end{pmatrix}, \tag{C.26}$$

where the normalization term is $C(x,p) \doteq [(m+\varepsilon)/(2\varepsilon)]^{1/2}$.

Now, let us calculate the corresponding spin coupling matrix $\mathcal{U}(x,p)$. When comparing the expression for $D(x,p)$ and $\Xi(x,p)$ that are introduced here with those given in Sec. C.1, one can notice that both systems are identical when replacing $\pi_\mu$ with $\zeta_\mu$. In fact, when repeating the same procedure to obtain Eq. (C.7), one obtains the following expression for the spin-coupling matrix

$$\mathcal{U}(t,\mathbf{x},\mathbf{p}) = -\frac{\boldsymbol{\pi}\times\{\zeta_0,\boldsymbol{\zeta}\}}{2\varepsilon(m+\varepsilon)}\cdot\boldsymbol{\sigma} + \frac{1}{2\varepsilon}\left(\frac{\partial\boldsymbol{\zeta}}{\partial p_\mu}\times\frac{\partial\boldsymbol{\zeta}}{\partial x^\mu}\right)\cdot\boldsymbol{\sigma}. \tag{C.27}$$

Let us calculate the Poisson bracket in the first term of Eq. (C.27). One obtains

$$\begin{aligned}\{\zeta_0, \boldsymbol{\zeta}\} &= \frac{\partial \zeta_0}{\partial p_\mu}\frac{\partial \boldsymbol{\zeta}}{\partial x^\mu} - \frac{\partial \zeta_0}{\partial x^\mu}\frac{\partial \boldsymbol{\zeta}}{\partial p_\mu} \\ &= \frac{\partial}{\partial p_\mu}(\pi_0 + \eta K_0)\frac{\partial \boldsymbol{\zeta}}{\partial x^\mu} - \frac{\partial \zeta_0}{\partial x^\mu}\frac{\partial}{\partial p_\mu}(\boldsymbol{\pi} + \eta \mathbf{K}) \\ &= \frac{\partial \boldsymbol{\zeta}}{\partial t} + K_0 \frac{\partial \eta}{\partial p_\mu}\frac{\partial \boldsymbol{\zeta}}{\partial x^\mu} + \boldsymbol{\nabla}\zeta_0 - \mathbf{K}\frac{\partial \eta}{\partial p_\mu}\frac{\partial \zeta_0}{\partial x^\mu} \\ &= \frac{\partial \boldsymbol{\zeta}}{\partial t} - \frac{\eta K_0}{\pi \cdot K}(K^\mu \partial_\mu)\boldsymbol{\zeta} + \boldsymbol{\nabla}\zeta_0 + \frac{\eta \mathbf{K}}{\pi \cdot K}(K^\mu \partial_\mu)\zeta_0, \end{aligned} \tag{C.28}$$

where I used

$$\frac{\partial \eta}{\partial p_\mu} = \frac{\partial}{\partial p_\mu}\frac{q^2 \langle\!\langle A^2_{\text{osc}} \rangle\!\rangle}{2(\pi \cdot K)} = -\frac{q^2 \langle\!\langle A^2_{\text{osc}} \rangle\!\rangle}{2(\pi \cdot K)^2}\frac{\partial}{\partial p_\mu}(\pi \cdot K) = -\frac{\eta}{\pi \cdot K}K^\mu. \tag{C.29}$$

Similarly, the second term in Eq. (C.27) is calculated as follows:

$$\begin{aligned}\left(\frac{\partial \boldsymbol{\zeta}}{\partial p_\mu} \times \frac{\partial \boldsymbol{\zeta}}{\partial x^\mu}\right)^i &= \varepsilon^{ijk}\frac{\partial \zeta^j}{\partial p_\mu}\frac{\partial \zeta^k}{\partial x^\mu} \\ &= \varepsilon^{ijk}\frac{\partial}{\partial p_\mu}(\pi^j + \eta K^j)\frac{\partial \zeta^k}{\partial x^\mu} \\ &= -\varepsilon^{ijk}\frac{\partial \pi^j}{\partial p^l}\frac{\partial \zeta^k}{\partial x^l} + \varepsilon^{ijk}K^j \frac{\partial \eta}{\partial p_\mu}\frac{\partial \zeta^k}{\partial x^\mu} \\ &= -\varepsilon^{ijk}\frac{\partial \zeta^k}{\partial x^j} - \varepsilon^{ijk}\frac{\eta K^j}{\pi \cdot K}(K^\mu \partial_\mu)\zeta^k \\ &= -(\boldsymbol{\nabla} \times \boldsymbol{\zeta})^i - \frac{\eta}{\pi \cdot K}[\mathbf{K} \times (K^\mu \partial_\mu)\boldsymbol{\zeta}]^i. \end{aligned} \tag{C.30}$$

Substituting Eqs. (C.28) and (C.30) into Eq. (C.27) leads to

$$\begin{aligned}\mathcal{U}(x,p) = -\frac{1}{2\varepsilon}\boldsymbol{\sigma} \cdot \Bigg(&\boldsymbol{\nabla} \times \boldsymbol{\zeta} + \frac{\eta}{\pi \cdot K}\mathbf{K} \times (K^\mu \partial_\mu)\boldsymbol{\zeta} + \frac{\boldsymbol{\zeta} \times (\boldsymbol{\nabla}\zeta_0 + \partial_t \boldsymbol{\zeta})}{m + \varepsilon} \\ &- \frac{\eta}{\pi \cdot K}\frac{K^0 \boldsymbol{\zeta} \times (K^\mu \partial_\mu)\boldsymbol{\zeta} - (\boldsymbol{\zeta} \times \mathbf{K})(K^\mu \partial_\mu)\zeta_0}{m + \varepsilon}\Bigg). \end{aligned} \tag{C.31}$$

Let us cast Eq. (C.31) in terms of more familiar quantities. The first term inside the brackets can be written as follows:

$$\boldsymbol{\nabla} \times \boldsymbol{\zeta} = \boldsymbol{\nabla} \times (\boldsymbol{\pi} + \alpha \mathbf{K}) = -q\boldsymbol{\nabla} \times \mathbf{A}_{\text{bg}} - \mathbf{K} \times \boldsymbol{\nabla}\eta = -q\mathbf{B}_{\text{bg}} - \mathbf{K} \times \boldsymbol{\nabla}\eta, \tag{C.32}$$

where I used $\boldsymbol{\nabla}\times\mathbf{K}=\boldsymbol{\nabla}\times\boldsymbol{\nabla}\Theta=0$ and $\mathbf{B}_{\text{bg}}(t,\mathbf{x})=\boldsymbol{\nabla}\times\mathbf{A}_{\text{bg}}$ is the background magnetic field. The second term inside the brackets of Eq. (C.31) can be written as

$$\mathbf{K}\times(K^\mu\partial_\mu)\boldsymbol{\zeta}=\mathbf{K}\times(K^\mu\partial_\mu)\boldsymbol{\pi}+\eta\mathbf{K}\times(K^\mu\partial_\mu)\mathbf{K}=\mathbf{K}\times(K^\mu\partial_\mu)\boldsymbol{\pi},\tag{C.33}$$

where I used

$$(K_\mu\partial_\mu)K_\nu=-(K_\mu\partial_\mu)\partial_\nu\Theta=-K_\mu\partial_\nu\partial_\mu\Theta=K^\mu\partial_\nu K_\mu=(1/2)\partial_\nu(K\cdot K)=0.\tag{C.34}$$

Likewise, the numerator of the third term of Eq. (C.31) can be written as

$$\begin{aligned}\boldsymbol{\nabla}\zeta_0+\partial_t\boldsymbol{\zeta}&=\boldsymbol{\nabla}(\pi_0+\eta K_0)+\partial_t(\boldsymbol{\pi}+\eta\mathbf{K})\\&=-q\boldsymbol{\nabla}V_{\text{bg}}+K_0\boldsymbol{\nabla}\eta+\eta\boldsymbol{\nabla}K_0-q\partial_t\mathbf{A}_{\text{bg}}+\mathbf{K}\partial_t\eta+\eta\partial_t\mathbf{K}\\&=q\mathbf{E}_{\text{bg}}+\mathbf{K}\partial_t\eta+K_0\boldsymbol{\nabla}\eta,\end{aligned}\tag{C.35}$$

where $\boldsymbol{\nabla}K_0=-\boldsymbol{\nabla}\partial_t\Theta=-\partial_t\boldsymbol{\nabla}\Theta=-\partial_t\mathbf{K}$ and $\mathbf{E}_{\text{bg}}(t,\mathbf{x})\doteq-\boldsymbol{\nabla}V_{\text{bg}}-\partial_t\mathbf{A}_{\text{bg}}$ is the background electric field. Similarly, the numerator of the last term in Eq. (C.31) simplifies to

$$K^0\boldsymbol{\zeta}\times(K^\mu\partial_\mu)\boldsymbol{\zeta}-(\boldsymbol{\zeta}\times\mathbf{K})(K^\mu\partial_\mu)\zeta_0=\boldsymbol{\zeta}\times[K^0(K^\mu\partial_\mu)\boldsymbol{\pi}-\mathbf{K}(K^\mu\partial_\mu)\pi_0].\tag{C.36}$$

Hence, substituting Eqs. (C.32)–(C.36) into Eq. (C.31) leads to

$$\begin{aligned}\mathcal{U}(x,p)=&\frac{1}{2\varepsilon}\boldsymbol{\sigma}\cdot\left[q\mathbf{B}_{\text{bg}}+\frac{\boldsymbol{\zeta}\times q\mathbf{E}_{\text{bg}}}{m+\varepsilon}-\left(\mathbf{K}-\frac{K_0\boldsymbol{\zeta}}{m+\varepsilon}\right)\times\left(\boldsymbol{\nabla}\eta-\frac{\eta}{\pi\cdot K}(K^\mu\partial_\mu)\boldsymbol{\pi}\right)\right.\\&\left.-\frac{\boldsymbol{\zeta}\times\mathbf{K}}{m+\varepsilon}\left(\partial_t\eta+\frac{\eta}{\pi\cdot K}(K^\mu\partial_\mu)\pi_0\right)\right].\end{aligned}\tag{C.37}$$

As a final step, let us further simplify the terms inside the square brackets of Eq. (C.37). The terms inside the first brackets can be written as follows:

$$\begin{aligned}\boldsymbol{\nabla}\eta-\frac{\eta}{\pi\cdot K}(K^\mu\partial_\mu)\boldsymbol{\pi}&=\frac{q^2\boldsymbol{\nabla}\langle\!\langle A_{\text{osc}}^2\rangle\!\rangle}{2(\pi\cdot K)}-\frac{\eta}{\pi\cdot K}\left[\boldsymbol{\nabla}(\pi\cdot K)+(K^\mu\partial_\mu)\boldsymbol{\pi}\right]\\&=\frac{q^2\boldsymbol{\nabla}\langle\!\langle A_{\text{osc}}^2\rangle\!\rangle}{2(\pi\cdot K)}-\frac{\eta}{\pi\cdot K}\left[\boldsymbol{\nabla}(\pi^0K^0)-\boldsymbol{\nabla}(\boldsymbol{\pi}\cdot\mathbf{K})+K^0\partial_t\boldsymbol{\pi}+(\mathbf{K}\cdot\boldsymbol{\nabla})\boldsymbol{\pi}\right]\\&=\frac{q^2\boldsymbol{\nabla}\langle\!\langle A_{\text{osc}}^2\rangle\!\rangle}{2(\pi\cdot K)}-\frac{\eta}{\pi\cdot K}\left[qK^0\mathbf{E}_{\text{bg}}+q\mathbf{K}\times\mathbf{B}_{\text{bg}}-(\pi^\mu\partial_\mu)\mathbf{K}\right].\end{aligned}\tag{C.38}$$

Likewise,

$$\begin{aligned}
\partial_t\eta + \frac{\eta}{\pi\cdot K}(K^\mu\partial_\mu)\pi_0 &= \frac{q^2\partial_t\langle\!\langle A^2_{\text{osc}}\rangle\!\rangle}{2(\pi\cdot K)} - \frac{\eta}{\pi\cdot K}\big[\partial_t(\pi\cdot K) - (K^\mu\partial_\mu)\pi_0\big] \\
&= \frac{q^2\partial_t\langle\!\langle A^2_{\text{osc}}\rangle\!\rangle}{2(\pi\cdot K)} - \frac{\eta}{\pi\cdot K}\big[\partial_t(\pi^0K^0) - \partial_t(\boldsymbol{\pi}\cdot\mathbf{K}) - K^0\partial_t\pi_0 - (\mathbf{K}\cdot\boldsymbol{\nabla})\pi_0\big] \\
&= \frac{q^2\partial_t\langle\!\langle A^2_{\text{osc}}\rangle\!\rangle}{2(\pi\cdot K)} + \frac{\eta}{\pi\cdot K}\big[q\mathbf{K}\cdot\mathbf{E}_{\text{bg}} - (\pi^\mu\partial_\mu)K^0\big]. \qquad \text{(C.39)}
\end{aligned}$$

Substituting Eqs. (C.38) and (C.39) into Eq. (C.37) leads to

$$\begin{aligned}
\mathcal{U}(x,p) = \frac{q}{2\varepsilon}&\left(\mathbf{B}_{\text{bg}} - \frac{\boldsymbol{\zeta}\times\mathbf{E}_{\text{bg}}}{m+\zeta^0}\right)\cdot\boldsymbol{\sigma} \\
&+ \frac{q^2}{4\varepsilon(\pi\cdot K)}\left(\mathbf{K}\times\boldsymbol{\nabla}\langle\!\langle A^2_{\text{osc}}\rangle\!\rangle - \frac{(\boldsymbol{\zeta}\times\mathbf{K})\partial_t\langle\!\langle A^2_{\text{osc}}\rangle\!\rangle}{m+\zeta^0} - \frac{K^0\boldsymbol{\zeta}\times\boldsymbol{\nabla}\langle\!\langle A^2_{\text{osc}}\rangle\!\rangle}{m+\zeta^0}\right)\cdot\boldsymbol{\sigma} \\
&+ \frac{q^2\langle\!\langle A^2_{\text{osc}}\rangle\!\rangle}{4\varepsilon(\pi\cdot K)^2}\left[\left(\frac{K^0\boldsymbol{\zeta}}{m+\zeta^0} - \mathbf{K}\right)\times\big[K^0q\mathbf{E}_{\text{bg}} + \mathbf{K}\times q\mathbf{B}_{\text{bg}} - (\pi^\mu\partial_\mu)\mathbf{K}\big]\right. \\
&\qquad\left. - \frac{\boldsymbol{\zeta}\times\mathbf{K}}{m+\zeta^0}\big[\mathbf{K}\cdot q\mathbf{E}_{\text{bg}} - (\pi^\mu\partial_\mu)K^0\big]\right]\cdot\boldsymbol{\sigma}. \qquad \text{(C.40)}
\end{aligned}$$

## C.4 Applications to wave turbulence and zonal-flow formation

In this Section, I prove that in the case of isolated systems ($Q = 0$), the Wigner–Moyal and WKE models conserve the total enstrophy $\mathcal{Z}(t)$ and the total energy $\mathcal{E}(t)$, whose expressions are given by Eqs. (9.31) and (9.32).

### C.4.1 Wigner–Moyal model

First, let us consider the Wigner–Moyal model [Eqs. (9.29) and (9.30)]. To show conservation of enstrophy, one has the following:

$$\begin{aligned}
\frac{\mathrm{d}\mathcal{Z}}{\mathrm{d}t} &= \int \mathrm{d}y\,(\partial_y U)(\partial_y\partial_t U) + \frac{1}{2}\int \mathrm{d}y\,\mathrm{d}^2p\,\partial_t\overline{W} \\
&= \frac{1}{2}\int \mathrm{d}y\,\mathrm{d}^2p\left[2U'''\left(\frac{1}{p_{\mathrm{D}}^2}\star p_xp_y\overline{W}\star\frac{1}{p_{\mathrm{D}}^2}\right) + \{\!\{\mathcal{H},\overline{W}\}\!\} + [[\Gamma,\overline{W}]]\right] \\
&= \frac{1}{2}\int \mathrm{d}y\,\mathrm{d}^2p\left[2U'''\left(\frac{1}{p_{\mathrm{D}}^2}\star p_xp_y\overline{W}\star\frac{1}{p_{\mathrm{D}}^2}\right) + 2\Gamma\overline{W}\right], \qquad \text{(C.41)}
\end{aligned}$$

where I used Eq. (A.15) so that $\int \mathrm{d}y\,\mathrm{d}^2p\,\{\!\{A,B\}\!\} = 0$ and similarly $\int \mathrm{d}y\,\mathrm{d}^2p\,[[A,B]] = 2\int \mathrm{d}y\,\mathrm{d}^2p\,AB$. To evaluate the remaining terms, I use the Fourier representations of $\overline{W}(t,y,\mathbf{p})$ and $U(t,y)$:

$$\overline{W}(t,y,\mathbf{p}) = \frac{1}{2\pi}\int \mathrm{d}q\,\overline{W}_q(t,\mathbf{p})e^{iqy}, \qquad U(t,y) = \frac{1}{2\pi}\int \mathrm{d}q\,U_q(t)e^{iqy}. \qquad \text{(C.42)}$$

Substituting Eq. (9.30b) and the above Fourier representations leads to

$$\frac{\mathrm{d}\mathcal{Z}}{\mathrm{d}t} = \frac{1}{2(2\pi)^2}\int \mathrm{d}y\, \mathrm{d}^2p\, \mathrm{d}q\, \mathrm{d}k\, U_q \overline{W}_k \left[-2iq^3\left(\frac{1}{p_\mathrm{D}^2} \star p_x p_y e^{iky} \star \frac{1}{p_\mathrm{D}^2}\right) e^{iqy} - q^2 \{\{e^{iqy}, p_x p_\mathrm{D}^{-2}\}\} e^{iky}\right]. \tag{C.43}$$

The Moyal products and the sine bracket are evaluated by substituting Eq. (9.34) and the property $A(\mathbf{p}) \star e^{i\mathbf{q}\cdot\mathbf{x}} = A(\mathbf{p}+\mathbf{q}/2)e^{i\mathbf{q}\cdot\mathbf{x}}$. Upon introducing the notation $A_{\pm q} \doteq A(\mathbf{p} \pm \mathbf{e}_y q/2)$ for any arbitrary symbol $A(\mathbf{p})$ that depends on the momentum coordinate, one obtains

$$\begin{aligned}
\frac{\mathrm{d}\mathcal{Z}}{\mathrm{d}t} &= -\frac{i}{2(2\pi)^2}\int \mathrm{d}^2p\, \mathrm{d}q\, \mathrm{d}k\, U_q \overline{W}_k \left[\frac{2p_x p_y q^3}{p_{\mathrm{D},+k}^2 p_{\mathrm{D},-k}^2} - p_x q^2\left(\frac{1}{p_{\mathrm{D},-q}^2} - \frac{1}{p_{\mathrm{D},+q}^2}\right)\right]\int \mathrm{d}y\, e^{i(k+q)y} \\
&= -\frac{i}{2\pi}\int \mathrm{d}^2p\, \mathrm{d}q\, \mathrm{d}k\, U_q \overline{W}_k p_x p_y q^3\left(\frac{1}{p_{\mathrm{D},+k}^2 p_{\mathrm{D},-k}^2} - \frac{1}{p_{\mathrm{D},+q}^2 p_{\mathrm{D},-q}^2}\right)\delta(k+q) \\
&= 0.
\end{aligned} \tag{C.44}$$

To show conservation of energy, one has

$$\begin{aligned}
\frac{\mathrm{d}\mathcal{E}}{\mathrm{d}t} &= \int \mathrm{d}y\, U(\partial_t U) + \frac{1}{2}\int \mathrm{d}y\, \mathrm{d}^2p\, \frac{\partial_t \overline{W}}{p_\mathrm{D}^2} \\
&= \frac{1}{2}\int \mathrm{d}y\, \mathrm{d}^2p\left[2U\frac{\partial}{\partial y}\left(\frac{1}{p_\mathrm{D}^2} \star p_x p_y \overline{W} \star \frac{1}{p_\mathrm{D}^2}\right) + \frac{1}{p_\mathrm{D}^2}\{\{\mathcal{H}, \overline{W}\}\} + \frac{1}{p_\mathrm{D}^2}[[\Gamma, \overline{W}]]\right] \\
&= -\frac{1}{2}\int \mathrm{d}y\, \mathrm{d}^2p\left[2U'\left(\frac{1}{p_\mathrm{D}^2} \star p_x p_y \overline{W} \star \frac{1}{p_\mathrm{D}^2}\right) - \frac{1}{p_\mathrm{D}^2}\left\{\left\{p_x U + \left[\left[U'', p_x p_\mathrm{D}^{-2}\right]\right]/2, \overline{W}\right\}\right\}\right. \\
&\qquad\qquad \left. - \frac{1}{p_\mathrm{D}^2}[[\{\{U'', p_x p_\mathrm{D}^{-2}\}\}/2, \overline{W}]]\right],
\end{aligned} \tag{C.45}$$

where I used the fact that $\{\{\overline{W}, \beta p_x/p_\mathrm{D}^2\}\}/p_\mathrm{D}^2$ can be written in terms of total derivatives on $y$, so its integral over $y$ is zero. The other terms can be expressed as follows. First,

$$\begin{aligned}
&\int \mathrm{d}y\, \mathrm{d}^2p\left[2U'\left(\frac{1}{p_\mathrm{D}^2} \star p_x p_y \overline{W} \star \frac{1}{p_\mathrm{D}^2}\right) - \frac{1}{p_\mathrm{D}^2}\{\{p_x U, W\}\}\right] \\
&= \frac{1}{(2\pi)^2}\int \mathrm{d}y\, \mathrm{d}^2p\, \mathrm{d}q\, \mathrm{d}k\, U_q\left[2iq\overline{W}_k\left(\frac{1}{p_\mathrm{D}^2} \star p_x p_y e^{iky} \star \frac{1}{p_\mathrm{D}^2}\right) - \frac{1}{p_\mathrm{D}^2}\{\{p_x e^{iqy}, \overline{W}_k\}\}\right]e^{iky} \\
&= \frac{1}{(2\pi)^2}\int \mathrm{d}^2p\, \mathrm{d}q\, \mathrm{d}k\, U_q\left[i\overline{W}_k\frac{2p_x p_y q}{p_{\mathrm{D},+k}^2 p_{\mathrm{D},-k}^2} - \frac{p_x}{ip_\mathrm{D}^2}\left(\overline{W}_{k,-q} - \overline{W}_{k,+q}\right)\right]\int \mathrm{d}y\, e^{i(k+q)y} \\
&= \frac{i}{2\pi}\int \mathrm{d}^2p\, \mathrm{d}q\, \mathrm{d}k\, U_q \overline{W}_k\left[\frac{2p_x p_y q}{p_{\mathrm{D},+k}^2 p_{\mathrm{D},-k}^2} + p_x\left(\frac{1}{p_{\mathrm{D},+q}^2} - \frac{1}{p_{\mathrm{D},-q}^2}\right)\right]\delta(k+q) \\
&= 0.
\end{aligned} \tag{C.46}$$

From Eq. (A.18), one has $A(\mathbf{p})e^{i\mathbf{k}\cdot\mathbf{x}} \star B(\mathbf{p})e^{i\mathbf{q}\cdot\mathbf{x}} = A(\mathbf{p}+\mathbf{q}/2)B(\mathbf{p}-\mathbf{k}/2)e^{i(\mathbf{k}+\mathbf{q})\cdot\mathbf{x}}$ for any constant vectors $\mathbf{k}$ and $\mathbf{q}$. Upon using this property, one obtains

$$\begin{aligned}
&\int \mathrm{d}y\, \mathrm{d}^2p\, \frac{1}{2p_\mathrm{D}^2}\left(\{\{[[U'', p_x p_\mathrm{D}^{-2}]], \overline{W}\}\} + [[\{\{U'', p_x p_\mathrm{D}^{-2}\}\}, \overline{W}]]\right) \\
&= -\frac{1}{(2\pi)^2}\int \mathrm{d}y\, \mathrm{d}^2p\, \mathrm{d}q\, \mathrm{d}k\, \frac{U_q q^2}{2p_\mathrm{D}^2}\left(\{\{[[e^{iqy}, p_x p_\mathrm{D}^{-2}]], \overline{W}_k e^{iky}\}\} + [[\{\{e^{iqy}, p_x p_\mathrm{D}^{-2}\}\}, \overline{W}_k e^{iky}]]\right) \\
&= -\frac{1}{(2\pi)^2}\int \mathrm{d}y\, \mathrm{d}^2p\, \mathrm{d}q\, \mathrm{d}k\, \frac{U_q p_x q^2}{2p_\mathrm{D}^2}\left(\{\{(p_{\mathrm{D},+q}^{-2} + p_{\mathrm{D},-q}^{-2})e^{iqy}, \overline{W}_k e^{iky}\}\}\right. \\
&\qquad\left. - \frac{1}{i}[[(p_{\mathrm{D},+q}^{-2} - p_{\mathrm{D},-q}^{-2})e^{iqy}, \overline{W}_k e^{iky}]]\right) \\
&= \frac{1}{(2\pi)^2}\int \mathrm{d}y\, \mathrm{d}^2p\, \mathrm{d}q\, \mathrm{d}k\, \frac{U_q p_x q^2}{2ip_\mathrm{D}^2}\left(\frac{\overline{W}_{k,q}}{p_{\mathrm{D},+q-k}^2} - \frac{\overline{W}_{k,-q}}{p_{\mathrm{D},+q+k}^2} + \frac{\overline{W}_{k,q}}{p_{\mathrm{D},-q-k}^2} - \frac{\overline{W}_{k,-q}}{p_{\mathrm{D},-q+k}^2}\right. \\
&\qquad\left. + \frac{\overline{W}_{k,q}}{p_{\mathrm{D},+q-k}^2} + \frac{\overline{W}_{k,-q}}{p_{\mathrm{D},+q+k}^2} - \frac{\overline{W}_{k,q}}{p_{\mathrm{D},-q-k}^2} - \frac{\overline{W}_{k,-q}}{p_{\mathrm{D},-q+k}^2}\right) e^{i(k+q)y} \\
&= -\frac{i}{2\pi}\int \mathrm{d}^2p\, \mathrm{d}q\, \mathrm{d}k\, U_q \overline{W}_k p_x q^2 \left(\frac{1}{p_{\mathrm{D},-q}^2 p_{\mathrm{D},-k}^2} - \frac{1}{p_{\mathrm{D},+q}^2 p_{\mathrm{D},+k}^2}\right)\delta(k+q) \\
&= 0.
\end{aligned} \tag{C.47}$$

Upon substituting Eqs. (C.46) and (C.47) into Eq. (C.45), one obtains $\dot{\mathcal{E}}(t) = 0$.

### C.4.2 WKE model

Now let us consider the WKE model [Eqs. (9.43) and (9.44)]. To show conservation of enstrophy, one has the following:

$$\begin{aligned}
\frac{\mathrm{d}\mathcal{Z}}{\mathrm{d}t} &= \int \mathrm{d}y\, (\partial_y U)(\partial_y \partial_t U) + \frac{1}{2}\int \mathrm{d}y\, \mathrm{d}^2p\, \partial_t \overline{W} \\
&= \frac{1}{2}\int \mathrm{d}y\, \mathrm{d}^2p\, \left(\frac{2p_x p_y}{p_\mathrm{D}^4} U''' \overline{W} + \{\mathcal{H}, \overline{W}\} + 2\Gamma \overline{W}\right) \\
&= \frac{1}{2}\int \mathrm{d}y\, \mathrm{d}^2p\, \left(\frac{2p_x p_y}{p_\mathrm{D}^4} U''' + 2\Gamma\right)\overline{W} \\
&= 0,
\end{aligned} \tag{C.48}$$

where I used Eq. (9.44b) and the fact that the integral of the Poisson bracket over all phase space is zero. In contrast, the tWKE does not conserve total enstrophy because $\Gamma = 0$ [see Eq. (9.45)]. This is also understood as follows: the tWKE manifestly conserves $\mathcal{Z}_\mathrm{dw}$, whereas $\mathcal{Z}_\mathrm{zf}$ is obviously not conserved; thus, $\mathcal{Z}_\mathrm{dw} + \mathcal{Z}_\mathrm{zf}$ cannot be conserved either.

To show conservation of energy, one has

$$
\begin{aligned}
\frac{\mathrm{d}\mathcal{E}}{\mathrm{d}t} &= \int \mathrm{d}y\, U(\partial_t U) + \frac{1}{2}\int \mathrm{d}y\, \mathrm{d}^2p\, \frac{\partial_t \overline{W}}{p_{\mathrm{D}}^2} \\
&= -\frac{1}{2}\int \mathrm{d}y\, \mathrm{d}^2p\, \frac{1}{p_{\mathrm{D}}^2}\left(\frac{2p_xp_y}{p_{\mathrm{D}}^2}\, U'\overline{W} - \{\mathcal{H}, \overline{W}\} - 2\Gamma\overline{W}\right) \\
&= -\frac{1}{2}\int \mathrm{d}y\, \mathrm{d}^2p\, \frac{1}{p_{\mathrm{D}}^2}\left(\frac{2p_xp_y}{p_{\mathrm{D}}^2}\, U'\overline{W} - \left\{p_xU + p_xp_{\mathrm{D}}^{-2}U'', \overline{W}\right\} + \frac{2p_xp_y}{p_{\mathrm{D}}^4}U'''\overline{W}\right),
\end{aligned}
\tag{C.49}
$$

where the integral of $\{\overline{W}, \beta p_x/p_{\mathrm{D}}^2\}/p_{\mathrm{D}}^2$ over $y$ is zero because it can be written as a total derivative on $y$. Finally,

$$
\begin{aligned}
\frac{\mathrm{d}\mathcal{E}}{\mathrm{d}t} &= -\frac{1}{2}\int \mathrm{d}y\, \mathrm{d}^2p\, \left(\frac{2p_xp_y}{p_{\mathrm{D}}^4}U'\overline{W} - \frac{p_x}{p_{\mathrm{D}}^2}U'\partial_{p_y}\overline{W} - \frac{p_x}{p_{\mathrm{D}}^4}\, U'''\, \partial_{p_y}\overline{W} - \frac{2p_xp_y}{p_{\mathrm{D}}^6}\, U''\partial_y\overline{W} + \frac{2p_xp_y}{p_{\mathrm{D}}^6}\, U'''\overline{W}\right) \\
&= -\frac{1}{2}\int \mathrm{d}y\, \mathrm{d}^2p\, \left(\frac{2p_xp_y}{p_{\mathrm{D}}^4}\, U'\overline{W} - \frac{2p_xp_y}{p_{\mathrm{D}}^4}U'\overline{W} - \frac{4p_xp_y}{p_{\mathrm{D}}^6}\, U'''\overline{W} + \frac{2p_xp_y}{p_{\mathrm{D}}^6}\, U'''\overline{W} + \frac{2p_xp_y}{p_{\mathrm{D}}^6}\, U'''\overline{W}\right) \\
&= 0.
\end{aligned}
\tag{C.50}
$$

Note that repeating the analysis for the tWKE model leads to an expression similar to Eq. (C.50) with only the first two terms in the integrand. These terms cancel out, thus showing that the tWKE model also conserves the total energy.